\DocumentMetadata{ 
	pdfstandard = a-2b,
	pdfversion  = 1.7,
	lang		= en-US,
}
\documentclass[twoside]{mitthesis}
\usepackage[utf8]{inputenc}
\usepackage[T1]{fontenc}
\usepackage{amssymb} 
\usepackage{amsfonts}
\usepackage{pgfplots}
\usepackage{amsmath}
\pgfplotsset{compat=1.17}
\usepackage{tikz}
\usepackage{esint}  
\usepackage{float} 
\usepackage{graphicx}
\usepackage{epstopdf}
\usepackage{tikz-3dplot}
\usepackage{listings}
\usepackage{tabularx}
\usepackage[version=4]{mhchem}
\usepackage{lipsum}
\IfPackageAtLeastTF{lipsum}{2021/09/20}{\setlipsum{auto-lang=false}}{}
\newcommand{\dd}{\mathrm{d}}
\usepackage{booktabs}
\usepackage{array}
\usepackage{amsmath,amsfonts,amsthm,amssymb}
\usepackage{mathrsfs}
\usepackage{mathtools}
\usepackage{longtable}
\usepackage{enumitem}
\usepackage[normalem]{ulem}
\usepackage{bigints}
\usepackage{xparse}
\usepackage{xcolor}
\usepackage{physics}
\usepackage{verbatim}
\usepackage{tabularx}
\usepackage{minibox}
\usepackage{comment}
\usepackage{appendix}
\usepackage{slashed}
\usepackage{marginnote}
\usepackage[nice]{nicefrac}
 
\usepackage[colorinlistoftodos]{todonotes}
\usepackage{hepunits}
\usepackage{stackengine,scalerel}
\usepackage{tikz}
\usepackage[compat=1.1.0]{tikz-feynman}
\usepackage{pgfplots}
\newcommand\hstar[1]{\ThisStyle{\ensurestackMath{%
\setbox0=\hbox{$\SavedStyle#1$}%
\stackengine{0pt}{\copy0}{\kern.2\ht0\smash{\SavedStyle\star}}{O}{c}{F}{T}{S}}}}
\definecolor{darkblue}{RGB}{0,0,196}
\definecolor{darkgreen}{RGB}{0,120,0}
\definecolor {darkgreen}{rgb}{0.2,0.7,0.2}

\usepackage{float}

\newcommand{\di}{{\rm d}}

\def\spt{{\cal S}}
\def\wT{{\widehat T}}
\def\wj{{\widehat j}}
\def\wJ{{\widehat J}}

\def\wspt{{\widehat{\cal S}}} 

\def\wPhi{{\widehat{\Phi}}}

\def\wrho{{\widehat{\rho}}}
\def\wrhol{{\widehat{\rho}_{\rm LE}}}

\newcommand{\x}{{\rm x}}

\newcommand{\be}{\begin{equation}}
\newcommand{\ee}{\end{equation}}                         %
\def\bea{\begin{eqnarray}}
\def\eea{\end{eqnarray}}   

\newcommand{\rr}[1]{\textcolor{red}{#1}}
\newcommand{\empsh}[1]{\emph{#1}}
\newcommand{\mycite}[1]{\hyperref[#1]{[\textcolor{red}{#1}]}}
\newcommand{\myref}[1]{\hyperref[#1]{[#1]}}
\newcommand{\mymultiref}[2]{\hyperref[#1]{[#1}\textcolor{blue}{-}\hyperref[#2]{#2]}}

\newcommand{\myrefg}[1]{\hyperref[#1]{[\textcolor{violet}{#1}]}}
\newcommand{\myrefs}[1]{\hyperref[#1]{[\textcolor{purple}{#1}]}}
\newcommand{\myrefp}[1]{\hyperref[#1]{[\textcolor{teal}{#1}]}}
\newcommand{\myrefn}[1]{\hyperref[#1]{[\textcolor{brown}{#1}]}}
\newcommand{\longmultimyref}[2]{\hyperref[#1]{[\textcolor{teal}{#1}}-\hyperref[#2]{\textcolor{teal}{#2}]}}
\ExplSyntaxOn
\NewDocumentCommand{\multimyref}{ >{\SplitList{,}} m }
 {
  \seq_clear:N \l_tmpa_seq 
  \ProcessList{#1}{\addtoseq} 
  [ 
  \seq_use:Nn \l_tmpa_seq {,~} 
  ] 
 }
\newcommand{\addtoseq}[1]{%
  \seq_put_right:Nn \l_tmpa_seq {\hyperref[#1]{\textcolor{teal}{#1}}} 
}
\ExplSyntaxOff
\usepackage[style=ext-numeric-comp,giveninits=true,maxbibnames=10,sorting=none]{biblatex}
\usepackage{hyperref} 
\hypersetup{%
	pdfsubject={Template for writing MIT theses with the mitthesis class},
	pdfkeywords={Massachusetts Institute of Technology, MIT},
	pdfurl={},
	pdfcontactemail={},
	pdfauthortitle={},
}
\usepackage{bm} 
\renewcommand{\x}{\boldsymbol{x}}
\renewcommand{\v}{\boldsymbol{v}}
\begin{document}

\begin{titlepage}
\thispagestyle{empty}
\centering
{\Huge \bfseries Formulation of Relativistic Dissipative Spin Hydrodynamics\par}
\vspace{2cm}

{\LARGE \bfseries Asaad Daher\par}
\vspace{2cm}

{\large \bfseries Dissertation\par}
\vspace{0.3cm}
{\normalsize Submitted to the\par}
{\normalsize Henryk Niewodniczański Institute of Nuclear Physics\par}
{\normalsize Polish Academy of Sciences\par}
{\normalsize Department of Theory of Structure of Matter\par}
{\normalsize Division of Theoretical Physics\par}
{\normalsize in partial fulfillment of the requirements for the degree of\par}
{\normalsize Doctor of Philosophy in Physics\par}
\vspace{1cm}
\begin{figure}[H]
\centering
\includegraphics[height=3.5cm]{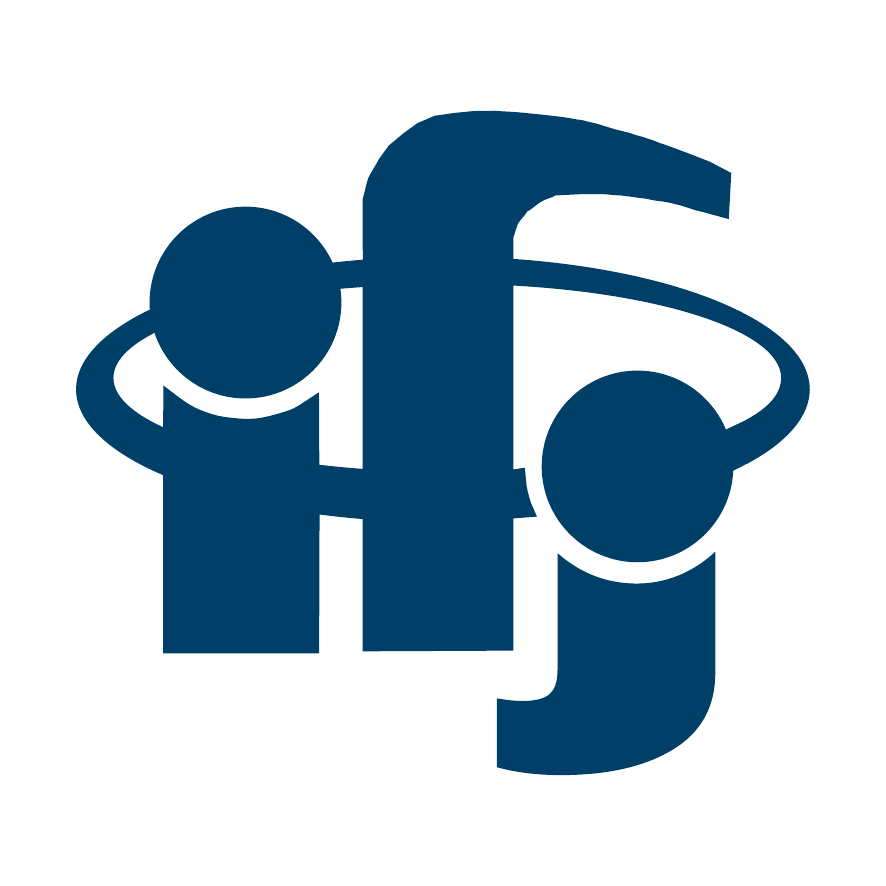}
\end{figure}
\vspace{0.5cm}
{\normalsize\bfseries Supervisor\par}
\vspace{0.2cm}
{\normalsize dr hab. Radoslaw Ryblewski, prof. IFJ PAN\par}
\vspace{2cm}

{\normalsize Kraków, Poland\par}
{\normalsize 2021--2025\par}
\end{titlepage}

\chapter*{STRESZCZENIE}
\pdfbookmark[0]{STRESZCZENIE}{STRESZCZENIE}

Głównym celem niniejszej pracy jest opracowanie spójnych podstaw teoretycznych hydrodynamiki dysypatywnej dla relatywistycznego płynu ze spinem, dalej określanej jako \textit{relatywistyczna dysypatywna hydrodynamika spinowa}. W ramach tej teorii dynamiczny opis relatywistycznego płynu wymaga wprowadzenia nowej wielkości makroskopowej — gęstości spinu, związanej z tensorem spinowym. Tensor ten, definiowany jako wartość średnia odpowiedniego operatora w kwantowej teorii pola, wnosi wkład do całkowitego momentu pędu układu. Potrzeba opracowania takiej teorii wynika z najnowszych pomiarów polaryzacji spinu hadron-ów produkowanych w niecentralnych relatywistycznych zderzeniach ciężkich jonów.
\medskip

W pracy zastosowano dwie różne metody sformułowania teorii. Pierwsza opiera się na kowariantnej termodynamice i stanowi rozszerzenie konwencjonalnych teorii \emph{Naviera--Stokesa} oraz \emph{Müllera--Israela--Stewarta} poprzez uwzględnienie tensora spinowego. Druga wywodzi się z ogólnych zasad relatywistycznej kwantowej mechaniki statystycznej i rozwija oraz uogólnia fundamentalne podejście \emph{Zubareva}. Obie metody mają na celu sformułowanie domkniętego układu równań ewolucji dla wielkości makroskopowych oraz — za pomocą analizy prądu entropii — identyfikację prądów dysypatywnych i odpowiadających im współczynników transportu.
\medskip

Oba podejścia oferują różne perspektywy dla przyszłych zastosowań w kontekście analizy pomiarów polaryzacji spinu. Poza swoim znaczeniem fenomenologicznym, teoria ta otwiera również szereg możliwości dalszych rozwinieć teoretycznych. Obejmują one m.in. weryfikację zależności termodynamicznych przyjętych w pierwszym podejściu na gruncie teorii mikroskopowej, a także głębsze zrozumienie wyłaniających się z tej teorii nowych współczynników transportu — w szczególności tych związanych z tensorem spinowym — poprzez modelowanie mikroskopowe lub dopasowanie do danych eksperymentalnych.

\renewcommand{\k}{\boldsymbol{k}}
%
%

\chapter*{ABSTRACT}

The primary objective of this thesis is to develop a consistent theoretical framework of dissipative hydrodynamics for a relativistic fluid with spin — hereafter referred to as \textit{relativistic dissipative spin hydrodynamics}. In this framework, the dynamical description of a relativistic fluid requires a new macroscopic variable, the spin density, which is associated with a spin tensor. This tensor, defined as the expectation value of a rank-3 tensor operator in quantum field theory, contributes to the system's total angular momentum. The need for such a theory is motivated by recent measurements of spin polarization of hadrons produced in non-central relativistic heavy-ion collisions.
\medskip

Two distinct formulation methods are employed. The first is grounded in covariant thermodynamics and extends the conventional \textit{Navier-Stokes} and \textit{Müller-Israel-Stewart} theories of relativistic hydrodynamics by incorporating a spin tensor. The second is based on principles of relativistic quantum statistical mechanics, building upon and generalizing the foundational \textit{Zubarev} approach. Both formulations aim to construct a closed system of evolution equations for the macroscopic variables and, via an entropy-current analysis, to identify the dissipative currents and their associated transport coefficients. 
\medskip

These two approaches provide different perspectives for future applications focused on spin polarization measurements. Beyond its phenomenological relevance, the theory also opens several avenues for further theoretical developments. These include the verification of the thermodynamic relations employed in the first formulation method using microscopic frameworks, as well as a deeper understanding of the emerging transport coefficients -- particularly those associated with the spin tensor -- through microscopic modeling or data-driven parameter extraction.



\chapter*{Acknowledgments}
\pdfbookmark[0]{Acknowledgments}{acknowledgments}

{\Huge \textit{To}} {\large\textit{my parents}}: this work is a tribute to your unwavering dedication and boundless patience. A special appreciation goes to my supervisor, Prof. Radoslaw Ryblewski, for his support and guidance, particularly for the academic freedom to explore my research interests. I have been very grateful to learn from and collaborate with Prof. Wojciech Florkowski at Jagiellonian University and Prof. Francesco Becattini and his group at INFN Florence. I also thank Dr. Arpan Das for his co-supervision during part of this dissertation, and Dr. Sushant Singh for discussions during the writing process. I am further grateful to the many outstanding collaborators and researchers with whom I had enriching interactions at national and international conferences. I extend my acknowledgments to the members of NO4 (NZ41) and KISD at IFJ PAN. Finally, to my partner, Patrycja Krekora — thank you for filling our time in Kraków with warmth.
\medskip

This research was supported by:
\begin{enumerate}
\item \textbf{2021–2025:}~{PhD student fellowship\newline Funded by: Krakow School of Interdisciplinary PhD Studies (KISD).}
\medskip

\item \textbf{2021–2024:}~{PhD student scholarship\newline Funded by: Polish National Science Centre SONATA BIS Grant No. 2018/30/E/ST2/00432 PI: Prof. Radoslaw Ryblewski.}
\medskip

\item{Visiting fellow at the Department of Physics INFN Florence \newline
Dates: May 18–June 18, Oct 1–Nov 1, 2023; June 9–July 9, 2024\newline
Funded by: Polish National Agency for Academic Exchange NAWA-STER Grant No. PPI/STE/2020/1/00020\newline
Host: Prof. Francesco Becattini.}
\end{enumerate}




\tableofcontents
\listoffigures
\listoftables
\chapter*{}
\section*{Notation}

\begin{enumerate}


\item Throughout this work, we adopt natural units, i.e., $c = \hbar = k_{B} = 1$, unless explicitly stated otherwise. Here, $c$, $\hbar$, and $k_{B}$ denote the speed of light, the reduced Planck constant, and the Boltzmann constant, respectively.
    
\item We employ a metric tensor with signature $g_{\mu\nu} = \operatorname{diag}(+1, -1, -1, -1)$, and use the totally antisymmetric Levi-Civita tensor $\epsilon^{\mu\nu\rho\sigma}$ with the sign convention $\epsilon^{0123} = -\epsilon_{0123} = +1$.

\item The Einstein summation convention is assumed unless stated otherwise. The dot product of four-vectors \( X = X^{\mu} e_{\mu} \) and \( Y = Y^{\nu} e_{\nu} \) in Minkowski spacetime is defined as 
$$ X \cdot Y = g_{\mu\nu} X^{\mu} Y^{\nu} = X^{0} Y^{0} - \boldsymbol{X} \cdot \boldsymbol{Y} ,$$
where bold symbols denote the corresponding spatial three-vectors.
    
\item The fluid four-velocity $u^{\mu}$ is normalized such that $u^{\mu} u_{\mu} = 1$. The projection operator orthogonal to $u^{\mu}$ is defined as 
$$\Delta^{\mu\nu} \equiv g^{\mu\nu} - u^{\mu} u^{\nu},$$
which, by construction, satisfies $\Delta^{\mu\nu} u_{\mu} = 0$.
    
\item For any tensor $X^{\mu\nu}$, we denote its symmetric and antisymmetric parts by 
$$X^{\mu\nu}_{S} \equiv X^{(\mu\nu)} \equiv \frac{1}{2}\left(X^{\mu\nu} + X^{\nu\mu}\right),\quad\text{and}\quad  X^{\mu\nu}_{A} \equiv X^{[\mu\nu]} \equiv \frac{1}{2}\left(X^{\mu\nu} - X^{\nu\mu}\right),$$
respectively. The projection of a tensor $X^{\alpha \beta \rho \delta \ldots}$ orthogonal to $u^{\rho}$ is denoted by 
$$X^{\alpha \beta\langle\gamma\rangle \delta \ldots} \equiv \Delta_\rho^\gamma X^{\alpha \beta \rho \delta \ldots}.$$
The traceless, symmetric projection orthogonal to $u^{\mu}$ of a tensor $X^{\alpha \beta}$ is defined as 
$$X^{\langle\mu\nu\rangle} \equiv \Delta^{\mu\nu}_{\alpha\beta} X^{\alpha\beta}$$
where
$$ \Delta^{\mu\nu}_{\alpha\beta}  \equiv \frac{1}{2}\Bigl(\Delta^{\mu}_{\ \alpha}\Delta^{\nu}_{\ \beta} + \Delta^{\mu}_{\ \beta}\Delta^{\nu}_{\ \alpha}\Bigr) - \frac{1}{3}\Delta^{\mu\nu}\Delta_{\alpha\beta}.$$
Similarly, the antisymmetric projection orthogonal to $u^{\mu}$ is defined by 
$$X^{\langle[\mu\nu]\rangle} \equiv \Delta^{[\mu\nu]}_{[\alpha\beta]} X^{\alpha\beta} $$
where
$$\Delta^{[\mu\nu]}_{[\alpha\beta]} \equiv \frac{1}{2}\Bigl(\Delta^{\mu}_{\ \alpha}\Delta^{\nu}_{\ \beta} - \Delta^{\mu}_{\ \beta}\Delta^{\nu}_{\ \alpha}\Bigr) .$$
For $X^{\alpha\beta}=\nabla^{\alpha}u^{\beta}$, the traceless symmetric projection orthogonal to $u^{\mu}$ yields the shear tensor $\sigma^{\mu\nu}=\nabla^{\langle\mu}u^{\nu\rangle}=\nabla^{(\mu}u^{\nu)}-\frac{1}{3}\theta\Delta^{\mu\nu}, $ while the antisymmetric projection reads $\nabla^{\langle[\mu}u^{\nu]\rangle}=\frac{1}{2}(\nabla^{\mu}u^{\nu}-\nabla^{\nu}u^{\mu})\equiv T \Omega^{\mu\nu}$.
\item The partial derivative operator is decomposed into components parallel and orthogonal to the flow direction as
$$\partial_{\mu} = u_{\mu} D + \nabla_{\mu},$$ 
where the comoving derivative is defined as $D \equiv u^{\mu} \partial_{\mu}$, and the spatial gradient orthogonal to $u^{\mu}$ is given by $\nabla_{\mu} \equiv \Delta_{\mu}^{\ \alpha} \partial_{\alpha}$, satisfying $u_{\mu}\nabla^{\mu} = 0$. The expansion rate is defined as $\theta \equiv \partial_{\mu} u^{\mu}= \nabla_{\mu} u^{\mu}$.
\end{enumerate}
\newpage
\section*{Abbreviations}

\begin{center}
\renewcommand{\arraystretch}{1.5} 
\begin{tabular}{p{3cm} p{10cm}}
\textbf{} & \textbf{} \\
\hline
BNL & Brookhaven National Laboratory \\
RHIC & Relativistic Heavy-Ion Collider \\
STAR & Solenoidal Tracker At RHIC\\
CERN & Conseil Européen pour la Recherche Nucléaire \\
LHC & Large Hadron Collider \\
ALICE& A Large Ion Collider Experiment\\
SPS & Super Proton Synchrotron\\
QFT & Quantum Field Theory \\
QCD & Quantum Chromodynamics \\
QGP & Quark-Gluon Plasma \\
DOF & Degrees Of Freedom \\
RF & Rest Frame\\
NS & Navier-Stokes \\
MIS & Müller-Israel-Stewart \\
BR & Belinfante-Rosenfeld \\
GLW & de Groot, van Leeuwen, van Weert \\
HW & Hilgevoord-Wouthuysen \\
\end{tabular}
\end{center}

\chapter{Introduction}
\label{Introduction}
%

%
Relativistic heavy-ion collision physics is a field of research dedicated to studying strongly interacting matter under extreme conditions of temperature and density. The characteristic features of a new state of matter -- quark-gluon plasma -- created in these conditions align with predictions of relativistic hydrodynamics with exceptionally low viscosity to entropy density ratio, suggesting its nearly perfect fluidity and vicinity of equilibrium.
Recent measurements of spin polarization of particles emitted in these collisions now tie the QGP’s concept of vorticity to particle spins. The surprises arising from facing theory predictions with experimental data, motivate new developments in relativistic hydrodynamics for relativistic fluids with spin.
%
%
\section{{Relativistic heavy-ion collisions and spin polarization measurements}}
\label{Spinpolarization}
%
%
%
\subsubsection{Quark-gluon plasma and relativistic hydrodynamics}

\textit{Quantum chromodynamics (QCD)}, the gauge theory of the strong interaction, has the distinctive property that its running coupling, $\alpha_s(Q)$, decreases with increasing momentum scale~\cite{Workman:2022ynf}. This phenomenon, known as \textit{asymptotic freedom}~\cite{Gross:1973id, Politzer:1973fx}, implies that at very short distances, quarks and gluons behave almost as free partons. At larger distances, however, they are confined within hadrons, entering the nonperturbative regime of \emph{confinement}. 

Understanding how confinement and asymptotic freedom emerge dynamically is one of the central goals of QCD research \cite{Quigg:2013ufa}. In particular, it motivates the study of QCD under conditions of finite temperature and density \cite{Kapusta:2006pm}, where the properties of strongly interacting matter can undergo qualitative transformations. Heating or compressing the medium weakens the confining force via \emph{color screening} \cite{Shuryak:1980tp}, leading to the expectation of a \emph{phase transition} from hadronic matter to a deconfined state of quarks and gluons \cite{Collins:1974ky,Cabibbo:1975ig}.

Modern \emph{lattice QCD} calculations \cite{Muroya:2003qs,Philipsen:2012nu,Petreczky:2012rq,Aoki:2012tk,Ratti:2018ksb} support this expectation: for vanishing baryon chemical potential, they reveal a \emph{smooth crossover} transition to a deconfined phase at a \emph{pseudo-critical temperature} $T_c \approx 155$ MeV \cite{Karsch:2001cy,Borsanyi:2010bp,HotQCD:2014kol,Bazavov:2011nk} -- remarkably close to the \emph{Hagedorn temperature}, initially inferred from the exponential growth of the hadron spectrum \cite{Hagedorn:1965st}. Above this threshold, the relevant degrees of freedom form the \emph{quark-gluon plasma (QGP)}, a state of matter first theorized in the late 1970s and early 1980s \cite{Shuryak:1980tp, Kalashnikov:1979dp, Kapusta:1979fh}, and believed to have filled the Universe a few microseconds after its creation \cite{Kolb:1990vq}.

To investigate the predicted thermodynamic properties of deconfined QCD matter experimentally, it is necessary to recreate the extreme conditions of temperature and density under which color confinement breaks down. The only practical and reproducible method to achieve this in the laboratory is through ultra-relativistic heavy-ion collisions~\cite{Schmidt_1993,Yagi:2005yb,Stock:2008ru,Florkowski:2010zz,Friman:2011zz,Satz:2018oiz,Busza:2018rrf,Elfner:2022iae}. These collisions briefly generate temperatures well above $T_c$ in approximate local equilibrium allowing to produce and study properties of strongly interacting matter in a deconfined state~\cite{Rischke:2003mt}. 

Early indications of deconfinement were observed already at CERN's Super Proton Synchrotron (SPS) in Pb–Pb collisions at $\sqrt{s_{\rm{NN}}}$= 17.3 GeV \cite{Heinz:2000bk}. However, compelling evidence emerged in 2005 when Au–Au collisions at $\sqrt{s_{\rm{NN}}}$= 200 GeV at the Relativistic Heavy-Ion Collider (RHIC) at Brookhaven National Laboratory (BNL) revealed temperatures significantly above $T_c$ and, in particular, strong suppression of high-transverse-momentum hadrons (jet quenching), signaling substantial parton energy loss in a colored medium \cite{STAR:2005gfr, PHENIX:2004vcz, PHOBOS:2004zne, BRAHMS:2004adc,Gyulassy:2004zy}. Since 2010, CERN's Large Hadron Collider (LHC) has extended the heavy-ion program to Pb–Pb collisions at $\sqrt{s_{\rm{NN}}}=$2.76 and 5.02 TeV per nucleon pair, confirming and extending these observations
\cite{ALICE:2010suc, ATLAS:2010isq, CMS:2011iwn}.

A variety of observables measured in these experiments, including jet quenching \cite{Mrowczynski:1991da,Casalderrey-Solana:2007knd,dEnterria:2009xfs,Mehtar-Tani:2013pia,Qin:2015srf,Carrington:2021dvw}, collective flow patterns \cite{Teaney:2000cw,Kolb:2003dz,Ollitrault:2007du,Heinz:2013th,Gale:2013da,Jaiswal:2016hex}, strangeness enhancement~\cite{Koch:2017pda}, and quarkonium suppression~\cite{Mocsy:2013syh,Rothkopf:2019ipj}, provide signatures of the formation of a new state of strongly interacting matter with properties qualitatively different from those of hadronic matter \cite{Hwa:2004yg,Hwa:2010npa,Wang:2016opj,Wang:2025qdg,Muller:2012zq,Braun-Munzinger:2015hba,Andronic:2017pug}. Intriguingly, these properties differ markedly from early theoretical expectations based on asymptotic freedom \cite{Shuryak:2003xe,Heinz:2009xj}. Rather than behaving as a weakly interacting gas of quasi-free quarks and gluons, the medium exhibits strong collective behavior characteristic of a strongly coupled, nearly perfect fluid \cite{Muller:2007rs}. 

The pronounced azimuthal anisotropy (quantified by the \emph{elliptic flow} coefficient $v_2$) observed in hadron momentum distributions, together with the hierarchy of other Fourier coefficients $v_n$ \cite{Bozek:2010bi,Gale:2013da,}, indicates that the medium thermalizes rapidly and displays significant collective flow \cite{Heinz:2013th}. Quantitative agreement with experimental data is achieved only when the dominant part of space–time evolution is modeled using relativistic dissipative hydrodynamics with a small shear-viscosity-to-entropy-density ratio $\eta/s \sim 0.1-0.2$~\cite{Teaney:2003kp,Romatschke:2007mq,Mrowczynski:1994xv,Jas:2007rw,Ollitrault:1992bk,Huovinen:2001cy,Hirano:2002ds,Kolb:2002ve,Hama:2005dz,Teaney:2009qa,Florkowski:2010cf,Florkowski:2013lya,Dusling:2015gta}, a value close to the conjectured \emph{Kovtun-Son-Starinets (KSS) lower bound} $(\eta/s)_{\rm KSS} \sim 1/(4\pi)$ \cite{Kovtun:2004de}, and further constrained through global Bayesian analyses \cite{Parkkila:2021tqq,Bernhard:2016tnd}. Despite these advances, the microscopic mechanisms responsible for the early thermalization of matter in ultra-relativistic heavy-ion collisions remain the subject of active investigation \cite{Janik:2006gp,Ryblewski:2013eja,Bozek:2010aj,Ryblewski:2010tn,Heller:2011ju,Zhu:2017oei,Heller:2011ju,Strickland:2013uga,Baier:2000sb,El:2007vg,Jankowski:2014lna,Carrington:2020ssh,Carrington:2024hhf}.  

Owing to its empirical success, dissipative hydrodynamics now forms the foundation of nearly all comprehensive models describing the evolution of matter created in relativistic heavy-ion collisions ~\cite{Bass:2000ib,Teaney:2001av,Kolb:2003dz,Bozek:2009ty,Huovinen:2003fa,Nonaka:2006yn,Broniowski:2008vp,Heller:2014wfa,Jaiswal:2016hex,Romatschke:2017ejr,Florkowski:2017olj}.
State-of-the-art simulations incorporate fluctuating, event-by-event initial conditions, second-order causal relativistic hydrodynamics for the deconfined phase, and hadronic transport transport models for the late-stage evolution in the confined phase \cite{Qiu:2011iv,Petersen:2008dd,Shen:2011eg,Song:2008si,Schenke:2010nt,Niemi:2015qia,Shen:2014vra,Schenke:2012wb,Karpenko:2013wva,Denicol:2018wdp,Noronha-Hostler:2014dqa,Bozek:2013ska,Bozek:2013uha,Bozek:2009dw,Blaschke:2022lqb}. Within this framework, transport coefficients and thermodynamic properties may be extracted by matching theoretical predictions to a broad range of experimental data such as heavy-flavor dynamics, electromagnetic emissivity, and critical fluctuations. Hydrodynamics thus provides a basis of a unified approach allowing quantitatively connect QCD to a wide array of observables in relativistic heavy-ion collision experiments. 
 

\subsubsection{Global spin polarization}
In fluid dynamics, an important property of the flow is its \emph{vorticity}, which quantifies the local rotational structure of the velocity field. In the nonrelativistic limit, vorticity is defined as \( \boldsymbol{\omega} = \nabla \times \boldsymbol{v} \)~\cite{acheson1990}. For a fluid undergoing rigid rotation with constant angular velocity, the vorticity is given by twice its angular velocity, allowing its magnitude to be estimated~\cite{Becattini:2016gvu}. In more general scenarios, where the fluid is not globally rotating, shear flows can also generate vorticity as the curl of the velocity field is nonzero in regions with velocity gradients~\cite{acheson1990}.
%
%

In the relativistic regime, several generalizations of vorticity arise, tailored to different physical contexts~\cite{Becattini:2015ska,Wu:2019eyi}. The direct extension of classical vorticity is the \textit{kinematical vorticity}, $\omega_K^{\mu\nu}=\partial^{[\nu}u^{\mu]}$, defined as the antisymmetric gradient of the four-velocity $u^{\mu}$~\cite{Gourgoulhon:2006bn}. Other important generalizations include \textit{T-vorticity}, $\omega_T^{\mu\nu}=\partial^{[\nu}(Tu^{\mu]})$, involving antisymmetric gradients of the temperature-weighted four-velocity, and the \textit{thermal vorticity}, $\varpi^{\mu\nu}=\partial^{[\nu}(u^{\mu]}/T)$, involving the antisymmetric gradient of the inverse-temperature-weighted four-velocity~\cite{Becattini:2015ska}; this concept will be discussed in Chapter~\ref{Quantum-statistical formulation}. Among these, thermal vorticity, as we qualitatively discuss later, is the suitable relativistic quantity for the study of polarization. Moreover, at global equilibrium with rotation, it becomes a constant field directly proportional to the system's angular velocity~\cite{Becattini:2012tc}, making it an accessible quantity in hydrodynamic modeling. Until recently, vorticity effects in the QGP received relatively little attention compared to symmetric flow gradients, primarily because no direct experimental signatures had been detected.

The interplay between macroscopic orbital angular momentum and the spin orientation of constituent particles was first demonstrated over a century ago in the \textit{Barnett effect}~\cite{Barnett:1915uqc,RevModPhys.7.129}, where the rotation of an uncharged, initially unmagnetized solid metal body induces a net alignment of spins along the rotation axis as a consequence of total angular momentum conservation~\footnote{An opposite effect, known as the \emph{Einstein–de Haas effect}, occurs when the alignment of spins due to an applied magnetic field induces mechanical rotation of the body to conserve total angular momentum \cite{Richardson1908,Einstein1915}.}. More recently, a connection between local vortical structures in a fluid and spin polarization was experimentally observed by Takahashi et al. in condensed matter systems~\cite{Takahashi2015, matsuo2017,maekawa2017spin}. In these experiments, mechanical shear generated vorticity within the conduction electron fluid inside metallic samples. Through the conservation of total angular momentum, spin-vorticity coupling produced measurable electron spin polarization, providing a direct experimental realization of the mechanism linking vorticity to spin polarization \cite{matsuo2017}.

A similar mechanism has been proposed for high-energy nuclear collisions~\cite{Becattini:2007sr}. However, unlike in condensed matter systems, where macroscopic rotation can be externally controlled, in heavy-ion collisions the magnitude and direction of the initial angular momentum fluctuate event by event, and the emitted particles are highly relativistic~\cite{Becattini:2021wqt}.
Furthermore, the short lifetime and extreme conditions of the QGP present additional challenges for the observation of spin polarization phenomena.

The flow structure of the QGP suggests that local vortical structures are generated due to the shear present in the initial velocity field, with significant event-by-event fluctuations~\cite{Becattini:2020ngoaa,Becattini:2015ska}. On average, the vorticity is expected to align with the system's global orbital angular momentum, $\boldsymbol{L} = \boldsymbol{b} \times \boldsymbol{p}_{\rm beam}$, where $\boldsymbol{b}$ is the impact parameter and $\boldsymbol{p}_{\rm beam}$ is the beam momentum~\footnote{The direction of $\boldsymbol{L}$ is orthogonal to the reaction plane (RP) which is spanned by the beam ($z$) direction and the direction of impact parameter $\boldsymbol{b}$ in the plane transverse to the beam ($x$ direction).}~\cite{Becattini:2007sr,Becattini:2021wqt}. Hence, although direct detection of these vortical structures remains experimentally not accessible, indirect evidence comes from the observed alignment of the average spin polarization vector of emitted particles with the direction of $\boldsymbol{L}$ due to the angular momentum conservation in the system.

Assuming local thermodynamical equilibrium, Refs.~\cite{Becattini:2007nd,Becattini:2013fla} established a relation between the average spin polarization vector of particles emitted at freeze-out and the average \emph{thermal vorticity}. This is achieved by a generalization of the Cooper-Frye prescription~\cite{Cooper:1974mv} -- a standard method for converting hydrodynamic fields into particle distributions. This \emph{spin-vorticity} coupling suggests that the fluid's vorticity imprints itself onto the final-state spin polarization patterns. 
Projecting the average spin polarization along $\boldsymbol{L}$ defines the so-called \textit{global spin polarization}. Notably, thermal vorticity can be extracted directly from the space-time hydrodynamic evolution of the QGP without relying on complicated perturbative QCD calculations~\cite{Liang:2004ph, Voloshin:2004ha, Gao:2007bc}.

\begin{figure}[H]
    \centering
    \begin{minipage}[t]{0.48\textwidth}
        \centering
        \includegraphics[height=6cm]{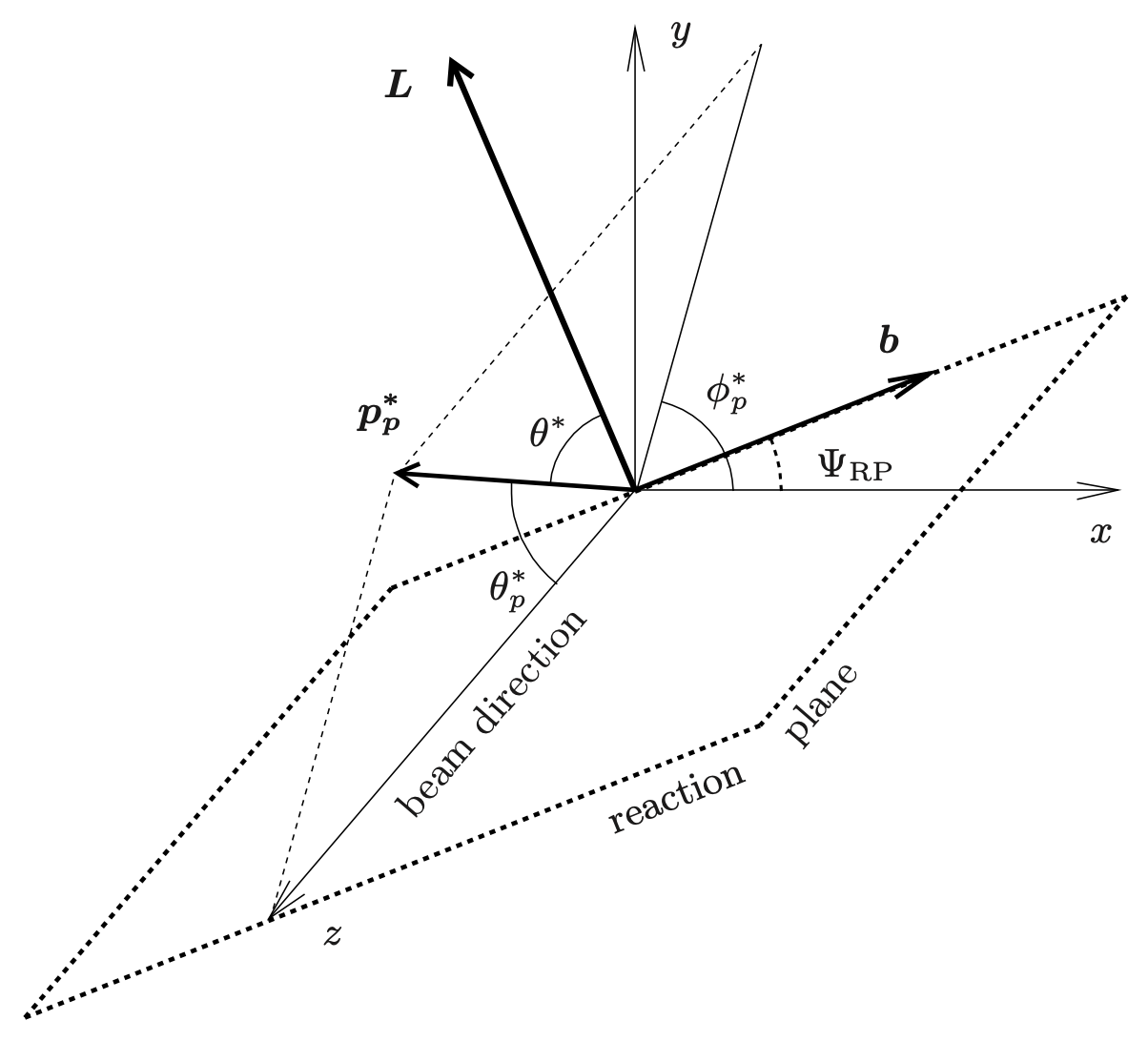}
    \end{minipage}
    \hfill
    \begin{minipage}[t]{0.48\textwidth}
        \centering
        \includegraphics[height=6.5cm]{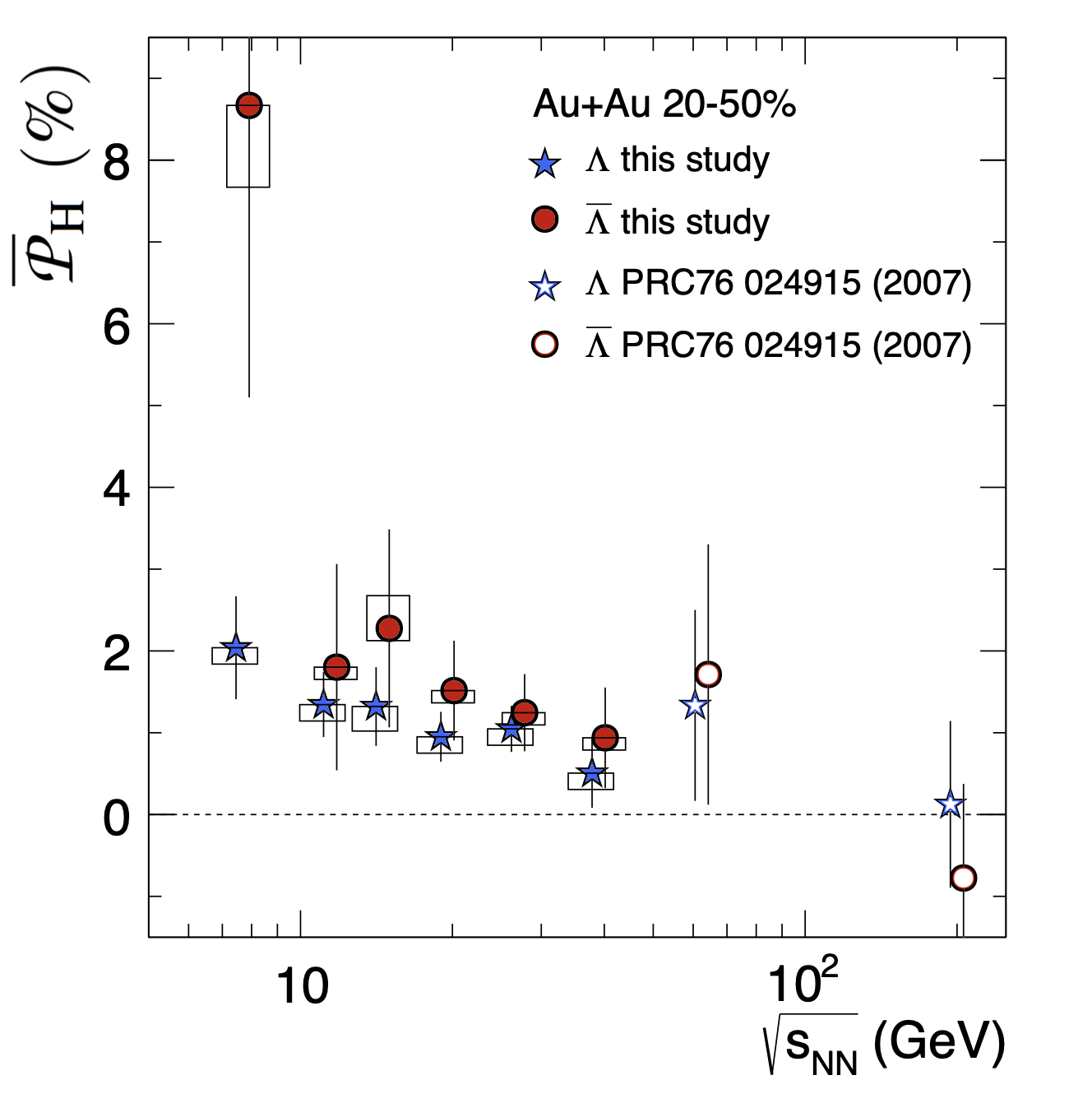}
    \end{minipage}
    \caption{Left: Schematic illustration of the hyperon decay parameters relative to the reaction plane~\cite{STAR:2007ccu}. The transformation $\boldsymbol{L} \rightarrow -\boldsymbol{L}$ corresponds to a shift $\Psi_{\rm RP} \rightarrow \Psi_{\rm RP} + \pi$. Right: Average global spin polarization of $\Lambda(\bar{\Lambda})$ as a function of collision energy~\cite{STAR:2017ckg}. Hollow stars and dots represent previously published data~\cite{STAR:2007ccu}. Boxes and vertical lines indicate systematic and statistical uncertainties, respectively.
}
    \label{fig:geometry_and_data}
\end{figure}

In 2017, the STAR Collaboration at RHIC~\cite{STAR:2017ckg} provided the first experimental evidence for the global polarization of $\Lambda$ hyperons (see~Fig.~\ref{fig:geometry_and_data}). The $\Lambda$ hyperon is particularly well-suited for polarization measurements because it undergoes a parity-violating weak decay, $\Lambda \rightarrow p + \pi^-$ (or $\bar{\Lambda} \rightarrow \bar{p} + \pi^+$), leading to a preferential emission of the proton ($p$) along the $\Lambda$ polarization vector in its rest frame~\cite{Pondrom:1985aw}. Consequently, an ensemble with polarization vector $\boldsymbol{P}^{*}_{\rm H}$ exhibits a measurable correlation with the proton momentum direction $\widehat{\boldsymbol{p}}^{\,*}_{p}$ via the angular distribution of daughter protons over the solid angle $\dd\Omega^{*}$ in the hyperon rest frame:
\begin{align}\label{exp1f}
\frac{\dd \rm{N}}{\dd\Omega^{*}}=\frac{1}{4\pi}\left(1+\alpha_{\rm H}\boldsymbol{P}^{*}_{\rm H}\cdot\widehat{\boldsymbol{p}}^{\,*}_{p}\right).
\end{align}
Here, the subscript H denotes $\Lambda$ or $\bar{\Lambda}$, and quantities with an asterisk are evaluated in the hyperon rest frame. The quantity $\alpha_{\Lambda}=-\alpha_{\bar{\Lambda}}\simeq 0.75$ is the weak decay parameter~\cite{ParticleDataGroup:2014cgo,BESIII:2018cnd}.

To obtain the \emph{global spin polarization}, denoted by $\overline{\mathcal{P}}_{\rm H}$, one must extract the projection of the polarization vector $\boldsymbol{P}^{*}_{\rm H}$ along the direction of the global orbital angular momentum $\widehat{\boldsymbol{L}}$~\footnote{Strictly speaking, the polarization measured in the hyperon rest frame should be projected along the direction of the total angular momentum, which must first be transformed to the same frame and only then averaged over 
$\Lambda$'s with different momenta in the center-of-mass frame~\cite{Florkowski:2021pkp}. The effect of frame-dependence of the total angular momentum direction is however significant only for the most energetic $\Lambda$'s.}. To this end, let $\widehat{\boldsymbol{n}}$ be an arbitrary unit vector.  Then, using Eq.~\eqref{exp1f}, the ensemble-averaged projection of the proton momentum along $\widehat{\boldsymbol{n}}$ yields~\cite{Becattini:2020ngoaa}
\begin{align}\label{generall}
\boldsymbol{P}^{*}_{\rm H} \cdot \widehat{\boldsymbol{n}} 
= \frac{3}{\alpha_{\rm H}} \left\langle \widehat{\boldsymbol{p}}^{\,*}_{p} \cdot \widehat{\boldsymbol{n}} \right\rangle.
\end{align}
Setting  $\widehat{\boldsymbol{n}}=\widehat{\boldsymbol{L}}$, the global spin polarization is obtained as
\begin{align}\label{GPinitial}
\overline{\mathcal{P}}_{\rm H}=\frac{3}{\alpha_{\rm H}}\langle\cos\theta^{*}\rangle 
=\frac{8}{\pi\alpha_{\rm H}}{\langle\sin(\Psi_{\rm RP}-\phi^{*}_{p})\rangle},
\end{align}
where $\theta^{*}$ is the angle between the orbital angular momentum $\boldsymbol{L}$ and  the proton momentum $\boldsymbol{p}^{*}_{p}$, $\Psi_{\rm RP}$ is the reaction plane angle, and $\phi_{p}^{*}$ is the azimuthal angle of the proton momentum in the hyperon rest frame; see left panel in Fig.~\ref{fig:geometry_and_data}.

Remarkably, observed global spin polarization is found to be consistent with theoretical predictions derived under the assumption of local thermodynamic equilibrium, utilizing a spin polarization formula that relates the spin polarization vector to the thermal vorticity at freeze-out~\cite{Karpenko:2016jyx,Li:2017slc,Vitiuk:2019rfv,Sun:2017xhx,Ivanov:2019ern}.

The experimental data reveal that the magnitude of the global polarization decreases with increasing collision energy (see right panel in Fig.~\ref{fig:geometry_and_data}). This trend is typically attributed to larger vorticity values at lower collision energies, offering deeper insight into the vortical structure of the QGP. Furthermore, the polarization magnitudes of the \(\Lambda\) and \(\bar{\Lambda}\) hyperons at a given collision energy are observed to differ slightly. The mechanisms underlying this splitting remain an open question, under active theoretical and experimental \cite{STAR:2023nvo,Hu:2024suz} investigation, and lie beyond the scope of the present discussion. Global spin polarization~\cite{STAR:2017ckg,STAR:2018gyt} has been observed over a wide range of collision energies, from low energies of 
Heavy Ion Synchrotron SIS18~\cite{Kornas:2020qzi,HADES:2022enx,STAR:2021beb} to the highest energies available at the LHC~\cite{ALICE:2019onw,ALICE:2021pzu}. Moreover, in addition to $\Lambda$ hyperons, the STAR Collaboration has also reported global spin polarization for other hyperons, such as $\Xi$ and $\Omega$~\cite{STAR:2020xbm}.


%
%

%
\subsubsection{Local spin polarization}
\begin{figure}[H]
    \centering
    \begin{subfigure}[t]{0.48\textwidth}
        \centering
        \includegraphics[height=6.5cm]{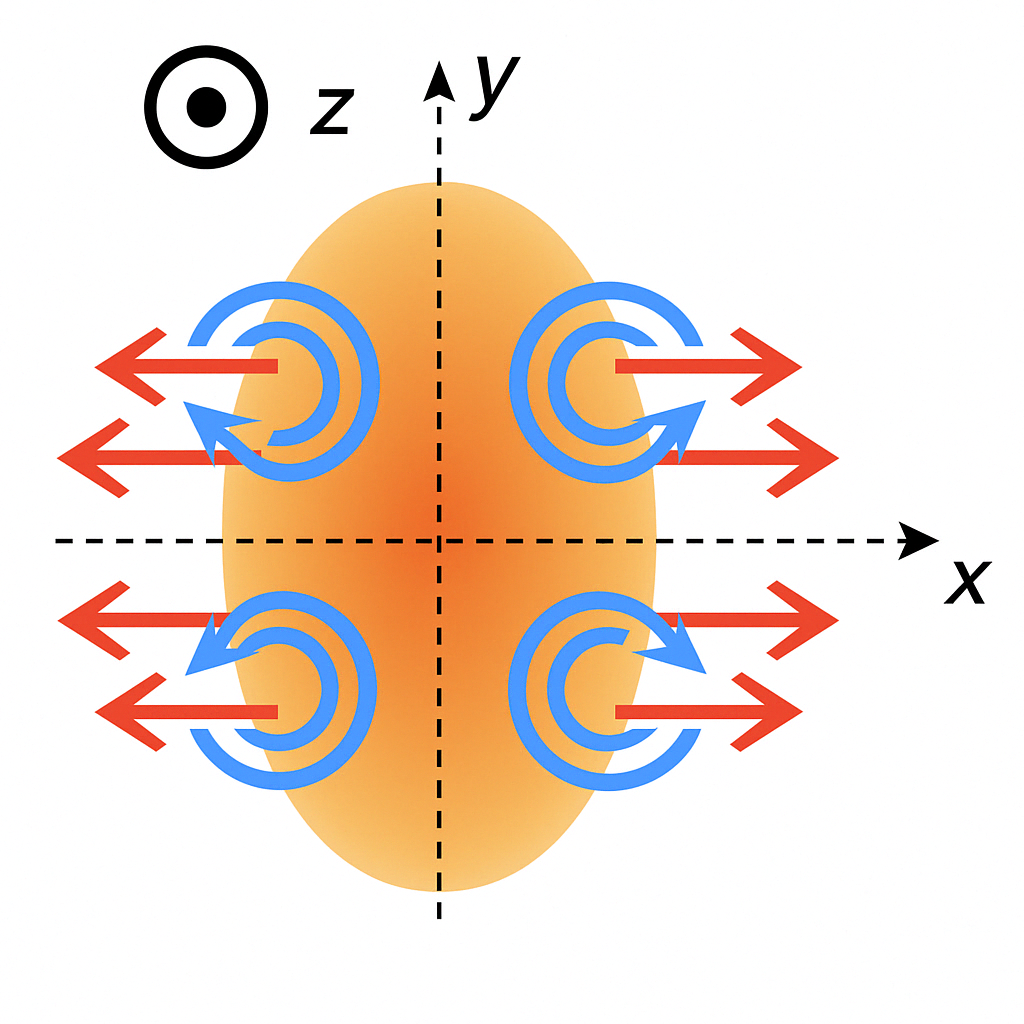}
    \end{subfigure}
    \hfill
    \begin{subfigure}[t]{0.48\textwidth}
        \centering
        \includegraphics[height=6cm]{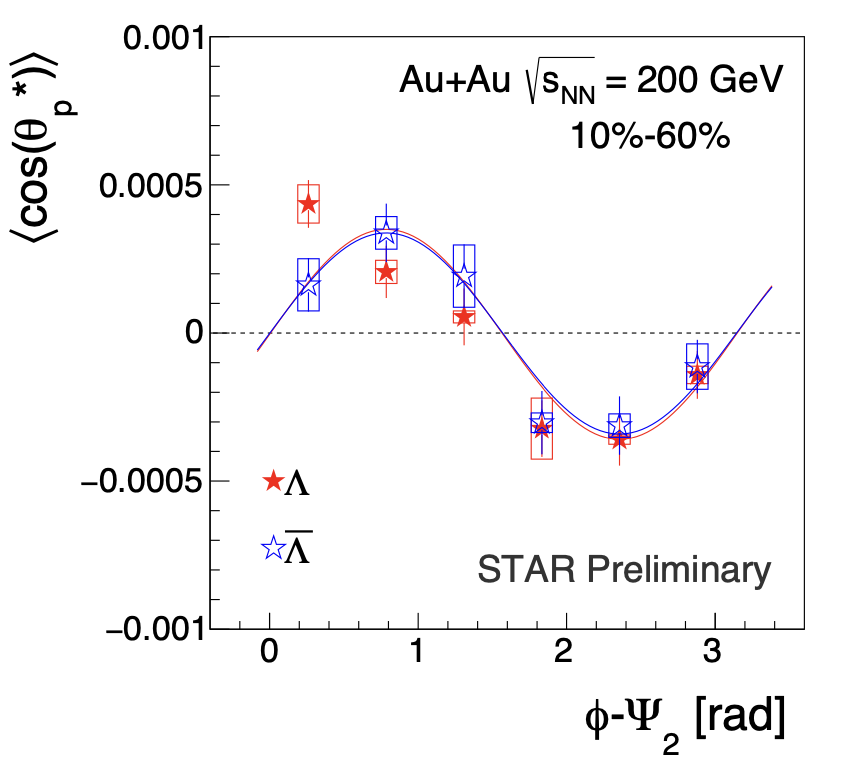}
    \end{subfigure}
    \caption{Left: Schematic illustration of the $z$-component of the nonrelativistic vorticity, \( \sim (\nabla \times \boldsymbol{v})_z \), in the transverse plane (blue arrows), generated by anisotropic flow expansion (red arrows)~\cite{STAR:2019erd}. Right: $\langle\cos\theta^{*}_{p}\rangle$ of $\Lambda$ and $\bar{\Lambda}$ as a function of the hyperons’ azimuthal angle relative to the second-order event plane $\Psi_{2}$~\cite{STAR:2019erd,Niida:2018hfw}.
}
    \label{fig:combined}
\end{figure}

The same spin-vorticity coupling mechanism leading to global spin polarization, when considered locally, may also give rise to nontrivial components of spin polarization along other directions than that of global orbital angular momentum. In particular, the four-velocity gradients related to anisotropic collective expansion in the transverse ($x-y$) plane to the beam are expected to generate vorticity components along the beam ($z$) direction, with a dependence on the azimuthal angle relative to the reaction plane~\cite{Pang:2016igs, Becattini:2017gcx, Voloshin:2017kqp,Xia:2018tes, Niida:2018hfw}; see Fig.~\ref{fig:combined}.  Such vortical structures are anticipated to develop at later stages of the system's evolution, near freeze-out, unlike the vortical structures leading to global spin polarization, which originate from the initial velocity fields. This implies the presence of a nontrivial component of the polarization vector $\boldsymbol{P}_{\rm H}^{*}$ along the beam ($z$) direction. Projecting the polarization vector along the $z$-axis thus defines the \textit{local spin polarization}. However, since the system's global orbital angular momentum defines a preferred average vorticity direction in an event, it follows that the vortical structures in the transverse plane -- and hence, the $z$-component of the polarization -- must vanish upon averaging over all azimuthal angles. 

By setting $\widehat{\boldsymbol{n}} = \widehat{\boldsymbol{p}}_{\rm beam}$ in Eq.~\eqref{generall}, the expression for the local spin polarization becomes:
\begin{align}
\mathcal{P}_{z}=\frac{1}{\alpha_{\rm H}}\frac{\langle\cos\theta^{*}_{p}\rangle}{\langle\cos^{2}\theta^{*}_{p}\rangle},
\end{align}
where $\theta^{*}_{p}$ is the polar angle of the proton momentum in
the $\Lambda(\textrm{or}~\bar{\Lambda})$ rest frame.
The local spin polarization of $\Lambda( \textrm{and}~\bar{\Lambda})$ hyperons was observed by the STAR Collaboration at RHIC~\cite{STAR:2019erd} and by the ALICE Collaboration at the LHC~\cite{ALICE:2021pzu}. The results for $\langle\cos\theta^{*}_{p}\rangle$ for $\Lambda$ and $\bar{\Lambda}$ as a function of the hyperons’ azimuthal angle relative to $\Psi_{2}$ measured by STAR Collaboration~\cite{STAR:2019erd} are shown in the right panel of Fig.~\ref{fig:combined}.

In contrast to global spin polarization, the theoretical model based on local thermodynamic equilibrium correctly reproduces the magnitude of the local spin polarization, but predicts the opposite sign -- a discrepancy known as the \textit{sign puzzle}~\cite{Becattini:2021suc, Becattini:2021iol, Fu:2021pok, Liu:2021uhn}. This mismatch between theory and experiment suggests that the assumption of local equilibrium may be insufficient to fully describe spin dynamics in the QGP. To address this, a new theoretical framework -- relativistic dissipative spin hydrodynamics -- is currently being developed, incorporating spin degrees of freedom through a spin tensor under nonequilibrium conditions, and is expected to yield a revised formula for the average spin polarization vector at freeze-out. Such a framework can be employed in numerical investigations and is expected to provide more accurate predictions. 
\medskip

In this thesis, we aim at constructing a consistent formalism of relativistic dissipative spin hydrodynamics; however, the derivation of the average spin polarization vector will be addressed in future work. 
%
%
\section{Relativistic hydrodynamics: methods and developments}
\subsubsection{Nonrelativistic fluid}
In modern physics, a \textit{fluid} is understood as a system composed of a large number of particles, where the characteristic scale of microscopic interactions is much smaller than the macroscopic size of the system~\cite{osti_5096753,landau2013fluid,rezzolla2013relativistic,Romatschke:2017ejr,Rischke:1998fq,Denicol:2014loa,Romatschke:2009im,Andersson:2020phh}. This scale separation allows the system to be treated as a continuum, described by a set of coarse-grained macroscopic variables, such as energy density, momentum density, and mass or number density. The evolution of these variables is governed by fundamental conservation laws. 

When the system is in local thermodynamic equilibrium, its behavior is described by \textit{ideal} (or \textit{perfect}) \textit{hydrodynamics}, governed by the \textit{Euler equations}~\cite{Euler1757}. Deviations from equilibrium necessitate the inclusion of dissipative effects, leading to \textit{dissipative hydrodynamics}, typically modeled by the \textit{Navier-Stokes equations}~\cite{Navier1822, Stokes1851}. 

The microscopic constituents of the fluid may be governed by quantum mechanics even in the nonrelativistic regime. In such cases, the fluid is described as a quantum-statistical ensemble, and the macroscopic quantities are interpreted as expectation values of corresponding quantum operators. These expectation values serve as the dynamical variables of the system and determine its macroscopic evolution.

As discussed in Sec.~\ref{Spinpolarization}, the discovery of a connection between local vortical structures in a fluid and the spin polarization of electrons in condensed matter systems~\cite{Takahashi2015, maekawa2017spin, matsuo2017} highlighted the relevance of spin degrees of freedom in fluid dynamics. This observation established a conceptual link between hydrodynamic vorticity and spin current, and motivated the development of nonrelativistic \textit{spin hydrodynamics}, where spin is treated on the same footing as other macroscopic hydrodynamic variables.
%
%
\subsubsection{Conventional relativistic fluid}

A \textit{relativistic fluid} is a continuous medium whose internal dynamics obeys the principles of special (or general) relativity \footnote{In this work we will limit ourselves to considerations following special relativity principles.}. The macroscopic state of such a fluid is described by covariant tensorial quantities: the \textit{energy-momentum tensor} $T^{\mu\nu}$ and the conserved \textit{number current} $N^{\mu}$. Their evolution is dictated by conservation laws, providing the foundation of \textit{relativistic hydrodynamics}. The relativstic generalization of ideal fluid dynamics leads to the \textit{relativistic Euler equations}. However, a direct extension of the Navier-Stokes (NS) equations into the relativistic domain is known to suffer from \emph{acausality} and \emph{instability}~\cite{landau2013fluid,Eckart:1940te,Hiscock:1985zz}. These deficiencies were addressed by the development of second-order dissipative theories, most notably the \textit{M\"uller-Israel-Stewart (MIS)} framework~\cite{Muller1967, Israel:1979wp, Hiscock:1987zz, Olson:1990rzl}. In recent years, several alternative causal and stable second-order~\cite{Denicol:2010xn, Denicol:2012cn, Baier:2007ix, Jaiswal:2012qm, Jaiswal:2013npa, Jaiswal:2013fc} as well as first-order  \cite{Bemfica:2017wps,Bemfica:2020zjp,Kovtun:2019hdm,Pandya:2021ief} theories were derived.

An important distinction from the nonrelativistic case is that in relativistic theory, the fields -- and hence $T^{\mu\nu}$ and $N^{\mu}$ -- are constructed as expectation values of quantum field operators, reflecting the field-theoretic nature of particle production and interactions in relativistic systems.
%
%
\subsubsection{Relativistic fluid with spin}
A \emph{relativistic fluid with spin}  consists of particles that not only carry energy, momentum, and quantum numbers, but also possess intrinsic angular momentum (spin). In this case, an additional macroscopic variable -- the \textit{spin density} -- must be introduced. The state of the fluid is then described by the \textit{spin tensor} \( S^{\lambda\mu\nu} \) (in addition to the energy-momentum tensor \( T^{\mu\nu} \) and the number current \( N^{\mu} \)), which encodes the density and flux of intrinsic angular momentum in spacetime. The evolution of these quantities is governed by conservation laws, which now include total angular momentum conservation. The spin tensor is not conserved independently; instead, it contributes to the total angular momentum tensor \( J^{\lambda\mu\nu} = L^{\lambda\mu\nu} + S^{\lambda\mu\nu} \), where \( L^{\lambda\mu\nu} \) is the  orbital angular momentum tensor. This decomposition will be derived from field-theoretic principles later in this work. 

Since spin dynamics is inherently dissipative, there is no analog of ideal spin hydrodynamics. Consequently, the task is to construct a dissipative theory of relativistic spin hydrodynamics, in particular by generalizing the structure of the relativistic Navier-Stokes equations to incorporate a spin tensor. Given the known issues of instability and acausality in first-order theories, this approach must ultimately lead to further generalizations, potentially extending the Müller-Israel-Stewart framework to include spin degrees of freedom. As in conventional relativistic hydrodynamics, the tensors \( N^{\mu} \), \( T^{\mu\nu} \), and \( S^{\lambda\mu\nu} \) are interpreted as expectation values of corresponding quantum field operators.
\medskip

The study of fluids composed of particles with intrinsic spin has a long and pioneering history~\cite{Weyssenhoff:1947iua, BohmVigier1958, Halbwachs1960, HessWaldmann1966}. Recent works~\cite{Becattini:2007sr,Becattini:2007nd, Becattini:2013fla} formalized the relation between the average spin polarization of produced hadrons and the thermal vorticity, calculated at freeze-out on top of a conventional hydrodynamic evolution in local equilibrium -- similarly to how momentum spectra are obtained. These studies have been instrumental in interpreting global spin polarization measurements and are closely related to seminal nonrelativistic investigations~\cite{Barnett:1915uqc, RevModPhys.7.129, Richardson1908, Einstein1915, Takahashi2015, maekawa2017spin, matsuo2017}. 

Prompted by local polarization measurements, a hydrodynamic framework describing the space-time evolution of spin via a spin tensor locally out-of-equilibrium, referred to as \textit{relativistic spin hydrodynamics}, was introduced in Refs.~\cite{Florkowski:2017ruc, Florkowski:2017dyn}. This framework is expected to modify the standard relation between average spin polarization and thermal vorticity. At the time of writing this thesis, several theoretical approaches to spin hydrodynamics had emerged:
\begin{itemize}
\item covariant thermodynamics~\cite{Hattori:2019lfp,Fukushima:2020ucl,Li:2020eon,She:2021lhe,Cao:2022aku,Hu:2022azy,Abbasi:2022rum,Kiamari:2023fbe,Ren:2024pur,Yang:2024duc,Florkowski:2024bfw,Drogosz:2024gzv,Drogosz:2024lkx,Dey:2024cwo}
\item relativistic kinetic theory~\cite{Florkowski:2018ahw,Florkowski:2019qdp,Hattori:2019ahi,Gao:2019znl,Weickgenannt:2019dks,Wang:2019moi,Li:2019qkf,Kapusta:2019sad,Liu:2019krs,Yang:2020hri,Liu:2020flb,Shi:2020htn,Bhadury:2020puc,Bhadury:2020cop,Speranza:2020ilk,Weickgenannt:2020aaf,Peng:2021ago,Singh:2021man,Sheng:2021kfc,Weickgenannt:2021cuo,Weickgenannt:2022zxs,Weickgenannt:2022qvh,Hidaka:2022dmn,Weickgenannt:2022jes,Weickgenannt:2023bss,Wagner:2024fhf,Chiarini:2024cuv,Weickgenannt:2024esg,Weickgenannt:2024ibf}
\item relativistic quantum-statistical approaches~\cite{Becattini:2018duy,Hu:2021lnx,Tiwari:2024trl,Buzzegoli:2025zud,She:2025qri}
\item holographic duality and gravity with torsion~\cite{Gallegos:2020otk,Hongo:2021ona,Floerchinger:2021uyo,Gallegos:2021bzp,Gallegos:2022jow}
\item Lagrangian effective field theory \cite{Montenegro:2017rbu,Montenegro:2017lvf,Montenegro:2018bcf,Montenegro:2020paq,Torrieri:2022ogj}
\end{itemize}  
with other related developments~\cite{Becattini:2009wh,Becattini:2011zz,Florkowski:2019gio,Becattini:2012pp,Garbiso:2020puw,Cartwright:2021qpp,Hu:2021pwh,Florkowski:2021wvk,Ambrus:2022yzz,Das:2022azr,Bhadury:2022ulr,Sarwar:2022yzs,Dey:2023hft,Bhadury:2023vjx,Weickgenannt:2023btk,Wang:2024afv,Fang:2024hxa,Banerjee:2024xnd,Drogosz:2024rbd,Lin:2024svh,Buzzegoli:2024mra,Lapygin:2025zhn,Singh:2024cub,Bhadury:2024ckc,Choi:2025tql,Giacalone:2025bgm,Sapna:2025yss,Funaki:2025isu,Armas:2025fvo}. For reviews on the recent progress in this field see Refs.~\cite{Florkowski:2018fap,Bhadury:2021oat,Becattini:2024uha,Huang:2024ffg}.
%
%
\section{Objectives and overview of this thesis}
\label{Objectives and overview of this thesis}
The central objective of this dissertation is to develop a mathematically consistent and physically complete theory of relativistic dissipative spin hydrodynamics using two methods. The first~\multimyref{AD1, AD2, AD3, AD4, AD6, AD7} in Chapters~\ref{Navier-Stokes limit} and \ref{MISMIS} extends the \textit{Navier-Stokes}~\cite{landau2013fluid, Eckart:1940te} and \textit{Müller-Israel-Stewart}~\cite{Muller1967, Israel:1979wp} frameworks by incorporating a spin tensor into the structure of covariant thermodynamics. The second~\multimyref{AD5, AD8} in Chapter~\ref{Quantum-statistical formulation} is grounded in relativistic quantum statistical mechanics and thermal \emph{quantum field theory (QFT)}, adapting and extending the foundational \textit{Zubarev} formalism~\cite{Zubarev, ChGvanWeert}.

%
\medskip

The thesis is organized as follows:
\begin{itemize}
    \item Chapter~\ref{From fields to fluids} provides a field-theoretical foundation for relativistic spin hydrodynamics in Minkowski spacetime. We examine the underlying conservation laws and their associated conserved quantities. 
    
    \item Chapter~\ref{Navier-Stokes limit} develops the theory in the Navier-Stokes (NS) limit by deriving the evolution equations that govern the system. Starting from the conservation laws and covariant thermodynamics, we derive the entropy current and the corresponding entropy production rate. This approach facilitates the identification of dissipative currents and their associated transport coefficients, ultimately leading to a closed system of evolution equations. Moreover, using linear-mode analysis, we demonstrate that the resulting equations exhibit linear instabilities in addition to their well-known causality violations arising from their parabolic structure. We then solve the equations under Bjorken flow conditions and find that, under certain circumstances, the numerical solutions display nonphysical behavior. All of the above analyses employ a phenomenological form of the spin tensor that is antisymmetric in its last two indices. For comparison, we also introduce the canonical spin tensor, which is fully antisymmetric, and show that the two approaches are equivalent under specific conditions. Both formulations are firmly rooted in field-theoretic principles, as outlined in Chapter~\ref{From fields to fluids}. 

    \item
    Motivated by the issues of stability and causality, in Chapter~\ref{MISMIS} we extend the formulation to the M\"uller-Israel-Stewart (MIS) limit by deriving second-order evolution equations. Specifically, we generalize the entropy current by including second-order terms in the hydrodynamic gradient expansion. Using entropy-current analysis, we identify the relevant transport coefficients and treat the dissipative currents as independent dynamical variables governed by relaxation-type evolution equations. Finally, we present a truncated version of these equations and perform a stability and causality analysis using linear-mode analysis in the rest frame, considering both high- and low-wavenumber limits.
\item

In Chapter~\ref{Quantum-statistical formulation}, we approach the formulation of spin hydrodynamics from a different theoretical perspective. We begin by revisiting the key concepts from nonrelativistic quantum statistical mechanics, which sets the stage for a brief introduction to the \textit{Zubarev} formulation of conventional relativistic hydrodynamics. We then extend this framework to derive the density operator describing a system of particles with spin. Using this operator, we derive the entropy current and entropy production rate. Additionally, we introduce a novel method -- presented in Appendix~\ref{AppendixC} -- for deriving the irreducible decomposition of a rank-$n$ tensor under rotations. This method enables us to extract the irreducible structure of dissipative currents at first-order in hydrodynamic gradient expansion. We conclude by summarizing these currents along with the corresponding transport coefficients. 

\item In Chapter~\ref{summaryandoutlook}, we summarize the key results of the thesis and outline directions for future work. 
\end{itemize}
\medskip

%
\subsection{List of papers contributing to this thesis}
\medskip

\begin{enumerate}[{start=1,font={\bfseries},label=[AD\arabic*]}]
\item \label{AD1}
\textbf{A. Daher}, A. Das, W. Florkowski,  R. Ryblewski,\\
\textit{Canonical and phenomenological formulations of spin hydrodynamics}, \\
Physical Review C \textbf{108} (2023) 2, 024902\\ arXiv: 2202.12609 [nucl-th]
\item \label{AD2}
\textbf{A. Daher}, A. Das, R. Ryblewski\\
\textit{Stability studies of first-order spin-hydrodynamic frameworks},\\
Physical Review D \textbf{107} (2023) 5, 054043 \\  arXiv: 2209.10460 [nucl-th]
\item \label{AD3}
R. Biswas, \textbf{A. Daher}, A. Das, W. Florkowski,  R. Ryblewski\\
\textit{Boost invariant spin hydrodynamics within the first order in derivative expansion},\\
Physical Review D \textbf{107} (2023) 9, 094022\\ arXiv: 2211.02934 [nucl-th]
\item \label{AD4}
R. Biswas, \textbf{A. Daher}, A. Das, W. Florkowski,  R. Ryblewski\\
\textit{Relativistic second-order spin hydrodynamics: An entropy-current analysis},\\
Physical Review D \textbf{108} (2023) 1, 014024 \\ arXiv: 2304.01009 [nucl-th]
\item \label{AD5}
F. Becattini, \textbf{A. Daher}, X.L. Sheng\\
\textit{Entropy current and entropy production in relativistic spin hydrodynamics},\\
Physics Letters B \textbf{850} (2024) 138533 \\
arXiv: 2309.05789 [nucl-th]
\item \label{AD6}
\textbf{A. Daher}, W. Florkowski, R. Ryblewski\\
\textit{Stability constraint for spin equation of state},\\
Physics Review D \textbf{110} (2024) 3, 034029 \\ arXiv: 2401.07608 [hep-ph]
\item\label{AD7}
\textbf{A. Daher}, W. Florkowski, R. Ryblewski, F. Taghinavaz\\
\textit{Stability and causality of rest frame modes in second-order spin hydrodynamics},\\
Physics Review D \textbf{109} (2024) 11, 114001 \\ arXiv: 2403.04711 [hep-ph]
%
%
%
%
%
%
\item\label{AD8}
\textbf{A. Daher}, X.L. Sheng, D. Wagner, F.Becattini\\
\textit{Dissipative currents and transport coefficients in relativistic spin hydrodynamics},\\ This work was submitted for peer review by the time this dissertation was completed\\
arXiv: 2503.03713 [nucl-th] 
\end{enumerate}
Further publications beyond the scope of this dissertation:
\begin{enumerate}
\item[]\label{}
\textbf{A. Daher}, L. Tinti, A. Jaiswal, R. Ryblewski\\
\textit{Quasiparticle second-order dissipative hydrodynamics at finite chemical potential},\\
Physics Review D \textbf{111} (2025) 7, 074011 \\ arXiv: 2412.06024 [hep-ph]
\end{enumerate}

\chapter{Relativistic spin hydrodynamics from field theory viewpoint}
\label{From fields to fluids}
This chapter derives conservation laws and their associated conserved quantities relevant to relativistic spin hydrodynamics, using field-theoretic methods in Minkowski spacetime. These conservation laws serve as fundamental tools for the formulation of spin hydrodynamics presented in the subsequent chapters.

Section~\ref{Global continuous transformations and conservation laws} demonstrates how global continuous symmetries of an underlying field theory give rise -- via Noether's theorem -- to conservation laws that remain valid in the hydrodynamic limit, provided the conserved quantities are identified with thermal expectation values of renormalized operators. Sections~\ref{translations} through \ref{U1globalfermion} present explicit derivations of the conserved quantities and corresponding conservation equations for scalar, vector, and spinor fields, covering the energy-momentum tensor, total angular momentum (including spin contributions), and U(1) charge current. These calculations allows us to identify the conservation laws governing relativistic spin hydrodynamics, which are summarized in Sec.~\ref{Relativisticspinhydrodynamics}. Finally, Sec.~\ref{Pseudo-gauge transformation concept} introduces pseudo-gauge transformations, which will employed in later chapters.

The material in this chapter is primarily based on established results, following Refs.~\cite{Ryder:1985wq,Weinberg:1995mt,Peskin:1995ev,Zee:2003mt,gilmore2006lie,kleinert2016particles}, with further citations included where appropriate.
%
%
\section{Global continuous transformations and conservation laws}
\label{Global continuous transformations and conservation laws}
We begin by considering the action functional constructed from a Lagrangian density describing relativistic classical or quantum fields in Minkowski spacetime. This action is invariant under global continuous transformations of the Poincaré group, understood here as the group of spacetime translations combined with proper, orthochronous Lorentz transformations \( \mathrm{SO}(1,3) \) which encompasses both spatial rotations and boosts~\footnote{The full Poincaré group \( \mathrm{ISO}(1,3) \) is defined as the semidirect product of spacetime translations and the full Lorentz group \( \mathrm{O}(1,3) \). Here, we are interested in the continuous symmetries of the action preserving both orientation and direction of time and hence focus on the subgroup \( \mathrm{SO}^{+}(1,3) \). In this thesis, we denote this subgroup -- the proper, orthochronous Lorentz group \( \mathrm{SO}^{+}(1,3) \) -- by \( \mathrm{SO}(1,3) \), omitting the \( + \) superscript for notational simplicity.}~\cite{Weinberg:1995mt,Zee:2003mt}.

In addition to spacetime symmetries, the action for complex field is also invariant under global internal 
 \( \mathrm{U}(1) \) transformations, corresponding to phase rotations in the complex fields space~\footnote{By \emph{global}, we mean that the transformation parameter is spacetime-independent; \emph{continuous} refers to the existence of a Lie group with continuous parameters; \emph{internal} indicates that the transformation acts on fields rather than coordinates.} \cite{Ryder:1985wq}. Table \ref{symmetry_table} summarizes these symmetries.
\begin{table}[H]
\renewcommand{\arraystretch}{1.5}
\begin{tabularx}{\textwidth}{ 
| >{\raggedright\arraybackslash}X 
| >{\raggedright\arraybackslash}X 
| >{\raggedright\arraybackslash}X | }
\hline
\textbf{Transformation type} & \textbf{Description} & \textbf{Mathematical form} \\ 
\hline
Spacetime translation & Shifts spacetime coordinates & \(x^{\prime\mu} = x^\mu \pm a^\mu\) \\ 
\hline
Proper,
orthochronous Lorentz transformation & Rotations and boosts & \(x^{\prime\mu} = \Lambda^\mu{}_\nu x^\nu\) \\ 
\hline
{Spatial rotation} & Rotations in spatial components & \(x^{\prime \, 0} =  x^0,\)  \(x^{\prime\, i} = R^i_j x^j\) \\
\hline
{Boost} & Changes relative velocity & \shortstack{$x^{\prime \, 0} = \gamma (x^0 - v x^i),$\\  $x^{\prime\,i} = \gamma (x^i - v x^0)$} for fixed spatial index $i$\\ 
\hline
{U(1) transformation} & Global phase rotation of fields & \(\psi^\prime = e^{i\alpha} \psi\) \\ 
\hline
\end{tabularx}
\caption{Summary of global continuous transformations.}
\label{symmetry_table}
\end{table}
\noindent
According to \emph{Noether's theorem}, each global continuous symmetry gives rise to a locally conserved quantity (the \emph{Noether current}) and an associated globally conserved charge (the \emph{Noether charge}) \cite{Noether:1918zz,Peskin:1995ev}.

As discussed in Chapter~\ref{Introduction}, a \emph{relativistic fluid} can be understood as the macroscopic continuum limit of a quantum many-body system. Its dynamics must respect the same global continuous symmetries as the underlying microscopic theory. Consequently, both the hydrodynamic and field-theoretic descriptions must exhibit the same conservation laws, provided the correspondence between conserved quantities is established via thermal expectation values of the relevant field-theoretic operators.

Specifically, let $\widehat{\mathcal{A}}^\mu$ be a conserved current in the microscopic theory and $\widehat{\rho}$ the stationary density operator in the Heisenberg picture. Then the corresponding hydrodynamic quantity is given by the thermal expectation value~\cite{Kapusta:2006pm,Bellac:2011kqa,Rindori:2020qqa,},
\begin{align}\label{TQF}
\mathcal{A}^\mu \equiv \langle \widehat{\mathcal{A}}^\mu \rangle = \text{Tr}[\widehat{\rho} \widehat{\mathcal{A}}^\mu]_{\text{ren}},
\end{align}
where the subscript "ren" indicates that this expectation value is appropriately renormalized~\footnote{Chapter~\ref{Quantum-statistical formulation} will discuss the explicit form of $\widehat{\rho}$ and the renormalization expectation values for relevant systems, such as the quark-gluon plasma.}.

If the operator-level conservation law $\partial_{\mu}\widehat{\mathcal{A}}^{\mu}=0$ holds, then the thermal expectation value satisfies, 
\begin{align}\label{TTQFT}
\partial_{\mu}\mathcal{A}^{\mu}=\partial_{\mu}\text{Tr}[\widehat{\rho}\mathcal{\widehat{A}}^{\mu}]_{\text{ren}}=\text{Tr}[\widehat{\rho}\,\partial_{\mu}\mathcal{\widehat{A}}^{\mu}]_{\text{ren}}=0,
\end{align}
where we have used $\partial_{\mu}\widehat{\rho}=0$ for a stationary state. Even when the trace $\text{Tr}[\widehat{\rho}\mathcal{\widehat{A}}^{\mu}]_{\text{ren}}$ cannot be computed analytically, the conservation law still holds, ensuring consistency between the microscopic and macroscopic descriptions. 
\begin{table}[H]
\renewcommand{\arraystretch}{1.5}
\begin{tabularx}{\textwidth}{ 
| >{\raggedright\arraybackslash}X 
| >{\centering\arraybackslash}>{\hsize=0.5\hsize}X 
| >{\raggedright\arraybackslash}>{\hsize=1.5\hsize}X 
| >{\raggedright\arraybackslash}X | }
\hline
\textbf{Field type} & \textbf{Spin} & \textbf{Examples} & \textbf{Mathematical representation} \\ 
\hline
Scalar  & \(0\) & Higgs boson & \(\phi(x)\) \\ 
\hline
Vector  & \(1\) & Photon, W/Z bosons, gluons & \(A^\mu(x)\) \\ 
\hline
Spinor  & \(\frac{1}{2}\) & Leptons, quarks & \(\psi(x)\) \\ 
\hline
\end{tabularx}
\caption{Field types considered in this chapter, with examples and mathematical representation.}
\label{fields_table}
\end{table}
\noindent
In the following, we systematically analyze several field theories, listed in Table~\ref{fields_table},  deriving their conserved quantities and conservation laws under global continuous transformations. This analysis enables us to identify the subset of conservation laws that underlie relativistic spin hydrodynamics. Since global continuous symmetries are defined via Lie group actions, their structure is independent of whether the fields are number-valued functions (as in classical field theory) or operators (as in quantum field theory). For this reason, we omit operator notations throughout Sections~\ref{translations}--\ref{U1globalfermion} for simplicity.

%
\section{Spacetime translation: scalar, vector, and spinor fields}
\label{translations}
\subsection*{Energy-momentum tensor and local conservation law}
Consider a Lagrangian density \(\mathcal{L}(\phi, \partial_\mu\phi)\) for a real scalar field \(\phi(x^\mu)\) that does not depend explicitly on the spacetime coordinates \(x^\mu\). The associated action functional is
\begin{align}
S[\mathcal{L}]=\int \di^{4}x\,\mathcal{L}.
\end{align}
Let us perform an infinitesimal spacetime translation of the coordinates,
\begin{align}
x^{\mu}\quad \longrightarrow \quad x^{\prime\,\mu}=x^{\mu}-\epsilon^{\mu},
\end{align}
where \(\epsilon^{\mu}\) is a constant infinitesimal four-vector that parametrizes the translation. The field evaluated at the translated point can be expanded using a Taylor series,
\begin{align}\label{phitranslation}
\phi(x^{\prime\,\mu})=\phi(x^{\mu})-\epsilon^{\nu}\partial_{\nu}\phi(x^{\mu})+\mathcal{O}(\epsilon^{2}),
\end{align}
As a scalar field, $\phi$ transforms under coordinate changes in such a way that it retains the same value at the same physical point,
\begin{align}\label{phiabsolutepoint}
\phi^{\prime}(x^{\prime})=\phi(x).
\end{align}
Combining Eqs.~\eqref{phitranslation} and \eqref{phiabsolutepoint}, the variation of the field under the infinitesimal translation is
\begin{align}
\delta\phi(x^{\prime})&=\phi^{\prime}(x^{\prime})-\phi(x^{\prime})=\epsilon^{\nu}\partial_{\nu}\phi(x).
\end{align}
Next, consider the variation of the Lagrangian density due to this field variation, 
\begin{align}\label{deltaL}
\delta \mathcal{L}&=\mathcal{L}\left(\phi^{\prime}(x^{\prime}),\partial_\mu\phi^{\prime}(x^{\prime})\right)-\mathcal{L}\left(\phi(x^{\prime}),\partial_\mu\phi(x^{\prime})\right).
\end{align}
Expanding to first order in $\epsilon$, and using the chain rule, one has
\begin{align}\label{deltaL}
\delta \mathcal{L}&=\epsilon^{\nu}\left(\frac{\partial \mathcal{L}}{\partial\phi(x)}\partial_{\nu}\phi(x)+\frac{\partial \mathcal{L}}{\partial(\partial_{\mu}\phi(x))}\partial_{\mu}  \partial_{\nu}\phi(x) \right).
\end{align}
We can now write the variation of the action
\begin{align}\label{deltaSforP}
\delta S&=\int \dd^{4}x\,\delta \mathcal{L}=\int \dd^{4}x\,\left[\epsilon^{\nu}\partial_{\nu}\phi\left( \frac{\partial\mathcal{L}}{\partial\phi}
-\partial_{\mu}\frac{\partial\mathcal{L}}{\partial(\partial_{\mu}\phi)}\right)+
\partial_{\mu}\left(\epsilon^{\nu}\partial_{\nu}\phi\frac{\partial\mathcal{L}}{\partial(\partial_{\mu}\phi)}\right)\right]\nonumber\\
&=\int \dd^{4}x\, \epsilon^{\nu}\partial_{\nu} \mathcal{L}.
\end{align}
Applying the \emph{Euler-Lagrange equations},
\begin{align}
 \frac{\partial\mathcal{L}}{\partial\phi}
-\partial_{\mu}\frac{\partial\mathcal{L}}{\partial(\partial_{\mu}\phi)}=0,
\end{align}
from Eq.~\eqref{deltaSforP} we obtain,
\begin{align}
\delta S=\int \label{EMTscalarfieldzero}\dd^4x\, \epsilon_\nu \partial_\mu\left(\frac{\partial \mathcal{L}}{\partial(\partial_\mu \phi)} \partial^\nu \phi - g^{\mu\nu} \mathcal{L}\right).
\end{align}
Since $\delta S\!=\!0$ for arbitrary $\epsilon$, we identify the conserved current associated with spacetime translations. The integrand defines the Noether current known as the \textit{canonical energy-momentum tensor}~\footnote{The derived Noether current is, strictly speaking, a current density, as it satisfies local conservation law. However, throughout this thesis, we refer to it simply as the ``current'' by convention.},
\begin{align}\label{EMTscalarfield}
T^{\mu\nu}=\frac{\partial \mathcal{L}}{\partial(\partial_{\mu}\phi)}\partial^{\nu}\phi-g^{\mu\nu}\mathcal{L},
\end{align}
which satisfies the \emph{local conservation law},
\begin{align}\label{localconservationEMToperator}
\partial_{\mu}T^{\mu\nu}=0.
\end{align}
This derivation follows the standard Noether procedure for spacetime symmetries cf. Refs.~\cite{Noether:1918zz,Peskin:1995ev,Baez:1995sj, Weinberg:1995mt}, with canonical energy-momentum tensors as discussed in Refs.~\cite{Srednicki:2007qs,Ryder:1985wq}.

\medskip
\noindent For a \emph{real scalar field} with Lagrangian density
\begin{equation}\label{realscalarlagrangian}
\mathcal{L} = \frac{1}{2} (\partial_\mu\phi )(\partial^\mu \phi) - V(\phi),
\end{equation}
where \(V(\phi)\) represents the potential term, the canonical energy-momentum tensor takes the explicit form
\begin{align}
T^{\mu\nu}=(\partial^{\mu}\phi)(\partial^{\nu}\phi)-g^{\mu\nu}\mathcal{L}.
\end{align}

\medskip
\noindent For a \emph{complex scalar field}, the general Lagrangian density depends on $\phi, \phi^{*}$, and their derivatives, $\mathcal{L}(\phi,\phi^{*},\partial_\mu\phi,\partial_\mu\phi^{*})$. The corresponding canonical energy-momentum tensor is
\begin{align}
T^{\mu\nu}=\frac{\partial \mathcal{L}}{\partial(\partial_{\mu}\phi)}\partial^{\nu}\phi+\partial^{\nu}\phi^{*}\frac{\partial\mathcal{L}}{\partial(\partial_{\mu}\phi^{*})}-g^{\mu\nu}\mathcal{L}.
\end{align}
\noindent For the commonly used form of the Lagrangian density
\begin{equation}\label{complexscalarfieldlagrangian}
\mathcal{L} = (\partial_\mu \phi^*)( \partial^\mu \phi) - V(\phi\,\phi^*),
\end{equation}
the energy-momentum tensor becomes
\begin{align}
T^{\mu\nu}=\partial^{\mu}\phi^{*}\partial^{\nu}\phi+\partial^{\nu}\phi^{*}\partial^{\mu}\phi-g^{\mu\nu}\mathcal{L}.
\end{align}

\medskip
\noindent Next, consider a \emph{vector field} \(\phi^{\mu}(x)\). Following the same procedure, the canonical energy-momentum tensor for a vector field is,
\begin{align}
T^{\mu\nu}=\frac{\partial \mathcal{L}}{\partial(\partial_{\mu}\phi^{\kappa})}\partial^{\nu}\phi^{\kappa}-g^{\mu\nu}\mathcal{L}.
\end{align}
A physically important example is the noninteracting electromagnetic field, described by the Lagrangian density,
\begin{align}\label{electromagneticfieldlagrangian}
\mathcal{L}=-\frac{1}{4}F^{\mu\nu}F_{\mu\nu},
\end{align}
where \(F_{\mu\nu} = \partial_{\mu}A_{\nu} - \partial_{\nu}A_{\mu}\) is the electromagnetic field strength (also known as \emph{Faraday}) tensor, and \(A_{\mu}\) is the electromagnetic four-potential. The canonical energy-momentum tensor becomes, 
\begin{align}\label{EMEM}
T^{\mu\nu}=-F^{\mu\lambda}\partial^{\nu}A_{\lambda}+\frac{1}{4}g^{\mu\nu}F^{\rho\sigma}F_{\rho\sigma}.
\end{align}
Notice that the above form is not symmetric\rr{~\footnote{One should also note that the canonical energy-momentum tensor in Eq.~\eqref{EMEM} is gauge dependent.}}. To obtain a symmetric energy-momentum tensor suitable for applications in general relativity, one applies the so-called \textit{Belinfante-Rosenfeld} symmetrization procedure~\cite{belinfante1939spin,cbelinfante1940current,rosenfeld1940}. This method systematically modifies the canonical energy-momentum tensor by adding a total derivative term, ensuring symmetry without affecting the local conservation law or the global conserved charges. The Belinfante-Rosenfeld symmetrization is an example of a so-called pseudo-gauge transformation -- the topic to be discussed in greater detail in Sec.~\ref{Pseudo-gauge transformation concept}.

\noindent
Finally, consider the \emph{free Dirac spinor} field with Lagrangian density
\begin{align}\label{Diraclagrangian}
\mathcal{L} = \Bar{\psi}\left(\frac{i}{2}\gamma^{\mu}\overset{\leftrightarrow}{\partial_\mu}-m\right)\psi.
\end{align}
Here, $\Bar{\psi}=\psi^{\dagger}\gamma^{0}$ is the Dirac adjoint with $\psi^{\dagger}$ denoting the hermitian conjugate and $\overset{\leftrightarrow}{\partial_\mu}$ is the symmetric derivative defined as $\Bar{\psi}(\gamma^{\mu}\overset{\leftrightarrow}{\partial_\mu})\psi\equiv \Bar{\psi}\gamma^{\mu}\partial_{\mu}\psi-\partial_{\mu}\Bar{\psi}\gamma^{\mu}\psi$. The Dirac matrices $\gamma^{\mu}$ satisfy the Clifford algebra 
\begin{align}
&\gamma^{\mu}\gamma^{\nu}+\gamma^{\nu}\gamma^{\mu}=\{\gamma^{\mu},\gamma^{\nu}\}=2g^{\mu\nu}, 
\end{align}
and the identity
\begin{align}
&\gamma^{\mu}\gamma^{\nu}\gamma^{\lambda}=g^{\mu\nu}\gamma^{\lambda}+g^{\nu\lambda}\gamma^{\mu}-g^{\mu\lambda}\gamma^{\nu}-i\epsilon^{\sigma\mu\nu\lambda}\gamma_{\sigma}\gamma^{5},\label{gamma5prop}
\end{align}
with $\gamma^{5}=i\gamma^{0}\gamma^{1}\gamma^{2}\gamma^{3}$. The corresponding canonical energy-momentum tensor for the Dirac field is,
\begin{align}
T^{\mu\nu}=\frac{\partial \mathcal{L}}{\partial(\partial_{\mu}\psi)}\partial^{\nu}\psi+\partial^{\nu}\Bar{\psi}\frac{\partial \mathcal{L}}{\partial(\partial_{\mu}\Bar{ \psi})}-g^{\mu\nu}\mathcal{L},
\end{align}
which, upon using Eq.~\eqref{Diraclagrangian}, evaluates to
\begin{align}
 T^{\mu\nu}&=\frac{i}{2}\left(\Bar{\psi}\gamma^{\mu}\partial^{\nu}\psi-\partial^{\nu}\Bar{\psi}\gamma^{\mu}\psi\right)-g^{\mu\nu}\mathcal{L}.
\end{align}
See also Refs.~\cite{Peskin:1995ev,Srednicki:2007qs,Ryder:1985wq,Weinberg:1995mt}
for detailed derivations and discussions of energy-momentum tensors in quantum field theory.
%
%
\subsection*{Total four-momentum and global conservation law}
\label{scalarFour-momentum and the global conservation law}
The local conservation of the energy-momentum tensor,
\begin{align}\label{locconsT}
\partial_{\mu}T^{\mu\nu}
&=0,
\end{align}
implies the existence of globally conserved quantities, provided appropriate boundary and geometric conditions are satisfied.

We define the conserved charge associated with the energy-momentum tensor or the \emph{total four-momentum} as an integral over a general spacelike hypersurface
\begin{align}\label{Ptotcons}
P^{\nu}=\int_{\Sigma_{\text{}}}\dd \Sigma_{\mu}T^{\mu\nu}.
 \end{align}
Here, $\dd \Sigma_{\mu}=n_\mu \dd \Sigma$, where $n_\mu$ is the future-directed unit vector normal to $\Sigma$ satisfying $n \cdot n =1$, and $\dd \Sigma$ is a volume element on $\Sigma$~\footnote{%
The three-dimensional hypersurface \( \Sigma \) is assumed to be spacelike, orientable, and smooth, with a globally defined future-directed unit normal vector \( n^\mu \) satisfying \( n^\mu n_\mu = 1 \). The directed surface element \( \dd\Sigma_\mu = n_\mu\, \dd\Sigma \) defines a directed three-form on \( \Sigma \), where \( \dd\Sigma \) is the proper volume element. This construction ensures that the integral \( \int_\Sigma \dd\Sigma_\mu\, T^{\mu\nu} \) is covariant and independent of the coordinate system used. For this integral to represent a globally conserved quantity (i.e., independent of the hypersurface), \( \Sigma \) should be also a \textit{Cauchy surface} -- a spacelike hypersurface intersecting every inextendible timelike or lightlike curve exactly once -- so that it captures the entire support of the field configuration. In flat spacetime, constant-time slices \( t = \text{const} \) are simple examples of such surfaces. More generally, hypersurfaces can be understood as level sets of a smooth global time function \( \tau(x) \), with \( \partial_\mu \tau \) timelike everywhere. This generalizes the notion of simultaneity to a coordinate-free setting and ensures that energy-momentum integrals are well-defined across arbitrary spacelike slicings of spacetime.%
}.

To show that $P^{\nu}$ is conserved, consider a compact spacetime region $\cal{V}$ bounded by two spacelike hypersurfaces, $\Sigma_{1}$ and $\Sigma_{2}$, and a timelike boundary $\Sigma_{\rm sides}$ at spatial infinity. Applying the Gauss (divergence) theorem to Eq.~\eqref{locconsT} over $\cal{V}$, \footnote{%
In curvilinear coordinates, the divergence theorem must be written in the form:
$
\int_{\mathcal{V}} \sqrt{-g}\, \dd^4x\, D_\mu T^{\mu\nu}
= \oint_{\partial \mathcal{V}} \dd\Sigma_\mu\, T^{\mu\nu},
$
where \( g = \det(g_{\mu\nu}) \) is the determinant of the spacetime metric expressed in the chosen coordinate system, and \( D_\mu \) is the covariant derivative compatible with \( g_{\mu\nu} \).  The inclusion of the factor \( \sqrt{-g} \) ensures that the integration measure \( \sqrt{-g}\, \dd^4x \) transforms as a scalar under coordinate changes, and the covariant divergence \( D_\mu T^{\mu\nu} \) ensures geometric invariance. In flat Cartesian coordinates, this reduces to the standard form \( \partial_\mu T^{\mu\nu} \) with \( \sqrt{-g} = 1 \) and \( D_\mu = \partial_\mu \).%
}
\begin{equation}
\int_{\mathcal{V}} \dd^4x\, \partial_\mu T^{\mu\nu} 
= \oint_{\partial \mathcal{V}} \dd\Sigma_\mu\, T^{\mu\nu},
\end{equation}
we obtain
\begin{align}
\oint \dd\Sigma_{\mu} \, T^{\mu\nu} = 0.
\end{align}
Here, \( \mathcal{V} \) denotes a finite spacetime volume, and \( \partial \mathcal{V} \) is its boundary.
 
Splitting the boundary into the three parts, $\partial \mathcal{V} = \Sigma_2 \cup (-\Sigma_1) \cup \Sigma_{\text{sides}}$, yields
\begin{align}\label{sigma12sides}
\int_{\Sigma_{2}}\dd\Sigma_{\mu}T^{\mu\nu}-\int_{\Sigma_{1}}\dd\Sigma_{\mu}T^{\mu\nu}+\int_{\Sigma_{\text{sides}}}\dd\Sigma_{\mu}T^{\mu\nu}=0,
 \end{align}
where the minus sign on \( \Sigma_1 \) indicates that its orientation is reversed with respect to the future-directed normal.

Assuming that the energy-momentum tensor falls off sufficiently rapidly at spatial infinity, the flux through $\Sigma_{\text{sides}}$ vanishes. This yields:
\begin{align}\label{sigma12sides2}
\int_{\Sigma_{2}}\dd\Sigma_{\mu}T^{\mu\nu}=\int_{\Sigma_{1}}\dd\Sigma_{\mu}T^{\mu\nu},
 \end{align}
demonstrating that the total four-momentum, as defined in Eq.~\eqref{Ptotcons}, is conserved between any two spacelike hypersurfaces satisfying the stated conditions.

In the special case where $\Sigma$ is a hypersurface of constant time $t$ in Minkowski spacetime, the unit normal is $n^\mu=(1,0,0,0)$ and the surface element reduces to $\dd\Sigma_\mu=(1,0,0,0) \dd^3x$. Substituting into Eq.~\eqref{Ptotcons}, we recover the standard expression
\begin{align}\label{Ptotconsspec}
P^{\nu}=\int_{}\,\dd ^3 x\, T^{0\nu},
 \end{align}
which defines the conserved energy $(\nu=0)$ and spatial momentum components $(\nu=1,2,3)$.

This covariant construction ensures that the conserved quantities associated with spacetime translation symmetry are independent of the specific family of hypersurfaces (foliation) used, provided the hypersurfaces are spacelike and the boundary fluxes vanish.

More generally, let $\{\Sigma_\tau\}$ be a foliation of spacetime defined by smooth, spacelike Cauchy hypersurfaces of global time function $\tau(x)$, with timelike gradient $D_\mu\tau(x)$ \cite{Wald:1984rg}~\footnote{{Milne coordinates (see also Sec.~\ref{boostinvariant}), used in modeling of ultrarelativistic heavy-ion collisions within relativistic hydrodynamics, provide a concrete example of this foliation.}}, where $D_\mu$ denotes the covariant derivative compatible with the spacetime metric. The total four-momentum evaluated on each hypersurface is given by $P^\nu\left[\Sigma_\tau\right]=\int_{\Sigma_\tau} \dd \Sigma_\mu T^{\mu \nu}(x)$, where $\dd \Sigma_{\mu}$ is defined as above. If the energy-momentum tensor satisfies the local conservation law $D_\mu T^{\mu\nu}=0$, and the flux through spatial infinity vanishes, then $P^\nu\left[\Sigma_\tau\right]$ is independent of the foliation parameter $\tau$. That is, the total four-momentum is conserved along the evolution 
\begin{align}\label{}
\frac{\dd P^{\nu}[\Sigma_\tau]}{\dd \tau}&=0. 
\end{align}
This expression generalizes the conservation law $\dd P^\mu(t)/\dd t=0$ from flat spacetime with Cartesian coordinates to arbitrary spacelike Cauchy foliations in globally hyperbolic spacetimes. It matches the geometric formulation of conserved charges used, for example, in canonical formulations of general relativity such as the ADM (Arnowitt–Deser–Misner) formalism \cite{Arnowitt:1962hi,Carroll:2004st}.  
%
%
\section{Lorentz transformation: scalar field}
\label{LTscalarr}
%
%
For a scalar field $\phi(x^{\mu})$, the Lorentz transformation of the spacetime coordinates is given by, 
\begin{align}\label{coordinateslorentz}
x^{\mu}\longrightarrow x^{\prime\,\mu\,}=\Lambda^{\mu}_{~\nu}x^{\nu},
\end{align}
where $\Lambda^{\mu}_{~\nu} \in \mathrm{SO}(1,3)$ is an element of the Lorentz group representing spatial rotations and boosts, as summarized in Table~\ref{symmetry_table}. The Minkowski metric $g^{\mu\nu}$ is invariant under Lorentz transformations as $\Lambda^{\mu}_{~\nu}$ satisfies
\begin{align}\label{conditionLambda}
\Lambda^{\mu}_{~\sigma}g^{\sigma\tau}\Lambda^{\nu}_{~\tau}=g^{\mu\nu},
\end{align}
along with the condition of having unit determinant. For an infinitesimal Lorentz transformation near the identity, we write 
\begin{align}\label{C17}
\Lambda^{\mu}_{~\nu} = \delta^{\mu}_{~\nu} + w^{\mu}_{~\nu},
\end{align}
where \(w^{\mu}_{~\nu}\) is infinitesimal. Substituting this into Eq.~\eqref{conditionLambda} and retaining only first-order terms, one finds that $w^{\mu}_{~\nu}$ is antisymmetric, $w^{\mu\nu}=-w^{\nu\mu}$.
%
%
%
%
The tensor \(w^{\mu}_{~\nu}\) belongs to the Lie algebra of the Lorentz group and can be expanded in terms of the six generators of $\mathrm{SO}(1,3)$ as
\begin{align}\label{wliealgebra}
w^{\mu}_{~\nu}&=\frac{1}{2}\Omega_{\alpha\beta}(A^{\alpha\beta})^{\mu}_{~\nu},
\end{align}
where \(\Omega_{\alpha\beta}\) is an antisymmetric matrix that represents the parameters of the infinitesimal Lorentz transformation, and \(A^{\alpha\beta}\) represents the generators of the Lorentz transformations. Explicitly, the generators can be grouped into spatial rotations \((L_k)^{\mu}_{~\nu}\) and boosts \((K_i)^{\mu}_{~\nu}\) so that
\begin{align}
w^{\mu}_{~\nu} = \frac{\delta \theta^k}{2} (L_k)^{\mu}_{~\nu} + \frac{\delta \beta^i}{2} (K_i)^{\mu}_{~\nu}.
\end{align}
where \(\delta \theta^k=(\delta\theta^{x},\delta\theta^{y},\delta\theta^{z})\) and \(\delta \beta^i=(\delta\beta^{x},\delta\beta^{y},\delta\beta^{z})\) are the infinitesimal spatial rotation and boost parameters, respectively. 
Under an infinitesimal transformation~\eqref{C17}, the scalar field transforms as
\begin{align}\label{phiL}
\phi(x^{\mu\,\prime})&=\phi(x^{\mu}+w^{\mu}_{~\nu}x^{\nu})=\phi(x)+w^{\mu}_{~\nu}x^{\nu}\partial_{\mu}\phi(x)+\mathcal{O}(w^{2}).
\end{align}
Since a scalar field is invariant under coordinate transformations, we have
\begin{align}\label{scalarabsolutepoint}
\phi^{\prime}(x^{\prime})=\phi(x).
\end{align}
Therefore, the variation of the field at a fixed point $x^{\prime}$ is, 
\begin{align}
\delta \phi(x^{\prime})&=\phi^{\prime}(x^{\prime})-\phi(x^{\prime}) 
 =-w^{\mu}_{~\nu}x^{\nu}\partial_{\mu}\phi(x).
\end{align}
This leads to the variation of the Lagrangian density, 
\begin{align}\label{C77}
\delta \mathcal{L}&=\mathcal{L}\left(\phi^{\prime}(x^{\prime}),\partial\phi^{\prime}(x^{\prime})\right)-\mathcal{L}\left(\phi(x^{\prime}),\partial\phi(x^{\prime})\right)\nonumber\\
&=-w^{\mu}_{~\nu}x^{\nu}\partial_{\mu}{\phi}\frac{\partial\mathcal{L}}{\partial\phi}-\partial_{{\lambda}}(w^{\mu}_{~\nu}x^{\nu}\partial_{\mu}\phi)\frac{\partial\mathcal{L}}{\partial(\partial_{\lambda}\phi)},
\end{align}
and, consequently, to the variation of the action, 
%
%
%
%
%
%
%
\begin{align}\label{2.23}
\delta S&=\int \dd^{4}x\,\delta \mathcal{L}\nonumber\\
&=\int \dd^{4}x\,\bigg\{- w^{\mu}_{~\nu}x^{\nu}\partial_{\mu}{\phi}\left[ \frac{\partial\mathcal{L}}{\partial\phi}-\partial_{\lambda}\frac{\partial\mathcal{L}}{\partial(\partial_{\lambda}\phi)}\right]
-\partial_{\lambda}\left(w^{\mu}_{~\nu}x^{\nu}\partial_{\mu}\phi\frac{\partial\mathcal{L}}{\partial(\partial_{\lambda}\phi)}\right)\bigg\}\nonumber\\
&=-\int \dd^{4}x\, \partial_{\mu}(x^{\nu}\mathcal{L})w^{\mu}_{~\nu}.
\end{align}
If the scalar field satisfies the Euler-Lagrange equations, then the action is invariant, and from Eq.~\eqref{2.23}, we obtain
\begin{align}
\int \dd^{4}x\, \partial_{\lambda}\left[x^{\mu}\left(\frac{\partial \mathcal{L}}{\partial(\partial_{\lambda}\phi)}\partial^{\nu}\phi {-g^{\lambda\nu}\mathcal{L}}\right)-x^{\nu}\left(\frac{\partial \mathcal{L}}{\partial(\partial_{\lambda}\phi)}\partial^{\mu}\phi-g^{\lambda\mu}\mathcal{L}\right)
\right]w_{\mu\nu}=0.
\end{align}
For each $w_{\mu\nu}$, we identify the conserved Noether current known as the \emph{orbital angular momentum tensor},
\begin{align}\label{OAMnoethercurrentdensity}
L^{\lambda\mu\nu}&=x^{\mu}\left(\frac{\partial \mathcal{L}}{\partial(\partial_{\lambda}\phi)}\partial^{\nu}\phi-g^{\lambda\nu}\mathcal{L}\right)-x^{\nu}\left(\frac{\partial \mathcal{L}}{\partial(\partial_{\lambda}\phi)}\partial^{\mu}\phi-g^{\lambda\mu}\mathcal{L}\right),
\end{align}
which satisfies the local conservation law,
\begin{align}\label{Lfieldconservation}
\partial_{\lambda}L^{\lambda\mu\nu}=0.
\end{align}
This tensor is antisymmetric in the last two indices: $L^{\lambda\mu\nu}=-L^{\lambda\nu\mu}$.

Using the expression for the canonical energy-momentum tensor~\eqref{EMTscalarfield}, the orbital angular momentum can be expressed as,
\begin{align}\label{LintermsofT}
L^{\lambda\mu\nu}=x^{\mu}T^{\lambda\nu}-x^{\nu}T^{\lambda\mu}.
\end{align}
Finally, note that the local conservation of the orbital angular momentum tensor $L^{\lambda\mu\nu}$ implies
\begin{align}\label{Lconservation}
\partial_{\lambda}L^{\lambda\mu\nu}&=\partial_{\lambda}\left(x^{\mu}T^{\lambda\nu}-x^{\nu}T^{\lambda\mu}\right) 
=T^{\mu\nu}-T^{\nu\mu} 
=0.
\end{align}
Thus, for scalar fields, the canonical energy-momentum tensor is symmetric, which is consistent with the conservation of the orbital angular momentum tensor. 

Following the methodology outlined in Sec.~\eqref{translations}, we can define the total orbital angular momentum as a conserved charge associated with the angular momentum tensor over a general spacelike hypersurface with specific boundary conditions as,
\begin{align}\label{totalorb}
L^{\mu\nu} = \int_{\Sigma} \dd \Sigma_\lambda \, L^{\lambda\mu\nu}.
\end{align}
If the hypersurface is chosen to be a constant time slice in Minkowski spacetime, we recover
\begin{align}\label{xyzss}
L^{\mu\nu} = \int \dd^{3}x \, L^{0\mu\nu},
\end{align}
such that it satisfies the global conservation law, 
\begin{align}\label{xyzssss}
\frac{\dd L^{\mu\nu}}{\dd t}=0.
\end{align}
%

%

\section{Lorentz transformation: vector and spinor fields}
\label{LTvectorandfermion}
In contrast to scalar field theories, fields with intrinsic spin -- such as vector fields (e.g., the electromagnetic field \( A^{\mu}(x) \) described by Eq.~\eqref{electromagneticfieldlagrangian}) or fermionic spinor fields (described by Eq.~\eqref{Diraclagrangian}) -- involve a more intricate derivation of the Noether current associated with Lorentz symmetry. In such cases, the total angular momentum tensor includes both orbital and spin contributions.
%
%
\subsection*{Vector field and the spin tensor antisymmetric in last two indices}
\medskip
Let us first consider a general vector field \(\phi^{\mu}(x)\) governed by an action of the form
\begin{align}\label{actionofvectorfield}
S[\mathcal{L}] = \int \dd^{4}x \, \mathcal{L}(\phi^{\mu}, \partial_\nu\phi^{\mu}).
\end{align}
Under an infinitesimal Lorentz transformation, the spacetime coordinates transform as  
\begin{align}\label{coordinateslorentz4}
x^{\nu} \longrightarrow x^{\prime\,\nu} = (\delta^{\nu}_{~\lambda}+w^{\nu}_{~\lambda}) x^{\lambda}.
\end{align}
The field evaluated at the new coordinates is, 
\begin{align}
\phi^{\mu}(x^{\prime})&=\phi^{\mu}(x)+w^{\nu}_{~\lambda}x^{\lambda}\partial_{\nu}\phi^{\mu}(x)+\mathcal{O}(w^{2}).
\end{align}
However, unlike scalar fields, vector fields also transform in their components under Lorentz transformations,
\begin{align}\label{vectorabsolutepoint}
\phi^{\prime\mu}(x^{\prime}) &= \Lambda^{\mu}_{~\nu}\phi^{\nu}(x) 
~\Longleftrightarrow~ 
\phi^{\prime\mu}(x) = \Lambda^{\mu}_{~\nu}\phi^{\nu}(\Lambda^{-1}x).
\end{align}
For infinitesimal transformations, this becomes
\begin{align}
\phi^{\prime\mu}(x^{\prime}) 
&= (g^{\mu}_{~\nu} + w^{\mu}_{~\nu})\phi^{\nu}(x) = \phi^{\mu}(x) + w^{\mu}_{~\nu}\phi^{\nu}(x).
\end{align}
Thus, the total variation of the vector field at the transformed coordinates is 
\begin{align}
\delta\phi^{\mu}(x^{\prime}) 
&= \phi^{\prime\mu}(x^{\prime}) - \phi^{\mu}(x^{\prime})= w^{\mu}_{~\nu}\phi^{\nu}(x) 
- w^{\nu}_{~\lambda}x^{\lambda}\partial_{\nu}\phi^{\mu}(x).
\end{align}
Compared to the scalar case in Eq.~\eqref{C77}, additional terms appear due to the internal transformation of the field components,
\begin{align}
\delta \mathcal{L}&=\mathcal{L}\left(\phi^{\prime}(x^{\prime}),\partial\phi^{\prime}(x^{\prime})\right)-\mathcal{L}\left(\phi(x^{\prime}),\partial\phi(x^{\prime})\right)\\
&=w^{\mu}_{~\delta}\phi^{\delta}(x)\frac{\partial\mathcal{L}}{\partial\phi^{\mu}}-w^{\nu}_{~\alpha}x^{\alpha}\partial_{\nu}\phi^{\mu}\frac{\partial\mathcal{L}}{\partial\phi^{\mu}}+\partial_{\lambda}(w^{\mu}_{~\delta}\phi^{\delta})\frac{\partial\mathcal{L}}{\partial(\partial_{\lambda}\phi^{\mu})}-\partial_{\lambda}(w^{\nu}_{~\alpha}x^{\alpha}\partial_{\nu}\phi^{\mu})\frac{\partial\mathcal{L}}{\partial(\partial_{\lambda}\phi^{\mu})}.
\nonumber
\end{align}
Assuming that the field equations hold (on-shell), the variation of the action leads to 
\begin{align}\label{Jcurrentdensityintegral}
\int \dd^{4}x\,\partial_{\lambda}\bigg[&x^{\mu}\left(\frac{\partial \mathcal{L}}{\partial(\partial_{\lambda}\phi^{\kappa})}\partial^{\nu}\phi^{\kappa}-g^{\lambda\nu}\mathcal{L}\right)-x^{\nu}\left(\frac{\partial \mathcal{L}}{\partial(\partial_{\lambda}\phi^{\kappa})}\partial^{\mu}\phi^{\kappa}-g^{\lambda\mu}\mathcal{L}\right)\nonumber\\
&+\left(\frac{\partial \mathcal{L}}{\partial(\partial_{\lambda}\phi_{\mu})}\phi^{\nu}-\phi^{\mu}\frac{\partial \mathcal{L}}{\partial(\partial_{\lambda}\phi_{\nu})}\right)\bigg]w_{\mu\nu}=0.
\end{align}
The two terms in the first line of the above equation form the orbital angular momentum tensor for a vector field, 
\begin{align}\label{C43}
L^{\lambda\mu\nu}&=x^{\mu}\left(\frac{\partial \mathcal{L}}{\partial(\partial_{\lambda}\phi^{\kappa})}\partial^{\nu}\phi^{\kappa}-g^{\lambda\nu}\mathcal{L}\right)-x^{\nu}\left(\frac{\partial \mathcal{L}}{\partial(\partial_{\lambda}\phi^{\kappa})}\partial^{\mu}\phi^{\kappa}-g^{\lambda\mu}\mathcal{L}\right)=x^{\mu}T^{\lambda\nu}-x^{\nu}T^{\lambda\mu},
\end{align}
where $T^{\lambda\nu}$ is the canonical energy-momentum tensor for the vector field. The term in the second line of Eq.~\eqref{Jcurrentdensityintegral} defines the \emph{spin angular momentum tensor},  
\begin{align}
S^{\lambda\mu\nu}=\frac{\partial \mathcal{L}}{\partial(\partial_{\lambda}\phi_{\mu})}\phi^{\nu}-\phi^{\mu}\frac{\partial \mathcal{L}}{\partial(\partial_{\lambda}\phi_{\nu})},
\end{align}
which is antisymmetric in the last two indices.

For each choice of $w_{\mu\nu}$, we identify the full integrand inside the brackets in Eq.~\eqref{Jcurrentdensityintegral} as the \emph{total angular momentum tensor} \(J^{\lambda\mu\nu}\)
\begin{align}\label{J=L+Sfields}
J^{\lambda\mu\nu}
&=L^{\lambda\mu\nu}+S^{\lambda\mu\nu},
\end{align}
which serves as the conserved Noether current associated with Lorentz symmetry.

In conclusion, while the scalar field theories yield symmetric canonical energy-momentum tensors and possess no intrinsic spin contributions, fields with internal Lorentz structure require both orbital and spin parts in the total angular momentum tensor. This decomposition plays a crucial role in understanding such fields' conserved charges and symmetry properties.  We close with a note that the conserved charge associated to the total angular momentum tensor follows the same definitions as in Eqs.~\eqref{totalorb}-\eqref{xyzssss}.

\subsection*{Spinor field and the totally antisymmetric spin tensor}

We now focus on the free fermion field described by the Dirac Lagrangian density in Eq.~\eqref{Diraclagrangian}. Under the infinitesimal Lorentz transformation in Eq.~\eqref{C17}, the fermion field evaluated at the transformed coordinates is, 
\begin{align}
\psi(x^{\prime})=\psi(x)+w^{\mu}_{~\nu}x^{\nu}\partial_{\mu}\psi.
\end{align}
Unlike the scalar field in Eq.~\eqref{scalarabsolutepoint} and the vector field in Eq.~\eqref{vectorabsolutepoint}, the spinor field transforms as
\begin{align}
\psi^{\prime\mu}(x^{\prime}) = D(\Lambda)^{\mu}_{~\nu}\psi^{\nu}(x) ~\Longleftrightarrow~ \psi^{\prime\mu}(x) = D(\Lambda)^{\mu}_{~\nu}\psi^{\nu}(\Lambda^{-1}x),
\end{align}
where \(D(\Lambda)\) is a \(4 \times 4\) spinor representation matrix of the Lorentz group. For an infinitesimal Lorentz transformation, the spinor representation becomes:
\begin{align}
D(I_{4} + w) = 1 - \frac{i}{4}w_{\mu\nu}\sigma^{\mu\nu},
\end{align}
where \(\sigma^{\mu\nu} = \frac{i}{2}[\gamma^{\mu}, \gamma^{\nu}]\) are the generators of the Lorentz algebra in spinor space. 

\noindent
Hence, the total transformation of the spinor field is
\begin{align}
\psi^\prime(x^\prime)=\psi(x)-\frac{i}{4}w_{\mu\nu}\sigma^{\mu\nu}\psi(x).
\end{align}
The total variation of the spinor field is, 
\begin{align}
\delta \psi(x^{\prime})&=\psi^{\prime}(x^{\prime})- \psi(x^{\prime})=-\frac{i}{4}w_{\mu\nu}\sigma^{\mu\nu}\psi(x)-w^{\mu}_{~\nu}x^{\nu}\partial_{\mu}\psi(x).
\end{align}
As in the case of vector fields, we construct the conserved Noether current associated with Lorentz symmetry. The total angular momentum tensor is given by:
\begin{align}
J^{\lambda\mu\nu}=L^{\lambda\mu\nu}+S^{\lambda\mu\nu}
\end{align}
where
\begin{align}\label{LDIRAC}
L^{\lambda\mu\nu}&=x^{\mu}\left(\frac{\partial \mathcal{L}}{\partial(\partial_{\lambda}\psi)}\partial^{\nu}\psi+\partial^{\nu}\Bar{\psi}\frac{\partial \mathcal{L}}{\partial(\partial_{\lambda}\Bar{\psi})}-g^{\lambda\nu}\mathcal{L}\right)\nonumber\\
&~~~-x^{\nu}\left(\frac{\partial \mathcal{L}}{\partial(\partial_{\lambda}\psi)}\partial^{\mu}\psi+\partial^{\mu}\Bar{\psi}\frac{\partial \mathcal{L}}{\partial(\partial_{\lambda}\Bar{\psi})}-g^{\lambda\mu}\mathcal{L}\right),
\end{align}
and 
\begin{align}
S^{\lambda\mu\nu}=-\frac{i}{2}\frac{\partial\mathcal{L}}{\partial(\partial_{\lambda}\psi)}\sigma^{\mu\nu}\psi+\frac{i}{2}\Bar{\psi}\sigma^{\mu\nu}\frac{\partial\mathcal{L}}{\partial(\partial_{\lambda}\Bar{\psi})},
\end{align}
are the orbital and spin angular momentum tensors, respectively.

Substituting the explicit form of the Dirac Lagrangian from Eq.~\eqref{Diraclagrangian}, we find: 
\begin{align}
L^{\lambda\mu\nu}=x^{\mu}\left(\frac{i}{2}\Bar{\psi}\gamma^{\lambda}\partial^{\nu}\psi-\frac{i}{2}\partial^{\nu}\Bar{\psi}\gamma^{\lambda}\psi\right)-x^{\nu}\left(\frac{i}{2}\Bar{\psi}\gamma^{\lambda}\partial^{\mu}\psi-\frac{i}{2}\partial^{\mu}\Bar{\psi}\gamma^{\lambda}\psi\right),
\end{align}
and 
\begin{align}\label{Spin tensor Dirac}
S^{\lambda\mu\nu}&=\frac{1}{4}\Bar{\psi}\{\gamma^{\lambda},\sigma^{\mu\nu}\}\psi.
\end{align}
Using the identity in Eq.~\eqref{gamma5prop}, we find that $S^{\lambda\mu\nu}$ is totally antisymmetric and takes the form,
\begin{align}\label{Totalantisymmspin}
S^{\lambda\mu\nu}=-\frac{1}{2}\epsilon^{\lambda\mu\nu\sigma}\Bar{\psi}\gamma_{\sigma}\gamma^{5}\psi.
\end{align}
The total, orbital, and spin angular momentum tensors derived above are commonly referred to as the canonical tensors for Dirac field~\cite{Florkowski:2018fap,Speranza:2020ilk}.
\section{$\mathrm{U}(1)$ global transformation: spinor field}
\label{U1globalfermion}
%
%
When examining transformations in field theory, it is essential to distinguish between spacetime coordinate transformations and internal transformations. Under spacetime transformations, the coordinates \(x^{\mu}\) themselves change. However, for internal transformations -- such as those involving global $\mathrm{U}(1)$ symmetry -- the coordinates remain fixed, and only fields transform. 

Consider a free Dirac spinor field $\psi(x)$ described by the Lagrangian density in Eq.~\eqref{Diraclagrangian}. The global $\mathrm{U}(1)$ transformation acts on the fields as follows
\begin{align}
\psi(x) \longrightarrow \psi^{\prime}(x)&=e^{-i\alpha}\,\psi(x)=\left(1-i\alpha+\mathcal{O}(\alpha^{2})\right)\psi(x)=\psi(x)-i\alpha\psi(x),
\end{align}
and 
\begin{align}
\Bar{\psi}(x) \longrightarrow \Bar{\psi}^{\prime}(x)=e^{i\alpha}\,\psi(x)=\Bar{\psi}(x)+i\alpha\Bar{\psi}(x),
\end{align}
where $e^{-i\alpha}$ and $e^{i\alpha}$ are elements of the unitary group $\mathrm{U}(1)$ and $\alpha$ is a constant. 

\noindent Consequently, variation of the Lagrangian density evaluated on the transformed fields becomes
\begin{align}
\delta\mathcal{L}&=\mathcal{L}\left(\psi^{\prime}(x), \partial\psi^{\prime}(x),\Bar{\psi}^{\prime}(x),\partial\Bar{\psi}^{\prime}(x)\right)-\mathcal{L}\left(\psi(x),\partial\psi(x),\Bar{\psi}(x),\partial\Bar{\psi}(x)\right)\nonumber\\
&= - i\alpha\psi \frac{\partial \mathcal{L}}{\partial \psi} - \partial_{\mu}(i\alpha \psi) \frac{\partial \mathcal{L}}{\partial (\partial_{\mu} \psi)}+ i\alpha\Bar{\psi} \frac{\partial \mathcal{L}}{\partial \Bar{\psi} } + \partial_{\mu}(i\alpha \Bar{\psi} ) \frac{\partial \mathcal{L}}{\partial (\partial_{\mu} \Bar{\psi} )}.
\end{align}
As with previous symmetry arguments, if \( \psi(x) \) and \( \Bar{\psi}(x) \) satisfy the Euler-Lagrange equations (i.e., if the action is invariant under this infinitesimal global $\mathrm{U}(1)$ transformation), we obtain
\begin{align}
\int \dd^{4}x\, \alpha\, \partial_{\mu}\left[-\frac{\partial \mathcal{L}}{\partial (\partial_{\mu} \psi)}i\psi + i\Bar{\psi} \frac{\partial \mathcal{L}}{\partial (\partial_{\mu} \Bar{\psi})}\right]=0.
\end{align}
From this, we identify the charge current (the integrand in square brackets) as
\begin{align}
N^{\mu} =  -\frac{\partial \mathcal{L}}{\partial (\partial_{\mu} \psi)}i\psi+i\Bar{\psi} \frac{\partial \mathcal{L}}{\partial (\partial_{{\mu}} \Bar{\psi})}.
\end{align}
This current satisfies the local conservation law
\begin{align}
\partial_{\mu} N^{\mu} = 0.
\end{align}
Specializing to the Dirac Lagrangian~\eqref{Diraclagrangian}, the variation yields the well-known result
\begin{align}\label{Diracparticlenumber}
N^{\mu}=\Bar{\psi}\gamma^{\mu}\psi.
\end{align}
This current is usually referred to as the \emph{particle number current}, or in this context, the \emph{fermion number current} for a spin-half Dirac field. 

The conserved charge associated with the particle number current can be obtained by integration over a general spacelike hypersurface with specific boundary conditions (see Sec.~\eqref{translations})
\begin{align}\label{}
N = \int_{\Sigma} \dd \Sigma_{\mu} N^{\mu}.
\end{align}
If we select the hypersurface to be a constant time slice in Minkowski spacetime, we get
\begin{align}\label{}
N= \int \dd^{3}x \, N^{0}=\int \dd^{3}x\,\bar{\psi}\gamma^{0}\psi=\int \dd^{3}x\, \psi^{\dagger}\psi,
\end{align}
such that 
\begin{align}
\frac{\dd N}{\dd t}=0.
\end{align}
In quantum field theory, the fermion field \( \psi(x) \) can be expanded in terms of creation and annihilation operators for particles and antiparticles. Consequently, \( N^{0} \) is found to be equal to the number of particles minus the number of antiparticles. Thus, \( N^{0} \) corresponds to the net fermion number density, and the conserved charge \( N \) denotes the \textit{net fermion number}. By ``net'' we refer to the difference between the number of particles and antiparticles in the system. For notational simplicity in later chapters, we omit the word ``net''.

\section{Relativistic spin hydrodynamics}
\label{Relativisticspinhydrodynamics}
In the preceding sections, we established that the presence or absence of spin degrees of freedom determines whether the energy–momentum tensor is asymmetric and whether a separate spin tensor must appear in the total angular momentum. Only the total angular momentum tensor is strictly conserved, and this principle carries over to the hydrodynamic regime through thermal expectation values of the underlying field operators. 

Concretely, the total angular momentum tensor obeys the local conservation law
\begin{align}
\partial_{\lambda} J^{\lambda\mu\nu}(x) = 0,
\end{align}
where
\begin{align} 
J^{\lambda\mu\nu}&=L^{\lambda\mu\nu}+S^{\lambda\mu\nu}
\end{align}
is composed of orbital, $L^{\lambda\mu\nu}$, and spin, $S^{\lambda\mu\nu}$, contributions. The orbital part may be written as 
\begin{align}
L^{\lambda\mu\nu}=x^{\mu}T^{\lambda\nu}-x^{\nu}T^{\lambda\mu},
\end{align}
which implies 
\begin{align}
\partial_{\lambda}J^{\lambda\mu\nu}&=\partial_{\lambda}L^{\lambda\mu\nu}+\partial_{\lambda}S^{\lambda\mu\nu}\nonumber\\
&=\partial_{\lambda}\left(x^{\mu}T^{\lambda\nu}-x^{\nu}T^{\lambda\mu}\right)+\partial_{\lambda}S^{\lambda\mu\nu}\nonumber\\
&=T^{\mu\nu}-T^{\nu\mu}+\partial_{\lambda}S^{\lambda\mu\nu}.
\end{align}
Hence, the spin tensor is not independently conserved; instead, it satisfies the \emph{spin continuity equation},
\begin{align}
\partial_{\lambda}S^{\lambda\mu\nu}=T^{\nu\mu}-T^{\mu\nu},
\end{align}
where \(T^{\mu\nu}\) is, in general, asymmetric. This illustrates the phenomenon of so-called ``spin-orbit'' interaction, in which angular momentum can transfer between \(L^{\lambda\mu\nu}\) and \(S^{\lambda\mu\nu}\), while the total \(J^{\lambda\mu\nu}\) remains conserved. In the free-fermion case, the spin tensor is totally antisymmetric, as shown in Eq.~\eqref{Totalantisymmspin}, while for vector field it is antisymmetric only in the last two indices.

Meanwhile, the energy-momentum tensor is always locally conserved,
\begin{align}
&\partial_{\mu}{T}^{\mu\nu}(x) = 0, \label{EMTspinhydro}
\end{align}
and each additional global symmetry (such as a $\textrm{U}(1)$ symmetry) yields a conserved current
\begin{align}
\partial_{\mu}{N}^{\mu}(x)=0.\label{Nspinhydro}
\end{align}
%
%


Using Eqs.~\eqref{TQF}-\eqref{TTQFT} to relate field-theoretic operators to their thermal expectation values, one arrives at the continuum-level conservation laws in \emph{relativistic spin hydrodynamics}. Thus, in essence, relativistic spin hydrodynamics is a large-scale effective description of an ensemble of particles described microscopically by quantum fields. 
%

Each locally conserved quantity has a corresponding globally conserved charge: the total four-momentum, total angular momentum, and net particle (fermion) number:
\begin{align}
&{P}^{\nu}(t) = \int \dd^{3}x \, {T}^{0\nu}~,~~~~\frac{\dd{P}^{\nu}(t)}{\dd t} = 0, \label{observablePspinhydro}\\
&{J}^{\mu\nu}(t) = \int \dd^{3}x \, {J}^{\,0\mu\nu}~,~~~~\frac{\dd{J}^{\mu\nu}(t)}{\dd t} = 0,\label{observableJspinhydro}\\
&{N}(t)=\int \dd^{3}x\, {N}^{\,0}~,~~~~\frac{\dd{N}(t)}{\dd t}=0.\label{observableNspinhydro}
\end{align}
\medskip 

A fermion field can be free (noninteracting) or coupled to other fields, such as the electromagnetic field in quantum electrodynamics (QED), the gluon field in QCD, or the weak gauge boson fields in the Standard Model. While we have not explicitly treated interacting fermions here, the same reasoning applies to strongly interacting systems such as the QGP. Although the QCD Lagrangian is significantly more elaborate than that of the free fermions or the electromagnetic field, the same global Poincaré invariance ensures the same set of local conservation laws and globally conserved charges -- albeit with net baryon number replacing net fermion number. Hence, relativistic spin hydrodynamics remains a valid effective description even for complex systems like the QGP, leaving detailed treatments of interacting fermions to future work. 
\medskip

Finally, \emph{conventional relativistic hydrodynamics} arises as the limiting case of the relativistic spin hydrodynamics in which there is no spin tensor. In that situation, only orbital angular momentum is present, and the energy–momentum tensor becomes symmetric by virtue of the absence of spin. Consequently, in standard relativistic hydrodynamics, one has
\begin{align}
\partial_{\lambda}J^{\lambda\mu\nu}&=\partial_{\lambda}L^{\lambda\mu\nu}=\partial_{\lambda}\left(x^{\mu}T^{\lambda\nu}-x^{\nu}T^{\lambda\mu}\right)=T^{\mu\nu}-T^{\nu\mu}=0.
\end{align}
%
  
\section{Pseudo-gauge transformation}
\label{Pseudo-gauge transformation concept}
In quantum field theory, it is important to emphasize that the energy-momentum $\widehat{T}^{\mu\nu}$ and spin $\widehat{S}^{\lambda\mu\nu}$ tensor operators are not uniquely defined \cite{halbwachs1960,Coleman:2018mew}. 
In general, it is possible to construct alternative pairs \(\{\widehat{T}^{\prime \mu\nu}, \widehat{S}^{\prime \lambda\mu\nu}\}\) via a \emph{pseudo-gauge transformation}~\footnote{The pseudo-gauge transformation can be further generalized by adding the term $\partial_{\alpha}\widehat{Z}^{\alpha\lambda\mu\nu}$ to the right-hand side of Eq.~\eqref{spinpsuedogauge} (where $\widehat{Z}$ satisfies the properties $\widehat{Z}^{\alpha\lambda\mu\nu} = -\widehat{Z}^{\lambda\alpha\mu\nu}$ and $\widehat{Z}^{\alpha\lambda\mu\nu} = -\widehat{Z}^{\alpha\lambda\nu\mu}$)~\cite{Hehl:1976vr,belinfante1939spin,cbelinfante1940current,rosenfeld1940,Speranza:2020ilk}. This is justified as the conditions~\eqref{psuedocon1}-\eqref{phiboundaryconditions2} will also be satisfied.}~\cite{Hehl:1976vr,Becattini:2011ev}, 
\begin{align}
&\widehat{T}^{\,\prime \mu\nu}=\widehat{T}^{\mu\nu}+\frac{1}{2}\partial_{\alpha}\left(\widehat{\Phi}^{\alpha\mu\nu}-\widehat{\Phi}^{\mu\alpha\nu}-\widehat{\Phi}^{\nu\alpha\mu}\right),\label{Tpsuedogauge}\\
&\widehat{S}^{\,\prime \lambda\mu\nu}=\widehat{S}^{\lambda\mu\nu}-\widehat{\Phi}^{\lambda\mu\nu} ,\label{spinpsuedogauge}
\end{align}
such that
\begin{align}\label{Jpsuedogauge}
\widehat{J}^{\, \prime \lambda\mu\nu}=\widehat{J}^{\,\lambda\mu\nu}+\frac{1}{2}\partial_{\alpha}\left[x^{\mu}\left(\widehat{\Phi}^{\alpha\lambda\nu}-\widehat{\Phi}^{\lambda\alpha\nu}-\widehat{\Phi}^{\nu\alpha\lambda}\right)-x^{\nu}\left(\widehat{\Phi}^{\alpha\lambda\mu}-\widehat{\Phi}^{\lambda\alpha\mu}-\widehat{\Phi}^{\mu\alpha\lambda}\right)\right].
\end{align}
Here, \( \widehat{\Phi} \) is an arbitrary differentiable rank-3 tensor operator -- often called a \emph{superpotential} -- antisymmetric in its last two indices, which depends on the underlying field. Such transformations are valid because the transformed pair \( \{\widehat{T}^{\,\prime \mu\nu}, \widehat{S}^{\,\prime \lambda\mu\nu}\} \):  
\begin{enumerate}
\item Satisfies the same continuity equations as \( \{\widehat{T}^{\mu\nu}, \widehat{S}^{\lambda\mu\nu}\} \). Indeed, from Eqs.~\eqref{Tpsuedogauge}-\eqref{Jpsuedogauge}, one can show: 
\begin{align}
&\partial_{\mu}\widehat{T}^{\,\prime\mu\nu}=\partial_{\mu}\widehat{T}^{\mu\nu}=0,\label{psuedocon1}\\
&\partial_{\lambda}\widehat{J}^{\,\prime\lambda\mu\nu}=\partial_{\lambda}\widehat{J}^{\,\lambda\mu\nu}=0.
\end{align}
\item Yields the same total conserved charges as \( \{\widehat{T}^{\mu\nu}, \widehat{S}^{\lambda\mu\nu}\} \), namely the same total four-momentum and total angular momentum. To see why, recall that the total conserved charges are defined as,   
\begin{align}\label{psuedocon2}
&\widehat{P}^{\, \prime \nu}(t) = \int_{V} \dd^{3}x \, \widehat{T}^{\, \prime 0\nu}~,~~\text{so that}~\int_{S} \dd S\, \widehat{T}^{\, \prime i\nu}n_{i}=\int_{V} \dd^{3}x\, \partial_{i}\widehat{T}^{\, \prime i\nu}=0,\\
&\widehat{J}^{\, \prime\mu\nu}(t)=\int_{V} \dd^{3}x\, \widehat{J}^{\, \prime 0\mu\nu}~,~~\text{so that} \int_{S}\dd S\,\widehat{J}^{\, \prime i\mu\nu}n_{i}=\int_{V}\dd^{3}x\, \partial_{i}\widehat{J}^{\,\prime i\mu\nu}=0, 
\end{align}
where $V$ a three-dimensional region bounded by the surface $\partial V=S$ with the surface normal $\boldsymbol{n}$. 

One has $\widehat{P}^{\,\prime\nu}=\widehat{P}^{\,\nu}$ and $\widehat{J}^{\,\prime \mu\nu}=\widehat{J}^{\mu\nu}$ if and only if \( \widehat{\Phi} \) satisfies the following conditions: 
\begin{align}
&\int_{S} \dd S\, \left(\widehat{\Phi}^{\,i0\nu}-\widehat{\Phi}^{\,0i\nu}-\widehat{\Phi}^{\,\nu i0}\right)n_{i}=0,\label{phiboundaryconditions1}\\
&\int_{S} \dd S\, \left[x^{\mu}\left(\widehat{\Phi}^{\,i0\nu}-\widehat{\Phi}^{\,0i\nu}-\widehat{\Phi}^{\,\nu i0}\right)-x^{\nu}\left(\widehat{\Phi}^{\,i0\mu}-\widehat{\Phi}^{\,0i\mu}-\widehat{\Phi}^{\,\mu i0}\right)\right]n_{i}=0.\label{phiboundaryconditions2}
\end{align}
\end{enumerate}
Therefore, a pseudo-gauge transformation of the form in Eqs.~\eqref{Tpsuedogauge}-\eqref{Jpsuedogauge} is always possible provided the boundary conditions~\eqref{phiboundaryconditions1}-\eqref{phiboundaryconditions2} are ensured. For example, starting with the Dirac Lagrangian for free fermions in Eq.~\eqref{Diraclagrangian}, one obtains the canonical tensors \( \{\widehat{T}^{\mu\nu}, \widehat{S}^{\lambda\mu\nu}\} \). Then for a particular choice of \( \widehat{\Phi}^{\lambda\mu\nu} \), one obtains new pairs \( \{\widehat{T}^{\, \prime \mu\nu}, \widehat{S}^{\, \prime \lambda\mu\nu}\} \) associated with different pseudogauges, such as: \emph{Belinfante-Rosenfeld (BR)} pseudogauge~\cite{belinfante1939spin,cbelinfante1940current,rosenfeld1940,Cohen1979}, \emph{Hilgevoord-Wouthuysen (HW)} pseudogauge~\cite{hilgevoord1963spin,hilgevoord1965energy,fradkin1961tensor}, or \emph{de Groot-van Leeuwen-van Weert (GLW)} pseudogauge~\cite{DeGroot:1980dk}. Notably, the BR choice \( \widehat{\Phi}^{\lambda\mu\nu} = \widehat{S}^{\lambda\mu\nu} \), yields a symmetric energy-momentum tensor and a vanishing spin tensor. Further discussion of various forms of \( \widehat{\Phi} \) within different theoretical frameworks appears in Refs.~\cite{Florkowski:2018fap, Speranza:2020ilk}.

The same reasoning applies to the expectation-value version of the pseudo-gauge transformation. Applying Eqs.~\eqref{TQF} to both sides of Eqs.~\eqref{Tpsuedogauge}-\eqref{Jpsuedogauge}, yields
\begin{align}
&{T}^{\prime \mu\nu}={T}^{\mu\nu}+\frac{1}{2}\partial_{\alpha}\left({\Phi}^{\alpha\mu\nu}-
{\Phi}^{\mu\alpha\nu}-{\Phi}^{\nu\alpha\mu}\right),\label{Tpseudogaugeaveragevalues}\\
&{S}^{\prime \lambda\mu\nu}={S}^{\lambda\mu\nu}-{\Phi}^{\lambda\mu\nu},\label{Spseudogaugeaveragevalues}
\end{align}
and
\begin{align}
    {J}^{\prime \lambda\mu\nu}={J}^{\lambda\mu\nu}+\frac{1}{2}\partial_{\alpha}\left[x^{\mu}({\Phi}^{\alpha\lambda\nu}-{\Phi}^{\lambda\alpha\nu}-{\Phi}^{\nu\alpha\lambda})-x^{\nu}({\Phi}^{\alpha\lambda\mu}-{\Phi}^{\lambda\alpha\mu}-{\Phi}^{\mu\alpha\lambda})\right].
\end{align}
Thus, relativistic spin hydrodynamics is also subject to pseudo-gauge transformations, meaning that the energy-momentum and the spin tensors are not uniquely defined and can be changed to a new couple of tensors fulfilling the same dynamical equations and providing the same integrated conserved charges. The fundamental role of such a property is under intensive investigation in recent literature~\cite{Speranza:2020ilk,Buzzegoli:2024mra}, extending beyond the scope of this thesis.  
\medskip

Given the material presented in this chapter, we now possess the fundamental tools—the conservation laws—that enable us to further develop the theory of relativistic spin hydrodynamics in Chapters~\ref{Navier-Stokes limit}, \ref{MISMIS}, and \ref{Quantum-statistical formulation}.
\chapter{Relativistic spin hydrodynamics: Navier-Stokes limit}
\label{Navier-Stokes limit}
In this chapter we formulate relativistic spin hydrodynamics in the Navier–Stokes (first-order) limit of the gradient expansion and derive the associated evolution equations. For simplicity, we confine the discussion to vanishing chemical potential.

Section~\ref{Thermodynamic relations} presents the covariant thermodynamic relations for a relativistic fluid with spin and explains why we do not begin from the ideal-fluid limit. By employing the conservation laws within the entropy-current analysis, we determine the dissipative currents and transport coefficients that feed into the evolution equations in Sections~\ref{eevolutionequationss}-\ref{Dissipative currents and transport coefficients : entropy-current analysis}. Next, we test the resulting equations through a linear-stability analysis in a specified background configuration (Section~\ref{stability}) and solve them for Bjorken flow in Section~\ref{boostinvariant}.  Throughout we adopt a phenomenological formulation, in which the spin tensor is antisymmetric in its last two indices. For comparison, Section~\ref{Canonical framework} develops a canonical formulation with a totally antisymmetric spin tensor and demonstrates that, under suitable conditions, the two approaches are equivalent. Both formulations are firmly grounded in field theory, as reviewed in Chapter~\ref{From fields to fluids} and revisited qualitatively in Section~\ref{Canonical framework}. Finally, in Section~\ref{Summary and outlookNS}, we summarize our results, assess their limitations, and motivate developments pursued in the next chapter. 

The derivation of the dissipative currents follows Ref.~\cite{Hattori:2019lfp}, whereas the remainder of the chapter is based on our work reported in Refs.~\multimyref{AD1,AD2,AD3,AD6}; additional references are cited where appropriate.
%

\section{Covariant thermodynamics}
\label{Thermodynamic relations}
\subsection{Conventional relativistic fluid}
In thermodynamic equilibrium and the absence of external forces, a conventional relativistic fluid at rest is characterized by a set of macroscopic quantities known as \emph{thermodynamic variables}. These variables can be categorized into two types: \emph{extensive} and \emph{intensive}. \emph{Extensive} variables scale with system size, meaning their magnitudes are additive for subsystems. These include \emph{internal energy} $U$, \emph{volume} $V$, \emph{entropy} $S$, and \emph{particle number} $N$. In contrast, \emph{intensive} variables are independent of system size. Among them are \emph{temperature} $T$, \emph{pressure} $P$, and \emph{chemical potential} $\mu$.
\noindent 
%
%

The thermodynamic variables obey the \emph{first law of thermodynamics}:
\begin{align}\label{firstlaw}
\dd{U} = {T} \dd{S} - {P} \dd{V} + \mu\,\dd{N},
\end{align}
which represents a change in the system's internal energy due to heat transfer, mechanical work, and particle exchange~\footnote{Here, \( \dd N \) represents the change in the number of particles, and \( \mu \) denotes the corresponding chemical potential. If the system consists of multiple particle species, the term \( \mu\,\dd N \) generalizes to the summation \( \sum_{i} \mu_{i}\,\dd N_{i} \) for $i=1,....,N_{\rm species}$. }.

In general, a function is said to be \emph{homogeneous of degree \( k > 0 \)} if scaling all inputs by a factor \( \lambda \) results in the output being scaled by \( \lambda^k \)~\cite{loomis1968advanced}, i.e., 
\begin{align*}
f(\lambda x_{1}, \ldots, \lambda x_{n}) = \lambda^{k} f(x_{1}, \ldots, x_{n}).
\end{align*}
Since the internal energy ${U}$ is a state function of extensive variables, as seen from Eq.~\eqref{firstlaw}, i.e,\, ${U}({S}, {V}, {N})$, it follows that ${U}$ is a homogeneous function of degree one (\(k=1\)). Furthermore, \emph{Euler theorem on homogeneous functions} states that if $f(x_{1}, \ldots, x_{n})$ is a homogeneous function of degree \( k \), then it satisfies the following relation~\cite{loomis1968advanced}:
\begin{align*}
k f(x_{1}, \ldots, x_{n}) = \sum_{i=1}^{n} x_{i} \frac{\partial f(x_{1}, \ldots, x_{n})}{\partial x_{i}}.
\end{align*}
Hence, by applying Euler theorem to the internal energy ${U}({S}, {V}, {N})$ and assuming it being a differentiable function, we obtain:
\begin{align}
{U}({S}, {V}, {N}) &= {S} \bigg(\frac{\partial {U}}{\partial {S}} \bigg)_{{V},\, {N}} + {V} \bigg(\frac{\partial {U}}{\partial{V}} \bigg)_{{S},\, {N}} + {N} \bigg(\frac{\partial {U}}{\partial {N}} \bigg)_{{S},\, {V}},\label{Eulerformula0}
\end{align}
which, recognizing that the partial derivatives are just the definitions of the intensive variables, simplifies to:
\begin{align}
{U} &=  {T}{S} -{P}{V} + \mu {N}.\label{Eulerformula}
\end{align}
This result is commonly referred to as the \emph{Euler thermodynamic equation}.

By taking the total differential of Eq.~\eqref{Eulerformula} and employing the first law of thermodynamics~\eqref{firstlaw}, we obtain the \emph{Gibbs–Duhem equation}:
\begin{align}\label{Gibbs–Duhem}
{V}\dd{P}={S}\dd{T}+{N}\dd\mu.
\end{align}

A conventional relativistic fluid is commonly conceptualized as a collection of \emph{fluid elements}, each of which represents a thermodynamic subsystem of the whole system. These elements are large enough to contain a vast number of particles yet remain small compared to the macroscopic scale of the fluid. Each fluid element can independently attain equilibrium and is thus characterized by a set of thermodynamic variables that vary gradually between neighboring elements. In this state, the fluid is said to be in \emph{local equilibrium}. When the entire fluid is in equilibrium state described by a single set of thermodynamic variables, the system is said to be in \textit{global equilibrium}. Consequently, it is more practical to express the first law of thermodynamics, as well as the Euler and Gibbs-Duhem equations describing a fluid element, in in terms of densities:
\begin{align}
\dd\varepsilon=T\dd s+\mu \dd n,\label{firstlawdensity}\\
\varepsilon+P=Ts+\mu n,\label{Eulerformuladensity}\\
 \dd P=s\dd T+n \dd \mu,\label{Gibbs-Durhamdensity}
\end{align}
where \( \varepsilon = \frac{{U}}{{V}} \) is the \emph{energy density}, \( n = \frac{{N}}{{V}} \) is the \emph{particle number density}, and \( s = \frac{{S}}{{V}} \) is the \emph{entropy density}. 

To account for all thermodynamic variables in Eq.~\eqref{Eulerformuladensity}, we first note that temperature and chemical potential are independent variables. Consequently, based on Eqs.~\eqref{firstlawdensity}-\eqref{Gibbs-Durhamdensity}, we may express 
\begin{align}
\varepsilon = \varepsilon(T,\mu)~,~~n = n(T,\mu).
\end{align}
such that the entropy density \(s = s(\varepsilon,n)\). The pressure is then expressed as a function of the energy density $\varepsilon$ and number density $n$, and represents an \emph{equation of state},
\begin{align}\label{eosnospin}
P = P(\varepsilon,n),
\end{align}
 which entirely defines the material properties of the fluid in equilibrium.
 
In this description, we transition from a set of thermodynamic variables assigned to each fluid element to a set of thermodynamic \emph{fields} that continuously vary with spacetime. Note that in the relativistic context, as can be inferred from Chapter~\ref{From fields to fluids}, the particle number $N$ and the corresponding density $n$ are \emph{net} quantities, meaning they represent the difference between particles and antiparticles of a single species. To avoid confusion, throughout this thesis, we will refer to \(N\) as the particle number and \(n\) as the particle number density unless otherwise stated.
\medskip 

One of the defining properties of fluids is their inability to resist shear forces, which causes them to continuously deform rather than to keep a fixed shape. In an incompressible fluid, this process occurs while preserving the total volume. As a result of these forces, the fluid elements acquire flow velocities that vary continuously over spacetime. A thermodynamic description of these fluid elements applies in their local rest frames, which are defined with respect to a fixed inertial observer by their local four-velocities. 

When dealing with \emph{relativistic} fluids, we have to account for the fact that in special relativity, the laws of physics must take the same form in all inertial frames of reference. However, the derivation of the thermodynamic relations in Eqs.~\eqref{firstlawdensity}-\eqref{Gibbs-Durhamdensity} assumes a fixed frame of reference. To ensure their validity in all inertial frames, we implement the following principles:
\begin{enumerate}
\item \textbf{Conserved currents and the rest frame.} The only \emph{covariant} quantities available to describe a conventional relativistic fluid are, as discussed in Chapter~\ref{From fields to fluids}, the energy-momentum tensor $T^{\mu\nu}$ and the particle number current $N^{\mu}$. The covariant form of the thermodynamic relations must ultimately be expressed in terms of these currents. To achieve this, we first decompose these currents in terms of thermodynamic fields. In equilibrium and in the rest frame of a fluid element, these currents must reduce to their corresponding thermodynamic fields.

\item \textbf{Four-velocity $u^{\mu}$ and the projector $\Delta^{\mu\nu}$.} To decompose the conserved currents in terms of thermodynamic fields, we use two fundamental mathematical objects: the four-velocity of the fluid element $u^{\mu}$ and the tensor $\Delta^{\mu\nu}$,
\begin{align}
&\Delta^{\mu\nu}=g^{\mu\nu}-u^{\mu}u^{\nu},
\end{align}
which projects vectors and tensors onto the space orthogonal to $u^{\mu}$. Hence by construction, the projector $\Delta^{\mu\nu}$ is orthogonal to $u^{\mu}$, namely $\Delta^{\mu\nu}u_{\nu}=0$. The four-velocity of a fluid element is given by:
\begin{align}
u^{\mu} = \gamma(1, \boldsymbol{v}),
\end{align}
where \(\gamma = (1 - v^2)^{-1/2}\) is the Lorentz factor, \(\boldsymbol{v}\) is the three-velocity of the fluid element, and \( v = |\boldsymbol{v}| \) is the magnitude of the three-velocity. The four-velocity satisfies the normalization condition:
\begin{align}
u^{\mu} u_{\mu} = 1.
\end{align}
%

In the rest frame (RF) of a fluid element, these expressions reduce to:
\begin{align}
&u^{\mu}_{\text{RF}} = (1, 0, 0, 0),\\
&\Delta^{\mu\nu}_{\text{RF}}=\text{diag}(0,-1,-1,-1).
\end{align}
\item \textbf{Covariance of tensorial forms.} Any tensorial form that is constructed in the rest frame must be covariant, ensuring its validity in any other inertial frame, as required by the principle of relativity. 
\end{enumerate}
Using points (1) and (2) from the above discussion, we can decompose the particle number current in equilibrium as: 
\begin{align}
N^{\mu}_{\text{RF}} &= n u^{\mu}_{\text{RF}}.\label{Nideal}
\end{align}
Similarly, we introduce the \emph{entropy current} and the \emph{thermal velocity current}:
\begin{align}
S^{\mu}_{\text{RF}} &= s u^{\mu}_{\text{RF}},\label{sidela}\\
\beta^{\mu}_{\text{RF}}&= \beta u^{\mu}_{\text{RF}}=\frac{u^{\mu}_{\text{RF}}}{T}\label{betaideal}.
\end{align}
Finally, for the energy density $\varepsilon$ and isotropic pressure $P$, we decompose the energy-momentum tensor in equilibrium as:
\begin{align}
T^{\mu\nu}_{\text{RF}} = \varepsilon u^{\mu}_{\text{RF}}u^{\nu}_{\text{RF}} - P\Delta^{\mu\nu}_{\text{RF}} =
\begin{pmatrix}
\varepsilon & 0 & 0 & 0 \\
0 & P & 0 & 0 \\
0 & 0 & P & 0 \\
0 & 0 & 0 & P
\end{pmatrix},
\end{align}
where $T^{00}=\varepsilon$ and $T^{ii}=P$. For a detailed derivation of this decomposition, we refer to Appendix~\ref{Appendix A}, which provides a more general form that applies beyond equilibrium.  

Using point (3), we emphasize that these decompositions hold in all inertial frames, allowing us to write:
\begin{align}
&N^{\mu}= n u^{\mu},\\
&S^{\mu}= s u^{\mu},\\
&\beta^{\mu}=\frac{u^{\mu}}{T},\\
&T^{\mu\nu} = \varepsilon u^\mu u^\nu - P \Delta^{\mu\nu},
\label{EMTandNdecompositions}
\end{align}
which are obtained from the previous forms by applying the Lorentz transformation. Using the decompositions presented above one may obtain the covariant form of the thermodynamic relations~\eqref{firstlawdensity}-\eqref{Gibbs-Durhamdensity} in the form: 
\begin{align}
    &\dd S^{\mu} = \beta_{\nu} \dd T^{\mu\nu} - \mu\beta \dd N^{\mu}, \label{covariantfirstlaw}\\
    &T^{\mu\nu} \beta_{\nu} + P \beta^{\mu} = S^{\mu} + \mu\beta N^{\mu}, \label{covarintEuler}\\
    &\dd (\beta^{\mu} P) = -T^{\mu\nu} \dd \beta_{\nu} + N^{\mu} \dd \left(\beta{\mu}\right).\label{CovariantGibbsDuhem}
\end{align}
One can verify, that in rest frame the above formulas reduce to their scalar counterparts~\eqref{firstlawdensity}-\eqref{Gibbs-Durhamdensity}.
\medskip 

Finally, we anticipate that in equilibrium, entropy is not produced, meaning that the divergence of the entropy current must vanish: 
\begin{align}\label{entropyconservationequilibriumnospin}
\partial_{\mu}S^{\mu} = \beta_{\nu} \partial_{\mu}T^{\mu\nu} - \mu\beta \partial_{\mu}N^{\mu}=0.
\end{align}
§This follows directly from the local conservation of the energy-momentum tensor \( T^{\mu\nu} \) and the particle number current \( N^{\mu} \).
%
\subsection{Relativistic fluid with spin}
\label{Relativistic fluid with spinnn}
In the case of a relativistic \emph{fluid with spin}, any change in the spin degrees of freedom is expected to contribute to the change in internal energy within a fluid element \cite{landau1884electrodynamics,de2013non}. Therefore, a new current must be introduced to account for the spin contribution in the covariant thermodynamic relations. Indeed, in Chapter~\ref{From fields to fluids}, we demonstrated that a relativistic fluid with spin satisfies the conservation of total angular momentum, in addition to the conservation of energy-momentum and particle number currents. Furthermore, we established that the total angular momentum tensor, $J^{\lambda\mu\nu}$ (which is antisymmetric in its last two indices) consists of both orbital and spin contributions:
\begin{align}
J^{\lambda\mu\nu}=L^{\lambda\mu\nu}+S^{\lambda\mu\nu}.
\end{align}
To incorporate the spin tensor $S^{\lambda\mu\nu}$ into the covariant thermodynamic relations~\eqref{covariantfirstlaw}-\eqref{CovariantGibbsDuhem}, we introduce a new (intensive) thermodynamic variable, the \emph{spin potential}~\footnote{The term \emph{spin potential} is used interchangeably with \emph{spin chemical potential} in the literature of spin hydrodynamics. In this thesis, we adopt the term \emph{spin potential} for simplicity of notation.}
, which satisfies the antisymmetric condition:
$$\omega_{\mu\nu} = -\omega_{\nu\mu}.$$
The spin potential is conjugate to the spin tensor \(S^{\lambda\mu\nu}\), in analogy with the chemical potential \(\mu\) which is conjugate to the particle number \(N^{\mu}\). 

Consequently,
the covariant form of the thermodynamic relations, now including the spin tensor, is given by:
\begin{align}
    &\dd S^{\mu}=\beta_{\nu}\dd T^{\mu\nu}-\mu\beta \dd N^{\mu}-\beta\omega_{\lambda\nu}\dd S^{\mu\lambda\nu},\label{covariantgeneralizedfirstlaw}\\
    &T^{\mu\nu}\beta_{\nu}+P\beta^{\mu}=S^{\mu}+\mu\beta N^{\mu}+\beta \omega_{\lambda\nu}S^{\mu\lambda\nu},\label{covariantgeneralizedEulerequation}\\
    &\dd (\beta^{\mu}P)=-T^{\mu\nu}\dd \beta_{\nu}+N^{\mu}\dd (\beta\mu)+S^{\mu\lambda\nu}\dd (\beta\omega_{\lambda\nu}).\label{covariantgeneralizedGibbsDuhem}
\end{align}
It is essential to emphasize that the incorporation of the spin tensor in the above equations is based on a heuristic arguments. In the absence of a rigorous derivation from field theory or kinetic theory, the validity of this formulation remains an open question.
\medskip

Similar to the case of a relativistic fluid without spin tensor, we expect that entropy is not produced in equilibrium. However, using Eq.~\eqref{covariantgeneralizedfirstlaw}, we observe that:  
\begin{align}\label{noidealspin}
\partial_{\mu}S^{\mu} = \beta_{\nu}\partial_{\mu}T^{\mu\nu} - \mu\beta \partial_{\mu}N^{\mu} - \beta\omega_{\mu\nu}\partial_{\lambda}S^{\lambda\mu\nu}\neq 0,
\end{align}
as, strictly speaking, the spin tensor, derived from field theory calculations in  Chapter~\ref{From fields to fluids}, regardless of the equilibrium condition, is not conserved separately:
\begin{align}
\partial_{\lambda}S^{\lambda\mu\nu}=T^{\nu\mu} - T^{\mu\nu}.
\end{align}
Consequently, the spin tensor contributes to the system in a dissipative manner. Therefore, in this thesis, when referring to equilibrium in spin hydrodynamics, we consider it only in the context of the leading-order term in the hydrodynamic gradient expansion. 
\medskip

Following the decomposition rules outlined in the previous subsection, we may express the spin tensor as: 
\begin{align}\label{WyRaa}
S^{\lambda\mu\nu}=u^{\lambda}S^{\mu\nu}+S^{\lambda\mu\nu}_{(1)}.
\end{align}
Here, we introduce a new variable called \emph{spin density} $S^{\mu\nu}$, which can be traced back to the seminal works of Weyssenhoff and Raabe~\cite{Weyssenhoff:1947iua}~\footnote{In the following, we adopt the symbol \( S^{\mu\nu} \) to represent both the spin density and the spin angular momentum. To prevent any potential confusion, the specific meaning of the symbol will be explicitly stated in the context prior to its use.
}. Moreover, we introduced the term $S^{\mu\lambda\nu}_{(1)}$ which encodes the spin dissipation and satisfies the orthogonality condition 
$$u_{\lambda}S^{\lambda\mu\nu}=0.$$
While one could argue that the spin tensor may contain additional terms, in here we treat this specific decomposition as an ansatz.

Taking into account the form of the spin density in Eq.~\eqref{WyRaa}, in the rest frame of the fluid element the covariant thermodynamic relations~\eqref{covariantgeneralizedfirstlaw}-\eqref{covariantgeneralizedGibbsDuhem} reduce to~\footnote{Analogous contribution to $\omega_{\mu\nu}S^{\mu\nu}$ in Eq.~\eqref{generalizedEulerequation}, albeit proportional to a single (scalar) combination of spin potential components, $\sim \zeta w\sim \sqrt{\omega^{\mu\nu}\omega_{\mu\nu}}\, w$ where $w$ is a spin density, was derived within kinetic theory framework for spin-half particles in Ref.~\cite{Florkowski:2017ruc}}~\cite{Hattori:2019lfp}: 
\begin{align}
&\dd \varepsilon=T\dd s+\mu \dd n+\omega_{\mu\nu}\dd S^{\mu\nu}\label{generalizedfirstlaw},\\
&\varepsilon+P=Ts+\mu n+\omega_{\mu\nu}S^{\mu\nu}\label{generalizedEulerequation},\\    
&\dd P=s\dd T+n\dd \mu+S^{\mu\nu}\dd \omega_{\mu\nu}\label{generalizedGibbsDuhem}.
\end{align}
Notably, the term \(S^{\lambda\mu\nu}_{(1)}\) does not appear in the rest frame expressions above, as it is of higher order in the hydrodynamic gradient expansion relative to the other variables. The topic of hydrodynamic gradient ordering will be discussed in Sec.~\ref{Dissipative currents and transport coefficients : entropy-current analysis}. 

Similarly to the conventional fluids, to account for the full set of thermodynamic variables in Eq.~\eqref{generalizedEulerequation}, we note that apart from temperature \( T \) and chemical potential \( \mu \), the independent variables now include also spin potential \( \omega^{\mu\nu} \). Consequently, we can express
\begin{align}\label{EOSS}
\varepsilon=\varepsilon\,(T,\mu,\omega^{\mu\nu})~,~~n=n\,(T,\mu,\omega^{\mu\nu})~,~~S^{\mu\nu}=S^{\mu\nu}\,(T,\mu,\omega^{\mu\nu}),
\end{align}
such that the entropy density reads $s = s\,(\varepsilon,n, S^{\mu\nu})$. The pressure is then expressed as a function of the energy, particle number, and spin densities, serving as an equation of state,  
\begin{align}
P = P \,(\varepsilon,n, S^{\mu\nu}).
\end{align}
We note that in the following sections the function \( S^{\mu\nu} = S^{\mu\nu}(T, \mu, \omega^{\mu\nu}) \) will be referred to as the \emph{spin equation of state}.
%
\section{Conservation equations}
\label{eevolutionequationss}
%
%
\subsection{Conventional relativistic fluid}
\label{Relativistic fluid}
In Sec.~\ref{Thermodynamic relations}, we found that the state of a fluid element in local equilibrium is characterized by the energy-momentum tensor \( T^{\mu\nu} \) and the particle number current \( N^{\mu} \), which have the following forms:
\begin{align}
&T^{\mu\nu} = \varepsilon u^{\mu} u^{\nu} - P \Delta^{\mu\nu},\label{Tequilibrium}\\
&N^{\mu} = n u^{\mu}. \label{Nequilibrium} 
\end{align}
Together, they completely encode, in a covariant way, the state of each fluid's cell.  
\begin{table}[H]
\renewcommand{\arraystretch}{2}
\begin{tabularx}{1\textwidth} { 
| >{\raggedright\arraybackslash}X 
| >{\raggedright\arraybackslash}X
| >{\raggedright\arraybackslash}X
| >{\raggedright\arraybackslash}X
| >{\centering\arraybackslash}X 
| >{\raggedleft\arraybackslash}X | }
\hline
\textbf{Dynamical variables} &\textbf{Constraints} &\textbf{Degrees of freedom}\\
\hline
Energy density: $\varepsilon(x)$ & 
&1\\
\hline
Four-velocity: $u^{\mu}(x)$ &$u_{\mu}u^{\mu}=1$& 3\\
\hline
Number density: $n(x)$ & 
&1 \\
\hline
\end{tabularx}
\caption{Primary dynamical variables in conventional relativistic hydrodynamics of perfect fluid.}
\label{Dynamical variables in relativistic hydrodynamics}
\end{table}
These currents, as demonstrated in Chapter~\ref{From fields to fluids}, are locally conserved, namely, they satisfy the following equations
\begin{align}
\label{lcl1}
&\partial_{\mu}T^{\mu\nu}=0,\\
\label{lcl2}
&\partial_{\mu}N^{\mu}=0.
\end{align}
By substituting expressions for the currents, Eqs.~\eqref{Tequilibrium} and \eqref{Nequilibrium}, into the local conservation laws~\eqref{lcl1} and \eqref{lcl2}, one obtains a closed set of five evolution equations for five \emph{primary} dynamical variables: energy density, particle number density, and four-velocity, see Table~\eqref{Dynamical variables in relativistic hydrodynamics}. If supplemented with proper initial conditions, they can be used to determine the dynamics of these variables in spacetime. 

By projecting the resulting conservation equations along and orthogonal to the four-velocity, one can obtain their equivalent forms, also known as the \emph{relativistic Euler equations} (see Table~\ref{Relativistic evolution equations in local equilibrium}).
\begin{table}[H]
\renewcommand{\arraystretch}{2}
\begin{tabularx}{1\textwidth} { 
| >{\raggedright\arraybackslash}X 
| >{\raggedright\arraybackslash}X
| >{\raggedright\arraybackslash}X
| >{\raggedright\arraybackslash}X
| >{\centering\arraybackslash}X 
| >{\raggedleft\arraybackslash}X | }
\hline
\textbf{Local conservation laws} & \textbf{Evolution equations} & \textbf{Number of equations}\\
\hline
{Conservation} of energy-momentum tensor projected along flow velocity $u_{\nu}\partial_{\mu}T^{\mu\nu}=0$ &  $D\varepsilon+(\varepsilon+P)\theta=0$& 1\\
\hline
Conservation of energy-momentum tensor projected orthogonal to the flow velocity $\Delta_{\nu}^{\alpha}\partial_{\mu}T^{\mu\nu}=0$& $(\varepsilon+P)D u^{\alpha}-\nabla^{\alpha}P=0$& 3\\
\hline
Conservation of particle number current  $\partial_{\mu}N^{\mu}=0$ & $Dn+n\theta=0$& 1\\
\hline
\end{tabularx}
\caption{Relativistic Euler evolution equations.}
\label{Relativistic evolution equations in local equilibrium}
\end{table}
\noindent
Here, we define the comoving derivative as \( D = u^{\mu} \partial_{\mu} \) and the spacelike (or transverse) gradient as 
\(\nabla_{\mu} = \Delta_{\mu\alpha} \partial^{\alpha} = \partial_{\mu} - u_{\mu} D\). Obviously, the latter is orthogonal to the fluid's element four-velocity, i.e., \(u_{\mu} \nabla^{\mu} = 0.\)
The expansion scalar is given by \( \theta = \partial_{\mu} u^{\mu} \). The pressure is related to energy density and particle number density through the equation of state \eqref{eosnospin}.
\medskip 

For a more realistic fluid dynamical description of physical systems, it is necessary to incorporate into the formalism the effects of dissipation. For that, we can extend the decompositions of the energy-momentum tensor~\eqref{Tequilibrium} and the particle number current~\eqref{Nequilibrium} by adding terms encoding the dissipation, 
\begin{align}
&T^{\mu\nu} = \varepsilon u^{\mu} u^{\nu} - P \Delta^{\mu\nu}+T^{\mu\nu}_{(1S)},\label{Toutofequilibrium}\\
&N^{\mu} = n u^{\mu}+N_{(1)}^{\mu}, \label{Noutofequilibrium} 
\end{align}
Here, \( N_{(1)}^{\mu} \) represents the particle number dissipation (or \emph{particle diffusion current}) and is defined such that \( u_{\mu}N_{(1)}^{\mu} = 0 \), while \( T_{(1S)}^{\mu\nu} \) encodes the dissipative components of the energy-momentum tensor. In here, both  \( T^{\mu\nu}_{(1S)} \) and \( T^{\mu\nu} \)  remain purely symmetric, as they must be, based on field theory considerations for scalar fields; see Chapter~\ref{From fields to fluids}. The dissipative terms $T^{\mu\nu}_{(1S)}$ and $N^{\mu}_{(1)}$ encode deviations from local equilibrium. Their magnitude can be understood through a systematic gradient expansion of the dynamic variables. Their gradient ordering in this scheme is listed in Table \ref{tab:gradient_expansionnospin}.
\begin{table}[H]
\renewcommand{\arraystretch}{2}
\begin{tabularx}{1\textwidth} { 
| >{\raggedright\arraybackslash}X 
| >{\raggedright\arraybackslash}X | }
\hline
\textbf{Quantity} & \textbf{Gradient Expansion} \\
\hline
Equilibrium quantities: \(\varepsilon, n, u^\mu\) &\text{Leading-order}, \(\mathcal{O}(\partial^0)\equiv\mathcal{O}(1)\) \\
\hline
Dissipative terms: \(T_{(1S)}^{\mu\nu}, N_{(1)}^\mu\) &\text{First-order}, \(\mathcal{O}(\partial^1)\) \\
\hline
\end{tabularx}
\caption{Gradient expansion order in conventional relativistic hydrodynamics.}
\label{tab:gradient_expansionnospin}
\end{table}
\noindent
Substituting Eqs.~\eqref{Toutofequilibrium}-\eqref{Noutofequilibrium} into the local conservation laws yields dissipative analogs of the Euler equations from Table~\ref{Relativistic evolution equations in local equilibrium}; see Table \ref{tab:RelativisticNavier-Stokesevolutionequations}.
\begin{table}[H]
\centering
\renewcommand{\arraystretch}{2}
\begin{tabularx}{\textwidth} { 
| >{\centering\arraybackslash}X 
| >{\centering\arraybackslash}X
| >{\centering\arraybackslash}X
| >{\centering\arraybackslash}X
| >{\centering\arraybackslash}X 
| >{\centering\arraybackslash}X | }
\hline
\textbf{Evolution equation} \\
\hline
$D\varepsilon+(\varepsilon+P)\theta=-u_{\nu}\partial_{\mu}T^{\mu\nu}_{(1S)} $ \\
\hline
$(\varepsilon+P)D u^{\alpha}-\nabla^{\alpha}P=-\Delta_{\nu}^{\alpha}\partial_{\mu}T^{\mu\nu}_{(1S)} $ \\
\hline
$Dn+n\theta=-\partial_{\mu}N_{(1)}^{\mu}$ \\
\hline
\end{tabularx}
\caption{Relativistic Navier-Stokes evolution equations.}
\label{tab:RelativisticNavier-Stokesevolutionequations}
\end{table}
\noindent
It is important to note that the dissipative currents in Eqs.~\eqref{Toutofequilibrium}-\eqref{Noutofequilibrium} introduce new degrees of freedom, whose dynamics, at this point, is unknown. As a result, the system of equations in Table \ref{tab:RelativisticNavier-Stokesevolutionequations} becomes underdetermined. 
In Sec.~\ref{Dissipative currents and transport coefficients : entropy-current analysis} we will show that this issue can be resolved by expressing the dissipative currents in terms of the gradients of the primary hydrodynamic variables listed in Table~\ref{Dynamical variables in relativistic hydrodynamics}. In this case, at first order in gradients, equations in Table~\ref{tab:RelativisticNavier-Stokesevolutionequations} are known as \emph{relativistic Navier-Stokes equations}.
%
%
\subsection{Relativistic fluid with spin}
In a relativistic fluid with spin, the dynamics involves eleven primary dynamical variables characterizing the state of a fluid element in local equilibrium, see Table \ref{Dynamical variables in relativistic spin hydrodynamics}.
\begin{table}[H]
\renewcommand{\arraystretch}{2}
\begin{tabularx}{1\textwidth} { 
| >{\raggedright\arraybackslash}X 
| >{\raggedright\arraybackslash}X
| >{\raggedright\arraybackslash}X
| >{\raggedright\arraybackslash}X
| >{\centering\arraybackslash}X 
| >{\raggedleft\arraybackslash}X | }
\hline
\textbf{Dynamical variables} &\textbf{Constraints} &\textbf{Degrees of freedom}\\
\hline
Energy density: $\varepsilon(x)$ & 
&1\\
\hline
Four-velocity: $u^{\mu}(x)$ & $u_{\mu}u^{\mu}=1$& 3\\
\hline
Number density: $n(x)$ &
&1 \\
\hline
Spin density: $S^{\mu\nu}(x)$ &
$S^{\mu\nu}=-S^{\nu\mu}$ 
&6\\
\hline 
\end{tabularx}
\caption{Primary dynamical variables in relativistic spin hydrodynamics.}
\label{Dynamical variables in relativistic spin hydrodynamics}
\end{table}
\noindent

In Sec.~\ref{Thermodynamic relations} we showed that these dynamical variables constitute the energy-momentum tensor \( T^{\mu\nu} \), the particle number current density \( N^{\mu} \), and the spin tensor $S^{\lambda\mu\nu}$. If dissipation is present, we supplement equilibrium currents with their dissipative parts
\begin{align}
T^{\mu\nu}&=\varepsilon u^{\mu}u^{\nu}-P\Delta^{\mu\nu}+T^{\mu\nu}_{(1S)}+T^{\mu\nu}_{(1A)},\label{Tleadingorder}\\
N^{\mu}&=nu^{\mu}+N_{(1)}^{\mu},\label{Nleadingorder}\\
S^{\mu\alpha\beta}&=u^{\mu}S^{\alpha\beta}+S^{\mu\alpha\beta}_{(1)}\label{Sleadingorder}.
\end{align}
Here, the dissipative part of the energy-momentum tensor is decomposed into symmetric $T^{\mu\nu}_{(1S)}$ and antisymmetric $T^{\mu\nu}_{(1A)}$ parts, since, following the formulation of relativistic spin hydrodynamics from field theory in Chapter~\ref{From fields to fluids}, the full energy-momentum tensor \( T^{\mu\nu} \) can be in general asymmetric. Apart from the presence of the spin tensor, the asymmetric structure of the energy-momentum tensor is one of the key features distinguishing the conventional relativistic hydrodynamics from relativistic spin hydrodynamics, and especially surprising for general relativity practitioners, given the fact that its Belinfante-Rosenfeld-symmetrized form is crucial for its applicability in Einstein's equations and reproducing the Hilbert definition~\cite{belinfante1939spin,cbelinfante1940current,rosenfeld1940}.

The terms $N^{\mu}_{(1)}$ and $S^{\mu\alpha\beta}_{(1)}$ encode the particle number and spin dissipation and are orthogonal to the four-velocity in their first index, i.e, $u_{\mu}N^{\mu}_{(1)}=0$ and $u_{\mu}S^{\mu\alpha\beta}_{(1)}=0$. Their magnitudes are of gradient order, see Table \ref{tab:gradient_expansion}.
\begin{table}[H]
\renewcommand{\arraystretch}{2}
\begin{tabularx}{1\textwidth} { 
| >{\raggedright\arraybackslash}X 
| >{\raggedright\arraybackslash}X | }
\hline
\textbf{Quantity} & \textbf{Gradient Expansion} \\
\hline
Equilibrium quantities: \(\varepsilon, n, u^\mu, S^{\mu\nu}\) & \text{Leading-order},~\(\mathcal{O}(\partial^0)\equiv\mathcal{O}(1)\) \\
\hline
Dissipative terms: \(T_{(1S)}^{\mu\nu}, T_{(1A)}^{\mu\nu}, N_{(1)}^\mu, S^{\lambda\mu\nu}_{(1)}\) & \text{First-order},~\(\mathcal{O}(\partial^1)\) \\
\hline
\end{tabularx}
\caption{Gradient expansion order in relativistic spin hydrodynamics.}
\label{tab:gradient_expansion}
\end{table}
\noindent
The spin potential, as defined in the thermodynamic relation Eq.~\eqref{covariantgeneralizedEulerequation}, is by construction assumed in this work to be of first-order in the hydrodynamic gradient expansion, i.e.,  
\begin{align}  
\omega^{\mu\nu} \sim \mathcal{O}(\partial).  
\end{align}  
The physical motivation for this gradient counting arises from the fact that the contribution of the spin tensor to the change in energy, see Eq.~\eqref{covariantgeneralizedEulerequation}, is subdominant compared to that of the standard hydrodynamic current. Indeed, the latter is a matter of "choice", meaning that the formulation of spin hydrodynamics inherently depends on this choice \footnote{An alternative of $\omega^{\mu\nu} \sim \mathcal{O}(1)$ was considered for instance in Ref.~\cite{She:2021lhe}}.

As demonstrated in Chapter~\ref{From fields to fluids}, for fluids with spin energy-momentum tensor and particle number current are locally conserved
\begin{align}
\label{TNconeqspinhydro1}
&\partial_{\mu}T^{\mu\nu}=0,\\
&\partial_{\mu}N^{\mu}=0,
\label{TNconeqspinhydro2}
\end{align}
while the spin tensor satisfies the following continuity equation, 
\begin{align}\label{spinconeqspinhydro}
\partial_{\lambda}S^{\lambda\mu\nu}&=T^{\nu\mu}-T^{\mu\nu} =-2T^{\mu\nu}_{(1A)}.
\end{align}
By substituting the expressions for the currents \eqref{Tleadingorder}-\eqref{Sleadingorder} into the local conservation laws \eqref{TNconeqspinhydro1}-\eqref{TNconeqspinhydro2} and the spin continuity equation \eqref{spinconeqspinhydro}, the evolution equations for the primary hydrodynamic variables are obtained. Similarly to the previous discussion, we may perform their projections along and orthogonal to $u$; the resulting forms are shown in Table~\ref{tab:Relativistic Navier-Stokes evolution equations with spin current}.
\begin{table}[H]
\centering
\renewcommand{\arraystretch}{2}
\begin{tabularx}{\textwidth}{ 
| >{\centering\arraybackslash}X 
| >{\centering\arraybackslash}X 
| >{\centering\arraybackslash}X 
| >{\centering\arraybackslash}X 
| >{\centering\arraybackslash}X 
| >{\centering\arraybackslash}X | }
\hline
 \textbf{Evolution equations} \\
\hline
 $D\varepsilon + (\varepsilon + p)\theta = -u_{\nu}\partial_{\mu}T^{\mu\nu}_{(1S)} - u_{\nu}\partial_{\mu}T^{\mu\nu}_{(1A)}$ \\
\hline
 $(\varepsilon + p) Du^{\alpha} - \nabla^{\alpha} p = -\Delta^{\alpha}_{\nu}\partial_{\mu}T^{\mu\nu}_{(1S)} - \Delta_{\nu}^{\alpha}\partial_{\mu}T^{\mu\nu}_{(1A)}$ \\
\hline
 $Dn + n\theta = -\partial_{\mu}N_{(1)}^{\mu}$ \\
\hline
 $DS^{\mu\nu} + S^{\mu\nu} \theta + \partial_{\lambda}S^{\lambda\mu\nu}_{(1)} = -2 T^{\mu\nu}_{(1A)}$ \\
\hline
\end{tabularx}
\caption{Relativistic Navier-Stokes evolution equations with spin tensor.}
\label{tab:Relativistic Navier-Stokes evolution equations with spin current}
\end{table}
\noindent
As in the case of conventional relativistic hydrodynamics, the dissipative currents introduce new dynamical degrees of freedom. When supplemented with the first-order expressions for the dissipative currents, the resulting evolution equations presented in Table~\ref{tab:Relativistic Navier-Stokes evolution equations with spin current} may be regarded as \emph{relativistic Navier-Stokes equations for fluids with spin}. The problem of establishing the forms of these currents will be addressed in the next section.
%
\section{Dissipative currents and transport coefficients: entropy-current analysis}
\label{Dissipative currents and transport coefficients : entropy-current analysis}
The objective of this section is to determine the dissipative currents for the relativistic Navier-Stokes evolution equations with spin tensor (see Table~\ref{tab:Relativistic Navier-Stokes evolution equations with spin current}). This is done through the entropy-current analysis, which, using the second law of
thermodynamics, will allow us to express them entirely in terms of the primary dynamical variables: energy density \( \varepsilon \), fluid four-velocity \( u^{\mu} \), and spin density \( S^{\mu\nu} \), and their gradients. As a result, (i) the system of evolution equations becomes mathematically closed, and (ii) the transport coefficients in the system are identified. For simplicity, throughout this chapter, we assume the limit of zero chemical potential, \(\mu = 0\). Methods discussed in this section will be employed in the subsequent section.
\medskip 

We start with the leading-order entropy current obtained in Eq.~\eqref{covariantgeneralizedEulerequation}, 
\begin{align}\label{Smu0}
S^{\mu}_{(0)} = T^{\mu\nu}_{(0)}\beta_{\nu} + P\beta^{\mu} - \beta\omega_{\alpha\beta}{S^{\mu\alpha\beta}_{(0)}} 
\end{align}
where we add the label "(0)" to indicate leading order quantities. In the presence of dissipation in the system, the entropy current receives contributions from the dissipative parts of the energy-momentum and spin tensors. Hence, the ansatz for the out-of-equilibrium entropy current at the first-order in gradients can be written in the following form,
\begin{align}\label{NSentropy}
S^{\mu} = S^{\mu}_{(0)} + T^{\mu\nu}_{(1)}\beta_{\nu} - \beta\omega_{\alpha\beta}S^{\mu\alpha\beta}_{(1)}.
\end{align}
Here, $T^{\mu\nu}_{(1)}$ represents the sum of its symmetric and antisymmetric parts, $T^{\mu\nu}_{(1)} = T^{\mu\nu}_{(1S)} + T^{\mu\nu}_{(1A)}$. Within the gradient counting chosen previously, the term $\beta\omega_{\alpha\beta}S^{\mu\alpha\beta}_{(1)}$ is $\mathcal{O}(\partial^{2})$, and therefore can be neglected. Consequently, the divergence of the entropy current reads as follows
\begin{align}\label{notfinalentropyproductionXu}
\partial_{\mu}S^{\mu} = \beta\left[T\partial_{\mu}S^{\mu}_{(0)} + u_{\nu}\partial_{\mu}T^{\mu\nu}_{(1)}\right] + T^{\mu\nu}_{(1)}\partial_{\mu}\beta_{\nu}.
\end{align}
In the presence of dissipation, the divergence of the leading-order entropy current~\eqref{Smu0} can be expressed as 
\begin{align}
\partial_{\mu}S^{\mu}_{(0)} = -\beta_{\nu}\partial_{\mu}T^{\mu\nu}_{(1)} - \beta\omega_{\alpha\beta}\partial_{\mu}(u^{\mu}S^{\alpha\beta})-\beta\omega_{\alpha\beta}\partial_{\mu}S^{\mu\alpha\beta}_{(1)}.
\end{align}
In the above equation we used the conservation law of the energy-momentum tensor $\partial_{\mu}T^{\mu\nu}_{(0)}=-\partial_{\mu}T^{\mu\nu}_{(1)}$ and similarly for the spin tensor. By combining this result with the spin continuity equation~\eqref{spinconeqspinhydro}, and considering terms up to second-order in gradient expansion, the divergence of the entropy current in Eq.~\eqref{notfinalentropyproductionXu} can be rewritten as,  
\begin{align}
\partial_{\mu}S^{\mu} &= {T^{\alpha\beta}_{(1S)}}\partial_{\alpha}\beta_{\beta} + {T^{\alpha\beta}_{(1A)}}\big[\partial_{\alpha}\beta_{\beta} + 2\beta\omega_{\alpha\beta}\big]. \label{Xuentropyproduction}
\end{align}

The symmetric $T^{\alpha\beta}_{(1S)}$ and antisymmetric $T^{\alpha\beta}_{(1A)}$ parts of the dissipative energy-momentum tensor can be decomposed with respect to the fluid four-velocity as follows,  
\begin{align} 
& T^{\alpha\beta}_{(1S)} = 2h^{(\alpha}u^{\beta)}+\pi^{\alpha\beta}+\Pi\Delta^{\alpha\beta},\label{XudecompsymmetricEMT}\\ 
&T^{\alpha\beta}_{(1A)} = 2q^{[\alpha}u^{\beta]}+\phi^{\alpha\beta}\label{Xudecompantisymmetric},
\end{align}
where we use the notation $X^{(\mu\nu)}=\frac{1}{2}\left(X^{\mu\nu} + X^{\nu\mu}\right)$ and $X^{[\mu\nu]} \equiv \frac{1}{2}\left(X^{\mu\nu} - X^{\nu\mu}\right)$. Here, the vectors \( h^{\mu} \) and \( q^{\mu} \) are orthogonal to the fluid four-velocity, namely \( h \cdot u = 0 \) and \( q \cdot u = 0 \). The tensor \( \pi^{\mu\nu} \) is symmetric, traceless and orthogonal to the flow four-vector $u_{\mu}\pi^{\mu\nu}=0$, \( \Pi \) is a scalar, and \( \phi^{\mu\nu} \) is antisymmetric and orthogonal to the flow velocity $u_{\mu}\phi^{\mu\nu}=0$. Consequently, the energy-momentum tensor can be expressed as, 
\begin{align}\label{EMTGENREALDECOM}
T^{\mu\nu}=&\,T^{\mu\nu}_{(0)}+T^{\mu\nu}_{(1S)}+T^{\mu\nu}_{(1A)}\nonumber\\
=&\,\varepsilon u^{\mu}u^{\nu}-P\Delta^{\mu\nu}+2h^{(\mu}u^{\nu)}+\pi^{\mu\nu}+\Pi\Delta^{\mu\nu}+2q^{[\mu}u^{\nu]}+\phi^{\mu\nu}.
\end{align}

The method leading to decomposition \eqref{XudecompsymmetricEMT}-\eqref{Xudecompantisymmetric} is shown in Appendix~\ref{Appendix A}. It is important to note that this decomposition holds only when the entropy production contains only gradients of the four-velocity and no gradients of the spin potential. This is justified within the counting scheme chosen here, namely when the gradients of the spin potential are considered to be of higher order in the gradient expansion (see Eq.~\eqref{Xuentropyproduction}). Moreover, the decomposition strictly holds only in the case of zero chemical potential, $\mu=0$, i.e., when there are no gradients of the chemical potential in the system~\footnote{In Chapter~\ref{Quantum-statistical formulation} and related Appendix~\ref{AppendixC} we irreducibly decompose these currents under rotation, taking into account all possible gradients associated with velocity, chemical potential, and spin potential.}.  
Substituting the decompositions~\eqref{XudecompsymmetricEMT}-\eqref{Xudecompantisymmetric} in the expression for the divergence of the entropy current~\eqref{Xuentropyproduction}, we get
\begin{align}\label{enproexp}
\partial_{\mu}S^{\mu}=&-\beta h^{\mu}(\beta \nabla_{\mu}T-Du_{\mu})+\beta\pi^{\mu\nu}\sigma_{\mu\nu}+\beta\Pi\theta-\beta q^{\mu}(\beta\nabla_{\mu}T+Du_{\mu}-4\omega_{\mu\nu}u^{\nu})\nonumber\\
&+\phi^{\mu\nu}(\Omega_{\mu\nu}+2\beta\Delta^{\alpha}_{\mu}\Delta^{\beta}_{\nu}\omega_{\alpha\beta}),
\end{align}
where $\sigma_{\mu\nu}=\nabla_{(\mu} u_{\nu)}-\frac{1}{3}\theta\Delta_{\mu\nu}$ is known as the \emph{shear tensor}, and $\Omega_{\mu\nu}=\Delta^{\alpha}_{~\mu}\Delta^{\beta}_{~\nu}\partial_{[\alpha}\beta_{\beta]} = \beta \nabla_{[\mu} u_{\nu]}$. 

Using the \emph{second law of thermodynamics, $\partial_{\mu}S^{\mu}\geq 0$,} we find that the dissipative currents have to take the forms
\begin{align} 
h^{\mu}&=-\kappa\left(Du^{\mu}-\beta\nabla^{\mu}T\right),\label{phenomenologicalheatflux}\\ \pi^{\mu\nu}&=2\eta\sigma^{\mu\nu},\label{}\\ \Pi &= \zeta \theta,\label{}\\ 
q^{\mu}&=\lambda \left(\beta\nabla^{\mu}T+Du^{\mu}-4\omega^{\mu\nu}u_{\nu}\right),\label{phemenologicalq}\\ 
\phi^{\mu\nu}&=\gamma\left(\Omega^{\mu\nu}+2\beta \Delta^{\mu}_{~\alpha}\Delta^{\nu}_{~\beta}\omega^{\alpha\beta}\right),\label{phenomenologicalphi}
\end{align}
where the \emph{transport coefficients} satisfy: $\kappa,~\eta,~\zeta,~\lambda,~\gamma~\geq 0$. In the formulas above, we readily identify the dissipative quantities also known in conventional first-order dissipative hydrodynamics. In particular, the four-vector $h^{\mu}$ is known as the \emph{heat flux} with $\kappa$ denoting the \emph{heat conductivity} coefficient, $\pi^{\mu\nu}$ is the \emph{shear stress tensor} with $\eta$ being the \emph{shear viscosity} coefficient, while $\Pi$ is the \emph{bulk pressure} with $\zeta$ denoting the \emph{bulk viscosity} coefficient. The current $q^{\mu}$ is the temporal projection $q^{\mu}=u_{\beta}\Delta^{\mu}_{\alpha}T^{\alpha\beta}_{(1A)}$ while $\phi^{\mu\nu}$ is the spatial projection $\Delta^{\alpha\mu}\Delta^{\beta\nu}T_{\alpha\beta(1A)}$ of an antisymmetric part of the energy-momentum tensor $T_{\alpha\beta(1A)}$ (see Table.~\ref{tab:EMTdecompSymbols} for details). Hence, we can interpret $\lambda$ as \emph{boost heat conductivity} and $\gamma$ as \emph{rotational viscosity}~\cite{Hattori:2019lfp}. The coefficients $\lambda$ and $\gamma$ are new transport coefficients in spin hydrodynamics. Note that the dissipative currents are expressed in terms of  $T$, $u^{\mu}$, and $\omega^{\alpha\beta}$ instead of $\varepsilon$, $u^{\mu}$, and $S^{\alpha\beta}$. However, as shown in Eq.~\eqref{EOSS}, we may always use the equation of state, and thermodynamic relations to transform between these two sets of quantities. 

\bigskip 

\noindent
Having the dissipative currents identified  in Eqs.~\eqref{phenomenologicalheatflux}-\eqref{phenomenologicalphi}, we can write the relativistic Navier-Stokes evolution equations with the spin tensor listed in Table~\ref{tab:Relativistic Navier-Stokes evolution equations with spin current} in their explicit forms, see Table \ref{tab:explicit relativistic Navier-Stokes evolution equations with spin current}.

\begin{table}[H]
\centering
\renewcommand{\arraystretch}{2}
\begin{tabularx}{\textwidth} { 
| >{\centering\arraybackslash}X 
| >{\centering\arraybackslash}X
| >{\centering\arraybackslash}X
| >{\centering\arraybackslash}X
| >{\centering\arraybackslash}X 
| >{\centering\arraybackslash}X | }
\hline
 \textbf{Relativistic Navier-Stokes hydrodynamics with spin ($\mu=0$)} \\
\hline
 $D\varepsilon+(\varepsilon+p)\theta = 2 \,h^{\mu}Du_{\mu} -\nabla \cdot (q+h)+\pi^{\mu \nu}  \partial_{\mu} u_{\nu}+\Pi\Delta^{\mu\nu}\partial_{\mu}u_{\nu}+\phi^{\mu \nu}  \partial_{\mu} u_{\nu}$ \\
\hline
 $(\varepsilon+p) Du^{\alpha} - \nabla^{\alpha}  p= 
-(q +h)\cdot \nabla u^{\alpha}+(q^{\alpha}-h^{\alpha})\theta
+\Delta^{\alpha}_{~\nu}D q^{\nu}-\Delta^{\alpha}_{~\nu}D h^{\nu}
-\Delta^{\alpha}_{~\nu} \partial_{\mu} (\pi^{\mu \nu}+\Pi\Delta^{\mu\nu}) -\Delta^{\alpha}_{~\nu} \partial_{\mu} \phi^{\mu \nu}$ \\
\hline
 $\partial_{\lambda}(u^{\lambda}S^{\mu\nu}) = -2(q^{\mu}u^{\nu}-q^{\nu}u^{\mu}+\phi^{\mu\nu})$ \\
\hline
\end{tabularx}
\caption{Explicit forms of the relativistic Navier-Stokes evolution equations with spin tensor in the zero chemical potential limit (\(\mu = 0\)).}
\label{tab:explicit relativistic Navier-Stokes evolution equations with spin current}
\end{table}

\noindent
Note that by taking the limit of zero spin tensor and, consequently, no antisymmetric dissipative part in the energy-momentum tensor, we can obtain the explicit forms of the conventional relativistic Navier-Stokes hydrodynamic equations, see Table \ref{tab:explicitRelativisticNavier-Stokesevolutionequations}.

\begin{table}[H]
\centering
\renewcommand{\arraystretch}{1.9}
\begin{tabularx}{\textwidth} { 
| >{\centering\arraybackslash}X 
| >{\centering\arraybackslash}X
| >{\centering\arraybackslash}X
| >{\centering\arraybackslash}X
| >{\centering\arraybackslash}X 
| >{\centering\arraybackslash}X | }
\hline
 \textbf{Conventional relativistic Navier-Stokes hydrodynamics ($\mu=0$)} \\
\hline
 $D\varepsilon+(\varepsilon+P)\theta+\partial_{\mu}h^{\mu}-h^{\nu}Du_{\nu}-\pi^{\mu\nu}\partial_{\mu}u_{\nu}-\Pi\theta=0$ \\
\hline
 $(\varepsilon+P)D u^{\alpha}-\nabla^{\alpha}P+h^{\mu}\partial_{\mu}u^{\alpha}+Dh^{\alpha}-u^{\alpha}u_{\nu}Dh^{\nu}+h^{\alpha}\theta+\Delta^{\alpha}_{\nu}\partial_{\mu}(\pi^{\mu\nu}+\Pi\Delta^{\mu\nu})=0$ \\
\hline
\end{tabularx}
\caption{{Explicit forms of the conventional relativistic Navier-Stokes evolution equations in the zero chemical potential limit (\(\mu = 0\)).}
}
\label{tab:explicitRelativisticNavier-Stokesevolutionequations}
\end{table}

\noindent
In Table \ref{Table1} we summarize the quantities involved in the decomposition of the energy-momentum tensor \(T^{\mu\nu}\) and the spin tensor \(S^{\lambda\mu\nu}\) (except for the term $S^{\lambda\mu\nu}_{(1)}$ which does not enter the final equations). For each quantity, we specify its mathematical type, the order in hydrodynamic gradient expansion, any conditions or constraints it satisfies, and the number of degrees of freedom it contributes to the system. 
\begin{table}[H]
\renewcommand{\arraystretch}{1.9}
\begin{tabularx}{1\textwidth} { 
| >{\raggedright\arraybackslash}X 
| >{\raggedright\arraybackslash}X
| >{\raggedright\arraybackslash}X
| >{\raggedright\arraybackslash}X
| >{\centering\arraybackslash}X 
| >{\raggedleft\arraybackslash}X | }
\hline
\textbf{Quantity} & \textbf{Mathematical object} &\textbf{Constraints}& \textbf{Hydrodynamic gradient order}& \textbf{Degrees of freedom}\\
\hline
$\varepsilon$  &  Scalar field & &$\mathcal{O}(\partial^{0})\equiv\mathcal{O}(1)$ & 1\\
\hline
$u^{\mu}$ & Four-vector &$u^{\mu}u_{\mu}=1$ &$\mathcal{O}(\partial^{0})\equiv\mathcal{O}(1)$ & 3\\
\hline
$ h^{\mu}$ & Four-vector& $h^{\mu}u_{\mu}=0$ &$\mathcal{O}(\partial)$ & 3 \\
\hline
$\pi^{\mu\nu}$ & Symmetric traceless tensor &\mbox{$\pi^{\mu\nu}u_{\mu}=0,$} $\pi^{\mu}_{\mu}=0$, $\pi^{\mu\nu}=\pi^{\nu\mu}$&$\mathcal{O}(\partial)$& 5\\
\hline
$\Pi$ & Scalar field& &$\mathcal{O}(\partial)$& 1  \\
\hline
$q^{\mu}$& Four-vector& $q^{\mu}u_{\mu}=0$& $\mathcal{O}(\partial)$& 3\\
\hline
$\phi^{\mu\nu}$& Antisymmetric tensor &$\phi^{\mu\nu}u_{\mu}=0$, $\phi^{\mu\nu}=-\phi^{\nu\mu}$& $\mathcal{O}(\partial)$& 3\\
\hline
$S^{\mu\nu}$& Antisymmetric tensor&$S^{\mu\nu}=-S^{\nu\mu}$ &$\mathcal{O}(\partial^{0})\equiv\mathcal{O}(1)$& 6\\
\hline
\end{tabularx} 
\caption{Number of independent components of various quantities used in the decomposition of the energy-momentum tensor $T^{\mu\nu}$ and the spin tensor $S^{\lambda\mu\nu}$.}
\label{Table1}
\end{table}

In general, the asymmetric energy-momentum tensor $T^{\mu\nu}$ in relativistic spin hydrodynamics can have sixteen independent components. From Table~\ref{Table1}, we observe that the sum of independent degrees of freedom entering \( T^{\mu\nu} \) is nineteen. The reason for this over-counting lies in the fact that, thus far, we were working in a general hydrodynamic frame, namely we did not define the four-velocity of the fluid. In the literature, two main hydrodynamic frames are commonly used which have different physical interpretations, namely the \emph{Eckart frame}~\cite{Eckart:1940te} and the \emph{Landau-Lifshitz frame}~\cite{landau2013fluid}. In these two cases, either the dissipative particle number current, \( N^{\mu}_{(1)} \), or the heat flux, \( h^{\mu} \), is zero, respectively. This implies that either number density \( n \) or energy density \( \varepsilon \) behave as if they were in local equilibrium. 

The four-velocity in the Landau frame is defined as
\begin{align}
T^{\mu\nu}u_{\nu} = \varepsilon u^{\mu} \quad \implies \quad h^{\mu} = 0,
\end{align}
which is equivalent to stating that 
\begin{align}
u_{\nu}T^{\mu\nu}_{(1S)}=0\quad \implies \quad h^{\mu}=0.
\end{align}
In the Eckart frame, the four-velocity is defined as
\begin{align}
N^{\mu} = n u^{\mu} \quad \implies \quad N^{\mu}_{(1)} = 0.
\end{align}
In the presence of an antisymmetric part of the energy-momentum tensor, the use of the Landau frame is nontrivial as, 
\begin{align}
T^{\mu\nu}u_{\nu} = \varepsilon u^{\mu} \quad \implies \quad h^{\mu}+q^{\mu} = 0,
\end{align}
or
\begin{align}\label{generalizedLandueframe}
u_{\mu}\left(T^{\mu\nu}_{(1S)}+T^{\mu\nu}_{(1A)}\right)=0\quad\implies\quad h^{\mu}+q^{\mu}=0.
\end{align}
We refer to the above frame choice as \emph{generalized Landau frame}. However, in both cases, either \( h^{\mu} = 0 \) or \( h^{\mu} + q^{\mu} = 0 \), the number of independent components is reduced from nineteen to sixteen.
%
%
\section{Canonical formulation of spin hydrodynamics}
\label{Canonical framework}
The formulation of relativistic spin hydrodynamics presented so far has been based on an asymmetric energy-momentum tensor, \( T^{\mu\nu} \), and a spin tensor, \( S^{\lambda\mu\nu} \), antisymmetric in its last two indices,  $S^{\lambda\mu\nu} = -S^{\lambda\nu\mu}$.
In the literature, the formulation of spin hydrodynamics based on this particular choice of currents is commonly called the \emph{phenomenological} formulation. The justification for the above given Lorentz index symmetries of phenomenological currents is grounded in field theory. However, as discussed in Chapter~\ref{From fields to fluids}, when constructing the framework of spin hydrodynamics for fermions, it is necessary
to consider an asymmetric energy-momentum tensor and a totally antisymmetric spin tensor. The choice of canonical currents constitutes a basis of what is known as a \emph{canonical} formulation of spin hydrodynamics.  
\medskip

With the motivation given above, in this section, we reformulate a first-order relativistic spin hydrodynamics using canonical forms of energy-momentum and spin tensors. We show that the phenomenological formulation and the canonical formulation are equivalent, namely, they can be directly connected by a pseudo-gauge transformation, provided the canonical formulation is appropriately modified. This section is fully based on Ref.~\multimyref{AD1}.
\subsection{Canonical formulation}
We define the canonical energy-momentum and spin tensors as,
\begin{align}
&T^{\mu\nu}_{\text{can}}=T^{\mu\nu}_{(0)}+T^{\mu\nu}_{\text{can}(1)},\label{EMTcanonical}\\
&S^{\mu\alpha\beta}_{\text{can}}=u^{\mu}S^{\alpha\beta}+u^{\beta}S^{\mu\alpha}+u^{\alpha}S^{\beta\mu}+S^{\mu\alpha\beta}_{\text{can}(1)}.\label{Spincanonical}
\end{align}
where $T^{\mu\nu}_{(0)}=\varepsilon u^{\mu}u^{\nu}-P \Delta^{\mu\nu}$ is the equilibrium part of the energy-momentum tensor. One can notice that here the leading order part of the spin tensor, $S^{\mu\alpha\beta}_{\text{can}(0)}=u^{\mu}S^{\alpha\beta}+u^{\beta}S^{\mu\alpha}+u^{\alpha}S^{\beta\mu}$, as opposed to the phenomenological case discussed in the previous section, is totally antisymmetric.  

The local conservation laws for the canonical energy-momentum tensor \( T^{\mu\nu}_{\text{can}} \) and the total angular momentum tensor 
\begin{align}
J^{\mu\alpha\beta}_{\text{can}} = \left( x^{\alpha}T^{\mu\beta}_{\text{can}} - x^{\beta}T^{\mu\alpha}_{\text{can}} \right) + S^{\mu\alpha\beta}_{\text{can}},
\end{align}
as well as the continuity equation for the canonical spin tensor \( S^{\mu\alpha\beta}_{\text{can}} \), read as follows 
\begin{align}
&\partial_{\mu}T^{\mu\nu}_{\text{can}}=0,\label{EMTconservationcanonical}\\
&\partial_{\mu}J^{\mu\alpha\beta}_{\text{can}}=0,\\
& \partial_{\mu}(u^{\mu}S^{\alpha\beta})+\partial_{\mu}S^{\mu\alpha\beta}_{\text{can}(1)}= -2T^{\alpha\beta}_{\text{can} (1A)}-2 \partial_{\mu}\Phi^{\mu\alpha\beta}_{\text{can}(0)}.\label{continuityspincanonical}
\end{align}
Here, for the notational simplicity, we have introduced the tensor $\Phi^{\mu\alpha\beta}_{\text{can(0)}}\equiv u^{[\alpha}S^{\beta]\mu}$ which is antisymmetric in the last two indices. 
\medskip

Similar to the phenomenological formulation in Sec.~\ref{Dissipative currents and transport coefficients : entropy-current analysis}, to determine the dissipative currents and transport coefficients, we first write down the leading-order entropy current derived in Eq.~\eqref{covariantgeneralizedEulerequation} replacing the currents with the canonical ones,  
\begin{align}\label{leadingcanonnical}
S^{\mu}_{\text{can}(0)} = T^{\mu\nu}_{\text{can}(0)}\beta_{\nu} + P\beta^{\mu} - \beta\omega_{\alpha\beta}S^{\mu\alpha\beta}_{(0)}.
\end{align}
The leading-order part of the spin tensor in the above thermodynamic relation reads \( S^{\mu\alpha\beta}_{(0)} = u^{\mu}S^{\alpha\beta} \), ensuring that, in the comoving frame of the fluid element, Eq.~\eqref{leadingcanonnical} reduces to Eq.~\eqref{generalizedEulerequation}.

Introducing dissipation, the out-of-equilibrium canonical entropy current is modified to 
\begin{align}\label{canonicaldissipativeentropycurrent}
S^{\mu}_{\text{can}} = S^{\mu}_{\text{can}(0)} + T^{\mu\nu}_{\text{can}(1)}\beta_{\nu}+\mathcal{O}(\partial^{2}).
\end{align}
Consequently, the divergence of the canonical entropy current is expressed as  
\begin{align}\label{notfinalentropyproductioncanonical}
\partial_{\mu}S^{\mu}_{\text{can}} = \beta\left[T\partial_{\mu}S^{\mu}_{\text{can}(0)} + u_{\nu}\partial_{\mu}T^{\mu\nu}_{\text{can}(1)}\right] + T^{\mu\nu}_{\text{can}(1)}\partial_{\mu}\beta_{\nu}.
\end{align}
The divergence of the leading-order entropy current in the presence of dissipation can be obtained from Eq.~\eqref{leadingcanonnical}, which gives  
\begin{align}
\partial_{\mu}S^{\mu}_{\text{can} (0)} = -\beta_{\nu}\partial_{\mu}T^{\mu\nu}_{\text{can}(1)} - \beta\omega_{\alpha\beta}\partial_{\mu}(u^{\mu}S^{\alpha\beta}).
\end{align}
Using this result, together with the spin continuity equation~\eqref{continuityspincanonical}, Eq.~\eqref{notfinalentropyproductioncanonical} can be expressed as   
\begin{align}
\partial_{\mu}S^{\mu}_{\text{can}} &= T^{\alpha\beta}_{\text{can}(1S)}\partial_{\alpha}\beta_{\beta} + T^{\alpha\beta}_{\text{can}(1A)}\big[\partial_{\alpha}\beta_{\beta} + 2\beta\omega_{\alpha\beta}\big] + 2\beta\omega_{\alpha\beta}\partial_{\mu}\Phi^{\mu\alpha\beta}_{\text{can}(0)}. \label{finalentropyproductioncanonial}
\end{align}
Compared to the phenomenological entropy production rate~\eqref{Xuentropyproduction}, its canonical analogue includes an additional term, $2\beta\omega_{\alpha\beta}\partial_{\mu}\Phi^{\mu\alpha\beta}_{\text{can}(0)}$. Consequently, employing the same decomposition for the symmetric and antisymmetric components of the energy-momentum tensor as before, in~\eqref{XudecompsymmetricEMT} and~\eqref{Xudecompantisymmetric}, and, similarly, decomposing $\partial_{\mu}\Phi^{\mu\alpha\beta}_{\text{can}(0)}$ and $\omega^{\alpha\beta}$ as
\begin{align}
%
%
&\partial_{\mu}\Phi^{\mu\alpha\beta}_{\text{can(0)}}=\delta q^{\alpha}u^{\beta}-\delta q^{\beta}u^{\alpha}+\delta\phi^{\alpha\beta},\label{}\\
&\omega^{\alpha\beta}=k^{\alpha}u^{\beta}-k^{\beta}u^{\alpha}+\lambda^{\alpha\beta},
\label{}
\end{align}
the entropy production~\eqref{finalentropyproductioncanonial} can be cast in the form
\begin{align}
\partial_{\mu}{S}^{\mu}_{\text{can}}= &-\beta h^{\mu}\left(\beta \nabla_{\mu}T-Du_{\mu}\right)+\beta\pi^{\mu\nu}\sigma_{\mu\nu}+\beta\Pi \theta-\beta q^{\mu}\left(\beta \nabla_{\mu}T+Du_{\mu}-4 \omega_{\mu\nu}u^{\nu}\right)\nonumber\\ & +\phi^{\mu\nu}\left(\Omega_{\mu\nu}+2\beta \Delta^{\alpha}_{~\mu}\Delta^{\beta}_{~\nu}\omega_{\alpha\beta}\right)+2\beta\left[2k_{\alpha}\delta q^{\alpha}+\lambda_{\alpha\beta}\delta\phi^{\alpha\beta}\right]. \label{}
\end{align}
Here, the new components $\delta q^{\alpha}$ and $k^{\alpha}$ as well as $\delta \phi^{\alpha\beta}$ and $\lambda^{\alpha\beta}$ have the same properties as $q^{\alpha}$ and $\phi^{\alpha\beta}$ respectively~\footnote{Note that here the terms $\delta q^{\alpha}$ and $\delta\phi^{\alpha\beta}$ should not be confused with perturbations of $q^{\alpha}$ and $\phi^{\alpha\beta}$.}.

Imposing second law of thermodynamics, $\partial_{\mu}S^{\mu}_{\rm can} \geq 0$, 
the dissipative contributions may be identified as follows
\begin{align}
&h^{\mu}=-\kappa\left(Du^{\mu}-\beta\nabla^{\mu}T\right),\label{canonicalheatflux}\\
&\pi^{\mu\nu}=2\eta\sigma^{\mu\nu},\label{}\\ 
&\Pi = \zeta \theta,\label{}\\ 
&q^{\mu}=\lambda \left(\beta\nabla^{\mu}T+Du^{\mu}-4\omega^{\mu\nu}u_{\nu}\right),\label{}\\ 
&\phi^{\mu\nu}=\gamma\left(\Omega^{\mu\nu}+2\beta \Delta^{\mu}_{~\alpha}\Delta^{\nu}_{~\beta}\omega^{\alpha\beta}\right),\label{canonicalphi}\\
&2 k_{\alpha}\delta q^{\alpha}+ \lambda_{\alpha\beta}\delta\phi^{\alpha\beta}\geq 0,\label{additionalconstriant}
\end{align}
where $\delta q^{\alpha}=\Delta^{\alpha}_{~\mu}u_{\nu}\partial_{\lambda}\Phi^{\lambda\mu\nu}_{\text{can(0)}}$ and $\delta\phi^{\alpha\beta}= \Delta^{\alpha}_{~[\mu}\Delta^{\beta}_{~\nu]}\partial_{\lambda}\Phi^{\lambda\mu\nu}_{\text{can(0)}}$~\cite{Daher:2022xon}. The dissipative currents~\eqref{canonicalheatflux}-\eqref{canonicalphi} are exactly the same as the phenomenological ones~\eqref{phenomenologicalheatflux}-\eqref{phenomenologicalphi}. However, due to $\Phi^{\mu\alpha\beta}_{\text{can(0)}}\equiv u^{[\alpha}S^{\beta]\mu}$, the additional condition given by Eq.~\eqref{additionalconstriant} imposes additional constraints on the spin density and, consequently, on the spin potential. These constraints make the resulting hydrodynamic framework not a well-defined initial value problem if an arbitrary set of the dynamical variables, $T$, $u^{\mu}$ and $\omega^{\mu\nu}$ is considered.
%
%
\subsection{Improved canonical formulation}
The issue of the additional constraint~\eqref{additionalconstriant} can be addressed by appropriately modifying the form of the energy-momentum tensor. In particular, one can add to it an additional divergence-free term, which does not affect the conservation law for the energy-momentum tensor, but it does affect the evolution of the spin tensor. With this modification, the energy-momentum and spin tensors read, 
\begin{align} 
& \widetilde{T}^{\mu\nu}_{\text{can}} = T^{\mu\nu}_{(0)}+T^{\mu\nu}_{\text{can}(1)} +\partial_{\lambda}\left (u^{\nu}S^{\mu\lambda}\right), \label{improvedcanonicalEMT}\\
&S^{\mu\alpha\beta}_{\text{can}}=u^{\mu}S^{\alpha\beta}+u^{\beta}S^{\mu\alpha}+u^{\alpha}S^{\beta\mu}+S^{\mu\alpha\beta}_{\text{can}(1)},\label{improvedspincanonicalEMT}
\end{align}
constituting the basis for the formulation of spin hydrodynamics known as the \emph{improved canonical} framework.
It is important to stress the fact that the additional divergence-free term is not a pseudo-gauge transformation as it does not follow the general definition of pseudo-gauge transformation~\eqref{Tpseudogaugeaveragevalues}-\eqref{Spseudogaugeaveragevalues}.
As a result, one can write 
\begin{align} 
&\partial_{\mu}\widetilde{T}^{\mu\nu}_{\text{can}}=0,\\
&\partial_{\mu}\widetilde{J}^{\mu\alpha\beta}=0,\\
&\partial_{\mu}(u^{\mu}S^{\alpha\beta})= -2T^{\alpha\beta}_{\text{can}(1A)}. \label{}
\end{align}

Similarly to the previous subsection, we can write the out-of-equilibrium improved canonical entropy current as,
\begin{align} 
\widetilde{{S}}^{\mu}_{\text{can}}={S}^{\mu}_{\text{can}(0)}+ \widetilde{T}^{\mu\nu}_{\text{can}(1)}\,\beta_{\nu}+{\mathcal{O}(\partial^2)}. \label{}
\end{align}
Here, compared to Eq.~\eqref{canonicaldissipativeentropycurrent}, the dissipative energy-momentum tensor $\widetilde{T}^{\mu\nu}_{(1)}$ includes an additional term $\partial_{\lambda}(u^{\nu}S^{\mu\lambda})$, see Eq.~\eqref{improvedspincanonicalEMT}. Therefore, the divergence of the entropy current is
\begin{align}
\partial_{\mu}\widetilde{{S}}^{\mu}_{\text{can}}= &T^{\alpha\beta}_{\text{can}(1S)}\partial_{\alpha}\beta_{\beta}+T^{\alpha\beta}_{\text{can}(1A)}\left[\partial_{\alpha}\beta_{\beta}+2\beta \omega_{\alpha\beta}\right]+\partial_{\mu}\beta_{\nu}\partial_{\lambda}(u^{\nu}S^{\mu\lambda}). \label{}
\end{align}
One should note, that the last term in the above equation is a total derivative, $\partial_{\mu}\left[\beta_{\nu}\partial_{\lambda}(u^{\nu}S^{\mu\lambda})\right]$. Hence, this term can be incorporated into the divergence of the entropy current on the left-hand side, resulting in the following equation,
\begin{align} 
& \partial_{\mu}\widetilde{{S}}^{\prime \mu}_{\text{can}}=T^{\alpha\beta}_{\text{can}(1S)}\partial_{\alpha}\beta_{\beta}+T^{\alpha\beta}_{\text{can}(1A)}\left[\partial_{\alpha}\beta_{\beta}+2\beta \omega_{\alpha\beta}\right], \label{}
\end{align}
where 
\begin{align} \widetilde{{S}}^{\prime\mu}_{\text{can}} & =\widetilde{{S}}^{\mu}_{\text{can}}-\beta_{\nu}\partial_{\lambda}\left(u^{\nu}S^{\mu\lambda}\right) = {S}^{\mu}_{(0)}+\beta_{\nu}T^{\mu\nu}_{\text{can}(1)} +\mathcal{O}(\partial^2)=S^{\mu}_{\text{can}}.\label{}
\end{align}
It is interesting to observe that the last equation indicates that $\widetilde{\mathcal{S}}^{\prime\mu}_{\text{can}}
= {\mathcal{S}}^\mu_{\rm can}$. Hence, the modification of the energy-momentum tensor proposed in Eq.~\eqref{improvedcanonicalEMT} does not change the original out-of-equilibrium entropy current~\eqref{canonicaldissipativeentropycurrent} in the canonical formulation. Therefore, the divergence of the out-of-equilibrium improved canonical entropy current $\partial_{\mu}\widetilde{S}^{\prime\mu}$ reads, 
\begin{align} \partial_{\mu}\widetilde{{S}}^{\prime \mu}_{\text{can}} = & -\beta h^{\mu}\left(\beta \nabla_{\mu}T-Du_{\mu}\right)+\beta\pi^{\mu\nu}\sigma_{\mu\nu}+\beta\Pi \theta-\beta q^{\mu}\left (\beta \nabla_{\mu}T+Du_{\mu}-4 \omega_{\mu\nu}u^{\nu}\right)\nonumber\\ &+\phi^{\mu\nu}\left(\Omega_{\mu\nu}+2\beta \Delta^{\alpha}_{~\mu}\Delta^{\beta}_{~\nu}\omega_{\alpha\beta}\right). \label{}
\end{align}
Note that, due to the use of an improved energy-momentum tensor, unlike Eq.~\eqref{finalentropyproductioncanonial}, this result is free of the potentially problematic terms. If we impose the second law of thermodynamics, we can identify the dissipative currents as in Eqs.~\eqref{canonicalheatflux}-\eqref{canonicalphi}, but without the constraint~\eqref{additionalconstriant}.
Therefore, the improved version of the canonical framework is well defined and the dissipative currents can be uniquely expressed by the hydrodynamic variables, i.e., $T$, $u^{\mu}$ and $\omega^{\mu\nu}$.
%
\subsection{Connection between improved canonical and phenomenological formulation}
In this section, we demonstrate that it is possible to explicitly construct a pseudo-gauge transformation that directly connects the improved canonical formulation, described by the tensors $\{\widetilde{T}^{\mu\nu}_{\text{can}}, S^{\mu\alpha\beta}_{\text{can}}\}$ in Eqs.~\eqref{improvedcanonicalEMT}-\eqref{improvedspincanonicalEMT}, and the phenomenological framework constructed with the pair $\{{T}^{\mu\nu}_{\text{ph}}, S^{\mu\alpha\beta}_{\text{ph}}\}$ in Sec.~\ref{Dissipative currents and transport coefficients : entropy-current analysis}, denoted here with the label ``ph'' for clarity. 
\medskip

Recalling the general form of the pseudo-gauge transformation introduced in Sec.~\ref{Pseudo-gauge transformation concept}, we define the phenomenological tensors in terms of the improved canonical tensors as follows
\begin{align} 
& T^{\mu\nu}_{\text{ph}}=\widetilde{T}^{\mu\nu}_{\text{can}}+\frac{1}{2}\partial_{\lambda}\left(\Sigma^{\lambda\mu\nu}-\Sigma^{\mu\lambda\nu}-\Sigma^{\nu\lambda\mu}\right),\label{EMTphcanpseudo-gauge}\\
& S^{\lambda\mu\nu}_{\text{ph}}=S^{\lambda\mu\nu}_{\text{can}}-\Sigma^{\lambda\mu\nu},\label{Spinphcanpseudo-gauge}
\end{align}
where the pseudo-gauge potential $\Sigma$ has the form
\begin{align} \Sigma^{\lambda\mu\nu}=2\Phi^{\lambda\mu\nu}_{\text{can(0)}}+\Sigma_{(1)}^{\lambda\mu\nu},
\label{equ30pap2}
\end{align}
and both $\Sigma^{\lambda\mu\nu}$ and $\Sigma^{\lambda\mu\nu}_{(1)}$ are antisymmetric only in the last two indices.
Here, we recall that $\Phi^{\lambda\mu\nu}_{\text{can}(0)}=u^{[\mu}S^{\nu]\lambda}$. Therefore, 
\begin{align} S^{\lambda\mu\nu}_{\text{ph}} & = u^{\lambda}S^{\mu\nu}+S_{\text{ph(1)}}^{\lambda\mu\nu} \label{}
\end{align}
where $S_{\text{ph(1)}}^{\lambda\mu\nu}=S^{\lambda\mu\nu}_{\text{can}(1)}-\Sigma_{(1)}^{\lambda\mu\nu}$, and
\begin{align} T^{\mu\nu}_{\text{ph}} & = \widetilde{T}^{\mu\nu}_{\text{can}}-\partial_{\lambda}(u^{\nu}S^{\mu\lambda})  = T^{\mu\nu}_{(0)}+T^{\mu\nu}_{\text{can}(1)}  =T^{\mu\nu}_{(0)}+T^{\mu\nu}_{\text{ph}(1)}. \label{equ32pap2}
\end{align}
We observe that the transformed currents \( \{T^{\mu\nu}_{\text{ph}}, S^{\mu\alpha\beta}_{\text{ph}}\} \) have the same forms as those used in Sec.~\eqref{Dissipative currents and transport coefficients : entropy-current analysis}. Hence, we anticipate that an entropy current analysis will yield identical dissipative currents and transport coefficients as those presented in Eqs.~\eqref{phenomenologicalheatflux}-\eqref{phenomenologicalphi}. Consequently, we conclude that starting from the improved canonical formulation of spin hydrodynamics, it is possible to recover the phenomenological one.
\medskip

Recall that, to achieve a well-defined description of dissipative canonical spin hydrodynamics, we have improved the canonical energy-momentum tensor by adding a total divergence $\partial_{\lambda}(u^{\nu}S^{\mu\lambda})$ to the standard dissipative part, as given in Eq.~\eqref{improvedcanonicalEMT}. Here, we explicitly show that such a term can arise when considering a pseudo-gauge transformation that originates from the phenomenological forms $\{T^{\mu\nu}_{\text{ph}}, S^{\mu\alpha\beta}_{\text{ph}}\}$, i.e.,
\begin{align} 
& \widetilde{T}^{\mu\nu}_{\text{can}} = T^{\mu\nu}_{\text{ph}}+\frac{1}{2}\partial_{\lambda}\bigg(\Psi^{\lambda\mu\nu}-\Psi^{\mu\lambda\nu}-\Psi^{\nu\lambda\mu}\bigg),\label{}\\ 
& S^{\mu\alpha\beta}_{\text{can}}=S^{\mu\alpha\beta}_{\text{ph}}-\Psi^{\mu\alpha\beta}.\label{}
\end{align} 
By choosing the pseudo-gauge potential $\Psi^{\mu\alpha\beta}$ in the form
\begin{align}
\Psi^{\mu\alpha\beta}=S_{\text{ph}}^{\alpha\mu\beta}-S_{\text{ph}}^{\beta\mu\alpha},
\label{}
\end{align}
we obtain the following, 
\begin{align} 
\widetilde{T}^{\mu\nu}_{\text{can}}&=T^{\mu\nu}_{\text{ph}}+\partial_{\lambda}(u^{\nu}S^{\mu\lambda})\label{},\\
S_{\text{can}}^{\mu\alpha\beta}&=S_{\text{ph}}^{\mu\alpha\beta}+S_{\text{ph}}^{\beta\mu\alpha}+S_{\text{ph}}^{\alpha\beta\mu}, \label{}
\end{align}
It should be emphasized that the term $\partial_{\lambda}\left(u^{\nu}S^{\mu\lambda}\right)$ was introduced in the previous subsection in a heuristic way to obtain a well-defined description within the canonical formalism. Interestingly, such a term is obtained here by applying the pseudo-gauge transformation to the phenomenological framework. 
%
%
\section{Linear mode analysis}
\label{stability}
Stability and causality are fundamental criteria for validating hydrodynamic evolution equations before their numerical implementation. In this section, we begin by reviewing some key concepts of linear mode analysis -- a technique that can be used to study stability and causality -- and provide a general discussion of the basic principles underlying it. With these preliminaries, we apply this method to the relativistic Navier-Stokes evolution equations with spin tensor presented in Table~\ref{tab:explicit relativistic Navier-Stokes evolution equations with spin current} to study their linear stability. Calculations are performed in the $h^{\mu} = 0$ case, also known as the Landau frame, and in the rest frame of the fluid element in the low-wavenumber limit. The first part of this section is loosely based on Refs.~\cite{Hiscock:1985zz,Hiscock:1987zz} and references therein. The rest of the section is based on~\multimyref{AD2,AD6}.

%
\subsection{Theoretical preliminaries}
To physically and technically address stability and causality criteria one commonly employs linear mode analysis~\footnote{Note that linear mode analysis assesses stability and causality within the linear regime of perturbations. However, for a more comprehensive understanding, a non-linear mode analysis should also be considered~\cite{Bemfica:2019knx}. Moreover, an alternative approach to addressing stability and causality was discussed in~\cite{Gavassino:2021kjm} and related works.
Nonetheless, these approaches are beyond the scope of this thesis and are considered potential directions for future work.}. In the following, we briefly describe the linear mode analysis scheme and show the criteria for stability and causality of the hydrodynamic equations. One usually proceeds as follows:
\begin{enumerate}
\item One starts by deriving the equations governing the evolution of linear perturbations of the dynamical variables with respect to their equilibrium values (also denoted as \emph{background}). For example, let \( X \) denote a dynamical variable which satisfies
\begin{align}
\partial_{t}X(t,\boldsymbol{x})-D_X \,\partial_{i}^{2}X(t,\boldsymbol{x})=0,
\end{align}
where $D_X$ is a constant and $i=1,2,3$. A linear perturbation of \( X \) with respect to a global equilibrium background reads,
\begin{align}
    X(t,\x) \longrightarrow X^\prime(t,\x)=X_{(0)} + \delta X(t,\x),
\end{align}
where \( X_{(0)} \) denotes the equilibrium value of \( X \) and \( \delta X(t,\x) \) represents a small deviation from it. The equation governing the dynamics of $\delta X(t,\x)$ then reads, 
\begin{align}\label{linearPDE}
\partial_{t}\delta X(t,\x)- D_X \partial_{i}^{2}\delta X(t,\x)=0,
\end{align}
which is a linear partial differential equation that we will refer to as \emph{perturbed equation}.
\item A \emph{plane-wave solution} ansatz of the perturbed equation is of the form,
\begin{align}
\delta X=\widetilde{\delta{X}} e^{-i\omega t+i\k\cdot\x},\label{PlanewaveHiscock}
\end{align}

where $\widetilde{\delta X}$ is constant in time and space, $\omega$ is the angular frequency, and $\k$ is the wave vector~\footnote{The proposed solution in Eq.~\eqref{PlanewaveHiscock} geometrically represents a plane wave propagating along the direction of the wave vector. However, it is also possible to consider a plane wave propagating opposite to  direction of the wave vector.}. Plugging the plane-wave solution~\eqref{PlanewaveHiscock} into Eq.~\eqref{linearPDE} yields the \emph{dispersion relation}, which defines the required connection between \( \omega \) and \( \k \), $\omega(\k)$, for the proposed solution to satisfy the perturbed equation. The dispersion relation $\omega(\k)$ is also referred to as a \emph{mode}.

\item For a more general solution ansatz, \( \delta X(t, \x) \) can be expressed as a superposition of plane waves or a \emph{wave packet}
\begin{align}\label{generalsolution}
\delta X(t, \x) \sim \int_{-\infty}^{+\infty} \int_{-\infty}^{+\infty} d\omega \, d\k \, \widetilde{ \delta X}(\omega, \k) e^{-i(\omega t - \k\cdot\x)}.
\end{align}
Similar to the plane wave solution, plugging the general solution ansatz into the perturbed equation allows us to express the differential perturbed equation as algebraic equation in Fourier space. This provides the required dispersion relation so that the proposed solution satisfies the perturbed equation.  

In case the perturbed hydrodynamic equations form a system of coupled linear differential equations, plugging the general solution~\eqref{generalsolution} gives back a system of algebraic equations that can be expressed in the matrix form,
\begin{equation}\label{matrixalgebraic}
\textbf{M}\,{\textbf{V}} = 0,
\end{equation}
where the vector \( {\textbf{V}} \) consists of the perturbed dynamical variables and the matrix $\textbf{M}$ depends on \( \omega \), \( \k \), and other coefficients specific to the equations under study. The nonzero solutions exist if and only if,
\begin{align}
    \text{det}(\textbf{M})=0.
\end{align}
%
The eigenvalues of the matrix $\textbf{M}$ are the dispersion relations. 
\item Linear mode analysis can be performed in various settings:  
\begin{itemize}  
\item Reference frame: The analysis can be carried out either in the rest frame of the fluid element or in a boosted frame.  
\item Wavenumber regimes: The study may consider different scales, including exact (all \(\k\)), low-\(\k\) , finite-\(\k\) , or high-\(\k\). The classification of these regimes depends on the complexity of the perturbed equations.  
\item Hydrodynamic frame: The choice of frame whether generic, Landau-Lifshitz, or Eckart, as described in Sec.~\ref{Dissipative currents and transport coefficients : entropy-current analysis}, also influences the analysis.  
\end{itemize}  
%
\item Naturally, the obtained dispersion relations should be constrained by physical requirements, which include:
\begin{itemize}
\item \textbf{Stability:} All perturbations of the dynamical variables must not grow exponentially with time. Therefore, we say that the perturbed hydrodynamic equations are \emph{stable} if and only if the imaginary part of each of the dispersion relations is negative, 
\begin{align}\label{Stabilitydef}
\mbox{Im}[\omega(\k)]<0.
\end{align}
\item \textbf{Causality:} All perturbations of the dynamical variables must not exceed the speed of light. Hence, we say that the perturbed hydrodynamic equations are \emph{causal} if and only if each dispersion relation satisfies: 
\begin{align}\label{Causalitydef}
\lim_{\k \to \infty} \bigg|\mbox{Re}\frac{w}{\k}\bigg|\leq 1~~\textrm{and}~~\lim_{\k \to \infty} \bigg|\frac{w}{\k}\bigg|~\text{is bounded.}
\end{align}
In the literature, the above formulas are also referred to as constraints for \emph{asymptotic causality}~\cite{Pu:2009fj, Xie:2023gbo}. Some comments on the above formulas defining causality are in place:

\textbf{(i)} Notice that we are using phase velocity and not group velocity, although the perturbations are wave packets (see Eq.~\eqref{generalsolution}). The reason is that in the high-$\k$ limit, the real parts of the dispersion relations approach a constant multiple of $\k$, and hence, both formulas are equivalent in that limit. 

\textbf{(ii)} If the system under consideration can be solved at exact-$\k$, then the above formula reduces to the standard one for group velocity. 

\textbf{(iii)} The second condition should be satisfied as pointed out in seminal work~\cite{KK} and then illustrated in Ref.~\cite{Xie:2023gbo}. Broadly speaking, this condition was introduced to account for all dispersion relations, including cases where they are purely imaginary. This is because $\omega$ can lead to an infinite propagation speed, even if it is purely imaginary, unless it satisfies the second condition. A straightforward example of this is the acausal nonrelativistic diffusion equation,
\begin{align}\label{diffusiontype}
\partial_t n - D_n \partial_{x}^2 n = 0,
\end{align}
where \( D_n \) represents the diffusion constant. By deriving its dispersion relation, one finds 
\begin{align}
\omega \sim i D_n \k^2,
\end{align}
which satisfies the first condition but does not fulfill the second condition in Eq.~\eqref{Causalitydef}. 

\textbf{(iv)} Finally, 
to avoid tedious causality calculations, it is sometimes sufficient to inspect the mathematical structure of the hydrodynamic equations. If they are parabolic (meaning the highest-order time derivative is first-order while the highest-order space derivative is second-order), then they will lead to acausal propagation of signals. A typical example is the nonrelativistic diffusion equation~\eqref{diffusiontype}.
\end{itemize}
\item For a given hydrodynamic frame, the investigation starts by checking stability in the fluid element rest frame at low-$\k$, followed by the finite-$\k$ regime, and finally at high-$\k$, where causality is also assessed. If the perturbations are unstable or acausal in the rest frame, further analysis in a boosted frame becomes unnecessary. However, if they are stable and causal in the rest frame, additional analysis in the boosted frame is required to draw a comprehensive conclusion.
\end{enumerate}
\medskip 

%
%
%
\subsection{Stability analysis: rest-frame low-wavenumber limit}
\label{Stability application: rest frame low-wavenumber limit}
The main goal of this section is to use the linear mode analysis reviewed in the previous section to check the stability of the relativistic Navier-Stokes evolution equations with spin tensor shown in Table~\ref{tab:explicit relativistic Navier-Stokes evolution equations with spin current}. The analysis will be conducted in the Landau frame, where $h^{\mu} = 0$, and in the fluid element's rest frame, in the low-wavenumber limit. It is important to note that a stability analysis of this formalism was also conducted in Ref.~\cite{Hattori:2019lfp}; however, our final results and conclusions differ, as we identify issues related to sign inconsistencies~\multimyref{AD2}.
\medskip

Following the scheme sketched above, we start by perturbing the dynamical variables in Table~\ref{Dynamical variables in relativistic spin hydrodynamics} on top of non-rotating homogeneous global equilibrium~\footnote{It is important to point out that the above perturbation scheme represents a particular choice, where the background three-velocity and the spin density are vanishing. In general, other choices, such as global equilibrium with finite thermal vorticity ~\cite{Becattini:2012tc} or other inhomogeneous global equilibrium configurations~\cite{Shokri:2023rpp}, are possible leading to different starting points of the analysis and, possibly, different conclusions.} [denoted with the label $(0)$]:    
\begin{align}
&\varepsilon\,(t,\x)\longrightarrow\varepsilon_{(0)}+\delta\varepsilon\,(t,\x),\label{eperturbationss}\\
&u^{\mu}\,(t,\x)\longrightarrow u^{\mu}_{(0)}+(0,\delta\v),\\
&S^{\mu\nu}\,(t,\x)\longrightarrow 0+\delta S^{\mu\nu}\,(t,\x),\label{Sperturbationss}
\end{align}
:where $u^{\mu}_{(0)}=(1,0,0,0)$. For the other hydrodynamic variables, we consider perturbations given by
\begin{align}
& P\longrightarrow P_{(0)}+\delta P,\\
& \omega_{\mu\nu}\longrightarrow 0+\delta\omega_{\mu\nu}.\label{omegaperpp}
\end{align}

To proceed, we need to calculate how the dissipative currents behave under such a global equilibrium configuration. The dissipative currents \( h^{\mu} \) and \( q^{\mu} \) as given in Eqs.~\eqref{phenomenologicalheatflux} and~\eqref{phemenologicalq}, can be further simplified using the relativistic Euler equation for \( u^{\mu} \), \( (\varepsilon+P)Du^{\alpha}-\nabla^{\alpha}P=0 \), see Table~\ref{Relativistic evolution equations in local equilibrium}. Using it one gets:
\begin{align}
h^{\mu} & = 0+\mathcal{O}(\partial^2),\\
q^{\mu} & =\lambda\bigg(2\frac{\nabla^{\mu}P}{\varepsilon+P}-4\omega^{\mu\nu}u_{\nu}\bigg)+\mathcal{O}(\partial^2).
\end{align}
Therefore, at the first order in gradients, the heat flux $h^{\mu}$ can be neglected. Moreover, the background part of various dissipative currents in Eqs.~\eqref{phenomenologicalheatflux}-\eqref{phenomenologicalphi} vanish, 
\begin{align}
\pi^{\mu\nu}_{(0)}=0~,~~ \Pi_{(0)}=0~,~~q^{\mu}_{(0)}=0~,~~\phi^{\mu\nu}_{(0)}=0.
\end{align}
Recall that $\pi^{\mu\nu}$, $\Pi$, $q^{\mu}$ and $\phi^{\mu\nu}$ are of first-order in hydrodynamic gradient expansion, and hence we consider $\delta\pi^{\mu\nu}$, $\delta \Pi$, $\delta q^{\mu}$ and $\delta \phi^{\mu\nu}$ up to second order in gradients and we neglect all higher order terms. Their respective perturbations read:
\begin{align}
    & \delta\pi^{\mu\nu}= \eta\bigg(\Delta^{\mu\alpha}_{(0)}\partial_{\alpha}\delta u^{\nu}+\Delta^{\nu\alpha}_{(0)}\partial_{\alpha}\delta u^{\mu}-\frac{2}{3}\Delta^{\mu\nu}_{(0)}\Delta^{\alpha\beta}_{(0)}\partial_{\beta}\delta u_{\alpha}\bigg),\\
    & \delta \Pi=\zeta (\partial_{\alpha}\delta u^{\alpha}),
    \label{}\\
    &  \delta q^{\mu}= \lambda\bigg(2\frac{\Delta^{\mu\alpha}_{(0)}\partial_{\alpha}\delta P}{\varepsilon_{(0)}+P_{(0)}}-4\delta \omega^{\mu\nu}u_{\nu}^{(0)}\bigg),
     \label{deltaqmu}\\
    &  \delta\phi^{\mu\nu}= \widetilde{\gamma}\bigg(\Delta^{\mu\alpha}_{(0)}\partial_{\alpha}\delta u^{\nu}-\Delta^{\nu\alpha}_{(0)}\partial_{\alpha}\delta u^{\mu}+4\Delta^{\mu}_{(0)\rho}\Delta^{\nu}_{(0)\lambda}\delta \omega^{\rho\lambda}\bigg).
\end{align}
such that $\delta \pi^{0i}=0$, $\delta q^{0}=0$, $\nabla_{(0)}^{\mu} = \Delta^{\mu\nu}_{(0)} \partial_{\nu}$ and $\delta \phi^{0i}=0$. Here, $\Delta^{\mu\nu}_{(0)}=g^{\mu\nu}_{(0)}-u^{\mu}_{(0)}u^{\nu}_{(0)}$ and $\widetilde{\gamma}=\beta\gamma/2$ where $\gamma$ is defined in Eq.~\eqref{phenomenologicalphi}.
\medskip 

%
Using Eq.~\eqref{deltaqmu}, the nonvanishing components of $\delta q^\mu$ can be further expressed as
\begin{align}
    \delta q^{i}  & = \lambda\bigg(2\frac{\nabla^{i}_{(0)}\delta P}{\varepsilon_{(0)}+P_{(0)}}-4\delta \omega^{i\nu}u_{\nu}^{(0)}\bigg)+\mathcal{O}(\partial^3)\nonumber\\
     & = \lambda^{\prime}c_s^2\partial^i\delta\varepsilon-\frac{4\lambda}{\chi_b}\delta S^{i0}+\mathcal{O}(\partial^3)\nonumber\\
     & = \lambda^{\prime}c_s^2\partial^i\delta\varepsilon-D_b\delta S^{i0}+\mathcal{O}(\partial^3).
     \label{}
\end{align}
Above, we used $\chi_{b} = {\partial S^{i0}}/{\partial \omega^{i0}}$ which along with $\chi_{s} = {\partial S^{ij}}/{\partial \omega^{ij}}$ originate from a relation between spin density and spin potential referred to as spin equation of state (to be discussed in detail in the next subsection).
We also defined the quantities $D_b = {4\lambda}/{\chi_b}$ and $\lambda' = {2\lambda}/({\varepsilon_{(0)} + P_{(0)})}$. Furthermore, we have utilized the definition of the (squared) speed of sound in a relativistic fluid, $c_{s}^{2}=\frac{\partial P}{\partial \varepsilon}$, which provides the relation between the energy density and the pressure.

For notational clarity, we introduce the variables $\delta \pi^{i}\equiv\delta T^{0i}$ such that $\delta T^{0i}$ can be calculated using the form of the energy-momentum tensor 
\begin{align}
     \delta\mathfrak{\pi}^i & =(\varepsilon_{(0)}+P_{(0)})\delta u^i-\lambda^{\prime}c_s^2\partial^i\delta\varepsilon+D_b\delta S^{i0}+\mathcal{O}(\partial^3).
\end{align}
By linearizing the evolution equations in Table~\ref{tab:explicit relativistic Navier-Stokes evolution equations with spin current} with respect to the perturbations in Eqs.~\eqref{eperturbationss}-\eqref{omegaperpp}, we obtain:
\begin{align}
&\partial_0\delta\varepsilon+\partial_i\delta\pi^i+2\bigg(\lambda^{\prime}c_s^2\partial_i\partial^i\delta\varepsilon-D_b\partial_i\delta S^{i0}\bigg)=0,\label{e-perturb-1}\\
&  \partial_0\delta\pi^i-c_s^2\partial^i\delta\varepsilon+(\gamma_{\perp}+\gamma^{\prime})(\delta^i_{~j}\partial^{k}\partial_k-\partial^i\partial_j)\delta\pi^j+\gamma_{||}\partial^i\partial_k\delta\pi^k+D_s\partial_k\delta S^{ki}=0,\label{u-perturb-1}\\
& \partial_{0}\delta S^{0i}= 2\lambda^{\prime}c_s^2\partial^i\delta\varepsilon-2D_b\delta S^{i0}+\mathcal{O}(\partial^3),\label{S0i-perturb-1}\\
&\partial_0\delta S^{ij}= -2D_s\delta S^{ij}-2\gamma^{\prime}(\partial^i\delta\pi^j-\partial^j\delta\pi^i)+\mathcal{O}(\partial^3),\label{Sij-perturb-1}
\end{align}
where $\delta^{i}_{~j}$ is the Kronecker delta, and we have used the following notation, 
\begin{align}\label{notationinstability}
    & 
    D_s\equiv \frac{4\widetilde{\gamma}}{\chi_s},~~\gamma^{\prime}\equiv\frac{\widetilde{\gamma}}{\varepsilon_{(0)}+P_{(0)}},
    ~~\gamma_{||}\equiv\frac{1}{\varepsilon_{(0)}+P_{(0)}}\left(\zeta+\frac{4}{3}\eta\right),~~\gamma_{\perp}\equiv \frac{\eta}{\varepsilon_{(0)}+P_{(0)}}.
\end{align}
We can notice from the perturbed equations that, in this framework, the perturbations of the standard hydrodynamic variables (\(\delta\varepsilon\), \(\delta u^i\)) are coupled with the spin density perturbation (\(\delta S^{0i}\), \(\delta S^{ij}\)).
\medskip

%
Using the ansatz from Eq.~\eqref{generalsolution}, we can express the perturbed dynamical variables as wave packets, allowing us to rewrite the perturbed differential equations~\eqref{e-perturb-1}-\eqref{Sij-perturb-1} as algebraic equations in Fourier space:
\begin{align}
& -i\omega\widetilde{\delta\varepsilon}+ik_{z}\widetilde{\delta\pi^z}+2\lambda^{\prime}c_s^2k_{z}^2\widetilde{\delta\varepsilon}+2D_bik_{z}\widetilde{\delta S^{0z}}=0,\label{number1}\\
& -i\omega\widetilde{\delta\pi^x}+(\gamma_{\perp}+\gamma^{\prime})k_{z}^2\widetilde{\delta\pi^x}+ik_{z}D_s\widetilde{\delta S^{zx}}=0,\label{}\\
& -i\omega\widetilde{\delta\pi^y}+(\gamma_{\perp}+\gamma^{\prime})k_{z}^2\widetilde{\delta\pi^y}+ik_{z}D_s\widetilde{\delta S^{zy}}=0,\label{}\\
& -i\omega\widetilde{\delta\pi^z}+ik_{z} c_s^2\widetilde{\delta\varepsilon}+\gamma_{||}k_{z}^2\widetilde{\delta\pi^z}=0,\label{equ33ver2}\\
& -i\omega\widetilde{\delta S^{xy}}+2D_s\widetilde{\delta S^{xy}}=0,\label{equ34ver2}\\
& -i\omega\widetilde{\delta S^{zx}}+2D_s \widetilde{\delta S^{zx}}-2\gamma^{\prime}(ik_{z})\widetilde{\delta \pi^x}=0,\label{equ35ver2}\\
& -i\omega\widetilde{\delta S^{yz}}+2D_s \widetilde{\delta S^{yz}}+2\gamma^{\prime}(ik_{z})\widetilde{\delta \pi^y}=0,\label{equ36ver2}\\
& -i\omega\widetilde{\delta S^{0x}}-2D_b\widetilde{\delta S^{0x}}=0,\label{equ37ver2}\\
& -i\omega\widetilde{\delta S^{0y}}-2D_b\widetilde{\delta S^{0y}}=0,\label{equ38ver2}\\
& -i\omega\widetilde{\delta S^{0z}}+2\lambda^{\prime}c_s^2(ik_{z})\widetilde{\delta\varepsilon}-2D_b\widetilde{\delta S^{0z}}=0.\label{equ39ver2}
\end{align}
Without loss of generality, according to the rotational symmetry of the system, above we considered $\k=(0,0,k_{z})$ . To determine the dispersion relations, we may express the algebraic equations for perturbations in the matrix form \eqref{matrixalgebraic} where $\mathbf{M}$ is a $10 \times 10$ matrix that depends on $\omega$ and $k_{z}$,
\begin{equation}
\resizebox{\textwidth}{!}{$
\mathbf{M}=
\begin{pmatrix}
i {k_{z}} & 2 i D_b k_{z} & 2 c_{s}^{2}\lambda^{'}k_{z}^2-i\omega  & 0 & 0 & 0 & 0 & 0 & 0 & 0 \\
 \gamma_{\parallel}{k_{z}}^2-i\omega  & 0 & i c_{s}^{2}k_{z} & 0 & 0 & 0 & 0 & 0 & 0 & 0 \\
 0 & -2 D_{b}-i \omega  & 2 i c_{s}^{2}\lambda^{'}{k_{z}} & 0 & 0 & 0 & 0 & 0 & 0 & 0 \\
 0 & 0 & 0 & iD_{s} k_{z} & k_{z}^2 (\gamma_{\perp}+\gamma^{\prime})-i \omega  & 0 & 0 & 0 & 0 & 0 \\
 0 & 0 & 0 & 2D_{s}-i \omega  & -2 i \gamma^{'}k_{z} & 0 & 0 & 0 & 0 & 0 \\
 0 & 0 & 0 & 0 & 0 & iD_{s}k_{z} & k_{z}^2 (\gamma _{\perp}+\gamma^{'})-i \omega  & 0 & 0 & 0 \\
 0 & 0 & 0 & 0 & 0 & 2 D_{s}-i \omega  & -2 i \gamma^{\prime}k_{z} & 0 & 0 & 0 \\
 0 & 0 & 0 & 0 & 0 & 0 & 0 & 2 D_{s}-i \omega  & 0 & 0 \\
 0 & 0 & 0 & 0 & 0 & 0 & 0 & 0 & -2 D_{b}-i \omega  & 0 \\
 0 & 0 & 0 & 0 & 0 & 0 & 0 & 0 & 0 & -2 D_{b}-i \omega  \\
\end{pmatrix}
$},
\end{equation}
and $\mathbf{V}$ is a vector of perturbations
\begin{align*}
{\textbf{V}}=\left(\widetilde{\delta\varepsilon},\widetilde{\delta\pi^x},\widetilde{\delta\pi^y},\widetilde{\delta\pi^z},\widetilde{\delta S^{0x}},\widetilde{\delta S^{0y}},\widetilde{\delta S^{0z}},\widetilde{\delta S^{xy}},\widetilde{\delta
S^{zy}},\widetilde{\delta S^{zx}}\right)^{\intercal}.
\end{align*}
As discussed before, nonzero solutions for the resulting matrix equation for $\textbf{V}$ exist if and only if $\text{det}(\textbf{M})=0$. Moreover, the block diagonal structure of the matrix simplifies calculations by allowing the calculation of the determinant of each block independently. Nevertheless, solving the determinant for the exact wavenumber is complicated. Instead, in the following, we limit ourselves to finding solutions in the low-wavenumber limit only. In this case, we obtain the following dispersion relations~\cite{Daher:2022wzf}:
\begin{align}
&\omega=-i\gamma_{\perp}k_{z}^{2}+\mathcal{O}(k_{z}^{4}) \quad(\text{two modes}),\label{shearmode}\\
    &\omega=+c_{s}k_{z}-\frac{i}{2}\gamma_{\parallel}k_{z}^{2}+\mathcal{O}(k_{z}^{3}),\label{sound+}\\
    &\omega=-c_{s}k_{z}-\frac{i}{2}\gamma_{\parallel}k_{z}^{2}+\mathcal{O}(k_{z}^{3}),\label{sound-}\\
    &\omega=-2iD_{s},\label{mode1}\\
    &\omega=+2iD_{b} \quad (\text{two modes}),\label{wrong1}\\
    &\omega=-2iD_{s}-i\gamma^{\prime}k_{z}^{2}+\mathcal{O}(k_{z}^{4})~\quad (\text{two modes}),\label{Ds2}\\
&\omega=2iD_{b}-2ic_{s}^{2}\lambda^{'}k_{z}^{2}+\mathcal{O}(k_{z}^{4})\label{wrong2}.
\end{align}
The first four dispersion relations in Eqs.~\eqref{shearmode}-\eqref{sound-} are the shear and sound modes present also in conventional relativistic hydrodynamics~\footnote{The classification of the modes into specific channels will be more clear in Sec.~\ref{Lmastabilityandcausality} where the calculation of dispersion relations will be performed explicitly employing the block structure of the characteristic matrix.}. In other words, given the same starting conditions, if we apply linear mode analysis for relativistic Navier-Stokes's equations without spin tensor in Table~\ref{tab:explicitRelativisticNavier-Stokesevolutionequations}, we would obtain the first four dispersion relations. The remaining six dispersion relations arise solely due to the presence of the spin tensor.
\medskip 
%
%

To analyze the stability, we evaluate the criterion from Eq.~\eqref{Stabilitydef} across all dispersion relations. To guarantee the stability of the entire framework, we demand that all the dispersion relations satisfy the constraint
\begin{align}
\mbox{Im}[\omega(k_{z})]<0.  
\end{align}
From Eqs.~\eqref{wrong1} and \eqref{wrong2}, it is evident that unstable modes emerge when \( D_b > 0 \). Throughout the calculations, \( \chi_b \) and \( \chi_s\) are assumed to be positive, which implies \( D_b>0 \) and \( D_s>0 \). Even if \( D_b \) and \( D_s \) were both assumed to be negative, the instability would still manifest, as indicated by Eqs.~\eqref{mode1} and \eqref{Ds2}. The only viable solution to remove the instability is to choose \( D_b<0 \) and \( D_s>0 \). However, one has to point out that the conclusions drawn in the above analysis hold if relation between the spin density and the spin potential has the following form (to be discussed in more detail in the next subsection),
\begin{align}
S^{\mu\nu} (T,\mu,\omega^{\mu\nu}) \sim S(T,\mu)\, \omega^{\mu\nu}.\label{StabilityspinEOSone}
\end{align}
This form (albeit the simplest nontrivial one) implies that \( \chi_b \) and \( \chi_s\)  and consequently \( D_b \) and \( D_s \) share the same sign, since the proportionality constant in the above relation should not alternate its sign depending on which components of the spin equation of state are under consideration. 
\medskip

In conclusion, we have shown that the perturbed Navier-Stokes equations~\eqref{e-perturb-1}-\eqref{Sij-perturb-1} with the spin equation of state of the form~\eqref{StabilityspinEOSone} exhibit unstable modes in Eqs.~\eqref{wrong1} and \eqref{wrong2}. A similar conclusion was found in Refs.~\cite{Xie:2023gbo,Sarwar:2022yzs}. 
In the next section, we will show that modifying the spin equation of state resolves the problem with unstable modes in the rest frame low-$k_{z}$ limit. 
%
%
\subsection{Stability constraints for the spin equation of state}
\label{Spin equation of state section}
In this section, we first discuss various forms of the spin equation of state. We then propose a specific spin equation of state, motivated by kinetic theory considerations~\cite{DeGroot:1980dk}, which resolves the particular stability issues discussed in the previous subsection.
The material discussed in this section is largely based on Ref.~\multimyref{AD6}.
\medskip

As discussed in Sec.~\ref{Relativistic fluid with spinnn}, the spin equation of state, \( S^{\mu\nu}(T, \mu, \omega^{\mu\nu}) \), is a necessary ingredient to close the set of hydrodynamic equations. As in the case of physical systems without spin, its particular form may be established when the underlying microscopic theory is specified. Until recently, such a theory was missing, and finding a correct form of the spin equation of state is currently a subject of intense research. Regardless of whether such a form can be derived from basic principles, one can attempt to construct it by referring to general requirements that it must fulfill, such as dimensional analysis.

In the previous section, we used the spin equation of state~\eqref{StabilityspinEOSone} as a linear function of the spin potential,
\begin{align}\label{XuEOS}
S^{\mu\nu}\,(T,\mu, \omega^{\mu\nu}) \sim S(T,\mu)\,\omega^{\mu\nu},
\end{align}
where \( S(T,\mu) \) is an unknown scalar function of temperature and chemical potential. In component form, this relation can be generally expanded as follows:
\begin{align}
    S^{i0}\,(T,\mu, \omega^{i0}) &= \chi_{b}(T,\mu)\,\omega^{i0} + C(T,\mu), \quad S^{ij}\,(T,\mu, \omega^{ij}) = \chi_{s}(T,\mu)\,\omega^{ij} + D(T,\mu),
\end{align}
where \( \chi_{b}(T,\mu) \) and \( \chi_{s}(T,\mu) \) represent coefficients defined as:
\begin{align}\label{sus}
\chi_b= \frac{\partial S^{i0}}{\partial \omega^{i0}}, \quad \chi_s = \frac{\partial S^{ij}}{\partial \omega^{ij}}.
\end{align}
Here, \( C(T,\mu) \) and \( D(T,\mu) \) are some functions of temperature and chemical potential. The form~\eqref{XuEOS} implies that the coefficients \( \chi_{b}(T) \) and \( \chi_{s}(T) \) share the same sign. 

In particular, in Ref.~\cite{Wang:2021ngp}, motivated by the hydrodynamic gradient expansion of the spin density, \( S^{\mu\nu} \sim \mathcal{O}(1) \), and the spin potential, \( \omega^{\mu\nu} \sim \mathcal{O}(\partial) \), it was proposed that in the high-temperature and vanishing chemical potential limit, the spin equation of state can take the following form
\begin{align}\label{ShipiEOS}
S^{\mu\nu} = a_{1}T^{2}\omega^{\mu\nu},   
\end{align}
where \( a_{1} \) is a constant. The underlying argument for this specific form is that, in the high-temperature regime, the large value of the temperature compensates for the small magnitude of \( \omega^{\mu\nu} \), which makes the gradient counting of \( S^{\mu\nu} \) and \( \omega^{\mu\nu} \) self-consistent. To maintain the consistency of the hydrodynamic gradient expansion on both sides of the equality away from the high-temperature regime, in the vanishing chemical potential limit, the following dependence can be proposed~\cite{Biswas:2022bht},
\begin{align}\label{boostinvariantEOS}
S^{\mu\nu}(T, \omega^{\mu\nu}) = S(T) \frac{\omega^{\mu\nu}}{\sqrt{\omega^{\mu\nu} \omega_{\mu\nu}}}.
\end{align}
Nevertheless, as discussed in Sec.~\ref{Stability application: rest frame low-wavenumber limit}, the form~\eqref{XuEOS} inevitably leads to the presence of unstable modes in the resulting evolution equations, making this form inadmissible in practical applications.

It turns out, that the stability issue discussed in Sec.~\ref{Stability application: rest frame low-wavenumber limit} may be simply resolved by observing that the signs of \( \chi_{b}(T,\mu) \) and \( \chi_{s}(T,\mu) \) are opposite to each other.
To see how this can be implemented, one can introduce another, more general, form of the spin equation of state, utilizing the decomposition of \( \omega^{\mu\nu} \) as a rank-2 antisymmetric tensor,  
\begin{align}\label{omegadecomp}
    \omega^{\gamma\delta} = k^{\gamma}u^{\delta} - k^{\delta}u^{\gamma} + \epsilon^{\gamma\delta\rho\sigma}u_{\rho}\omega_{\sigma},
\end{align}
where \( k \) and \( \omega \) are spacelike vectors satisfying the orthogonality conditions: \( k \cdot u = 0 \) and \( \omega \cdot u = 0 \). Substituting the decomposition~\eqref{omegadecomp} into Eq.~\eqref{XuEOS} we get, 
\begin{align}\label{ourspinEOS}
    S^{\gamma\delta}\,(T, \mu, \omega^{\gamma\delta})=S_{1}\,(T, \mu)\,(k^{\gamma}u^{\delta}-k^{\delta}u^{\gamma})+S_{2}\,(T, \mu)\, \epsilon^{\gamma\delta\rho\sigma}u_{\rho}\omega_{\sigma}.
\end{align}
where, this time, the scalar functions $S_{1}(T,\mu)$ and $S_{2}(T,\mu)$ are allowed to be different.

In the rest frame of the local fluid element, where \( u^{\lambda} = (1, \boldsymbol{0}) \), we find:  
\begin{align}
S^{i0} = S_{1}\,\omega^{i0}, \quad S^{ij} = S_{2}\,\omega^{ij}.
\end{align}
Hence, it is natural to identify  \( S_{1} = \chi_{b} \) and \( S_{2} = \chi_{s} \), as defined in Eq.~\eqref{sus}. 
\medskip 

As we saw, the primary motivation for the form~\eqref{ourspinEOS} is that it allows for \( \chi_{b} \) and \( \chi_{s} \) having different signs, a feature that is crucial in the stability analysis discussed in Sec.~\ref{stability}. However, there remains a question whether, given sensible underlying microscopic theory, the resulting scalar functions in Eq.~\eqref{ourspinEOS} indeed satisfy the conditions \( S_{1} < 0 \) and \( S_{2} > 0 \) (equivalently $\chi_{b} < 0$ and $\chi_{s}>0$). One widely used form of the spin tensor in equilibrium is the one introduced by de Groot, van Leeuwen, and van Weert in their seminal textbook on relativistic kinetic theory~\cite{DeGroot:1980dk}, hereafter referred to as the GLW form. In the case where the spin part of the angular momentum is separately conserved, the local equilibrium GLW spin tensor has the following structure~\cite{Florkowski:2019qdp}~\footnote{Indeed, the choice of such a spin tensor contradicts the narrative of this chapter, where we discussed that the spin tensor is inherently not conserved based on field theory arguments in Chapter~\ref{From fields to fluids}. However, the chosen form is merely one example, and in the future, we look forward to exploring other kinetic theory forms of the spin tensor for this specific calculation.},
\begin{equation}
S^{\alpha,\, \beta \gamma}_{\rm GLW} = A_1 u^\alpha \omega^{\beta \gamma} + A_2 u^\alpha u^{[\beta} k^{\gamma]}  
+ A_3 \left( u^{ [\beta} \omega^{\gamma ] \alpha}
+ g^{\alpha [ \beta} k^{\gamma ] }\right),
\label{eq:SGLW}
\end{equation}
where the coefficients $A_1$, $A_2$, and $A_3$ are functions of temperature and chemical potential. The  corresponding spin density tensor $S^{\beta \gamma}_{\rm GLW}=u_{\alpha}S^{\alpha\beta\gamma}_{\rm GLW}$ reads,
\begin{equation}
S^{\beta \gamma}_{\rm GLW} = \left(A_1 - \frac{A_2}{2}  -A_3 \right)
\left( k^\beta u^\gamma - k^\gamma u^\delta \right) + A_1 \epsilon^{\beta \gamma \rho \sigma} u_\rho \omega_\sigma,
\label{eq:SGLWden}
\end{equation}
hence,
\begin{equation}
S^{\alpha,\, \beta \gamma}_{\rm GLW} = u^\alpha S^{\beta \gamma}_{\rm GLW} +   A_3\left(  \Delta^{\alpha[\beta}k^{\gamma]}    +   \epsilon^{\alpha[\beta \lambda \chi} u_\lambda \omega_\chi u^{\gamma]} \right).
\label{eq:SGLW2}
\end{equation}
We note that the second term on the right-hand side of the equation above is explicitly orthogonal to $u_\alpha$; hence, it does not contribute to the thermodynamic relations~\eqref{covariantgeneralizedfirstlaw}-\eqref{covariantgeneralizedGibbsDuhem}. The coefficients $A_1, A_2$, and $A_3$ are given by the following expressions~\cite{Florkowski:2019qdp}
\begin{equation}
A_1 = C (n_0 - B_0), \quad A_2 = 2 C (n_0 - 3 B_0),  \quad A_3 = C B_0,  
\end{equation}
where $C = \cosh(\mu/T)$, $B_0 = -2 (T^2/m^2) s_0$, and  $n_0$ ($s_0$) is the equilibrium number density (entropy density) of spinless massive classical particles, and $m$ is the particle mass. 

It is trivial to see that, 
\begin{align}
    S_1 = A_1- \frac{1}{2}A_2- A_3  = C B_0 < 0~,~~S_2 = A_1 > 0.
\end{align}
Using explicit forms of the coefficients $S_1$ and $S_2$ one has
\begin{equation}\label{S11}
S_1 = - C\, \frac{T^3}{\pi^2} \left[4 K_2(x) + x K_1(x)  \right]<0
\end{equation}
and
\begin{equation}
S_2 = C\, \frac{T^3}{2 \pi^2} \left[(8+x^2) K_2(x) + 2 x K_1(x)  \right]>0 ,
\end{equation}
where $K_n(x)$'s are the modified Bessel functions of the second kind of order $n$, and $x=m/T$. Interestingly, in the ultra-relativistic limit ($m/T \ll1$), the two coefficients become exactly opposite~\footnote{The limit $m \to 0$ cannot be taken, as the formalism of Ref.~\cite{DeGroot:1980dk} is defined only for massive particles.},  
\begin{equation}
S_1 = - \frac{8 C T^5}{\pi^2 m^2} , \quad S_2=-S_1.
\end{equation}
On the other hand, in the large mass limit, we find
\begin{equation}
S_1 = - \frac{C T^3}{\pi^{3/2}} 
\sqrt{\frac{m}{2T}} e^{-m/T}, 
\quad S_2 =  \frac{C m T^2}{2 \pi^{3/2}} 
\sqrt{\frac{m}{2T}} e^{-m/T},
\end{equation}
which means that $S_1$ (boost-related) contribution is subleading.
\medskip 

One can check that replacing \eqref{XuEOS} with \eqref{ourspinEOS} does not affect the stability analysis performed in the previous subsection, except for changing the conclusions. With Eq.~\eqref{ourspinEOS}, we obtain the same structure for the dispersion relations given in Eqs.~\eqref{shearmode}-\eqref{wrong2}, with the only difference of the coefficients now satisfying the conditions $\chi_{b} < 0$ and $\chi_{s} > 0$. This, in turn, stabilizes the modes in Eqs.~\eqref{wrong1} and \eqref{wrong2}.

\medskip
In conclusion, at this stage of the analysis, we find that the perturbed Navier-Stokes equations with spin, Eqs.~\eqref{e-perturb-1}-\eqref{Sij-perturb-1}, in the rest frame low-$k_{z}$ limit and under the Landau frame choice, $h^{\mu}=0$, can be stable provided the spin equation of state has the form~\eqref{ourspinEOS}. However, to place this result in a broader context, some important comments are in place at this point:
\begin{itemize}
\item 
It is worth stressing that conclusions about stability drawn from the analysis presented above are in line with the conclusions of the early pioneering work~\cite{Hiscock:1985zz}, which demonstrated that the conventional relativistic Navier-Stokes equations without spin are stable in the fluid rest frame, assuming the Landau frame definition, but only in the low-$\k$ regime. The same analysis revealed that the theory is always unstable in both the general hydrodynamic frame and the Eckart frame. When analyzed in a boosted frame, all three hydrodynamic frame definitions lead to unstable modes. Additionally, the parabolic structure of the conventional relativistic Navier-Stokes equations in Table~\ref{tab:explicitRelativisticNavier-Stokesevolutionequations} implies that they are acausal.  
 
\item The results of Ref.~\cite{Hiscock:1985zz} also agree with those of Ref.~\cite{Xie:2023gbo}, where the stability analysis of relativistic Navier-Stokes equations with spin was extended to the high-$k_{z}$ limit, revealing that the system exhibits instabilities in the fluid rest frame, assuming the Landau frame. Moreover, as expected from the parabolic nature of the perturbed equations, acausal perturbations also emerge in the high-$k_{z}$ regime. Consequently, further analysis of stability and causality in the boosted frame is unnecessary.

\item The above-discussed results along with those presented in~\cite{Xie:2023gbo} prove that the relativistic Navier-Stokes evolution equations with spin, similarly to their conventional counterparts, are generally unstable and acausal. In the case of conventional relativistic hydrodynamics, these problems are usually overcome by extending the theory to the second-order in gradients. However, it is worth noting that the dependence of the stability on the structure of the spin equation of state revealed here is inherent to the relativistic Navier-Stokes evolution equations with spin and sustains its decisive role in the stability of second-order theory. The development of the second-order theory for systems with spin and its linear mode analysis will be the subject of study in Chapter~\ref{MISMIS}.

\item
The research on the stability and causality of relativistic hydrodynamics is currently a hot research topic. Recent results have revealed an intricate connection between these two aspects~\cite{Denicol:2008ha, Pu:2009fj, Bemfica:2020zjp, Gavassino:2021kjm}. The novelty of these findings lies in their ability to significantly reduce the complexity of calculations related to the linear mode analysis. This direction represents a compelling topic for future research.
\end{itemize}
%

\section{Solutions for systems subject to Bjorken flow}
\label{boostinvariant}

In this section, we implement the Bjorken flow expansion into the relativistic Navier-Stokes evolution equations in Table.~\ref{tab:explicit relativistic Navier-Stokes evolution equations with spin current} and solve the resulting equations numerically. The numerical results suggest that under certain conditions, the system develop nonphysical features. This observation is in line with the findings achieved within the stability analysis in previous sections. Presented results are mainly based on Ref.~\multimyref{AD3}
\medskip 

The Bjorken flow is a simple model of the expansion of matter formed in the collision of two heavy ions at extremely high energies~\cite{Bjorken:1983}. The main model assumption is that, in the center of mass frame, the system in the central rapidity region is transversely homogeneous and undergoes a boost-invariant expansion along the beam direction (the $z$-direction or the longitudinal direction). This assumption may be formulated by stating that the three-velocity of the fluid element in this region is given by:
\begin{align}
v^{z} &= \frac{z}{t}, \quad \text{for} \ |z| < t, \nonumber\\
v^{x} &= v^{y} = 0.
\end{align}
%
With the above assumption, the four-velocity (in Minkowski coordinates) reads, 
\begin{align}\label{Bjokenvelocityminkowski}
u^{\mu}=\left(u^{t},u^{x},u^{y},u^{z}\right)=\gamma\left(1,0,0,v^{z}\right)=\left(\frac{t}{\sqrt{t^{2}-z^{2}}},0,0,\frac{z}{\sqrt{t^{2}-z^{2}}}\right),
\end{align}
where, due to the normalization condition of four-velocity, $u \cdot u =1$, the Lorentz factor is $\gamma=1/\sqrt{1-(v^{z})^2}=t/\sqrt{t^{2}-z^{2}}$.

To investigate the fluid dynamics undergoing the Bjorken expansion, it is convenient to use the so-called Milne coordinate system $(\tau,x,y,\eta)$ ~\cite{milne1935relativity,Bjorken:1983} where the longitudinal proper time $\tau$ and the spacetime rapidity $\eta$ are given by the formulas
\begin{align}
\tau=\sqrt{t^{2}-z^{2}} \quad\textrm{and}\quad  \eta=\frac{1}{2}\ln\left(\frac{t+z}{t-z}\right),
\end{align}
respectively.  The inverse transformation to Minkowski coordinates $(t,x,y,z)$ is given by the expressions
%
%
%
\begin{align}\label{coordinatetransformation}
    t=\tau \cosh(\eta), \quad z=\tau\sinh(\eta),
\end{align}
Hence, the four-velocity may be expressed as
\begin{align}
    u^{\mu}&= (\cosh\eta,0,0,\sinh\eta),
\end{align} 
which, as can be easily shown, is static in Milne coordinates $u^{\mu\prime}= (1,0,0,0)$.
In Milne coordinates, the partial derivatives are, 
\begin{align}
&\partial_{t}=\cosh(\eta)\partial_{\tau}-\frac{1}{\tau}\sinh(\eta)\partial_{\eta},\\
&\partial_{z}=-\sinh(\eta)\partial_{\tau}+\frac{1}{\tau}\cosh(\eta)\partial_{\eta},
\end{align}
which allows us to express the comoving derivative  and the expansion scalar as follows 
\begin{align}\label{DUPD}
D&=u^{\mu}\partial_{\mu}=\partial_{\tau}, \qquad \theta=\partial_{\mu}u^{\mu}=\frac{1}{\tau}.
\end{align}
Consequently, the four-acceleration is
\begin{align}
Du^{\mu}=u^{\alpha}\partial_{\alpha}u^{\mu}=0,
\end{align}
while components of the spacelike gradient \( \nabla^{\mu} = \partial^{\mu} - u^{\mu} D \) read
\begin{align}
& \nabla_{0}=\partial_{0}-u_{0}D=-\frac{1}{\tau}\sinh(\eta)\partial_{\eta},\\
& \nabla_{i}=\partial_{i}-u_{i}D=\frac{1}{\tau}\cosh(\eta)\partial_{\eta}.
\end{align}
%

%
\subsection{Evolution equations}
We start implementing the Bjorken flow expansion in the relativistic Navier-Stokes equations with spin from Table~\ref{tab:explicit relativistic Navier-Stokes evolution equations with spin current} by analyzing how the dissipative currents in these equations behave subject to it. Using the formulas valid for Bjorken flow, the dissipative currents in Eqs.~\eqref{phenomenologicalheatflux}-\eqref{phenomenologicalphi} simplify to:
\begin{align}
 &h^{\mu}=0,\label{heatcurrentmilne}\\
&\pi^{\mu\nu}=\eta\left(\partial^{\mu}u^{\nu}+\partial^{\nu}u^{\mu}\right)-\frac{2\eta}{3\tau}\Delta^{\mu\nu},\label{pimunumilne}\\
&\Pi=\frac{\zeta}{\tau},\label{Pimilne}\\
&q^{\mu}=-4\lambda\omega^{\mu\nu}u_{\nu},\label{qMilne}\\
&\phi^{\mu\nu}=2\gamma\beta(\omega^{\mu\nu}+2u^{[\mu}\omega^{\nu]\beta}u_{\beta}).\label{phiMilne}
\end{align}
Here, we have used the fact that under Bjorken flow, all scalars, including thermodynamic quantities, are functions of the proper time $\tau$ only. Using decomposition of the spin potential from Eq.~\eqref{omegadecomp} in Eqs.~\eqref{qMilne}-\eqref{phiMilne}, we can further write
\begin{align}
& q^{\mu}=-4\lambda\kappa^{\mu},\label{qMilneomegak}\\
& \phi^{\mu\nu}=2\gamma\beta\epsilon^{\mu\nu \alpha \beta}u_{\alpha}\omega_\beta.
\label{phiMilneomega}
\end{align}

Since both $\kappa^{\mu}$ and $\omega_{\beta}$ are space-like four-vectors, without loss of generality, we can decompose them in the four-vector basis $(X^{\mu}, Y^{\mu}, Z^{\mu})$ which spans the space orthogonal to the four-velocity~\cite{Florkowski:2019qdp}, such that it satisfies the relation $\Delta^{\mu\nu}=-X^{\mu}X^{\nu}-Y^{\mu}Y^{\nu}-Z^{\mu}Z^{\nu}$~\cite{Martinez:2012tu} and the normalization conditions:
\begin{align}
    &u^{\mu}X_{\mu}=0, \quad u^{\mu}Y_{\mu}=0, \quad u^{\mu}Z_{\mu}=0,\\
    & X^{\mu}X_{\mu}=-1, \quad Y^{\mu}Y_{\mu}=-1, \quad Z^{\mu}Z_{\mu}=-1,\\
    & X^{\mu}Y_{\mu}=0, \quad X^{\mu}Z_{\mu}=0,\quad Y^{\mu}Z_{\mu}=0. 
\end{align} 
In the case of Bjorken flow, they may be defined as follows~\cite{Florkowski:2019qdp}:
\begin{align}
X^{\mu}&\equiv (0,1,0,0),\\
Y^{\mu}&\equiv (0,0,1,0),\\
Z^{\mu}&\equiv (\sinh\eta,0,0,\cosh\eta),
\end{align}
and, consequently, $\kappa^{\mu}$ and $\omega_{\beta}$ may be parametrized as:
\begin{align}
\kappa^{\mu} & \equiv C_{\kappa X}X^{\mu}+C_{\kappa Y}Y^{\mu}+C_{\kappa Z}Z^{\mu} =\left(C_{\kappa Z}\sinh\eta,C_{\kappa X},C_{\kappa Y},C_{\kappa Z}\cosh\eta\right),\label{}\\
\omega^{\mu} & \equiv C_{\omega X}X^{\mu}+C_{\omega Y}Y^{\mu}+C_{\omega Z}Z^{\mu}  =\left(C_{\omega Z}\sinh\eta,C_{\omega X},C_{\omega Y},C_{\omega Z}\cosh\eta\right).\label{}
\end{align}
Here, the coefficients $C_{\kappa X}, C_{\kappa Y}, C_{\kappa Z}, C_{\omega X}, C_{\omega Y}, C_{\omega Z}$ are scalar functions of the proper time $\tau$ only. Interestingly, we observe that the constitutive relation for $h^{\mu}$ in Eq.~\eqref{phenomenologicalheatflux} for the Bjorken flow $h^{\mu}$ vanishes identically. This, in turn, implies that, according to the generalized Landau condition~\eqref{generalizedLandueframe}, $q^{\mu}=0$. Therefore, from Eq.~\eqref{qMilne} we have $\omega^{\mu\nu}u_{\nu}=0$ and $\kappa^{\mu}=0$ as can be inferred from Eq.~\eqref{qMilneomegak}. Consequently, the spin potential $\omega^{\mu\nu}$ has only three independent components, as it is completely determined by the four-vector $\omega^{\mu}$. The contraction $\omega^{\mu\nu}\omega_{\mu\nu}$ thus reads 
\begin{align}
\omega^{\mu\nu}\omega_{\mu\nu}&=-2\omega^{\mu}\omega_{\mu} =  2C^{2}>0.
\label{}
\end{align}
where $C = \sqrt{C_{\omega X}^2+C_{\omega Y}^2+C_{\omega Z}^2}$ is the magnitude of the spin potential. 

We can now turn back to the analysis of the evolution equations. For the energy density evolution equation (first equation in Table~\ref{tab:explicit relativistic Navier-Stokes evolution equations with spin current}) under Bjorken flow, we obtain the following expression,
\begin{align}
    \frac{\dd\varepsilon}{\dd\tau}+\frac{\varepsilon+P}{\tau}-\frac{1}{\tau}\bigg(\frac{2}{3}\frac{\eta}{\tau}+\frac{\zeta}{\tau}\bigg)=0.
    \label{bjorkenendeneq}
\end{align}
Here, we utilize the expressions for the dissipative currents given in Eqs.~\eqref{heatcurrentmilne}-\eqref{phiMilne}. Although the current $\phi^{\mu\nu}$ explicitly appears in this equation before implementing Bjorken flow, it can be shown for the Bjorken flow that $u_{\nu}\partial_{\mu}\phi^{\mu\nu}=0$. Therefore, the antisymmetric parts of the energy-momentum tensor disappear from the evolution of the energy density. Note that $\eta/s$ and $\zeta/s$ (here $\eta$ should not be confused with the spacetime rapidity), where $s$ is the entropy density, are dimensionless quantities. In terms of these dimensionless variables, the above equation can be rewritten as
\begin{align}
\frac{\dd \varepsilon}{\dd\tau}+\frac{\varepsilon+P}{\tau}-\frac{s}{\tau^2}\bigg(\frac{2}{3}\frac{\eta}{s}+\frac{\zeta}{s}\bigg)=0.
\label{energydensityevobjor}
\end{align}
Now, let us consider the vector equation (also referred to as velocity or momentum equation) in Table~\ref{tab:explicit relativistic Navier-Stokes evolution equations with spin current}. We find that for  the Bjorken flow, this equation is trivially fulfilled due to the following properties:
\begin{align}
& \Delta^{\alpha\mu}\partial_{\mu}P=0,\\
& \Delta^{\alpha}_{~\nu}\partial_{\mu}\pi^{\mu\nu}=0,\\
& \Delta^{\alpha}_{~\nu}\partial_{\mu}(\Pi\Delta^{\mu\nu})=0,\\
& \Delta^{\alpha}_{~\nu}\partial_{\mu}\phi^{\mu\nu}=0.
\end{align}
To reproduce the above equations, one has to substitute the forms of the dissipative currents from Eqs.~\eqref{heatcurrentmilne}-\eqref{phiMilne}. 

The remaining equation to be considered is the spin density evolution equation in Table~\ref{tab:explicit relativistic Navier-Stokes evolution equations with spin current}. Under the Bjorken flow, this equation reads
\begin{align}
\frac{\partial S^{\mu\nu}}{\partial\tau}+\frac{S^{\mu\nu}}{\tau}=-2\phi^{\mu\nu}.\label{spindensitybjorkenfirst}
\end{align}
To proceed, we have to define the spin equation of state, which relates the spin density to temperature and spin potential $\omega^{\mu\nu}$. In the current analysis, we use the form already presented in Sec.~\ref{Spin equation of state section}, namely
\begin{align}\label{spinEOSdouble}
S^{\mu\nu}(T,\omega^{\mu\nu})=  S_{0}(T)\frac{\omega^{\mu\nu}}{\sqrt{\omega:\omega}},
\end{align}
where $S_{0}(T)$ is a scalar function of temperature only and $\omega:\omega\equiv\omega_{\mu\nu}\omega^{\mu\nu}$. The reasoning behind this choice is based on the equal hydrodynamic gradient order expansion given that $S^{\mu\nu}\sim \mathcal{O}(1)$ and $\omega^{\mu\nu}\sim\mathcal{O}(\partial)$. In consequence of Eq.\eqref{spinEOSdouble}, both the left-hand side and the right-hand side of Eq.~\eqref{spindensitybjorkenfirst} depend on \( \omega^{\mu\nu} \). As previously argued, since we are working in the generalized Landau frame, \( \omega^{\mu\nu} \) has three independent components. Thus, Eq.~\eqref{spindensitybjorkenfirst} can be rewritten as three differential equations governing the evolution of these three independent components of the tensor \( \omega^{\mu\nu} \):
\begin{align}
& \frac{\dd}{\dd\tau}\left(\frac{S_0}{\sqrt{2}C}C_{\omega X}\right)+\left(\frac{S_0}{\sqrt{2}C}C_{\omega X}\right)\frac{1}{\tau}=-4\beta\gamma C_{\omega X}, \label{}\\
& \frac{\dd}{\dd\tau}\left(\frac{S_0}{\sqrt{2}C}C_{\omega Y}\right)+\left(\frac{S_0}{\sqrt{2}C}C_{\omega Y}\right)\frac{1}{\tau}= -4\beta\gamma  C_{\omega Y}, \label{}\\
& \frac{\dd}{\dd\tau}\left(\frac{S_0}{\sqrt{2}C}C_{\omega Z}\right)+\left(\frac{S_0}{\sqrt{2}C}C_{\omega Z}\right)\frac{1}{\tau}=-4\beta\gamma C_{\omega Z}. \label{}
\end{align}
After several algebraic manipulations, from the above three evolution equations, we can obtain a scalar equation for the magnitude of the spin potential $C$,
\begin{align}\label{magspinpotevo}
C&=-\frac{1}{4\sqrt{2}\gamma\beta}\left(S_0^\prime(T) \frac{\dd T}{\dd\tau}+\frac{S_{0}}{\tau}\right),
\end{align}
where we defined $S_0^\prime(T)=\dd S_{0}(T)/\dd T$. 
%
%
%
%

To determine the evolution of the magnitude of the spin potential from Eq.~\eqref{magspinpotevo}, we need to specify the form of the temperature-dependent function $S_0(T)$ and the transport coefficient $\gamma$. Unfortunately, at the moment, we lack any microscopic insights into these functions. Consequently, in our calculations, we use the forms that follow from the reasoning that refers to dimensional analysis and the overall simplicity. For example, in natural units, the energy dimension of the energy density and temperature is 
\begin{align}
[\varepsilon]=[E]^{4}~,~[T]=[E].
\end{align}
Then, using the Euler thermodynamic equation~\eqref{generalizedEulerequation}, we can find the energy dimension for the pressure and entropy density as
\begin{align}
[P]=[E]^{4}~,~[s]=[E]^{3}.
\end{align}
Finally, given the fact that the total angular momentum is of energy dimension 3, so is the spin tensor and the spin density. Again, using the Euler thermodynamic relation, the energy dimension of the spin potential reads, 
\begin{align}
[\omega^{\mu\nu}]=[E]^{1}.
\end{align}
Therefore, from Eq.~\eqref{spinEOSdouble}, the constant $S_0(T)$ has energy dimension three. Hence, based solely on dimensional analysis, we propose,
\begin{align}
S_{0}(T)=\frac{\alpha}{\sqrt{2}}T^{a}M^{b}K_{n}\left(\frac{M}{T}\right),\label{s0}
\end{align}
where $a$ and $b$ are numerical constants satisfying the condition $a+b=3$ and $M$ is the particle mass. The parameter $\alpha$ is a pure numerical constant. The appearance of the modified Bessel function is rooted in the fact that later, for numerical purposes, we are going to relate $S_{0}(T)$ to entropy density and particle density. 

To determine a specific form of $\gamma$, we again use dimensional analysis. Since the energy-momentum tensor has energy dimension four, so does $\phi^{\mu\nu}$. Then, according to Eq.~\eqref{phiMilneomega}, the coefficient $\gamma$ has the same energy dimension as $\phi^{\mu\nu}$. Therefore, we propose, 
\begin{align}
\gamma=\tilde{\alpha} \, T^{c+1} M^{d} K_{m}\left(\frac{M}{T}\right)\label{gamma},
\end{align}
where $c$ and $d$ are numerical constants fulfilling \mbox{$c+d=3$}, and $\tilde{\alpha}$ is yet another pure numerical constant. 

By substituting the expressions for \( S_{0}(T) \) and \( \gamma \) into the evolution equation for the magnitude of the spin potential~\eqref{magspinpotevo}, we obtain
\begin{equation}\label{evolutionofCnewnew}
C(\tau)=C_{11}(\tau)\frac{\dd T(\tau)}{\dd\tau}+C_{12}(\tau),
\end{equation}
where
\begin{align}\label{constants1}
&C_{11} = -\frac{\alpha}{8\tilde{\alpha}}\left(\frac{T}{M}\right)^{a-c}\left(a  \frac{K_n}{T K_m}+\frac{K_n^{\prime}}{K_m}\right) 
\quad C_{12} = -\frac{\alpha}{8\tilde{\alpha}}\left(\frac{T}{M}\right)^{a-c}\frac{1}{\tau}\frac{K_n}{K_m}.
\end{align}
To improve readability, in the above equation, we skipped the argument $\frac{M}{T}$ of the Bessel functions.  

The last step to close the system of equations is to define the relation of the energy density to temperature and spin potential in Eq.~\eqref{EOSS}, which is required in Eq.~\eqref{energydensityevobjor}. From Euler thermodynamic equation~\eqref{generalizedEulerequation}, we know that $\varepsilon$ includes first-order corrections in $\omega^{\mu\nu}$. Therefore, we propose the following general relation respecting the dimensions, 
\begin{align}\label{epsilon(T,omega)}
\varepsilon(T,\omega^{\mu\nu})=\varepsilon_{0}(T)+T\,S_0^\prime(T) \sqrt{\omega:\omega},
\end{align}
where, for a massive Boltzmann gas (without the spin tensor), the explicit expression for $\varepsilon_{0}(T)$ is~\cite{Florkowski:2019qdp}
\begin{align}
\varepsilon_{0}=\frac{g_s}{2\pi^{2}}T^{4}\left(\frac{M}{T}\right)^{2}\left[3K_{2}\left(\frac{M}{T}\right)+\frac{M}{T}K_{1}\left(\frac{M}{T}\right)\right].
\end{align}
Here, $g_s$ is a factor that accounts for spin and particle-antiparticle degeneracy. 
Now, using the form of $S_{(0)}(T)$, Eq.\eqref{epsilon(T,omega)} can be written in terms of the magnitude of the spin potential
\begin{align}
\varepsilon(T,C)=\varepsilon_{0}(T)+a\alpha T^a M^b K_n^{}\left(\frac{M}{T}\right)C+\alpha T^{a+1} M^b K_n^{\prime}\left(\frac{M}{T}\right)C.
\label{}
\end{align}
Recall that both temperature $T=T(\tau)$ and magnitude of spin potential $C=C(\tau)$ are functions of the proper time $\tau$; then, by taking the derivative of the energy density with respect to the proper time $\tau$, we obtain, 
\begin{align}
\frac{\dd\varepsilon}{\dd\tau}=&\varepsilon_2C_{22}\frac{\dd^2T}{\dd\tau^2}+(\varepsilon_2C_{21}+\varepsilon_1 C_{11})\left(\frac{\dd T}{\dd\tau}\right)^2+\left(\frac{\dd\varepsilon_{0}}{\dd T}+C_{12}\varepsilon_1+C_{23}\varepsilon_2\right)\left(\frac{\dd T}{\dd\tau}\right)+\varepsilon_2 C_{24}.
\end{align}
Various temperature-dependent coefficients appearing in the above equation are given as:
\begin{align}\label{constants2}
& \varepsilon_1=a\alpha T^{a}M^{b}(aT^{-1}K_{n}+K^{\prime}_{n})+\alpha(a+1)T^a M^bK_n^{\prime}+\alpha T^{a+1}M^b K^{\prime\prime}_n,\nonumber\\
&\varepsilon_2=\alpha T^a M^b \left(a K_n+ T K_n^{\prime}\right),\nonumber\\
& C_{22} = -\frac{\alpha}{8\tilde{\alpha}}T^{a-c}M^{c-a}\left[a \frac{K_n}{T K_m}+\frac{K_n^{\prime}}{K_m}\right],\nonumber\\ 
& C_{21} = -\frac{\alpha M^{c-a}}{8\tilde{\alpha}}\bigg[a(a-c-1)T^{a-c-2}\frac{K_n}{K_m}+aT^{a-c-1}\frac{K_n^{\prime}}{K_m}-aT^{a-c-1}\frac{K_nK_m^{\prime}}{K_m^2}\nonumber\\
&~~~~~~~~~~~~~~~~~~~~~+(a-c)T^{a-c-1}\frac{K_n^{\prime}}{K_m}+T^{a-c}\frac{K_n^{\prime\prime}}{K_m}-T^{a-c}\frac{K^{\prime}_m K^{\prime}_n}{K_m^2}\bigg],\nonumber\\
& C_{23} = -\frac{\alpha}{8\tilde{\alpha}}T^{a-c}M^{c-a}\frac{1}{\tau}\left[(a-c) \frac{K_n}{T K_m}+ \frac{K_n^{\prime}}{K_m}-\frac{K_n K_m^{\prime}}{K_m^2}\right],\nonumber\\
& C_{24} = \frac{\alpha}{8\tilde{\alpha}}T^{a-c
}M^{c-a} \frac{1}{\tau^2}\frac{K_n}{K_m}.
\end{align}
For the second term in Eq.~\eqref{energydensityevobjor}, i.e., \(\varepsilon + P\), we require the expression for pressure in terms of temperature. However, from Eq.~\eqref{EOSS}, we know that pressure may be regarded as a function of energy density and spin density. Therefore, we may write
\begin{align}
P=P_{0}(T)+ S_0(T)\sqrt{\omega:\omega},
\end{align}
such that $\varepsilon_{0}+P_{0}=Ts_{0}$. For a massive Boltzmann gas the explicit expression for $P_{0}(T)$ is
\begin{align}
P_{0}=\frac{g_s}{2\pi^{2}}T^{2}M^{2}K_{2}\left(\frac{M}{T}\right).\label{}
\end{align}

Using the expression for $S_{0}$ in Eq.~\eqref{s0} we may express the pressure in terms of the magnitude of the spin potential $C$ as follows
\begin{align}
P(T,C)=P_{0}(T)+\alpha T^a M^b K_n\left(\frac{M}{T}\right)C. 
\end{align}

Consequently, the proper time evolution of the temperature is given by the following second-order ordinary differential equation,
\begin{align}\label{Temperatureequation3.267}
& \frac{\dd\varepsilon}{\dd\tau}+ \frac{\varepsilon+P}{\tau}-\frac{s_{0}}{\tau^2}\left(\frac{2}{3}\frac{\eta}{s_{0}}+\frac{\zeta}{s_0}\right)=0\nonumber\\
\implies &     A(\tau) \frac{\dd^2T}{\dd\tau^2}+B(\tau)\left(\frac{\dd T}{\dd\tau}\right)^2+D(\tau)\frac{\dd T}{\dd\tau}+E(\tau)=0,
\end{align}
where the various coefficients are defined as follows:
\begin{align}\label{constant3}
& A(\tau)= \varepsilon_2~C_{22},\nonumber\\
& B(\tau)= \left(\varepsilon_2C_{21}+\varepsilon_1 C_{11}\right),\nonumber\\
& D(\tau)= \left(\frac{\dd\varepsilon_{0}}{\dd T}+C_{12}\varepsilon_1+C_{23}\varepsilon_2\right)+\frac{\varepsilon_2}{\tau}C_{11}+\frac{\alpha}{\tau}T^a M^{3-a}K_n~C_{11},\nonumber\\
& E(\tau)= \varepsilon_2 C_{24}+\frac{\varepsilon_0+P_0}{\tau}+\frac{\varepsilon_2~C_{12}}{\tau}+\frac{\alpha}{\tau}T^aM^{3-a}K_n C_{12}-\frac{s_{0}}{\tau^2}\left(\frac{2}{3}\frac{\eta}{s_{0}}+\frac{\zeta}{s_0}\right).
\end{align}
In summary, the final forms of the equations that need to be solved are those governing the evolution of temperature and the magnitude of the spin potential, as summarized in Table \ref{finaltablechpater2}.
\begin{table}[H]
\centering
\renewcommand{\arraystretch}{3}
\begin{tabularx}{\textwidth} { 
| >{\centering\arraybackslash}X 
| >{\centering\arraybackslash}X
| >{\centering\arraybackslash}X
| >{\centering\arraybackslash}X
| >{\centering\arraybackslash}X 
| >{\centering\arraybackslash}X | }
\hline
\textbf{{Relativistic Navier-Stokes equations with spin for Bjorken flow}} \\
\hline
$  A(\tau) \frac{\dd^2T}{\dd\tau^2}+B(\tau)\left(\frac{\dd T}{\dd\tau}\right)^2+D(\tau)\frac{\dd T}{\dd\tau}+E(\tau)=0 $ \\
\hline
$C(\tau)=C_{11}(\tau)\frac{\dd T}{\dd\tau}+C_{12}(\tau)$ \\
\hline
\end{tabularx}
\caption{Evolution equations for energy density and the magnitude of the spin potential in the Bjorken flow model.}
\label{finaltablechpater2}
\end{table}
%
%
\subsection{Numerical results}
By appropriately choosing the values of the constants in the expressions for $S_{0}(T)$ and $\gamma$ in Eqs.~\eqref{s0} and \eqref{gamma}, respectively, all the functions in Eqs.~\eqref{constants1}, \eqref{constants2} and \eqref{constant3} may be determined, allowing for solving evolution equations in Table~\ref{finaltablechpater2} for the temperature and the magnitude of the spin potential. 
For our numerical simulations, we propose two different cases:
\medskip

{\bf Case I}: In this case, we assume that the function $S_0(T)$ is given by the particle number density $n_0(T)$, while $\gamma(T)$ is proportional to the pressure $P_0(T)$,
\begin{align}
& S_0(T)\equiv n_0(T), ~~ \gamma(T) \equiv \mathcal{A}\,P_0(T),
\label{caseI}
\end{align}
where 
\begin{align}
~\alpha = g_s \sqrt{2}/(2\pi^2)\quad\text{and}\quad\mathcal{A}= 2\pi^{2} \tilde{\alpha}/g_s,
\end{align}
which corresponds to the following choice: $a=1, n=2, c=1, m=2$.
To be able to treat the spin effects as a small correction to the standard (spinless) dynamics, we assume a very small value of the parameter  $\tilde{\alpha}=0.001$. 
\medskip 

{\bf Case II}: In this case, we assume that the functions $S_0(T)$ and $\gamma(T)$ are both related to the entropy density $s_0(T)$, namely
\begin{align}
S_0(T)\equiv s_0(T), ~~ \gamma(T) \equiv T s_0(T).
\label{caseII}
\end{align}
This means that in such a case, we consider: $a=0, n=3, c=0, m=3$, as well as we choose
\begin{align}
\alpha=g_s\sqrt{2}/(2\pi^2)~\text{and}~\tilde{\alpha}=g_s/(2\pi^2).
\end{align} 
It is important to note that only a consistently developed kinetic theory for particles with spin (or some other microscopic theory) can uniquely determine $S_0$ and $\gamma$. 
\medskip 

The temperature evolution can be determined uniquely if the initial conditions are provided at \( \tau = \tau_0 \) for both the temperature function \( T(\tau) \) and its derivative \( \dd T(\tau)/\dd \tau \). The initial temperature gradient can be determined from Eq.~\eqref{evolutionofCnewnew} if the initial values of \( T(\tau) \) and \( C(\tau) \) are specified. Hence, as expected, the dynamics of the system is determined by the initial values of the temperature and magnitude of the spin potential. We also note that it might also be possible to find analytic solutions for some specific choices of the coefficients in Eqs.~\eqref{constants1}, \eqref{constants2}, and \eqref{constant3}; however, here we limit ourselves to full numerical solutions only. Our choice of initial conditions is as follows, 
\begin{align}
&T_{0}=T(\tau_0)=200\,\text{MeV}\nonumber\\
&C_{0}=C(\tau_0)=50\,\text{MeV} 
\end{align}
where $\tau_0=0.5\,\text{fm}$. The internal degeneracy factor is $g_s=4$ (particles and antiparticles with spin $1/2$), and the effective particle mass is $M=500$. Moreover, we use the KSS-bound value of the specific shear viscosity $\eta/s_0=1/(4\pi)$. We ignore the effect of bulk viscosity since, as one can infer from Eq.~\eqref{bjorkenendeneq}, in the case of Bjorken flow expansion, its effect cannot be distinguished from that of shear viscosity.
\medskip 

\begin{figure}[H]
\centering
	\includegraphics[scale=0.42]{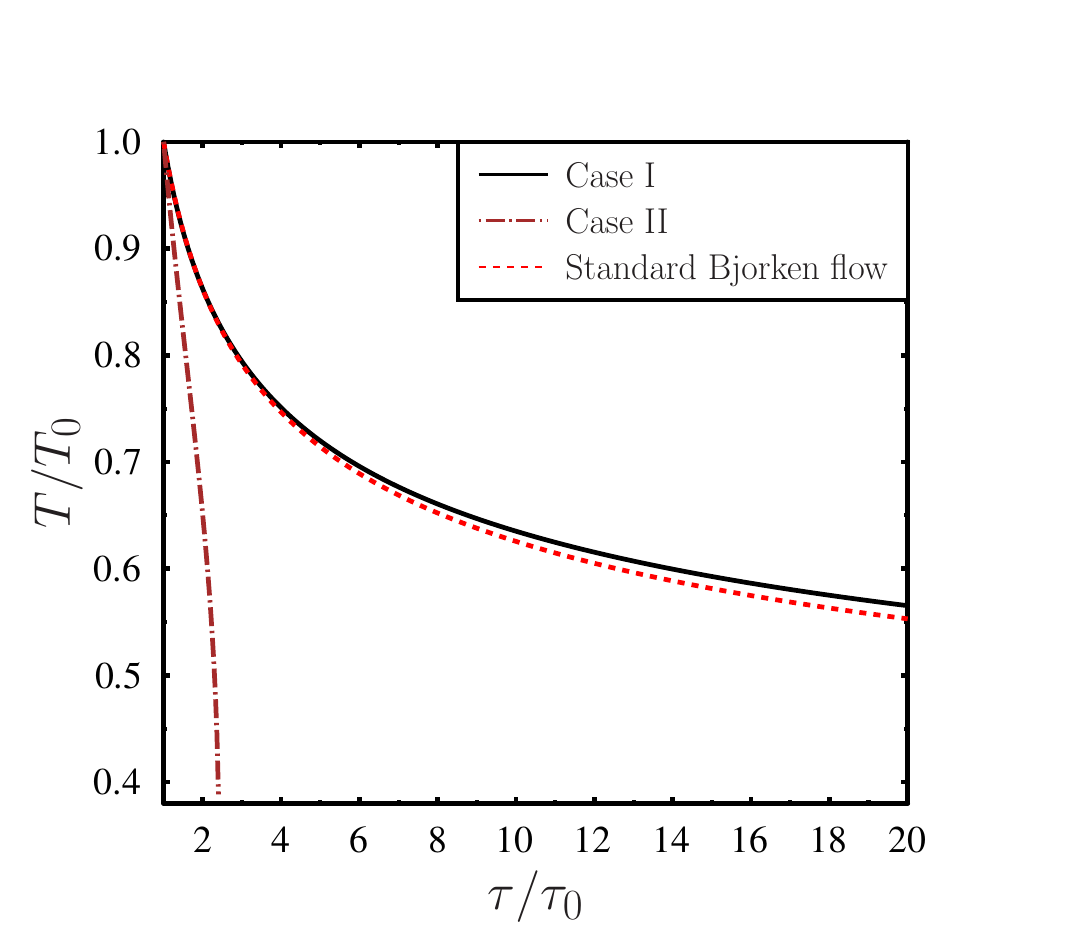}
\includegraphics[scale=0.43]{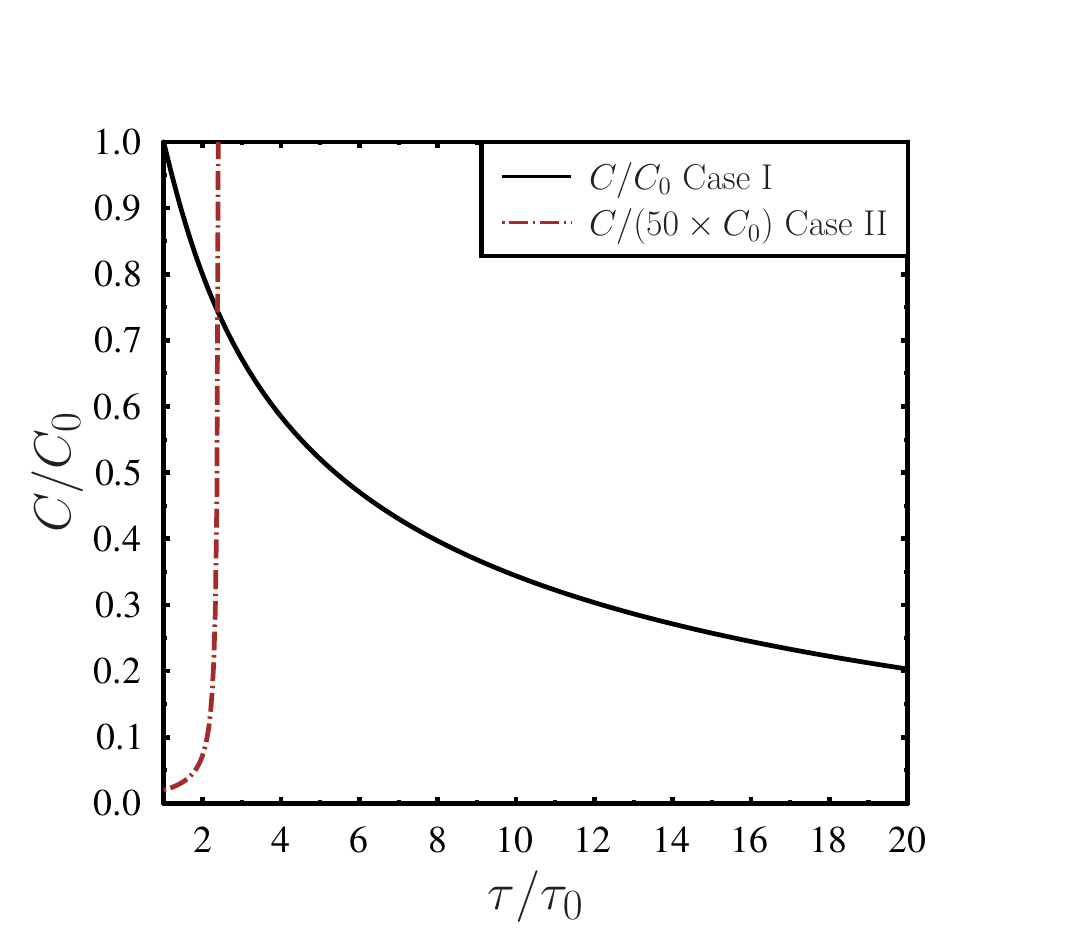}
	\caption{Left panel: Proper time evolution of temperature scaled by its initial value. The solid black line represents the temperature evolution for the \textbf{Case I}. The dashed-dotted brown line represents the temperature evolution for the \textbf{Case II}. The dashed red line represents the evolution of temperature for the standard Bjorken flow without spin. Here we used $T_0=200$ MeV at $\tau_0=0.5$ fm. Right panel: Proper time evolution of the magnitude of the spin potential $C$ normalized to its initial value.  Here, we consider $C_0=50$ MeV at $\tau_0=0.5$ fm. The factor 50 in the legend is a scaling factor introduced to fit all curves in one figure.}
	\label{fig:1}
\end{figure}

In the left panel in Fig.~\ref{fig:1}, we show the proper-time evolution of temperature normalized to its initial value. The solid black line and the dashed-dotted brown line represent the solutions of Eq.~\eqref{Temperatureequation3.267} for \textbf{Case I} and \textbf{Case II}, respectively. The dashed red line represents the standard Bjorken flow solution (no spin tensor included). Note that for the standard Bjorken flow solution, the temperature evolution equation is a first-order differential equation. We emphasize that one cannot set $C=0$ in Eq.~\eqref{Temperatureequation3.267} to obtain the standard Bjorken flow without spin since to obtain Eq.~\eqref{Temperatureequation3.267}, we used Eq.~\eqref{evolutionofCnewnew}, which assumes $C\neq0$. In \textbf{Case I}, the temperature profile for the evolution is very close to the standard Bjorken solution, while for the the \textbf{Case II}, we observe a rapid drop in temperature. For larger evolution times, for the \textbf{Case II}, the temperature becomes negative, which suggests that this solution cannot be accepted as physically meaningful.

In the right panel in Fig.~\ref{fig:1}, we present the proper time evolution of the magnitude of the spin potential $C(\tau)$ normalized to its initial value. We observe that $C$ monotonically decreases with proper time for the \textbf{Case I}. However, for the \textbf{Case II}, we observe that the spin potential rapidly grows with time, indicating a singular behavior in this case. 
\medskip

In conclusion, the equation of state given by Eq.~\eqref{s0}, which is based on dimensional analysis, along with the form of the coefficient $\gamma$ in Eq.~\eqref{gamma} for the \textbf{Case II}, clearly leads to nonphysical behavior in the system. At this stage, we cannot definitively rule out either \textbf{Case I} or \textbf{Case II}, as only a consistently developed kinetic theory for particles with spin can uniquely determine $S_0$ and $\gamma$. However, the behavior of the system observed in the \textbf{Case II} may either suggest that its setup is unphysical or indicate the instability of equations found in Sec.~\ref{stability}.
%
\section{Summary and outlook}
\label{Summary and outlookNS}
Before we proceed to the next chapter, we briefly review what we have accomplished here. We have formulated relativistic spin hydrodynamics in the Navier–Stokes (first-order), deriving ten coupled evolution equations for ten dynamical variables. For simplicity, we set the chemical potential to zero; incorporating a nonvanishing chemical potential is left for future work. 

By ``formulation'' we mean that we have:

\begin{itemize}
    \item established covariant thermodynamic relations for a relativistic fluid with spin,
    \item identified the dissipative currents and transport coefficients through conservation laws and an entropy-current analysis,
    \item performed a linear-stability study of the resulting equations, and
    \item obtained quasi-analytic solutions under Bjorken flow.
\end{itemize}   
As in conventional relativistic hydrodynamics, first-order (Navier–Stokes) equations are parabolic, hence acausal and generically unstable. A notable new feature is that -- in the local rest frame at small wavenumber, and within the Landau-frame choice $h^\mu=0$ -- stability depends sensitively on the spin equation of state. The Bjorken-flow solutions exhibit the same dependence. Because no microscopic theory is yet available, the functional form of the spin equation of state remains an educated guess.
%

The cornerstone of the present construction is the set of covariant thermodynamic relations involving the spin tensor, Eqs.~\eqref{covariantgeneralizedfirstlaw}-\eqref{covariantgeneralizedGibbsDuhem}. Confirming or refining these relations from an underlying microscopic theory -- such as relativistic kinetic theory or quantum field theory -- is an important task for the future. Likewise, our gradient-ordering assignment for the spin potential was a choice; adopting a different counting scheme could yield modified evolution equations. Clarifying the appropriate gradient counting therefore remains an open problem.
%

In the next chapter, we extend the analysis to the M\"uller-Israel-Stewart (second-order) framework for dissipative spin hydrodynamics, with the goal of deriving evolution equations that are both causal and stable. 
%

\chapter{Relativistic spin hydrodynamics: M\"uller-Israel-Stewart limit}
\label{MISMIS}

In this chapter, we develop relativistic spin hydrodynamics in the Müller-Israel–Stewart (MIS) framework, i.e., at second-order in the gradient expansion, under the simplifying assumption of vanishing chemical potential. We analyze the theory's stability and causality in the rest frame in the limiting cases of low- and high-wavenumber and within the Landau-frame choice $h^\mu=0$.

Our chief objective is to derive the corresponding evolution equations aiming to remedy the acausality and generic instabilities of the Navier–Stokes equations discussed in the previous chapter. To address these issues, in Sec.~\ref{Formulation} we introduce the second-order entropy current and identify all the dissipative currents  (dynamical variables) of the theory. Subsequently, in Sec.~\ref{relaxation-typeIS}, employing entropy-current analysis, we determine the associated transport coefficients and construct relaxation-type dynamical equations of the dissipative currents. In Sec.~\ref{Lmastabilityandcausality}, we present a truncated set of these equations and subject them to a linear-mode analysis that allows us to examine their stability and causality. Finally, in Sec.~\ref{Summary and outlookIS}, we summarize our findings and outline future perspectives.

Our derivation is inspired by the seminal works of Müller, Israel, and Stewart on second-order (conventional) relativistic hydrodynamics~\cite{Muller1967,Israel:1979wp}. The presentation is based on Refs.~\multimyref{AD4,AD7}. 

\section{Second-order theory and spin dissipative currents}
\label{Formulation}
We begin by revisiting the leading-order entropy current~\eqref{Smu0}, which results from the covariant formulation of the Euler thermodynamic equation incorporating a spin tensor~\eqref{covariantgeneralizedEulerequation},
\begin{align}\label{Smu0IS}
S^{\mu}_{(0)} = T^{\mu\nu}_{(0)}\beta_{\nu} + P\beta^{\mu} - \beta\omega_{\alpha\beta} S^{\mu\alpha\beta}_{(0)}.
\end{align}
Here, $\beta_{\nu} = u_{\nu}/T$ represents the thermal velocity four-vector, while $T^{\mu\nu}_{(0)} = \varepsilon u^{\mu}u^{\nu} - P\Delta^{\mu\nu}$ and $S^{\mu\alpha\beta}_{(0)} = u^{\mu}S^{\alpha\beta}$ denote the leading-order energy-momentum and spin tensors, respectively.
\medskip 

In the Navier-Stokes (NS) limit, the entropy current takes the form shown in Eq.~\eqref{NSentropy},
\begin{align}\label{SNSINIS}
S^{\mu}_{\text{NS}} &= T^{\mu\nu}\beta_{\nu}+P\beta^{\mu}-\beta\omega_{\alpha\beta}S^{\mu\alpha\beta}\nonumber\\
&=S^{\mu}_{(0)} + T^{\mu\nu}_{(1)}\beta_{\nu} - \beta\omega_{\alpha\beta}S^{\mu\alpha\beta}_{(1)},
\end{align}
where $T_{(1)}^{\mu\nu}$ represents the dissipative part of the energy-momentum tensor and $S^{\mu\alpha\beta}_{(1)}$ denotes the dissipative part of the spin tensor. Recall that, in our gradient counting scheme, the spin potential $\omega_{\alpha\beta}$ is a first-order quantity, $\mathcal{O}(\partial)$, and, consequently, the term $\beta\omega_{\alpha\beta}S_{(1)}^{\mu\alpha\beta}$ is of second order. As a result, this term is considered being of higher order in the NS entropy current, namely
\begin{align}
S^{\mu}_{\text{NS}} =S^{\mu}_{(0)} + T^{\mu\nu}_{(1)}\beta_{\nu}+\mathcal{O}(\partial^{2}),
\end{align}
which prevents us from determining the spin dissipative currents in the NS limit (for a detailed discussion, see Sec.~\ref{Dissipative currents and transport coefficients : entropy-current analysis}).

Decomposing $T^{\mu\nu}_{(1)}$ into symmetric and antisymmetric parts, the divergence of the NS entropy current may be expressed as follows  
\begin{align}
\partial_{\mu}S^{\mu}_{\text{NS}} &= {T^{\alpha\beta}_{(1S)}}\partial_{\alpha}\beta_{\beta} + {T^{\alpha\beta}_{(1A)}}\big[\partial_{\alpha}\beta_{\beta} + 2\beta\omega_{\alpha\beta}\big], \label{XuentropyproductionIS}
\end{align}
where the symmetric and antisymmetric parts of the dissipative energy-momentum tensor can be decomposed as:
\begin{align} 
& T^{\mu\nu}_{(1S)} = 2h^{(\mu}u^{\nu)}+\pi^{\mu\nu}+\Pi\Delta^{\mu\nu},\\ 
&T^{\mu\nu}_{(1A)} = 2q^{[\mu}u^{\nu]}+\phi^{\mu\nu}.
\end{align}
The dissipative currents in the above decomposition were determined in Chapter \ref{Navier-Stokes limit}; however, it was shown that the resulting evolution equations are, in general, unstable and acausal.
%
\medskip

It is well known~\cite{Hiscock:1985zz,Hiscock:1987zz} that, in the case of conventional relativistic hydrodynamics, the problems with stability and causality of NS equations may be resolved within the so-called Müller-Israel-Stewart (MIS) theory~\cite{Muller1967,Israel:1979wp}. The MIS formalism generalizes the NS form of entropy current by incorporating second-order gradient corrections to the dissipative currents. Here, we follow this method and, hence, write  
\begin{align}
S^{\mu}_{\text{MIS}}&=\beta_{\nu}T^{\mu\nu}_{}+\beta^{\mu}P-\beta\omega_{\alpha\beta}S^{\mu\alpha\beta}+Q^{\mu}\nonumber\\
&=\beta_{\nu}T^{\mu\nu}_{(0)}+\beta^{\mu}P-\beta\omega_{\alpha\beta}S^{\mu\alpha\beta}_{(0)}+\beta_{\nu}T^{\mu\nu}_{\rm(1)}-\beta\omega_{\alpha\beta}S^{\mu\alpha\beta}_{(1)}+Q^{\mu}\nonumber\\
&=S^{\mu}_{\rm NS}-\beta\omega_{\alpha\beta}S_{(1)}^{\mu\alpha\beta}+Q^{\mu}.
\label{eq16}
\end{align}
Here, the current $Q^{\mu}$, in general, includes all terms up to second order, $\mathcal{O}(\partial^2)$, and, as we show later, is composed of the currents: $h^{\mu}$, $\pi^{\mu\nu}$, $\Pi$, $q^{\mu}$, $\phi^{\mu\nu}$, and $S^{\mu\alpha\beta}_{(1)}$. Obviously, the term $\beta\omega_{\alpha\beta}S_{(1)}^{\mu\alpha\beta}$ in $S^{\mu}_
{\text{MIS}}$ can no longer be neglected. Therefore, before presenting the most general explicit expression for $Q^{\mu}$, we need to first define $S^{\mu\alpha\beta}_{(1)}$.
\medskip

The dissipative contribution to the spin tensor, \( S^{\mu\alpha\beta}_{(1)} \), is a rank-3 tensor that is antisymmetric in its last two indices and is orthogonal to the four-velocity in its first index, \( u_{\mu} S^{\mu\alpha\beta}_{(1)} = 0 \), hence it may be decomposed with respect to four-velocity as follows
\begin{align}
S^{\mu\alpha\beta}_{(1)}=2u^{[\alpha}\Delta^{\mu\beta]}\Phi+2u^{[\alpha}\tau^{\mu\beta]}_{(s)}+2u^{[\alpha}\tau^{\mu\beta]}_{(a)}+\Theta^{\mu\alpha\beta}.\label{eq17}
\end{align}
The procedure leading to the above result is presented in detail in Appendix~\ref{Appendix B}. The new dissipative currents are all of first order in the hydrodynamic gradient expansion, and satisfy the following properties: $u_{\mu} \tau_{(s)}^{\mu \beta}=$ $u_{\mu} \tau_{(a)}^{\mu \beta}=u_{\mu} \Theta^{\mu \alpha \beta}=0; \tau_{(s)}^{\mu \beta}=\tau_{(s)}^{\beta \mu}, \tau_{(a)}^{\mu \beta}=$ $-\tau_{(a)}^{\beta \mu}$, $\tau_{(s)\mu}^{~~\mu}=0$, $\Theta^{\mu \alpha \beta}=-\Theta^{\mu \beta \alpha}$, $u_{\mu}\Theta^{\mu \alpha \beta}=0$, $u_{\alpha}\Theta^{\mu \alpha \beta}=0$, and $u_{\beta}\Theta^{\mu \alpha \beta}=0$.
\medskip

Let us now examine the number of independent dynamical variables and the number of evolution equations in the system. According to the discussion so far, the energy-momentum and spin tensors can be written as:
\begin{align}
&T^{\mu\nu}=T^{\mu\nu}_{(0)}+T^{\mu\nu}_{(1S)}+T^{\mu\nu}_{(1A)}=\varepsilon u^{\mu}u^{\nu}-P\Delta^{\mu\nu}+2h^{(\mu}u^{\nu)}+\pi^{\mu\nu}+\Pi\Delta^{\mu\nu}+2q^{[\mu}u^{\nu]}+\phi^{\mu\nu},\\
&S^{\mu\alpha\beta}=S^{\mu\alpha\beta}_{(0)}+S^{\mu\alpha\beta}_{(1)}=u^{\mu}S^{\alpha\beta}+2u^{[\alpha}\Delta^{\mu\beta]}\Phi+2u^{[\alpha}\tau^{\mu\beta]}_{(s)}+2u^{[\alpha}\tau^{\mu\beta]}_{(a)}+\Theta^{\mu\alpha\beta}.
\end{align}
From Table~\ref{Table1}, it is evident that the energy-momentum tensor contributes 16 independent dynamical variables after imposing the Landau ($h^{\mu}=0$) or the generalized Landau ($h^{\mu}+q^{\mu}=0$) hydrodynamic frame, while the spin tensor (see Table \ref{Dynamical variables of the spin tensor  SIS.}) has 24 independent dynamical degrees of freedom, since, as we noted already, $S^{\lambda\mu\nu}_{(1)}$ this time cannot be neglected. On the other hand,  as shown in Sec.~\ref{Dissipative currents and transport coefficients : entropy-current analysis},  the total number of evolution equations derived from the conservation of the energy-momentum tensor and the conservation of total angular momentum is 10, namely, we have:
\begin{align}
&u_{\nu}\partial_{\mu} T^{\mu\nu}=0\implies D\varepsilon+(\varepsilon+P)\theta = 2 \,h^{\mu}Du_{\mu} -\nabla \cdot (q+h)+\pi^{\mu \nu}  \partial_{\mu} u_{\nu}\nonumber\\
&~~~~~~~~~~~~~~~~~~~~~~~~~~~~~~~~~~~~~~~~~~~~~~~+\Pi\Delta^{\mu\nu}\partial_{\mu}u_{\nu}+\phi^{\mu \nu}  \partial_{\mu} u_{\nu},\label{uTIS}\\
&\Delta^{\alpha}_{\nu}\partial_{\mu}T^{\mu\nu}=0\implies(\varepsilon+P) Du^{\alpha} - \nabla^{\alpha}  P= 
-(q +h)\cdot \nabla u^{\alpha}+(q^{\alpha}-h^{\alpha})\theta\nonumber\\
&~~~~~~~~~~~~~~~~~~~~~~~~~~~~~~~~~~~~~~~~~~~~~~~~~~~~~~~
+\Delta^{\alpha}_{~\nu}D q^{\nu}-\Delta^{\alpha}_{~\nu}D h^{\nu}-\Delta^{\alpha}_{~\nu} \partial_{\mu} \phi^{\mu \nu}
\nonumber\\
&~~~~~~~~~~~~~~~~~~~~~~~~~~~~~~~~~~~~~~~~~~~~~~~~~~~~~~~-\Delta^{\alpha}_{~\nu} \partial_{\mu} (\pi^{\mu \nu}+\Pi\Delta^{\mu\nu}),\label{DeltaTIS}\\
&\partial_{\lambda}S^{\lambda\mu\nu}=-2T^{\mu\nu}_{(1A)}=0\implies \partial_{\lambda}(u^{\lambda}S^{\mu\nu})+\partial_{\lambda} S^{\lambda\mu\nu}_{(1)} = -2(q^{\mu}u^{\nu}-q^{\nu}u^{\mu}+\phi^{\mu\nu}).\label{SpinISevo}
\end{align}
This means that we need to determine $16+24-10=30$ additional equations to close the system. To do so, in the following section, we derive the evolution equations for the dissipative currents.
%
%
\noindent
\begin{table}[H]
\renewcommand{\arraystretch}{2}
\begin{tabularx}{\textwidth} { 
| >{\raggedright\arraybackslash\hsize=0.7\hsize}X 
| >{\raggedright\arraybackslash\hsize=1.5\hsize}X
| >{\raggedright\arraybackslash}X
| >{\raggedright\arraybackslash}X
| >{\centering\arraybackslash\hsize=0.8\hsize}X | }
\hline
\textbf{Quantity} & \textbf{Mathematical object} &\textbf{Constraints}& \textbf{Hydrodynamic gradient order}& \textbf{Degrees of freedom}\\
\hline
$S^{\mu\nu}$& Antisymmetric tensor&$S^{\mu\nu}=-S^{\nu\mu}$ &$\mathcal{O}(\partial^{0})\equiv\mathcal{O}(1)$& 6\\
\hline
 $\Phi$ & Scalar field&&$\mathcal{O}(\partial)$&1\\
\hline
 $\tau^{\mu\nu}_{(s)}$ &Symmetric traceless tensor&$\tau_{(s)}^{\mu\nu}u_{\mu}=0,$ $\tau^{\mu}_{\mu(s)}=0$, $\tau^{\mu\nu}_{(s)}=\tau_{(s)}^{\nu\mu}$&$\mathcal{O}(\partial)$ &5\\
\hline 
 $\tau^{\mu\nu}_{(a)}$ & Antisymmetric tensor& $\tau^{\mu\nu}_{(a)}u_{\mu}=0$, $\tau^{\mu\nu}_{(a)}=-\tau^{\nu\mu}_{(a)}$& $\mathcal{O}(\partial)$&3\\
\hline
$\Theta^{\lambda\mu\nu}$&Antisymmetric tenor in the last 2 indices&$\Theta^{\mu \alpha \beta}=-\Theta^{\mu \beta \alpha}$, $u_{\mu}\Theta^{\mu \alpha \beta}=0$, $u_{\alpha}\Theta^{\mu \alpha \beta}=0$, $u_{\beta}\Theta^{\mu \alpha \beta}=0$&$\mathcal{O}(\partial)$&9\\
\hline
$S^{\lambda\mu\nu}_{(1)}$& Antisymmetric tensor in the last 2 indices &$u_{\lambda}S^{\lambda\mu\nu}_{(1)}=0$, $S^{\lambda\mu\nu}=-S^{\lambda\nu\mu}$ &$\mathcal{O}(\partial)$ & 18\\
\hline
\end{tabularx}
\caption{Number of independent degrees of freedom of various quantities in the decomposition of the spin tensor $S^{\lambda\mu\nu}$.}
\label{Dynamical variables of the spin tensor  SIS.}
\end{table}
%
%
\section{Relaxation-type equations and transport coefficients: entropy-current analysis}
\label{relaxation-typeIS}
To derive the evolution equations for the dissipative currents listed in Tables~\ref{Table1} and \ref{Dynamical variables of the spin tensor  SIS.}, we need to determine the explicit form of $Q^{\mu}$ in the entropy current~\eqref{eq16}. The term \( Q^{\mu} \) represents a general four-vector that includes contributions up to second-order \( \mathcal{O}(\partial^2) \), composed of all possible combinations of the dissipative currents. The form of $Q^{\mu}$ is constrained by the condition that leading-order entropy should be greater than the first-order $S^{\mu}_{\text{NS}}$ or the second-order $S^{\mu}_{\text{MIS}}$. By contracting Eq.~\eqref{eq16} with $u^{\mu}$ we obtain, 
\begin{align}
u_{\mu}S_{\text{MIS}}^{\mu}-u_{\mu}S^{\mu}_{(0)}=u_{\mu}Q^{\mu},
\end{align}
which leads to the condition $u_{\mu}Q^{\mu}\leq 0$. Now we can express $Q^{\mu}$ in terms of all possible second-order combinations of dissipative currents respecting the constraint $u\cdot Q\leq 0$,
\begin{align}
Q^{\mu} = & ~~u^{\mu}\left(a_{1}\Pi^2+a_{2}\pi^{\lambda\nu}\pi_{\lambda\nu}+a_{3}h^{\lambda}h_{\lambda}+a_{4}q^{\lambda}q_{\lambda}+a_{5}\phi^{\lambda\nu}\phi_{\lambda\nu}\right)\nonumber
\\
\nonumber
& + u^{\mu}\left(\tilde{a}_{1}\Phi^{2}+\tilde{a}_{2}\tau_{(s)}^{\lambda\nu}\tau_{(s)\lambda\nu}+\tilde{a}_{3}\tau_{(a)}^{\lambda\nu}\tau_{(a)\lambda\nu}+\tilde{a}_{4}\Theta^{\lambda\alpha\beta}\Theta_{\lambda\alpha\beta}\right)\nonumber
\\
& + \Big( b_{1}\Pi h^{\mu} + b_{2} \pi^{\mu\nu}h_{\nu}+b_{3}\phi^{\mu\nu}h_{\nu}+b_{4}\Pi q^{\mu}+ b_{5} \pi^{\mu\nu}q_{\nu}+b_{6}\phi^{\mu\nu}q_{\nu}\Big)\nonumber
\\
& + \left(\tilde{b}_{1}\Phi h^{\mu} + \tilde{b}_{2} \tau^{\mu\nu}_{(s)}h_{\nu}+\tilde{b}_{3}\tau^{\mu\nu}_{(a)}h_{\nu}+\tilde{b}_{4}\Phi q^{\mu}+ \tilde{b}_{5} \tau^{\mu\nu}_{(s)}q_{\nu}+\tilde{b}_{6}\tau^{\mu\nu}_{(a)}q_{\nu} \right)\nonumber
\\
&+\left(c_{1}\Theta^{\mu\alpha\beta}\phi_{\alpha\beta}+c_{2}\Theta^{\mu\alpha\beta}\tau_{(a)\alpha\beta}\right)\nonumber\\
& +\left(c_3 \Theta^{\alpha\beta\mu}\Delta_{\alpha\beta}\Pi+c_4 \Theta^{\alpha\beta\mu}\pi_{\alpha\beta}+c_5 \Theta^{\alpha\beta\mu}\Delta_{\alpha\beta}\Phi+c_6\Theta^{\alpha\beta\mu}\tau_{(s)\alpha\beta}\right)\nonumber\\
& +\left(c_7 \Theta^{\alpha\beta\mu}\phi_{\alpha\beta}+c_8\Theta^{\alpha\beta\mu}\tau_{(a)\alpha\beta}\right),
\label{eq19}
\end{align}
where $a_{i}, \tilde{a}_{i}, b_{i}, \Tilde{b}_{i}, $ and $c_{i}$ are dimensionful coefficients. It is clear that due to the condition $u\cdot Q\leq 0$, the $a$ and $\tilde{a}$ coefficients have to the have definite signatures: $a_1\leq 0$, $a_2\leq 0$, $a_3\geq 0$, $a_4\geq 0$, $a_5\leq 0$, $\tilde{a}_1\leq 0$, $\tilde{a}_2\leq 0$, $\tilde{a}_3\leq 0$, $\tilde{a}_4\geq 0$. However, there are no such sign constraints for $b_{i}, \Tilde{b}_{i}$, or $c_{i}$. Therefore, the divergence of the entropy current, as given in Eq.~\eqref{eq16}, is expressed as
\begin{align}
\label{enprodrat}
\partial_{\mu}S^{\mu}_{\text{MIS}} &= T^{\mu\nu} \partial_{\mu} \beta_{\nu} + \beta_{\nu} \partial_{\mu} T^{\mu\nu} + \partial_{\mu} (P\beta^{\mu}) - S^{\mu\alpha\beta} \partial_{\mu} (\beta\omega_{\alpha\beta}) - \beta\omega_{\alpha\beta} \partial_{\mu} S^{\mu\alpha\beta} + \partial_{\mu} Q^{\mu}.
\end{align}
Using the conservation of the energy-momentum tensor and the continuity equation of the spin tensor, i.e., 
\begin{align}
&\partial_{\mu}T^{\mu\nu}=0,\nonumber\\
&\partial_{\mu}S^{\mu\alpha\beta}=-2T^{\mu\nu}_{(1A)},\nonumber
\end{align}
equation~\eqref{enprodrat} can be further rewritten as
\begin{align}
\partial_{\mu}S^{\mu}_{\text{MIS}} &= \partial_{\mu} (P\beta^{\mu}) + T^{\mu\nu}_{(0)} \partial_{\mu} \beta_{\nu} - S^{\mu\alpha\beta}_{(0)} \partial_{\mu} (\beta\omega_{\alpha\beta}) \nonumber\\
&\quad + \left(\partial_{\mu} \beta_{\nu} + 2\beta\omega_{\mu\nu} \right) T^{\mu\nu}_{(1A)} + T^{\mu\nu}_{(1S)} \partial_{\mu} \beta_{\nu} - S^{\mu\alpha\beta}_{(1)} \partial_{\mu} (\beta\omega_{\alpha\beta}) + \partial_{\mu} Q^{\mu}.
\label{equ59ver2}
\end{align}
Furthermore, using the Gibbs-Duhem thermodynamic relation~\eqref{covariantgeneralizedGibbsDuhem}, the first line in Eq.~\eqref{equ59ver2} vanishes, leading to  
\begin{align}\label{FinalISentropypro}
\partial_{\mu}S^{\mu}_{\text{MIS}} &= \left(\partial_{\mu} \beta_{\nu} + 2\beta\omega_{\mu\nu} \right) T^{\mu\nu}_{(1A)} + T^{\mu\nu}_{(1S)} \partial_{\mu} \beta_{\nu} - S^{\mu\alpha\beta}_{(1)} \partial_{\mu} (\beta\omega_{\alpha\beta}) + \partial_{\mu} Q^{\mu}.
\end{align}

The goal now is to explicitly calculate each term in Eq.~\eqref{FinalISentropypro} in terms of various dissipative currents. The first two terms give exactly the NS part shown in Eq.~\eqref{XuentropyproductionIS}, and hence
\begin{align}\label{ISforNS}
\left(\partial_{\mu} \beta_{\nu} + 2\beta\omega_{\mu\nu} \right) T^{\mu\nu}_{(1A)} + T^{\mu\nu}_{(1S)} \partial_{\mu} \beta_{\nu}=&\beta h^{\mu}(\beta \nabla_{\mu}T-Du_{\mu})+\beta\pi^{\mu\nu}\sigma_{\mu\nu}+\beta\Pi\theta\nonumber\\
&-\beta q^{\mu}(\beta\nabla_{\mu}T+Du_{\mu}-4\omega_{\mu\nu}u^{\nu})\nonumber\\
&+\phi^{\mu\nu}(\Omega_{\mu\nu}+2\beta\Delta^{\alpha}_{\mu}\Delta^{\beta}_{\nu}\omega_{\alpha\beta}).
\end{align}
For the third term in Eq.~\eqref{FinalISentropypro}, we simply substitute the expression for $S^{\mu\alpha\beta}_{(1)}$ derived in Eq.~\eqref{eq17}, which gives
\begin{align}
-S^{\mu\alpha\beta}_{(1)}\partial_{\mu}(\beta\omega_{\alpha\beta}) =&-\left(2u^{[\alpha}\Delta^{\mu\beta]}\Phi+2u^{[\alpha}\tau_{(s)}^{\mu\beta]}+2u^{[\alpha}\tau_{(a)}^{\mu\beta]}+\Theta^{\mu\alpha\beta}\right)\partial_{\mu}(\beta\omega_{\alpha\beta})\nonumber\\
=&-2u^{[\alpha}\Delta^{\mu\beta]}\Phi \nabla_{\mu}(\beta\omega_{\alpha\beta})-2u^{[\alpha}\tau_{(s)}^{\mu\beta]}\nabla_{\mu}(\beta\omega_{\alpha\beta})\nonumber\\
 &-2u^{[\alpha}\tau_{(a)}^{\mu\beta]}\nabla_{\mu}(\beta\omega_{\alpha\beta})\Theta^{\mu\alpha\beta}\nabla_{\mu}(\beta\omega_{\alpha\beta})\nonumber\\
 =&-2\Phi u^{\alpha}\nabla^{\beta}(\beta\omega_{\alpha\beta})-2u^{\alpha}\tau^{\mu\beta}_{(s)}\nabla_{\mu}(\beta\omega_{\alpha\beta})-2u^{\alpha}\tau^{\mu\beta}_{(a)}\nabla_{\mu}(\beta\omega_{\alpha\beta})\nonumber\\
 &-\Theta^{\mu\alpha\beta}\nabla_{\mu}(\beta\omega_{\alpha\beta})\nonumber\\
 =&-\tau_{\mu\beta(s)}u^{\alpha}\left(\Delta^{\gamma\mu}\Delta^{\rho\beta}+\Delta^{\gamma\beta}\Delta^{\rho\mu}-\frac{2}{3}\Delta^{\gamma\rho}\Delta^{\mu\beta}\right)\nabla_{\gamma}(\beta\omega_{\alpha\rho})\nonumber\\
&-2\Phi u^{\alpha}\nabla^{\beta}(\beta\omega_{\alpha\beta})-2\tau_{\mu\beta(a)}u^{\alpha}\Delta^{\gamma[\mu}\Delta^{\beta]\rho}\nabla_{\gamma}(\beta\omega_{\alpha\rho})\nonumber\\
&-\Theta_{\mu\alpha\beta}\Delta^{\alpha\delta}\Delta^{\beta\rho}\Delta^{\mu\gamma}\nabla_{\gamma}(\beta\omega_{\delta\rho}).
\label{equ65ver2}
\end{align}
%
%
As the final step, we need to examine the term \( \partial_{\mu} Q^{\mu} \), which can be evaluated using the expression for \( Q^{\mu} \) given in Eq.~\eqref{eq19}. A straightforward calculation yields
\begin{align}
\partial_{\mu}Q^{\mu}&=h_{\alpha}\mathcal{A}^{\alpha}+q_{\alpha}\mathcal{B}^{\alpha}+\pi_{\alpha\beta}\mathcal{C^{\alpha\beta}}+\Pi \mathcal{D}+\phi_{\alpha\beta}\mathcal{E^{\alpha\beta}}\nonumber\\
&~~~+\Phi \mathcal{F}+\tau^{\alpha\beta}_{(s)}\mathcal{G_{\alpha\beta}}+\tau^{\alpha\beta}_{(a)}\mathcal{H_{\alpha\beta}}+\Theta_{\alpha\beta\gamma}\mathcal{I}^{\alpha\beta\gamma},
\label{equ23ver2}
\end{align}
which is equivalent to 
\begin{align}
\partial_{\mu}Q^{\mu}&=h_{\alpha}\mathcal{A}^{\langle\alpha\rangle}+q_{\alpha}\mathcal{B}^{\langle\alpha\rangle}+\pi_{\alpha\beta}\mathcal{C^{\langle\alpha\beta\rangle}}+\Pi \mathcal{D}+\phi_{\alpha\beta}\mathcal{E^{\langle[\alpha\beta]\rangle}}\nonumber\\
&~~~+\Phi \mathcal{F}+\tau^{\alpha\beta}_{(s)}\mathcal{G_{\langle\alpha\beta\rangle}}+\tau^{\alpha\beta}_{(a)}\mathcal{H_{\langle[\alpha\beta]\rangle}}+\Theta_{\alpha\beta\gamma}\mathcal{I}^{\langle\alpha\rangle\langle\beta\rangle\langle\gamma\rangle},
\label{eq33}
\end{align}
such that
\begin{align}
    & \mathcal{A}^{\langle\alpha\rangle}\equiv\Delta^{\alpha\beta}\mathcal{A}_{\beta}; \quad u_{\alpha}\mathcal{A}^{\langle\alpha\rangle}=0,\label{A}\\
    & \mathcal{B}^{\langle\alpha\rangle}\equiv\Delta^{\alpha\beta}\mathcal{B}_{\beta}; \quad u_{\alpha}\mathcal{B}^{\langle\alpha\rangle}=0,\label{B}\\
    & \mathcal{C}_{\langle\alpha\beta\rangle}\equiv\Delta_{\alpha\beta}^{\mu\nu}\mathcal{C}_{\mu\nu}=\left(\Delta^{(\mu}_{~\alpha}\Delta^{\nu)}_{~\beta}-\frac{\Delta_{\alpha\beta}}{3}\Delta^{\mu\nu}\right)\mathcal{C}_{\mu\nu}; \quad u^{\alpha}\mathcal{C}_{\langle\alpha\beta\rangle}=0; \quad g^{\alpha\beta}\mathcal{C}_{\langle\alpha\beta\rangle}=0,\label{C} \\
    & \mathcal{E}_{\langle[\alpha\beta]\rangle}\equiv \Delta^{[\mu\nu]}_{[\alpha\beta]}\mathcal{E}_{\mu\nu}\equiv\frac{1}{2}\left(\Delta^{\mu}_{~\alpha}\Delta^{\nu}_{~\beta}-\Delta^{\nu}_{~\alpha}\Delta^{\mu}_{~\beta}\right)\mathcal{E}_{\mu\nu}; \quad u_{\alpha}\mathcal{E^{\langle[\alpha\beta]\rangle}}=0, \label{E}\\
    & \mathcal{G_{\langle\alpha\beta\rangle}}\equiv\Delta^{\mu\nu}_{\alpha\beta}\mathcal{G_{\mu\nu}}; \quad u_{\alpha}\mathcal{G^{\langle\alpha\beta\rangle}}=0, \quad g_{\alpha\beta}\mathcal{G^{\langle\alpha\beta\rangle}}=0,\label{G}\\
    & \mathcal{H}_{\langle[\alpha\beta]\rangle}\equiv \Delta^{[\mu\nu]}_{[\alpha\beta]}\mathcal{H}_{\mu\nu}\equiv\frac{1}{2}\left(\Delta^{\mu}_{~\alpha}\Delta^{\nu}_{~\beta}-\Delta^{\nu}_{~\alpha}\Delta^{\mu}_{~\beta}\right)\mathcal{H}_{\mu\nu}; \quad u_{\alpha}\mathcal{H^{\langle[\alpha\beta]\rangle}}=0,\label{H}\\
    & \mathcal{I}^{\langle\alpha\rangle\langle\beta\rangle\langle\gamma\rangle}\equiv\Delta^{\alpha\mu}\Delta^{\beta\nu}\Delta^{\gamma\delta}\mathcal{I}_{\mu\nu\delta};\quad u_{\alpha}\mathcal{I}^{\langle\alpha\rangle\langle\beta\rangle\langle\gamma\rangle}=0; \quad u_{\beta}\mathcal{I}^{\langle\alpha\rangle\langle\beta\rangle\langle\gamma\rangle}=0;\quad u_{\gamma}\mathcal{I}^{\langle\alpha\rangle\langle\beta\rangle\langle\gamma\rangle}=0.
\label{I2}
\end{align}
%
Here, we used the notation where: the projection of a four-vector $X^{\mu}$ orthogonal to $u^{\mu}$ is denoted by $X^{\langle\mu\rangle}=\Delta^{\mu\nu} X_{\nu}$, the traceless and symmetric projection operator orthogonal to $u^{\mu}$ is given by $X^{\langle\mu\nu\rangle}=\Delta^{\mu\nu}_{\alpha\beta} X^{\alpha\beta} = \frac{1}{2}\Bigl(\Delta^{\mu}_{\ \alpha}\Delta^{\nu}_{\ \beta} + \Delta^{\mu}_{\ \beta}\Delta^{\nu}_{\ \alpha} - \frac{2}{3}\Delta^{\mu\nu}\Delta_{\alpha\beta}\Bigr) X^{\alpha\beta}$, and the antisymmetric projection operator orthogonal to $u^{\mu}$ is defined as $X^{\langle[\mu\nu]\rangle}= \Delta^{[\mu\nu]}_{[\alpha\beta]} X^{\alpha\beta} = \frac{1}{2}\Bigl(\Delta^{\mu}_{\ \alpha}\Delta^{\nu}_{\ \beta} - \Delta^{\mu}_{\ \beta}\Delta^{\nu}_{\ \alpha}\Bigr) X^{\alpha\beta}$. 

In the above equations, to obtain scalars $\mathcal{D}$ and $\mathcal{F}$, vectors $\mathcal{A}_{\beta}$ and  $\mathcal{B}_{\beta}$, and tensors $\mathcal{C}_{\mu\nu}$, $\mathcal{E_{\mu\nu}}$, $\mathcal{G_{\mu\nu}}$, $\mathcal{H}_{\mu\nu}$, and $\mathcal{I}_{\mu\nu\delta}$ we: \textbf{(i)} take the partial derivative of $Q^{\mu}$ in~Eq.~\eqref{eq19} (note that the partial derivative of the parameters $a_{i}, \tilde{a}_{i}, b_{i}, \tilde{b}_{i},$ and $c_{i}$ is not zero), \textbf{(ii)} collect all terms having common dissipative currents, keeping in mind that in such a process, some terms, such as `$\pi_{\mu\nu}h^{\nu}\nabla^{\mu}b_{2}$', will contribute to two different dissipative current sectors, in this case to $h^{\nu}$ and $\pi_{\mu\nu}$. For that reason, in such a case, we introduce a constant factor $l$ (similarly $\tilde{l}$) such that
\begin{align}
\pi_{\mu\nu}h^{\nu}\nabla^{\mu}b_{2}=l_{h\pi} \pi_{\mu\nu}h^{\nu}\nabla^{\mu}b_{2}+(1-l_{h\pi}) \pi_{\mu\nu}h^{\nu}\nabla^{\mu}b_{2}.
\end{align}
Following the above procedure, we obtain various scalars $\mathcal{D}$ and $\mathcal{F}$, vectors $\mathcal{A}_{\beta}$ and $\mathcal{B}_{\beta}$, and tensors $\mathcal{C}_{\mu\nu}$, $\mathcal{E_{\mu\nu}}$, $\mathcal{G_{\mu\nu}}$, $\mathcal{H}_{\mu\nu}$, and $\mathcal{I}_{\mu\nu\delta}$, whose explicit forms are listed below:\\
\resizebox{\textwidth}{!}{
\begin{minipage}{\textwidth}
\begin{align}
\mathcal{D}=&a_1\Pi\theta+\Pi Da_1+2a_1D\Pi+(1-l_{\Pi h})h^{\mu}\nabla_{\mu}b_1 
-b_1(1-\tilde{l}_{\Pi h})h^{\mu}Du_{\mu}+\frac{1}{2}b_1\nabla_{\mu}h^{\mu} \nonumber\\
&+\frac{1}{2}l_{\Pi q}q^{\mu}\nabla_{\mu}b_4
-\frac{1}{2}\tilde{l}_{\Pi q}b_4 q^{\mu}Du_{\mu}+\frac{1}{2}b_4\nabla_{\mu}q^{\mu}
+l_{\Theta\Pi}\Theta^{\alpha\mu\nu}\Delta_{\alpha\mu}\nabla_{\nu}c_3 \nonumber\\
&-\tilde{l}_{\Theta\Pi}c_3\Delta_{\alpha\mu}\Theta^{\alpha\mu\nu}Du_{\nu}
+\frac{1}{2}c_3\Delta_{\alpha\beta}\nabla_{\mu}\Theta^{\alpha\beta\mu},
\label{equ78ver2}
\end{align}
\end{minipage}
}
\resizebox{\textwidth}{!}{
\begin{minipage}{\textwidth}
\begin{align}
\mathcal{A}^{\mu}=& a_3h^{\mu}\theta+h^{\mu}Da_3+2a_3Dh^{\mu}
+l_{\Pi h}\Pi\nabla^{\mu}b_1+b_1\nabla^{\mu}\Pi 
-b_1\tilde{l}_{\Pi h}\Pi Du^{\mu}+l_{\pi h}\pi^{\lambda\mu}\nabla_{\lambda}b_2\nonumber\\
&+b_2\nabla_{\lambda}\pi^{\lambda\mu} 
-b_2\tilde{l}_{\pi h}\pi^{\lambda\mu}Du_{\lambda} 
+l_{\phi h}\phi^{\lambda\mu}\nabla_{\lambda}b_3+b_3\nabla_{\lambda}\phi^{\lambda\mu}-b_3\tilde{l}_{\phi h}\phi^{\lambda\mu}Du_{\lambda}   \nonumber\\
&+l_{\Phi h}\Phi\nabla^{\mu}\tilde{b}_1
+\tilde{b}_1\nabla^{\mu}\Phi -\tilde{b}_1\tilde{l}_{\Phi h}\Phi Du^{\mu} +l_{\tau_s h}\tau^{\lambda\mu}_{(s)}\nabla_{\lambda}\tilde{b}_2 
+\tilde{b}_2\nabla_{\lambda}\tau^{\lambda\mu}_{(s)} 
\nonumber\\
&-\tilde{b}_2\tilde{l}_{\tau_{s}h}\tau^{\lambda\mu}_{(s)}Du_{\lambda}+l_{\tau_a h}\tau^{\lambda\mu}_{(a)}\nabla_{\lambda}\tilde{b}_3+\tilde{b}_3\nabla_{\lambda}\tau^{\lambda\mu}_{(a)} 
-\tilde{b}_3\tilde{l}_{\tau_{a}h}\tau^{\lambda\mu}_{(a)}Du_{\lambda},
\label{equ79ver2}
\end{align}
\end{minipage}
}
\resizebox{\textwidth}{!}{
\begin{minipage}{\textwidth}
\begin{align}
\mathcal{B}^{\mu}=& a_4q^{\mu}\theta+q^{\mu}Da_4+2a_4Dq^{\mu}
+(1-l_{\Pi q})\Pi\nabla^{\mu}b_4+b_4\nabla^{\mu}\Pi 
-b_4(1-\tilde{l}_{\Pi q})\Pi Du^{\mu} \nonumber\\
&+(1-l_{\pi q})\pi^{\lambda\mu}\nabla_{\lambda}b_5
+b_5\nabla_{\lambda}\pi^{\lambda\mu} 
-b_5(1-\tilde{l}_{\pi q})\pi^{\lambda\mu}Du_{\lambda} 
+l_{\phi q}\phi^{\lambda\mu}\nabla_{\lambda}b_6 \nonumber\\
&+b_6\nabla_{\lambda}\phi^{\lambda\mu} 
-b_6\tilde{l}_{\phi q}\phi^{\lambda\mu}Du_{\lambda} 
+l_{\Phi q}\Phi\nabla^{\mu}\tilde{b}_4
+\tilde{b}_4\nabla^{\mu}\Phi 
-\tilde{b}_4\tilde{l}_{\Phi q}\Phi Du^{\mu} \nonumber\\
&+l_{\tau_s q}\tau^{\lambda\mu}_{(s)}\nabla_{\lambda}\tilde{b}_5 
+\tilde{b}_5\nabla_{\lambda}\tau^{\lambda\mu}_{(s)} 
-\tilde{b}_5\tilde{l}_{\tau_{s}q}\tau^{\lambda\mu}_{(s)}Du_{\lambda}+l_{\tau_a q}\tau^{\lambda\mu}_{(a)}\nabla_{\lambda}\tilde{b}_6  \nonumber\\
&
+\tilde{b}_6\nabla_{\lambda}\tau^{\lambda\mu}_{(a)} 
-\tilde{b}_6\tilde{l}_{\tau_{a}q}\tau^{\lambda\mu}_{(a)}Du_{\lambda},
\label{equ80ver2}
\end{align}
\end{minipage}
}
\resizebox{\textwidth}{!}{
\begin{minipage}{\textwidth}
\begin{align}
\mathcal{C}^{\mu\nu} =& a_2\theta \pi^{\mu\nu}+\pi^{\mu\nu}Da_2+2a_2 D\pi^{\mu\nu} 
+(1-l_{\pi h})h^{(\nu}\nabla^{\mu)}b_2 
-b_2(1-\tilde{l}_{\pi h})h^{(\nu}Du^{\mu)} \nonumber\\
&+b_2\nabla^{(\mu}h^{\nu)} 
+l_{\pi q}q^{(\nu}\nabla^{\mu)}b_5 
-\tilde{l}_{\pi q}b_5 q^{(\nu}Du^{\mu)} 
+b_5\nabla^{(\mu}q^{\nu)}+c_4\nabla_{\alpha}\Theta^{(\mu\nu)\alpha} \nonumber\\
&+l_{\Theta\pi}\Theta^{(\mu\nu)\alpha}\nabla_{\alpha}c_4 
-\tilde{l}_{\Theta\pi}c_4\Theta^{(\mu\nu)\alpha}Du_{\alpha} 
,
\label{equ81ver2}
\end{align}
\end{minipage}
}
\resizebox{\textwidth}{!}{
\begin{minipage}{\textwidth}
\begin{align}
\mathcal{E}^{\mu\nu}=& a_5\theta\phi^{\mu\nu}+\phi^{\mu\nu}Da_5+2a_5 D\phi^{\mu\nu} 
+(1-l_{\phi h})h^{[\nu}\nabla^{\mu]}b_3 
-b_3(1-\tilde{l}_{\phi h})h^{[\nu}Du^{\mu]} \nonumber\\
&+b_3\nabla^{[\mu}h^{\nu]} 
+(1-l_{\phi q})q^{[\nu}\nabla^{\mu]}b_6 
-b_6(1-\tilde{l}_{\phi q})q^{[\nu}Du^{\mu]} 
+b_6\nabla^{[\mu}q^{\nu]} \nonumber\\
&+l_{\Theta\phi}\Theta^{\lambda\mu\nu}\nabla_{\lambda}c_1 
-\tilde{l}_{\Theta\phi}c_1\Theta^{\lambda\mu\nu}Du_{\lambda} 
+c_3\nabla_{\lambda}\Theta^{\lambda\mu\nu}+c_7\nabla_{\lambda}\Theta^{[\mu\nu]\lambda} \nonumber\\
&+k_{\Theta\phi}\Theta^{[\mu\nu]\lambda}\nabla_{\lambda}c_7 
-\tilde{k}_{\Theta\phi}c_7\Theta^{[\mu\nu]\lambda}Du_{\lambda},
\label{equ82ver2}
\end{align}
\end{minipage}
}
\resizebox{\textwidth}{!}{
\begin{minipage}{\textwidth}
\begin{align}
\mathcal{F}=&\tilde{a}_1\theta\Phi+\Phi D\tilde{a}_1+2\tilde{a}_1D\Phi+(1-l_{\Phi h})h^{\mu}\nabla_{\mu}\tilde{b}_1-(1-\tilde{l}_{\Phi h})\tilde{b}_1h^{\mu}Du_{\mu}+\tilde{b}_1\nabla_{\mu}h^{\mu}\nonumber\\
&+(1-l_{\Phi q})q^{\mu}\nabla_{\mu}\tilde{b}_4-(1-\tilde{l}_{\Phi q})\tilde{b}_4 q^{\mu}Du_{\mu}+\tilde{b}_4\nabla_{\mu}q^{\mu}+l_{\Theta\Phi}\Theta^{\alpha\mu\nu}\Delta_{\alpha\mu}\nabla_{\nu}c_5\nonumber\\
&-\tilde{l}_{\Theta\Pi}c_5\Delta_{\alpha\mu}\Theta^{\alpha\mu\nu}Du_{\nu}+c_5\Delta_{\alpha\beta}\nabla_{\mu}\Theta^{\alpha\beta\mu},
\label{equ83ver2}
\end{align}
\end{minipage}
}
\resizebox{\textwidth}{!}{
\begin{minipage}{\textwidth}
\begin{align}
\mathcal{G}^{\mu\nu} =&\tilde{a}_2\theta \tau^{\mu\nu}_{(s)}+\tau^{\mu\nu}_{(s)}D\tilde{a}_2+2\tilde{a}_2 D\tau^{\mu\nu}_{(s)}+(1-l_{\tau_{s} h})h^{(\nu}\nabla^{\mu)}\tilde{b}_2-\tilde{b}_2(1-\tilde{l}_{\tau_s h})h^{(\nu}Du^{\mu)}\nonumber\\
&+\tilde{b}_2\nabla^{(\mu}h^{\nu)}+(1-l_{\tau_s q})q^{(\nu}\nabla^{\mu)}\tilde{b}_5-(1-\tilde{l}_{\tau_s q})\tilde{b}_5 q^{(\nu}Du^{\mu)}+\tilde{b}_5\nabla^{(\mu}q^{\nu)}\nonumber\\
&+l_{\Theta\tau_s}\Theta^{(\mu\nu)\lambda}\nabla_{\lambda}c_6-\tilde{l}_{\Theta\tau_s}c_6\Theta^{(\mu\nu)\lambda}Du_{\lambda}+c_6\nabla_{\lambda}\Theta^{(\mu\nu)\lambda},
\label{equ84ver2}
\end{align}
\end{minipage}
}
\resizebox{\textwidth}{!}{
\begin{minipage}{\textwidth}
\begin{align}
\mathcal{H}^{\mu\nu} =&\tilde{a}_3\theta\tau^{\mu\nu}_{(a)}+\tau^{\mu\nu}_{(a)}D\tilde{a}_3+2\tilde{a}_3 D\tau^{\mu\nu}_{(a)}+(1-l_{\tau_a h})h^{[\nu}\nabla^{\mu]}\tilde{b}_3-\tilde{b}_3(1-\tilde{l}_{\tau_a h})h^{[\nu}Du^{\mu]}\nonumber\\
&+\tilde{b}_3\nabla^{[\mu}h^{\nu]}+(1-l_{\tau_a q})q^{[\nu}\nabla^{\mu]}\tilde{b}_6-\tilde{b}_6(1-\tilde{l}_{\tau_a q})q^{[\nu}Du^{\mu]}+\tilde{b}_6\nabla^{[\mu}q^{\nu]}
\nonumber\\
&+l_{\Theta\tau_a}\Theta^{\lambda\mu\nu}\nabla_{\lambda}c_2-\tilde{l}_{\Theta\tau_a}c_2\Theta^{\lambda\mu\nu}Du_{\lambda}+c_2\nabla_{\lambda}\Theta^{\lambda\mu\nu}+c_8\nabla_{\lambda}\Theta^{[\mu\nu]\lambda}\nonumber\\
&+k_{\Theta\tau_a}\Theta^{[\mu\nu]\lambda}\nabla_{\lambda}c_8-\tilde{k}_{\Theta\tau_a}c_8\Theta^{[\mu\nu]\lambda}Du_{\lambda},
\label{equ85ver2}
\end{align}
\end{minipage}
}
\resizebox{\textwidth}{!}{
\begin{minipage}{\textwidth}
\begin{align}
\mathcal{I}^{\alpha\mu\nu}  =&\tilde{a}_4\theta\Theta^{\alpha\mu\nu}+ \Theta^{\alpha\mu\nu}D\tilde{a}_4+2\tilde{a}_4D\Theta^{\alpha\mu\nu} + (1-l_{\Theta \phi})\phi^{\mu\nu}\nabla^{\alpha}c_1-(1-\tilde{l}_{\Theta \phi})c_1\phi^{\mu\nu}Du^{\alpha}\nonumber\\
&+c_1\nabla^{\alpha}\phi^{\mu\nu}+ (1-l_{\Theta \tau_a})\tau_{(a)}^{\mu\nu}\nabla^{\alpha}c_2-(1-\tilde{l}_{\Theta \tau_a})c_2\tau^{\mu\nu}_{(a)}Du^{\alpha}+c_2\nabla^{\alpha}\tau^{\mu\nu}_{(a)}\nonumber\\
&+(1-l_{\Theta\Pi})\Pi\Delta^{\alpha[\mu}\nabla^{\nu]}c_3-(1-\tilde{l}_{\Theta\Pi})c_3\Pi\Delta^{\alpha[\mu}Du^{\nu]}+c_3\Delta^{\alpha[\mu}\nabla^{\nu]}\Pi
\nonumber\\
&+(1-l_{\Theta\Phi})\Phi\Delta^{\alpha[\mu}\nabla^{\nu]}c_5-(1-\tilde{l}_{\Theta\Phi})c_5\Phi\Delta^{\alpha[\mu}Du^{\nu]}+c_5\Delta^{\alpha[\mu}\nabla^{\nu]}\Phi\nonumber\\
&+(1-k_{\Theta\tau_a})\tau_{(a)}^{\alpha[\mu}\nabla^{\nu]}c_8-(1-\tilde{k}_{\Theta\tau_a})c_8\tau_{(a)}^{\alpha[\mu}Du^{\nu]}+c_8\nabla^{[\nu}\tau_{(a)}^{\alpha\mu]}\nonumber\\
&+(1-k_{\Theta\phi})\phi^{\alpha[\mu}\nabla^{\nu]}c_7-(1-\tilde{k}_{\Theta\phi})c_7\phi^{\alpha[\mu}Du^{\nu]}+c_7\nabla^{[\nu}\phi^{\alpha\mu]}\nonumber\\
&+(1-l_{\Theta\tau_s})\tau_{(s)}^{\alpha[\mu}\nabla^{\nu]}c_6-(1-\tilde{l}_{\Theta\tau_s})c_6\tau_{(s)}^{\alpha[\mu}Du^{\nu]}+c_6\nabla^{[\nu}\tau_{(s)}^{\alpha\mu]} \nonumber\\
&+(1-l_{\Theta\pi})\pi^{\alpha[\mu}\nabla^{\nu]}c_4-(1-\tilde{l}_{\Theta\pi})c_4\pi^{\alpha[\mu}Du^{\nu]}+c_4\nabla^{[\nu}\pi^{\alpha\mu]}.
\label{equ86ver2}
\end{align}
\end{minipage}
}
\medskip

By substituting all terms obtained above into Eq.~\eqref{FinalISentropypro}, we obtain the complete expression for the divergence of the entropy current at second-order. This formulation incorporates the contributions from both first-order and second-order corrections of dissipative currents. The full expression can be written as
\begin{align}
\partial_{\mu}S^{\mu}_{\rm MIS}=&-\beta h^{\mu}\left(\beta\nabla_{\mu}T-Du_{\mu}-T\mathcal{A}_{\langle\mu\rangle}\right)-\beta q^{\mu}\left(\beta\nabla_{\mu}T+Du_{\mu}-4\omega_{\mu\nu}u^{\nu}-T\mathcal{B}_{\langle\mu\rangle}\right)\nonumber\\
&+\beta\pi^{\mu\nu}\left(\sigma_{\mu\nu}+T\mathcal{C}_{\langle\mu\nu\rangle}\right)+\beta\Pi\left(\theta+T\mathcal{D}\right)+\phi^{\mu\nu}\left(\Omega_{\mu\nu}+2\beta\omega_{\langle\mu\rangle\langle\nu\rangle}+\mathcal{E}_{\langle[\mu\nu]\rangle}\right)\nonumber\\
&
+\Phi\left[-2 u^{\alpha}\nabla^{\beta}(\beta\omega_{\alpha\beta})+\mathcal{F}\right]
+\tau^{\mu\beta}_{(s)}\left[-2u^{\alpha}\Delta^{\gamma\rho}_{\mu\beta}\nabla_{\gamma}(\beta\omega_{\alpha\rho})+\mathcal{G}_{\langle\mu\beta\rangle}\right]\nonumber\\
&+\Theta^{\mu\alpha\beta}\left[-\Delta^{\delta}_{~\alpha}\Delta^{\rho}_{~\beta}\Delta^{\gamma}_{~\mu}\nabla_{\gamma}(\beta\omega_{\delta\rho})+\mathcal{I}_{\langle\mu\rangle\langle\alpha\rangle\langle\beta\rangle}\right]\nonumber\\
&
+\tau^{\mu\beta}_{(a)}\left[-2u^{\alpha}\Delta^{[\gamma\rho]}_{[\mu\beta]}\nabla_{\gamma}(\beta\omega_{\alpha\rho})
+\mathcal{H}_{\langle[\mu\beta]\rangle}\right].
\label{equ32ver2}
\end{align} 
Similarly to the NS theory, the second law of thermodynamics $\partial_{\mu}S^{\mu}_{\text{MIS}}\geq 0$ is guaranteed if the dissipative currents have the following forms:
\begin{align}
& h^{\mu}=-\kappa\left(Du^{\mu}-\beta\nabla^{\mu}T+T\mathcal{A^{\langle\mu\rangle}}\right),\label{equ34ver2}\\
& q^{\mu}= \lambda\left(Du^{\mu}+\beta\nabla^{\mu}T-4\omega^{\mu\nu}u_{\nu}-T\mathcal{B^{\langle\mu\rangle}}\right),\label{equ35ver2}\\
& \Pi=\zeta\big(\theta+T\mathcal{D}\big),\label{equ2}\\
& 
\pi^{\mu\nu}=2\eta\left(\sigma^{\mu\nu}+T\mathcal{C^{\langle\mu\nu\rangle}}\right),\label{equ36ver2}\\
& \phi^{\mu\nu}=\gamma\left(\Omega^{\mu\nu}+2\beta\omega^{\langle\mu\rangle\langle\nu\rangle}+\mathcal{E^{\langle[\mu\nu]\rangle}}\right),\label{equ37ver2}\\
& \Phi=\chi_{1}\left(-2 u^{\alpha}\nabla^{\beta}(\beta\omega_{\alpha\beta})+\mathcal{F}\right),\label{equ38ver2}\\
& \tau^{\mu\beta}_{(s)}=\chi_{2}\left[-u^{\alpha}\left(\Delta^{\gamma\mu}\Delta^{\rho\beta}+\Delta^{\gamma\beta}\Delta^{\rho\mu}-\frac{2}{3}\Delta^{\gamma\rho}\Delta^{\mu\beta}\right)\nabla_{\gamma}(\beta\omega_{\alpha\rho})+\mathcal{G}^{\langle\mu\beta\rangle}\right],\label{equ39ver2}\\
& \tau^{\mu\beta}_{(a)}=\chi_{3} \left[-u^{\alpha}(\Delta^{\gamma\mu}\Delta^{\rho\beta}-\Delta^{\gamma\beta}\Delta^{\rho\mu})\nabla_{\gamma}(\beta\omega_{\alpha\rho})
+\mathcal{H}^{\langle[\mu\beta]\rangle}\right],\label{equ40ver2}\\
& \Theta^{\mu\alpha\beta}= -\chi_4  \left[-\Delta^{\delta\alpha}\Delta^{\rho\beta}\Delta^{\gamma\mu}\nabla_{\gamma}(\beta\omega_{\delta\rho})+\mathcal{I}^{\langle\mu\rangle\langle\alpha\rangle\langle\beta\rangle}\right]\label{equ41ver2},
\end{align}
where the transport coefficients must satisfy: $\kappa\geq 0$, $\lambda\geq 0$, $\eta\geq0$, $\zeta\geq0$ $\gamma\geq0$, $\chi_1\geq 0$,
$\chi_2\geq 0$, $\chi_3\geq 0$, and $\chi_4\geq 0$. Here, as before, $\kappa$ is the heat conductivity, $\eta$ is the shear viscosity, and $\zeta$ is the bulk viscosity. The transport coefficients $\lambda$ and $\gamma$ are identified as boost heat conductivity and rotational viscosity, respectively. The remaining coefficients $\chi_1, \chi_2, \chi_{2},$ and $\chi_{4}$ are the new transport coefficients related to spin. 
\medskip 

The expressions in Eqs.~\eqref{equ34ver2}-\eqref{equ41ver2} of various transport coefficients are, in fact, relaxation-type dynamical equations. To see this, we substitute the explicit expressions of $\mathcal{D}$, $\mathcal{A^{\langle\mu\rangle}}$, $\mathcal{B^{\langle\mu\rangle}}$, $\mathcal{C^{\langle\mu\nu\rangle}}$ and $\mathcal{E^{\langle[\mu\nu]\rangle}}$ from Eqs.~\eqref{equ78ver2}-\eqref{equ86ver2} into Eqs.~\eqref{equ34ver2}-\eqref{equ41ver2}. In this way, we may write relaxation-type dynamical equations for the dissipative currents of the energy-momentum tensor as: 
\begin{align}
Dh^{\langle\mu\rangle}+\frac{h^{\mu}}{\tau_{h}} =&-\frac{1}{2a_3}\bigg[\beta(Du^{\mu}-\beta\nabla^{\mu}T)+a_3h^{\mu}\theta+h^{\mu}Da_3+l_{\Pi h}\Pi\nabla^{\mu}b_1+b_1\nabla^{\mu}\Pi\nonumber\\
&-\tilde{b}_1\tilde{l}_{\Phi h}\Phi Du^{\mu}+l_{\tau_s h}\tau^{\lambda\mu}_{(s)}\nabla_{\lambda}\tilde{b}_2+\tilde{b}_2\Delta^{\mu}_{~\nu}\nabla_{\lambda}\tau^{\lambda\nu}_{(s)}-\tilde{b}_3\tilde{l}_{\tau_{a}h}\tau^{\lambda\mu}_{(a)}Du_{\lambda}\nonumber\\
&-b_1\tilde{l}_{\Pi h}\Pi Du^{\mu}+l_{\pi h}\pi^{\lambda\mu}\nabla_{\lambda}b_2+b_2\Delta^{\mu}_{~\nu}\nabla_{\lambda}\pi^{\lambda\nu}-b_2\tilde{l}_{\pi h}\pi^{\lambda\mu}Du_{\lambda}\nonumber\\
&+l_{\phi h}\phi^{\lambda\mu}\nabla_{\lambda}b_3-b_3\tilde{l}_{\phi h}\phi^{\lambda\mu}Du_{\lambda}+l_{\Phi h}\Phi\nabla^{\mu}\tilde{b}_1+\tilde{b}_1\nabla^{\mu}\Phi\nonumber\\
&-\tilde{b}_2\tilde{l}_{\tau_{s}h}\tau^{\lambda\mu}_{(s)}Du_{\lambda}+l_{\tau_a h}\tau^{\lambda\mu}_{(a)}\nabla_{\lambda}\tilde{b}_3+\tilde{b}_3\Delta^{\mu}_{~\nu}\nabla_{\lambda}\tau^{\lambda\nu}_{(a)}\bigg],
\label{equ44ver2}
\end{align}
\resizebox{\textwidth}{!}{
\begin{minipage}{\textwidth}
\begin{align}
Dq^{\langle\mu\rangle}+\frac{q^{\mu}}{\tau_{q}} =&\frac{1}{2a_4}\bigg[\beta(\beta\nabla^{\mu}T+Du^{\mu}-4\omega^{\mu\nu}u_{\nu})-a_4 q^{\mu}\theta-q^{\mu}Da_4-b_4 \nabla^{\mu}\Pi-b_5\Delta^{\mu}_{~\nu}\nabla_{\lambda}\pi^{\lambda\nu}
\nonumber\\
& -(1-l_{\Pi q})\Pi\nabla^{\mu}b_4+b_4(1-\tilde{l}_{\Pi q}) \Pi Du^{\mu}-(1-l_{\pi q}) \pi^{\lambda\mu}\nabla_{\lambda}b_5-l_{\phi q} \phi^{\lambda\mu}\nabla_{\lambda}b_6\nonumber\\
&-\tilde{b}_4 \nabla^{\mu}\Phi+\tilde{b}_4\tilde{l}_{\Phi q} \Phi Du^{\mu}-l_{\tau_s q} \tau^{\lambda\mu}_{(s)}\nabla_{\lambda}\tilde{b}_5
\nonumber-\tilde{b}_5 \Delta^{\mu}_{~\nu}\nabla_{\lambda}\tau^{\lambda\nu}_{(s)}+\tilde{b}_5\tilde{l}_{\tau_{s}q} \tau^{\lambda\mu}_{(s)}Du_{\lambda}\nonumber\\
&+b_5(1-\tilde{l}_{\pi q}) \pi^{\lambda\mu}Du_{\lambda}-b_6 \Delta^{\mu}_{~\nu}\nabla_{\lambda}\phi^{\lambda\nu}+b_6\tilde{l}_{\phi q} \phi^{\lambda\mu}Du_{\lambda}-l_{\Phi q} \Phi\nabla^{\mu}\tilde{b}_4\nonumber\\
&-l_{\tau_a q} \tau^{\lambda\mu}_{(a)}\nabla_{\lambda}\tilde{b}_6-\tilde{b}_6 \Delta^{\mu}_{~\nu}\nabla_{\lambda}\tau^{\lambda\nu}_{(a)}+\tilde{b}_6\tilde{l}_{\tau_{a}q} \tau^{\lambda\mu}_{(a)}Du_{\lambda}\bigg],
\label{equ45ver2}
\end{align}
\end{minipage}
}
\resizebox{\textwidth}{!}{
\begin{minipage}{\textwidth}
\begin{align}
D\Pi+\frac{\Pi}{\tau_{\Pi}}=&-\frac{1}{2a_1}\bigg[~\beta\theta+a_1\Pi\theta+\Pi Da_1+(1-l_{\Pi h})h^{\mu}\nabla_{\mu}b_1-~b_1(1-\tilde{l}_{\Pi h})h^{\mu}Du_{\mu}\nonumber\\
&+b_1\nabla_{\mu}h^{\mu}+l_{\Pi q}q^{\mu}\nabla_{\mu}b_4-\tilde{l}_{\Pi q}b_4  q^{\mu}Du_{\mu}+b_4 \nabla_{\mu}q^{\mu}+l_{\Theta\Pi}\Theta^{\alpha\mu\nu}\Delta_{\alpha\mu}\nabla_{\nu}c_3\nonumber\\
&-\tilde{l}_{\Theta\Pi}c_3\Delta_{\alpha\mu}\Theta^{\alpha\mu\nu}Du_{\nu}+c_3\Delta_{\alpha\beta}\nabla_{\mu}\Theta^{\alpha\beta\mu}\bigg],
\label{equ43ver2}
\end{align}
\end{minipage}
}
\resizebox{\textwidth}{!}{
\begin{minipage}{\textwidth}
\begin{align}
D\pi^{\langle\mu\nu\rangle}+\frac{\pi^{\mu\nu}}{\tau_{\pi}}=&-\frac{1}{2a_2}\bigg[\beta\sigma^{\mu\nu}+a_2 \theta \pi^{\mu\nu}+ \pi^{\mu\nu}Da_2+(1-l_{\pi h}) h^{\langle\mu}\nabla^{\nu\rangle}b_2+b_2\nabla^{\langle\mu}h^{\nu\rangle}
\nonumber\\
&-b_2(1-\tilde{l}_{\pi h})h^{\langle\mu}Du^{\nu\rangle}
+l_{\pi q}q^{\langle\mu}\nabla^{\nu\rangle}b_5-\tilde{l}_{\pi q}b_5 q^{\langle\mu}Du^{\nu\rangle}+b_5 \nabla^{\langle\mu}q^{\nu\rangle}
\nonumber\\
&+l_{\Theta\pi}\Theta^{\langle\mu\nu\rangle\alpha}\nabla_{\alpha}c_4-\tilde{l}_{\Theta\pi}c_4\Theta^{\langle\mu\nu\rangle\alpha}Du_{\alpha}+c_4\nabla_{\alpha}\Theta^{\langle\mu\nu\rangle\alpha}\bigg],
\label{equ46ver2}
\end{align}
\end{minipage}
}
\resizebox{\textwidth}{!}{
\begin{minipage}{\textwidth}
\begin{align}
D\phi^{\langle[\mu\nu]\rangle}+\frac{\phi^{\mu\nu}}{\tau_{\phi}} =&-\frac{1}{2a_5}\bigg[\left(\Omega^{\mu\nu}+2\beta\omega^{\langle\mu\rangle\langle\nu\rangle}\right)+a_5\theta\phi^{\mu\nu}+\phi^{\mu\nu}Da_5+b_3\Delta^{[\mu\nu]}_{[\alpha\beta]}\nabla^{[\alpha}h^{\beta]}
\nonumber\\
&+(1-l_{\phi h})h^{[\nu}\nabla^{\mu]}b_3-b_3(1-\tilde{l}_{\phi h})h^{[\nu}Du^{\mu]}
+(1-l_{\phi q})q^{[\nu}\nabla^{\mu]}b_6
\nonumber\\
&-b_6(1-\tilde{l}_{\phi q})q^{[\nu}Du^{\mu]}
+b_6\Delta^{[\mu\nu]}_{[\alpha\beta]}\nabla^{[\alpha}q^{\beta]}+l_{\Theta\phi}\Theta^{\lambda\mu\nu}\nabla_{\lambda}c_1\nonumber\\
&-\tilde{l}_{\Theta\phi}c_1\Theta^{\lambda\mu\nu}Du_{\lambda}+c_3\Delta^{[\mu\nu]}_{[\alpha\beta]}\nabla_{\lambda}\Theta^{\lambda\alpha\beta}+k_{\Theta\phi}\Theta^{[\mu\nu]\lambda}\nabla_{\lambda}c_7
\nonumber\\
&-\tilde{k}_{\Theta\phi}c_7\Theta^{[\mu\nu]\lambda}Du_{\lambda}+c_7\Delta^{[\mu\nu]}_{[\alpha\beta]}\nabla_{\lambda}\Theta^{[\alpha\beta]\lambda}\bigg].
 \label{equ47ver2}
\end{align}
\end{minipage}
}
In the above equations, the relaxation times of various dissipative quantities are defined as, $\tau_{\Pi}=-2a_1\zeta T\geq 0$, $\tau_{h}=2a_3\kappa T\geq 0$, $\tau_{q}=2a_4\lambda T\geq 0$, $\tau_{\pi}=-4a_2\eta T\geq 0$, and $\tau_{\phi}=-2a_5\gamma\geq 0$. Moreover, $Dh^{\langle\mu\rangle}=\Delta^{\mu}_{~\nu}Dh^{\nu}$, $Dq^{\langle\mu\rangle}=\Delta^{\mu}_{~\nu}Dq^{\nu}$, $D\pi^{\langle\mu\nu\rangle}=\Delta^{\mu\nu}_{\alpha\beta}D\pi^{\alpha\beta}$, and $D\phi^{\langle[\mu\nu]\rangle}=\Delta^{[\mu\nu]}_{[\alpha\beta]}D\phi^{\alpha\beta}$. Similarly, the dynamical equations for dissipative currents appearing in the spin tensor read:

\resizebox{\textwidth}{!}{
\begin{minipage}{\textwidth}
\begin{align}
D\Phi+\frac{\Phi}{\tau_{\Phi}}=&-\frac{1}{2\tilde{a}_1}\bigg[-2 u^{\alpha}\nabla^{\beta}(\beta\omega_{\alpha\beta})+\tilde{a}_1\theta\Phi+\Phi D\tilde{a}_1+(1-l_{\Phi h})h^{\mu}\nabla_{\mu}\tilde{b}_1+\tilde{b}_1\nabla_{\mu}h^{\mu}\nonumber\\
&-(1-\tilde{l}_{\Phi h})\tilde{b}_1h^{\mu}Du_{\mu}+(1-l_{\Phi h})q^{\mu}\nabla_{\mu}\tilde{b}_4-(1-\tilde{l}_{\Phi q})\tilde{b}_4 q^{\mu}Du_{\mu}+\tilde{b}_4\nabla_{\mu}q^{\mu}
\nonumber\\
&+l_{\Theta\Phi}\Theta^{\alpha\mu\nu}\Delta_{\alpha\mu}\nabla_{\nu}c_5-\tilde{l}_{\Theta\Pi}c_5\Delta_{\alpha\mu}\Theta^{\alpha\mu\nu}Du_{\nu}+c_5\Delta_{\alpha\beta}\nabla_{\mu}\Theta^{\alpha\beta\mu}\bigg],
\label{equ48ver2}
\end{align}
\end{minipage}
}
\resizebox{\textwidth}{!}{
\begin{minipage}{\textwidth}
\begin{align}
D\tau^{\langle\mu\nu\rangle}_{(s)}+\frac{\tau^{\mu\nu}_{(s)}}{\tau_{\tau_s}}  =&-\frac{1}{2\tilde{a}_2}\bigg[-u^{\alpha}\left(\Delta^{\gamma\mu}\Delta^{\rho\nu}+\Delta^{\gamma\nu}\Delta^{\rho\mu}-\frac{2}{3}\Delta^{\gamma\rho}\Delta^{\mu\nu}\right)\nabla_{\gamma}(\beta\omega_{\alpha\rho})+\tilde{a}_2\theta \tau^{\mu\nu}_{(s)}\nonumber\\
&+\tau^{\mu\nu}_{(s)}D\tilde{a}_2+(1-l_{\tau_{s} h})h^{\langle\mu}\nabla^{\nu\rangle}\tilde{b}_2-\tilde{b}_2(1-\tilde{l}_{\tau_s h})h^{\langle\mu}Du^{\nu\rangle}+\tilde{b}_2\nabla^{\langle\mu}h^{\nu\rangle}
\nonumber\\
&+(1-l_{\tau_s q})q^{\langle\mu}\nabla^{\nu\rangle}\tilde{b}_5-(1-\tilde{l}_{\tau_s q})\tilde{b}_5 q^{\langle\mu}Du^{\nu\rangle}+\tilde{b}_5\nabla^{\langle\mu}q^{\nu\rangle}\nonumber\\
&+l_{\Theta\tau_s}\Theta^{\langle\mu\nu\rangle\lambda}\nabla_{\lambda}c_6-\tilde{l}_{\Theta\tau_s}c_6\Theta^{\langle\mu\nu\rangle\lambda}Du_{\lambda}+c_6\nabla_{\lambda}\Theta^{\langle\mu\nu\rangle\lambda}\bigg],
\label{equ49ver2}
\end{align}
\end{minipage}
}
\resizebox{\textwidth}{!}{
\begin{minipage}{\textwidth}
\begin{align}
D\tau^{\langle[\mu\nu]\rangle}_{(a)}+\frac{\tau^{\mu\nu}_{(a)}}{\tau_{\tau_a}}=&-\frac{1}{2\tilde{a}_3}\bigg[-u^{\alpha}(\Delta^{\gamma\mu}\Delta^{\rho\nu}-\Delta^{\gamma\nu}\Delta^{\rho\mu})\nabla_{\gamma}(\beta\omega_{\alpha\rho})   +\tilde{a}_3\theta\tau^{\mu\nu}_{(a)}+\tau^{\mu\nu}_{(a)}D\tilde{a}_3\nonumber\\
&+(1-l_{\tau_a h})h^{[\nu}\nabla^{\mu]}\tilde{b}_3-\tilde{b}_3(1-\tilde{l}_{\tau_a h})h^{[\nu}Du^{\mu]}+\tilde{b}_3\Delta^{[\mu\nu]}_{[\alpha\beta]}\nabla^{[\alpha}h^{\beta]}\nonumber\\
&+(1-l_{\tau_a q})q^{[\nu}\nabla^{\mu]}\tilde{b}_6-\tilde{b}_6(1-\tilde{l}_{\tau_a q})q^{[\nu}Du^{\mu]}+\tilde{b}_6\Delta^{[\mu\nu]}_{[\alpha\beta]}\nabla^{[\alpha}q^{\beta]}\nonumber\\
&+l_{\Theta\tau_a}\Theta^{\lambda\mu\nu}\nabla_{\lambda}c_2-\tilde{l}_{\Theta\tau_a}c_2\Theta^{\lambda\mu\nu}Du_{\lambda}+c_2\Delta^{[\mu\nu]}_{[\alpha\beta]}\nabla_{\lambda}\Theta^{\lambda\alpha\beta}\nonumber\\
&+k_{\Theta\tau_a}\Theta^{[\mu\nu]\lambda}\nabla_{\lambda}c_8-\tilde{k}_{\Theta\tau_a}c_8\Theta^{[\mu\nu]\lambda}Du_{\lambda}+c_8\Delta^{[\mu\nu]}_{[\alpha\beta]}\nabla_{\lambda}\Theta^{[\alpha\beta]\lambda}\bigg],
\label{equ50ver2}
\end{align}
\end{minipage}
}
\resizebox{\textwidth}{!}{
\begin{minipage}{\textwidth}
\begin{align}
D\Theta^{\langle\alpha\rangle\langle\mu\rangle\langle\nu\rangle}=&\frac{-\Theta^{\alpha\mu\nu}}{\tau_{\Theta}}-\frac{1}{2\tilde{a}_4}\bigg[-\Delta^{\delta\mu}\Delta^{\rho\nu}\Delta^{\gamma\alpha}\nabla_{\gamma}(\beta\omega_{\delta\rho})+\tilde{a}_4\theta\Theta^{\alpha\mu\nu}+ (1-l_{\Theta \phi})\phi^{\mu\nu}\nabla^{\alpha}c_1\nonumber\\
&+(1-k_{\Theta\tau_a})\tau_{(a)}^{\alpha[\mu}\nabla^{\nu]}c_8 -(1-\tilde{k}_{\Theta\tau_a})c_8\tau_{(a)}^{\alpha[\mu}Du^{\nu]}+c_8\Delta^{\alpha a}\Delta^{\mu b} \Delta^{\nu c}\nabla_{[c}\tau_{(a)ab]}\nonumber\\
&+(1-l_{\Theta\tau_s})\tau_{(s)}^{\alpha[\mu}\nabla^{\nu]}c_6 -(1-\tilde{l}_{\Theta\tau_s})c_6\tau_{(s)}^{\alpha[\mu}Du^{\nu]}+c_6\Delta^{\alpha a}\Delta^{\mu b} \Delta^{\nu c}\nabla_{[c}\tau_{(s)ab]}
\nonumber\\
& +(1-k_{\Theta\phi})\phi^{\alpha[\mu}\nabla^{\nu]}c_7-(1-\tilde{k}_{\Theta\phi})c_7\phi^{\alpha[\mu}Du^{\nu]}+c_7\Delta^{\alpha a}\Delta^{\mu b} \Delta^{\nu c}\nabla_{[c}\phi_{ab]}\nonumber\\
&+(1-l_{\Theta\pi})\pi^{\alpha[\mu}\nabla^{\nu]}c_4 -(1-\tilde{l}_{\Theta\pi})c_4\pi^{\alpha[\mu}Du^{\nu]}+c_4\Delta^{\alpha a}\Delta^{\mu b} \Delta^{\nu c}\nabla_{[c}\pi_{ab]}\nonumber\\
& + (1-l_{\Theta \tau_a})\tau_{(a)}^{\mu\nu}\nabla^{\alpha}c_2-(1-\tilde{l}_{\Theta \tau_a})c_2\tau^{\mu\nu}_{(a)}Du^{\alpha}+c_2\Delta^{\alpha a}\Delta^{\mu b}\Delta^{\nu  c}\nabla_a\tau_{bc(a)}\nonumber\\
&+(1-l_{\Theta\Pi})\Pi\Delta^{\alpha[\mu}\nabla^{\nu]}c_3 -(1-\tilde{l}_{\Theta\Pi})c_3\Pi\Delta^{\alpha[\mu}Du^{\nu]}+c_3\Delta^{\alpha[\mu}\nabla^{\nu]}\Pi\nonumber\\
&+(1-l_{\Theta\Phi})\Phi\Delta^{\alpha[\mu}\nabla^{\nu]}c_5 -(1-\tilde{l}_{\Theta\Phi})c_5\Phi\Delta^{\alpha[\mu}Du^{\nu]}+c_5\Delta^{\alpha[\mu}\nabla^{\nu]}\Phi
\nonumber\\
&+ \Theta^{\alpha\mu\nu}D\tilde{a}_4-(1-\tilde{l}_{\Theta \phi})c_1\phi^{\mu\nu}Du^{\alpha}+c_1\Delta^{\alpha a}\Delta^{\mu b}\Delta^{\nu  c}\nabla_{a}\phi_{bc}\bigg].
\label{equ51ver2}
\end{align}
\end{minipage}
}
In above equations, $D\tau^{\langle\mu\nu\rangle}_{(s)}\!\equiv\! \Delta^{\mu\nu}_{\alpha\beta}D\tau_{(s)}^{\alpha\beta}$, $D\tau^{\langle[\mu\nu]\rangle}_{(a)}\!\equiv\! \Delta^{[\mu\nu]}_{[\alpha\beta]}D\tau_{(a)}^{\alpha\beta}$,  $D\Theta^{\langle\alpha\rangle\langle\mu\rangle\langle\nu\rangle}\equiv\! \Delta^{\alpha a}\Delta^{\mu b}\Delta^{\nu  c}D\Theta_{abc} $. The spin-relaxation times can be identified as: $\tau_{\Phi}=-2\tilde{a_1}\chi_1\geq 0$, $\tau_{\tau_s}\equiv -2\tilde{a}_2\chi_2\geq 0$, $\tau_{\tau_a}\equiv -2\tilde{a}_3\chi_3\geq 0$, and $\tau_{\Theta}\equiv 2\tilde{a}_4\chi_4\geq 0$. 

Collectively, the relaxation-type dynamical equations governing the evolution of the dissipative currents, together with the evolution equations~\eqref{uTIS}-\eqref{SpinISevo} resulting from conservation laws, constitute a closed system of 16+24 coupled partial differential equations for 16+24 dynamical variables.
%

%
\section{Linear mode analysis: stability and causality} 
\label{Lmastabilityandcausality}
In this section, we use the linear mode analysis introduced in Sec.~\ref{stability} to check the stability and causality of the formulated evolution equations~\eqref{uTIS}--\eqref{SpinISevo} and \eqref{equ44ver2}--\eqref{equ51ver2} describing the dynamics of the variables characterizing energy-momentum tensor and the spin tensor shown in Tables~\ref{Table1} and \ref{Dynamical variables of the spin tensor SIS.}. To make the analysis feasible, we consider a truncated version of these equations. We perform the analysis in the Landau frame where $h^{\mu}=0$ and the fluid element rest frame. We examine linear perturbations of the dynamical variables around a homogeneous global equilibrium. We find that the truncated equations are stable in the low- and high-wavenumber limits with dependence on the spin equation of state. Moreover, they are found to be causal. The presentation is based primarily on Ref.~\multimyref{AD7}.
%
\subsection{Truncated second-order dissipative spin hydrodynamics}
\label{a ssspecial case}


To make the stability and causality assessment feasible, it is desirable to simplify the evolution equations~\eqref{uTIS}--\eqref{SpinISevo} and \eqref{equ44ver2}--\eqref{equ51ver2}. For this purpose, we adopt the expression for \(Q^{\mu}\) given in Eq.~\eqref{eq19}, however neglecting the cross terms proportional to \(\Pi\,q^{\mu}\), \(\pi^{\mu\nu}q_{\nu}\), \(\Theta^{\alpha\beta\mu}\phi_{\alpha\beta}\), and other similar contributions. Hence, we choose
\begin{align}
Q^{\mu} = & ~~u^{\mu}\left(a_{1}\Pi^2+a_{2}\pi^{\lambda\nu}\pi_{\lambda\nu}+a_{4}q^{\lambda}q_{\lambda}+a_{5}\phi^{\lambda\nu}\phi_{\lambda\nu}\right)\nonumber
\\
& + u^{\mu}\left(\tilde{a}_{1}\Phi^{2}+\tilde{a}_{2}\tau_{(s)}^{\lambda\nu}\tau_{(s)\lambda\nu}+\tilde{a}_{3}\tau_{(a)}^{\lambda\nu}\tau_{(a)\lambda\nu}+\tilde{a}_{4}\Theta^{\lambda\alpha\beta}\Theta_{\lambda\alpha\beta}\right).
\end{align}
Starting from Eq.~\eqref{FinalISentropypro} and following the methodology outlined in Secs.~\ref{Formulation}-\ref{relaxation-typeIS}, we find that the evolution equations governing the dynamics of the energy-momentum tensor are:
\begin{align}
D\varepsilon+(\varepsilon+P)\theta=\pi^{\mu\nu}\partial_{\mu}u_{\nu}+\Pi\theta-\nabla\cdot q+\phi^{\mu\nu}\partial_{\mu}u_{\nu}\,,\label{energyeq}
\end{align}
\vspace{-2em}
\begin{align}
(\varepsilon+P) Du^{\alpha} - \nabla^{\alpha}  P=
&-\Delta^{\alpha}_{\nu}\partial_{\mu}\pi^{\mu\nu}-\Delta^{\mu\alpha}\partial_{\mu}\pi+\pi D u^{\alpha}-q^{\mu}\partial_{\mu}u^{\alpha}\nonumber\\
&+\Delta^{\alpha}_{\nu}Dq^{\nu}+q^{\alpha}\theta-\Delta^{\alpha}_{\nu}\partial_{\mu}\phi^{\mu\nu}\, ,
\end{align}
\vspace{-2em}
\begin{align}
    &\tau_{\Pi}D\Pi+\Pi=\zeta\left[\theta+Ta_{1}\Pi\theta+T\Pi Da_{1}\right],
\end{align}
\vspace{-2em}
\begin{align}
\tau_{\pi}\Delta^{\mu\nu}_{\alpha\beta}D\pi^{\alpha\beta}+\pi^{\mu\nu}=2\eta\left[(\nabla^{(\mu}u^{\nu)}-\frac{1}{3}\theta\Delta^{\mu\nu})+Ta_{2}\theta\pi^{\mu\nu}+T\pi^{\mu\nu}Da_{2}\right],
\end{align}
\vspace{-2em}
\begin{align}
\tau_{q}\Delta^{\mu}_{\nu}Dq^{\nu}+q^{\mu}=\lambda\left[(\beta\nabla^{\mu}T+Du^{\mu}-4\omega^{\mu\nu}u_{\nu})-Ta_{4}q^{\mu}\theta-Tq^{\mu}Da_{4}\right],\label{qeq}
\end{align}

\vspace{-2em}
\begin{align}
\tau_{\phi}\Delta^{[\mu\nu]}_{[\alpha\beta]}D\phi^{\alpha\beta}+\phi^{\mu\nu}=\gamma\left[ (\beta\nabla^{[\mu}u^{\nu]}+2\beta\Delta^{\mu\alpha}\Delta^{\nu\beta}\omega_{\alpha\beta})+a_{5}\theta\phi^{\mu\nu}+\phi^{\mu\nu}Da_{5}\right]\label{phieq}.
\end{align}
%
Similarly, the evolution equations for the spin density and spin dissipative currents reduce to: 
\begin{align}
DS^{\alpha\beta}+S^{\alpha\beta}\theta+\partial_{\mu}S^{\mu\alpha\beta}_{1}=-2(q^{\alpha}u^{\beta}-q^{\beta}u^{\alpha}+\phi^{\alpha\beta}),
\end{align}
\vspace{-2em}
\begin{align}
    \tau_{\Phi}D\Phi+\Phi=\chi_{1}\left[-2u^{\alpha}\nabla^{\beta}(\beta\omega_{\alpha\beta})+\tilde{a}_{1}\theta\Phi+\Phi D\tilde{a}_{1}\right],\label{Phieq}
\end{align}
\vspace{-2em}
\begin{align}
\tau_{\tau_{s}}\Delta^{\mu\nu}_{\alpha\beta}D\tau^{\alpha\beta}_{(s)}+\tau^{\mu\nu}_{(s)}=\chi_{2}\bigg[&-u^{\alpha}(\Delta^{\gamma\mu}\Delta^{\rho\nu}+\Delta^{\gamma\nu}\Delta^{\rho\mu}-\frac{2}{3}\Delta^{\gamma\rho}\Delta^{\mu\nu})\nabla_{\gamma}(\beta\omega_{\alpha\rho})\nonumber\\
&+\tilde{a}_{2}\theta\tau^{\mu\nu}_{(s)}+\tau^{\mu\nu}_{(s)}D\tilde{a}_{2}\bigg],
\label{tauseq}
\end{align}
\vspace{-2em}
\begin{align}
\tau_{\tau_{a}}\Delta^{[\mu\nu]}_{[\alpha\beta]}D\tau_{(a)}^{\alpha\beta}+\tau^{\mu\nu}_{(a)}=\chi_{3}\left[-u^{\alpha}(\Delta^{\gamma\mu}\Delta^{\rho\nu}-\Delta^{\gamma\nu}\Delta^{\rho\mu})\nabla_{\gamma}(\beta\omega_{\alpha\rho})+\tilde{a}_{3}\theta\tau_{(a)}^{\mu\nu}+\tau^{\mu\nu}_{(a)}D\tilde{a}_{3}\right],
\label{tauaeq}
\end{align}
\vspace{-2em}
\begin{align}  \tau_{\Theta}\Delta^{\alpha}_{\lambda}\Delta^{\mu}_{\sigma}\Delta^{\nu}_{\beta}D\Theta^{\lambda\sigma\beta}+\Theta^{\alpha\mu\nu}=-\chi_{4}\left[-\Delta^{\delta\mu}\Delta^{\rho\nu}\Delta^{\gamma\alpha}\nabla_{\gamma}(\beta\omega_{\delta\rho})+\tilde{a}_{4}\theta\Theta^{\alpha\mu\nu}+\Theta^{\alpha\mu\nu}D\tilde{a}_{4}\right].
\label{Thetaeq}
\end{align}
%
%
Different transport coefficients and relaxation times in the above equations are identified in Sec.~\ref{relaxation-typeIS}.

\subsection{Dispersion relations: rest frame low- and high-wavenumber limits}

The linear perturbations of hydrodynamic fields in hydrodynamic equations in Section~\ref{a ssspecial case} are characterized by the following expressions:
\begin{align}
&\varepsilon(t,\x)\rightarrow\varepsilon_{(0)}+\delta\varepsilon(t,\x),\label{perturbationeeIS}\\
& P(t,\x)\longrightarrow P_{(0)}+\delta P(t,\x),\\
&u^{\mu}(t,\x)\rightarrow u^{\mu}_{(0)}+\delta u^{\mu}(t,\x)=(1,\boldsymbol{0})+(0,\delta \v),\\
&S^{\mu\nu}(t,\x)\rightarrow 0+\delta S^{\mu\nu}(t,\x),\\
&\omega^{\mu\nu}(t,\x)\rightarrow 0 +\delta \omega^{\mu\nu}(t,\x),\\
&X(t,\x)\rightarrow 0+\delta X(t,\x),
\label{perturbationsss}
\end{align}
where $X(t,\x)$ represents various dissipative currents of the energy-momentum and spin tensors subject to linear perturbation.
%

To proceed, we employ the following relations between perturbations of thermodynamic quantities:
\begin{align}
    &\delta P=c_{s}^{2}\delta \varepsilon,\\
    &\delta T=\frac{T_{(0)}\,c_{s}^{2}}{\varepsilon_{(0)}+P_{(0)}}\delta\varepsilon,
\end{align}
where $c_{s}^{2}$ is the speed of sound squared. 

Additionally, we employ the following spin equation of state (see Eq.~\eqref{ourspinEOS})  
\begin{align}\label{SOeFHERE}
    S^{\gamma\delta}(T, \omega^{\gamma\delta}) = S_1(T) (k^\gamma u^\delta - k^\delta u^\gamma) + S_2(T) \epsilon^{\gamma\delta\rho\sigma} u_\rho \omega_\sigma,
\end{align}
which, in the rest frame of the fluid element, yields  
\begin{align}
\chi_b = \frac{\partial S^{i0}}{\partial \omega^{i0}} < 0, \quad \chi_s = \frac{\partial S^{ij}}{\partial \omega^{ij}} > 0.
\end{align}
For a detailed discussion of the role of the above quantities, see Sec.~\eqref{Spin equation of state section}. Furthermore, we introduce the following notations for the transport coefficients, which will be utilized in subsequent discussions:
\begin{align}
\label{definition(s)s}
&D_{b}=\frac{4\lambda}{\chi_{b}}<0,~~D_{s}=\frac{4\gamma}{\chi_{s}}>0,~~\lambda^{\prime}=\frac{\lambda}{\varepsilon_{(0)}+P_{(0)}},~~\gamma^{\prime}=\frac{\gamma\beta_{(0)}}{2},\nonumber\\
&\tilde{\chi}_{1}=\frac{2\beta_{(0)}}{\chi_{b}}\chi_{1}<0,~~{\tilde{\chi}_{2}=\frac{\beta_{(0)}}{\chi_{b}}\chi_{2}}<0,~~{\tilde{\chi}_{3}=\frac{\beta_{(0)}}{\chi_{b}}\chi_{3}}<0,~~{\tilde{\chi}_{4}=\frac{\beta_{(0)}}{\chi_{s}}\chi_{4}}>0.
\end{align}
The resulting linearized evolution equations in Eqs.~\eqref{perturbationeeIS}-\eqref{perturbationsss} read~\cite{Daher:2024bah}:
\begin{align}
\partial_{0}\delta\varepsilon+(\varepsilon_{(0)}+P_{(0)})\partial_{i}\delta u^{i}+\partial_{i}\delta q^{i}=0,\label{energyperturbation}
\end{align}
\vspace{-2em}
\begin{align}
    (\varepsilon_{(0)}+P_{(0)})\partial_{0}\delta u^{i}-c_{s}^{2}\partial^{i}\delta\varepsilon-\partial_{0}\delta q^{i}+\partial_{\mu}\delta \pi^{\mu i}+\partial^{i}\delta\pi+\partial_{\mu}\delta\phi^{\mu i}=0,
\end{align}
\vspace{-2em}
\begin{align}
    \tau_{\Pi}\partial_{0}\delta\Pi+\delta\Pi-\zeta\partial_{i}\delta u^{i}=0\, ,
\end{align}
\vspace{-2em}
\begin{align}
    \tau_{\pi}\partial_{0}\delta\pi^{ij}+\delta \pi^{ij}-\eta(\partial^{i}\delta u^{j}+\partial^{j}\delta u^{i})+\frac{2}{3}\eta\Delta^{ij}_{(0)}\partial_{k}\delta u^{k}=0\, ,
\end{align}
\vspace{-2em}
\begin{align}
    \tau_{q}\partial_{0}\delta q^{i}+\delta q^{i}-\lambda^{\prime}c_{s}^{2}\partial^{i}\delta\varepsilon-\lambda\partial_{0}\delta u^{i}+D_{b}\delta S^{i0}=0 ,
\end{align}
\vspace{-2em}
\begin{align}
   \tau_{\phi}\partial_{0}\delta \phi^{ij}+\delta \phi^{ij}-\gamma^{\prime}(\partial^{i}\delta u^{j}-\partial^{j}\delta u^{i})-\frac{\beta_{(0)}}{2}D_{s}\delta S^{ij}=0, \,
\end{align}
\vspace{-2em}
\begin{align}
    \partial_{0}\delta S^{0i}+\Delta^{ki}_{(0)}\partial_{k}\delta\Phi+\partial_{k}\delta\tau_{(s)}^{ki}+\partial_{k}\delta\tau^{ki}_{(a)}-2\delta q^{i}=0\, ,
\end{align}
\vspace{-2em}
\begin{align}
    \partial_{0}\delta S^{ij}+\partial_{k}\delta\Theta^{kij}+2\delta\phi^{ij}=0\, ,
\end{align}
\vspace{-2em}
\begin{align}
    \tau_{\Phi}\partial_{0}\delta\Phi
+\delta\Phi-{\tilde{\chi}_{1}\partial_{i}\delta S^{i0}}=0\, ,
\end{align}
\vspace{-2em}
\begin{align}
\tau_{\tau_{s}}\partial_{0}\delta\tau_{(s)}^{ij}+\delta \tau_{(s)}^{ij}{-\tilde{\chi}_{2}(\partial^{i}\delta S^{j0}+\partial^{j}\delta S^{i0})+\frac{2}{3}\tilde{\chi}_{2}\Delta^{ij}_{(0)}\partial_{k}\delta S^{k0}}=0\, ,
\end{align}
\vspace{-2em}
\begin{align}
    \tau_{\tau_{a}}\partial_{0}\delta\tau_{(a)}^{ij}+\delta \tau_{(a)}^{ij}+{\tilde{\chi}_{3}(-\partial^{i}\delta S^{j0}+\partial^{j}\delta S^{i0})}=0\, ,
\end{align}
\vspace{-2em}
\begin{align}
\tau_{\Theta}\partial_{0}\delta\Theta^{ijk}+\delta\Theta^{ijk}-{\tilde{\chi}_{4}\partial^{i}\delta S^{jk}}=0.\label{Thetaperturbation}
\end{align}
%

Using the solution in the form~\eqref{generalsolution}, the perturbed dynamical variables can be represented as wave packets. This approach enables us to convert the differential perturbed equations~\eqref{energyperturbation}--\eqref{Thetaperturbation} into algebraic equations in Fourier space. Without loss of generality, we assume \( \k=(0,0,k_z) \), leveraging the system's rotational symmetry. The resulting algebraic equations can be rewritten in matrix form, which will be used to determine the required dispersion relations. The matrix equation reads
\begin{align}\label{matrixequation12}
\mathbf{M}_{40\times40} \, \mathbf{V}=0,
\end{align}
where $\mathbf{M}_{40\times40}$ is a $40\times 40$ block diagonal matrix given by 
\begin{align}
&\mathbf{M}_{40\times40}=\left(
\begin{array}{ccccc}
\mathbf{A}_{10\times10} & 0 & 0 & 0 & 0 \\
0 & \mathbf{B}_{9\times9} & 0 & 0 & 0 \\
0 & 0 & \mathbf{B}_{9\times9} & 0 & 0 \\
0 & 0 & 0 & \mathbf{C}_{3\times3}& 0\\
0 & 0 & 0 & 0 & \mathbf{D}_{9\times9}
\end{array}
\right), 
\end{align}
and the vector $\mathbf{V}$ is defined as 
\begin{align}
\mathbf{V}=(\mathbf{v}_{\mathbf{A}},\mathbf{v}_{\mathbf{B}_{\mathbf{x}}},\mathbf{v}_{\mathbf{B}_{\mathbf{y}}},\mathbf{v}_{\mathbf{C}},\mathbf{v}_{\mathbf{D}})^{\intercal}.
\end{align}
The components of $\mathbf{V}$ are given by the dynamical variables:
\begin{align}
&\mathbf{v}_{\mathbf{A}}=\left(\delta \tilde{\varepsilon},\delta \tilde{u}^{z},\delta \tilde{\pi}^{xx},\delta\tilde{\pi}^{yy},\delta \tilde{\Pi},\delta \tilde{q}^{z},\delta \tilde{\Phi},\delta\tilde{S}^{0z},\delta \tilde{\tau}^{xx}_{(s)},\delta \tilde{\tau}^{yy}_{(s)}\right),\nonumber\\
&\mathbf{v}_{\mathbf{B_{x}}}=\left(\delta \tilde{u}^{x},\delta \tilde{q}^{x},\delta\tilde{\pi}^{zx},\delta\tilde{\phi}^{zx},\delta \tilde{S}^{0x},\delta \tilde{\tau}^{zx}_{(s)},\delta \tilde{\tau}_{(a)}^{zx},\delta \tilde{S}^{xz},\delta \tilde{\Theta}^{zxz}\right),\nonumber\\
&\mathbf{v}_{\mathbf{B}_{\mathbf{y}}}=\left(\delta \tilde{u}^{y},\delta \tilde{q}^{y},\delta\tilde{\pi}^{zy},\delta\tilde{\phi}^{zy},\delta \tilde{S}^{0y},\delta \tilde{\tau}^{zy}_{(s)},\delta \tilde{\tau}_{(a)}^{zy},\delta \tilde{S}^{yz},\delta \tilde{\Theta}^{zyz}\right),\nonumber\\
&\mathbf{v}_{\mathbf{C}}=(\delta \tilde{S}^{xy},\delta \tilde{\phi}^{xy},\delta \tilde{\Theta}^{zxy}),\nonumber\\
&\mathbf{v}_{\mathbf{D}}=\left(\delta \tilde{\pi}^{xy},\delta \tilde{\tau}^{xy}_{(s)},\delta \tilde{\tau}_{(a)}^{xy},\delta\tilde{\Theta}^{xxy},\delta\tilde{\Theta}^{xxz},\delta\tilde{\Theta}^{yyz},\delta\tilde{\Theta}^{xyz},\delta\tilde{\Theta}^{yxz},\delta\tilde{\Theta}^{yxy}\right).
\label{fluctuations}
\end{align}

\noindent
The explicit forms for the matrices $\mathbf{A}_{10\times10}$, $\mathbf{B}_{9\times9}$, $\mathbf{C}_{3\times3}$, and $\mathbf{D}_{9\times9}$ are:
\smallskip

\begin{equation}
\resizebox{\textwidth}{!}{$
\mathbf{A}_{10\times10}=
\begin{pmatrix}
 -i \omega  & i k_{z} & 0 & i (\varepsilon_{(0)}+P_{(0)})k_{z} & 0 & 0 & 0 & 0 & 0 & 0 \\
 i c_{s}^{2} k_{z} & i \omega  & 0 & -i (\varepsilon_{(0)}+P_{(0)}) \omega  & -i k_{z} & -i k_{z} & -i k_{z} & 0 & 0 & 0 \\
 0 & 0 & 0 & -i k_{z} \zeta  & 1-i \tau_{\Pi} \omega  & 0 & 0 & 0 & 0 & 0 \\
 0 & 0 & 0 & \frac{-2}{3}i \eta k_{z} & 0 & 1-i\tau_{\pi} \omega  & 0 & 0 & 0 & 0 \\
 0 & 0 & 0 & -\frac{2}{3}i \eta k_{z} & 0 & 0 & 1-i \tau_{\pi} \omega  & 0 & 0 & 0 \\
 i c_{s}^{2}\lambda^{\prime} k_{z} & 1-i\tau_{q} \omega  & -D_{b} & i \lambda  \omega  & 0 & 0 & 0 & 0 & 0 & 0 \\
 0 & -2 & -i \omega  & 0 & 0 & 0 & 0 & -i k_{z} & -i k_{z} & -i k_{z} \\
 0 & 0 & i \tilde{\chi}_{1} k_{z} & 0 & 0 & 0 & 0 & 0 & 0 & 1-i \tau_{\Phi} \omega  \\
 0 & 0 & \frac{2}{3} i \tilde{\chi}_{2} k_{z} & 0 & 0 & 0 & 0 & 1-i\tau_{\tau_{s}} \omega  & 0 & 0 \\
 0 & 0 & \frac{2}{3} i \tilde{\chi_{2}} k_{z} & 0 & 0 & 0 & 0 & 0 & 1-i \tau_{\tau_{s}} \omega  & 0 \\
\end{pmatrix}
$}\nonumber,
\end{equation}
\begin{equation}
\resizebox{\textwidth}{!}{$
\mathbf{B}_{9\times9}=
\begin{pmatrix}
i \omega  & 0 & 0 & -i(\varepsilon_{(0)}+P_{(0)}) \omega  & 0 & i k_{z} & 0 & 0 & i k_{z} \\
 0 & 0 & 0 & i \eta k_{z} & 0 & 1-i\tau_{\pi} \omega  & 0 & 0 & 0 \\
 1-i \tau_{q} \omega  & -D_{b} & 0 & i \lambda  \omega  & 0 & 0 & 0 & 0 & 0 \\
 0 & 0 & -\frac{D_{s}}{2T_{(0)}} & -i \gamma^{'}k_{z} & 0 & 0 & 0 & 0 & -1+i\tau_{\phi} \omega  \\
 -2 & -i \omega  & 0 & 0 & 0 & 0 & ik_{z} & i k_{z} & 0 \\
 0 & 0 & -i \omega  & 0 & ik_{z} & 0 & 0 & 0 & -2 \\
 0 & -i \tilde{\chi}_{2} k_{z} & 0 & 0 & 0 & 0 & 0 & 1-i \tau_{\tau_{s}} \omega  & 0 \\
 0 & -i \tilde{\chi}_{3} k_{z} & 0 & 0 & 0 & 0 & 1-i \tau_{\tau_{a}} \omega  & 0 & 0 \\
 0 & 0 & i \tilde{\chi}_{4}k_{z} & 0 & 1-i \tau_{\Theta} \omega  & 0 & 0 & 0 & 0 \\
\end{pmatrix}
$}\nonumber,
\end{equation}
\begin{eqnarray*}
\mathbf{C}_{3\times3}=\left(
\begin{array}{ccc}
 -\frac{D_{s}}{2T_{(0)}} & 0 & 1-i \tau_{\phi} \omega  \\
 -i \omega  & i k_{z} & 2 \\
 i \tilde{\chi}_{4}k_{z} & 1-i\tau_{\Theta} \omega  & 0 \\
\end{array}
\right),\nonumber\\
\end{eqnarray*}
\begin{equation}
\resizebox{\textwidth}{!}{$
\mathbf{D}_{9\times9}=
\begin{pmatrix}
 1-i\tau_{\tau_{s}} \omega  & 0 & 0 & 0 & 0 & 0 & 0 & 0 & 0 \\
 0 & 1-i \tau_{\tau_{a}} \omega  & 0 & 0 & 0 & 0 & 0 & 0 & 0 \\
 0 & 0 & 1-i \tau_{\Theta} \omega  & 0 & 0 & 0 & 0 & 0 & 0 \\
 0 & 0 & 0 & 1-i \tau_{\Theta} \omega  & 0 & 0 & 0 & 0 & 0 \\
 0 & 0 & 0 & 0 & 1-i \tau_{\Theta} \omega  & 0 & 0 & 0 & 0 \\
 0 & 0 & 0 & 0 & 0 & 1-i \tau_{\Theta} \omega   & 0 & 0 & 0 \\
 0 & 0 & 0 & 0 & 0 & 0 & 1-i \tau_{\Theta} \omega  & 0 & 0 \\
 0 & 0 & 0 & 0 & 0 & 0 & 0 & 1-i \tau_{\Theta} \omega  & 0 \\
 0 & 0 & 0 & 0 & 0 & 0 & 0 & 0 & 1-i\tau_{\pi} \omega  \\
\end{pmatrix}
$}.
\end{equation}
\smallskip

\noindent
Nonzero solutions to the Eq.~\eqref{matrixequation12} exist if and only if $\det(\mathbf{M}) = 0$. Taking advantage of the block diagonal structure of $\mathbf{M}$, we can compute the determinant of each block independently, 
\begin{align}
\det(\mathbf{M}) &= \det(\mathbf{A}) \det(\mathbf{B})^{2} \det(\mathbf{C}) \det(\mathbf{D}) = 0. \label{determinanttt}
\end{align}
Each block matrix, with its associated vector, forms a \emph{channel} that can be treated independently. The physical interpretation of each channel depends mainly on the behavior of the perturbation of the fluid element velocity with respect to the wave vector in a rotation-invariant fluid~\cite{Kovtun:2019hdm}. In particular, the cases with $\delta \v \parallel \k$ specify the \emph{sound channel}, whereas if $\delta \v \perp \k$ we deal with the \empsh{shear channel}. Recall that, for our choice of $\k=(0,0,k_{z})$, we obtain the following distinct channels
\begin{itemize}  
\item The matrix \( \mathbf{A}_{10\times10} \) corresponds to the sound channel, as the associated vector \( \mathbf{v}_{\mathbf{A}} \) contains  \( \delta u^z = \delta v^z \parallel k^z \). In this work, we refer to it as the sound channel \( \mathbf{A} \).
\item The matrix \( \mathbf{B}_{9\times9} \) corresponds to the shear channel, as the associated vector \( \mathbf{v}_{\mathbf{B}}=(\mathbf{v}_{\mathbf{B}_{\mathbf{x}}},\mathbf{v}_{\mathbf{B}_{\mathbf{y}}}) \) contains \( \delta u^x = \delta v^x \) and \( \delta u^y = \delta v^y \), with both components orthogonal to \( k^z \). We designate it as the shear channel \( \mathbf{B} \). 
\item The matrix \( \mathbf{C}_{3\times3} \), associated with vector \( \mathbf{v}_{\mathbf{C}} \), contains perturbations that are exclusively within the domain of spin hydrodynamics. The precise physical interpretation of this channel warrants further investigation in future research. We refer to it as channel \( \mathbf{C} \). 
\item The matrix \( \mathbf{D}_{9\times9} \), associated with vector \( \mathbf{v}_{\mathbf{D}} \), represents a decoupled channel, as \( \mathbf{D}_{9\times9} \) is diagonal. We designate it as channel \( \mathbf{D} \).  
\end{itemize}
%
%
\subsubsection*{Low-wavenumber modes}
Calculating the determinant of each matrix in Eq.~\eqref{determinanttt} for an exact wavenumber is complicated. Therefore, we evaluate the determinant in the low- and high-wavenumber limits. We begin with the low-wavenumber approximation.
\medskip 

Up to second order in $k_z$, the dispersion relations corresponding to the determinant of the matrix $\mathbf{A}$ in Eq.~\eqref{determinanttt}, read:
\begin{align}
    &\omega_{1,2}=\pm c_{s}k_{z}-i\frac{(\frac{4}{3}\eta+\zeta)}{2(\varepsilon_{(0)}+P_{(0)})}k_{z}^{2},\label{oso12}\\
    &\omega_{3}=-\frac{i}{\tau_{\Phi}}-i\frac{\tilde{\chi}_{1}}{\Sigma_{\Phi}}k_{z}^{2},\\
    &\omega_{4}=-\frac{i}{\tau_{\pi}}+\frac{4}{3}i\frac{\eta}{(\varepsilon_{(0)}+P_{(0)})\Lambda_{\pi}}k_{z}^{2},\\
    &\omega_{5}=-\frac{i}{\tau_{\Pi}}+i\frac{\zeta}{(\varepsilon_{(0)}+P_{(0)})\Lambda_{\Pi}}k_{z}^{2},\\
    &\omega_{6}=-\frac{i}{\tau_{\tau_{s}}}-\frac{4}{3}i\frac{\tilde{\chi}_{2}}{\Sigma_{\tau_{s}}}k_{z}^{2},\\
    &\omega_{7,8}=-i\frac{\left(1\pm \sqrt{8D_{b}(\tau_{q}-\lambda^{\prime})+1}\right)}{2(\tau_{q}-\lambda^{\prime})}+\mathcal{O}(k_{z}^{2}),\label{oso11}
\end{align}
where we have introduced the following notation:
\begin{align}
&\Sigma_{\Phi}=1+\frac{2D_{b}\tau_{\Phi}^{2}}{(\tau_{\Phi}-\tau_{q})+\lambda^{\prime}},~~\Sigma_{\tau_{s}}=1+\frac{2D_{b}\tau_{\tau_{s}}^{2}}{\left(\tau_{\tau_{s}}-\tau_{q}\right)+\lambda^{\prime}},\nonumber\\
&\Lambda_{\pi}=1+\frac{\lambda^{\prime}}{2D_{b}\tau_{\pi}^{2}+\tau_{\pi}-\tau_{q}},~~\Lambda_{\Pi}=1+\frac{~\lambda^{\prime}}{2D_{b}\tau_{\Pi}^{2}+\tau_{\Pi}-\tau_{q}}.
\end{align}
For simplicity of the calculation, we neglect the contribution of the $k_{z}^{2}$ term in Eq.~\eqref{oso11}, as the respective expression is complicated. Moreover, we note that since the matrix $\mathbf{A}$ is ten-dimensional, it is expected to have 10 dispersion relations. However, in the above, we display only eight of them, as two purely damped modes are listed together with channel $\mathbf{D}$ (see Eqs.~\eqref{extra1nqm} and \eqref{extra2nqm}), which becomes eleven-dimensional in this way. The first two modes in Eq.~\eqref{oso12} are the sound modes present also in conventional relativistic hydrodynamics. 

\medskip

The dispersion relations obtained from the condition $\textrm{det}(\mathbf{B})^{2} = 0$, as indicated by the squared determinant, are doubly degenerate. Therefore, we obtain the following doubly degenerate dispersion relations:
\begin{align}
    &\omega_{1}=-i\frac{\eta}{(\varepsilon_{(0)}+P_{(0)})}k_{z}^{2},\label{osh1}\\
    &\omega_{2}=-\frac{i}{\tau_{\pi}}+i\frac{\eta}{(\varepsilon_{(0)}+P_{(0)})\Lambda_{\pi}}k_{z}^{2},\\
    &\omega_{3}=-\frac{i}{\tau_{\tau_{\Theta}}}+i\frac{\tilde{\chi}_{4}}{\Upsilon_{\Theta}}k_{z}^{2},\\
    &\omega_{4}=-\frac{i}{\tau_{\tau_{a}}}-i\frac{\tilde{\chi}_{3}}{\Sigma_{\tau_{a}}}k_{z}^{2},\\
    &\omega_{5}=-\frac{i}{\tau_{\tau_{s}}}-i\frac{\tilde{\chi}_{2}}{\Sigma_{\tau_{s}}}k_{z}^{2},\\
    &\omega_{6,7}=-i\frac{\left(1\pm\sqrt{1-4D_{s}\tau_{\phi}\beta_{(0)}}\right)}{2\tau_{\phi}}+\mathcal{O}(k_{z}^{2}),\\
    &\omega_{8,9}=-i\frac{\left(1\pm \sqrt{8D_{b}(\tau_{q}-\lambda^{\prime})+1}\right)}{2(\tau_{q}-\lambda^{\prime})}+\mathcal{O}(k_{z}^{2}).\label{osh89}
\end{align}
The first doubly degenerate mode is the shear mode in conventional relativistic hydrodynamics. Here, we have introduced the following notation,
\begin{align}
\Sigma_{\tau_{a}}=1+\frac{2D_{b}\tau_{\tau_{a}}^{2}}{\left(\tau_{\tau_{a}}-\tau_{q}\right)+\lambda^{\prime}},~~\Upsilon_{\Theta}=1+D_{s}\beta_{(0)}\left(\frac{\tau_{\Theta}^{2}}{\tau_{\phi}-\tau_{\Theta}}\right).
\end{align}

From the determinant of the matrix $\mathbf{C}$ in Eq.~\eqref{determinanttt}, we extract the following dispersion relations:
\begin{align}
    &\omega_{1}=-\frac{i}{\tau_{\Theta}}+i\frac{\tilde{\chi}_{4}}{\Upsilon_{\Theta}}k_{z}^{2},\\
    &\omega_{2}=-i\frac{\left(1-\sqrt{1-4D_{s}\tau_{\phi}\beta_{(0)}}\,\right)}{2\tau_{\phi}}\nonumber\\
    &~~~~~~~-i\frac{\tilde{\chi}_{4}\tau_{\phi}\left(1+\sqrt{1-4D_{s}\tau_{\phi}\beta_{(0)}}\right)}{-2\left(2D_{s}\tau_{\Theta}\beta_{(0)}-\sqrt{1-4D_{s}\tau_{\phi}\beta_{(0)}}\right)\tau_{\phi}+\left(1-\sqrt{1-4D_{s}\tau_{\phi}\beta_{(0)}}\right)\tau_{\Theta}}k_{z}^{2},\\
    &\omega_{3}=-i\frac{\left(1+\sqrt{1-4D_{s}\tau_{\phi}\beta_{(0)}}\,\right)}{2\tau_{\phi}}\nonumber\\
    &~~~~~~~-i\frac{\tilde{\chi}_{4}\tau_{\phi}\left(-1+\sqrt{1-4D_{s}\tau_{\phi}\beta_{(0)}}\right)}{2\left(2D_{s}\tau_{\Theta}\beta_{(0)}+\sqrt{1-4D_{s}\tau_{\phi}\beta_{(0)}}\right)\tau_{\phi}-\left(1+\sqrt{1-4D_{s}\tau_{\phi}\beta_{(0)}}\right)\tau_{\Theta}}k_{z}^{2}.
\end{align}

Finally, solving the determinant of the matrix $\mathbf{D}$ leads to:
\begin{align}
    &\omega_{1,2}=-\frac{i}{\tau_{\pi}},\label{extra1nqm}\\
    &\omega_{3\rightarrow 8}=-\frac{i}{\tau_{\Theta}},\\
    &\omega_{9}=-\frac{i}{\tau_{\tau_{a}}},\\
    &\omega_{10,11}=-\frac{i}{\tau_{\tau_{s}}}.\label{extra2nqm}
\end{align}
It can be noticed that the above dispersion relations are independent of the wavenumber, and they are purely imaginary, hence, they are all damped modes.
%
%
\subsubsection*{High-wavenumber limit }
We now turn to calculating the dispersion relations in the high-wavenumber limit. It is crucial to emphasize that the channel classification that we outlined after Eq.~\eqref{determinanttt} remains entirely valid in the high-wavenumber regime.
\medskip

From the determinant of matrix $\mathbf{A}$ in Eq.~\eqref{determinanttt} in the high-wavenumber limit, we obtain the following dispersion relations:
\begin{align}\label{eq:AC-HM}
    &{\omega_1 = -\frac{i ( \frac{4}{3} \eta + \xi)}{\frac{4}{3} \eta \tau_\Pi+ \xi \tau_\pi} 
    + \mathcal{O}\left(\frac{1}{k_z^2}\right) + \mathcal{O}\left(\frac{1}{k_z^4}\right),}\\
&{\omega_2 = - \frac{i(\frac{4}{3}\tilde{\chi}_2 + \tilde{\chi}_1)}{\frac{4}{3}\tilde{\chi}_2 \tau_\Phi + \tilde{\chi}_1 \tau_{\tau_s}} 
    +\mathcal{O}\left(\frac{1}{k_z^2}\right) + \mathcal{O}\left(\frac{1}{k_z^4}\right),}\\
    &{\omega_{3,4} =-\frac{i}{2 \tau_{\tau_s} \tau_\Phi} \frac{  \tau_{\tau_s}^2 \tilde{\chi}_1 + \frac{4}{3} \tau_\Phi ^2 \tilde{\chi}_2}{ \tau_{\tau_s} \tilde{\chi}_1 + \frac{4}{3} \tau_\Phi \tilde{\chi}_2} \pm \sqrt{-\frac{\frac{4}{3}\tilde{\chi}_2 \tau_\Phi + \tilde{\chi}_1 \tau_{\tau_s}}{\tau_{\tau_s} \tau_\Phi}}\, k_z + \mathcal{O}\left(\frac{1}{k_z}\right)},\\
    &{\omega_{s_{1},s_{2}} = -i \frac{\sum\limits_{n=0}^3 c_n v_{s_1,s_2}^{2n}}{\sum\limits_{n=0}^3 d_n v_{s_1,s_2}^{2n}} + v_{s_1,s_2} \, k_z + \mathcal{O}\left(\frac{1}{k_z}\right)}.
\end{align}
For the sake of simplicity, here, $\omega_{s_{1},s_{2}}$ represents 4 modes with $s_1 = \pm$ and $s_2 = \pm$. Hence, the four modes are:  $\omega_{+,+},\,\omega_{+,-},\,\omega_{-,+},\,\omega_{-,-}$, where:
\begin{align}
    &{v_{s_1,s_2} = s_1 \sqrt{\frac{\mathcal{A} + s_2 \mathcal{B}}{2 (\varepsilon_{(0)} + P_{(0)}) (\tau_q - \lambda^\prime) \tau_\pi \tau_\Pi}}},\nonumber\\
    &{\mathcal{A} \equiv c_s^2 (\varepsilon_{(0)} + P_{(0)}) (\tau_q + 3 \lambda^\prime) \tau_\pi \tau_\Pi + \tau_q \left(\frac{4}{3} \eta \tau_\Pi+ \xi \tau_\pi\right)},\nonumber\\
&{\mathcal{B}^2 \equiv \mathcal{A}^2 - 4 c_s^2 \tau_\pi \tau_\Pi \lambda  (\tau_q - \lambda^\prime) \left(\frac{4}{3} \eta \tau_\Pi+ \xi \tau_\pi\right)},\nonumber
\end{align}
along with the terms:
\begin{align}
{c_0 = -c_s^2 \lambda^\prime \left[\tilde{\chi}_1 \left(\frac{4\eta}{3} (\tau_\Pi + \tau_{\tau_s}) + \xi (\tau_\pi + \tau_{\tau_s})\right) + \frac{4 \tilde{\chi}_2}{3} \left(\frac{4\eta}{3} (\tau_\Pi + \tau_\Phi) + \xi (\tau_\pi + \tau_\Phi)\right)\right],} \nonumber
\end{align}
\vspace{-1em}
\begin{align}
    c_1 &= \frac{1}{9} \bigg\{ 
    -3 c_s^2 \bigg[\lambda^\prime \bigg( 4 \eta \big(\tau_\Pi \tau_\Phi + \tau_{\tau_s} (\tau_\Pi + \tau_\Phi) \big) 
    + 3 \xi \big(\tau_\pi \tau_\Phi + \tau_{\tau_s} (\tau_\pi + \tau_\Phi) \big)-3 (\varepsilon_{(0)} + P_{(0)}) \nonumber\\
    &\quad  \bigg( 
    \tau_\pi \tau_\Pi \big(3 \tilde{\chi}_1 + 4 \tilde{\chi}_2 \big) 
    + 4 \tau_\Phi \tilde{\chi}_2 (\tau_\pi + \tau_\Pi) 
    + 3 \tau_{\tau_s} \tilde{\chi}_1 (\tau_\pi + \tau_\Pi) 
    \bigg) \bigg] - (\varepsilon_{(0)} + P_{(0)})\nonumber\\
    &\quad  \bigg[ 
    \tau_q \bigg( 
    \tau_\pi \tau_\Pi (3 \tilde{\chi}_1 + 4 \tilde{\chi}_2) 
    + 4 \tau_\Phi \tilde{\chi}_2 (\tau_\pi + \tau_\Pi) 
    + 3 \tau_{\tau_s} \tilde{\chi}_1 (\tau_\pi + \tau_\Pi) 
    \bigg) \nonumber\\
    &\quad + \tau_\pi \tau_\Pi (3 \tau_{\tau_s} \tilde{\chi}_1 + 4 \tau_\Phi \tilde{\chi}_2) 
    \bigg]+ 4 \eta \bigg[ 
    \tau_q \bigg( \tau_\Pi (3 \tilde{\chi}_1 + 4 \tilde{\chi}_2)  + 3 \tau_{\tau_s} \tilde{\chi}_1\nonumber\\
    &\quad  + 4 \tau_\Phi \tilde{\chi}_2 \bigg) 
    + \tau_\Pi (3 \tau_{\tau_s} \tilde{\chi}_1 + 4 \tau_\Phi \tilde{\chi}_2) 
    \bigg] + 3 \xi \bigg[ 
    \tau_q \bigg( \tau_\pi (3 \tilde{\chi}_1 + 4 \tilde{\chi}_2) \nonumber\\
    &\quad + 3 \tau_{\tau_s} \tilde{\chi}_1 + 4 \tau_\Phi \tilde{\chi}_2 \bigg) 
    + \tau_\pi (3 \tau_{\tau_s} \tilde{\chi}_1 + 4 \tau_\Phi \tilde{\chi}_2) 
    \bigg] 
    \bigg\},\nonumber
\end{align}
\vspace{-1em}
\begin{align}
    c_2 &= \frac{1}{3} \bigg\{ (\varepsilon_{(0)} + P_{(0)}) \bigg[ 
    3 c_s^2 \bigg( 3 \lambda^\prime (\tau_\pi \tau_\Pi \tau_\Phi 
    + \tau_\pi \tau_{\tau_s} (\tau_\Pi + \tau_\Phi) 
    + \tau_\Pi \tau_{\tau_s} \tau_\Phi ) + \tau_\pi \tau_\Pi \tau_q (\tau_\Phi + \tau_{\tau_s}) \nonumber\\
    &\quad 
    + \tau_{\tau_s} \tau_\Phi (\tau_q (\tau_\pi + \tau_\Pi) + \tau_\pi \tau_\Pi ) \bigg) + \lambda^\prime \bigg( \tau_\pi (3 \tau_\Pi \tilde{\chi}_1 + 4 \tau_\Pi \tilde{\chi}_2 + 4 \tau_\Phi \tilde{\chi}_2 )  + 4 \tau_\Pi \tau_\Phi \tilde{\chi}_2 \nonumber\\
    &\quad 
   + 3 \tau_{\tau_s} \tilde{\chi}_1 (\tau_\pi + \tau_\Pi ) \bigg) 
   - \tau_\pi \tau_\Pi (4 \tau_\Phi \tilde{\chi}_2 + 3 \tau_q \tilde{\chi}_1) 
    - 4 \tau_q \tilde{\chi}_2 (\tau_\pi (\tau_\Pi + \tau_\Phi) + \tau_\Pi \tau_\Phi ) \nonumber\\
    &\quad - 3 \tau_{s} \tilde{\chi}_1 (\tau_q (\tau_\pi + \tau_\Pi ) + \tau_\pi \tau_\Pi ) \bigg] + 4 \eta ( \tau_\Pi \tau_q \tau_\Phi + \tau_q \tau_{\tau_s} (\tau_\Pi + \tau_\Phi ) + \tau_\Pi \tau_{\tau_s} \tau_\Phi ) \nonumber\\
    &\quad + 3 \xi ( \tau_\pi \tau_q \tau_\Phi + \tau_q \tau_{\tau_s} (\tau_\pi + \tau_\Phi ) + \tau_\pi \tau_{\tau_s} \tau_\Phi ) \bigg\},\nonumber
\end{align}
\vspace{-1em}
\begin{align}
c_3 =& -(\varepsilon_{(0)} + P_{(0)}) \bigg[-\lambda^\prime \bigg(\tau_\pi  \tau_\Pi  \tau_\Phi +\tau_{\tau_s} \tau_\Phi  (\tau_\pi +\tau_\Pi )+\tau_\pi  \tau_\Pi  \tau_{\tau_s}\bigg) + \tau_\Phi  \bigg(\tau_\Pi  \tau_q (\tau_\pi +\tau_{\tau_s})\nonumber\\
 &+\tau_\pi  \tau_q \tau_{\tau_s}+\tau_\pi  \tau_\Pi  \tau_{\tau_s}\bigg)+ \tau_\pi  \tau_\Pi  \tau_q \tau_{\tau_s}\bigg],\nonumber
\end{align}
and 
\begin{align}
&{d_0 = -2c_s^2 \lambda^\prime \left(\frac{4}{3} \eta \tau_\Pi+ \xi \tau_\pi\right) \left(\frac{4}{3}\tilde{\chi}_2 \tau_\Phi + \tilde{\chi}_1 \tau_{\tau_s}\right),}\nonumber\\ 
&d_1 = 4 \bigg[\tau_q \left(\frac{4}{3} \eta \tau_\Pi+ \xi \tau_\pi\right) \left(\frac{4}{3}\tilde{\chi}_2 \tau_\Phi + \tilde{\chi}_1 \tau_{\tau_s}\right) + c_s^2 \bigg(-\lambda^\prime \tau_{\tau_s} \tau_\Phi \left(\frac{4}{3} \eta \tau_\Pi+ \xi \tau_\pi\right)\nonumber\\
&~~~~~~~~~+ (\varepsilon_{(0)} + P_{(0)}) \tau_\pi \tau_\Pi(3\lambda^\prime + \tau_q) \left(\frac{4}{3}\tilde{\chi}_2 \tau_\Phi + \tilde{\chi}_1 \tau_{\tau_s}\right)\bigg)\bigg],\nonumber\\
&d_2 = 6 \tau_q \tau_{\tau_s} \tau_\Phi \left(\frac{4}{3} \eta \tau_\Pi+ \xi \tau_\pi\right) + 6 (\varepsilon_{(0)} + P_{(0)}) \tau_\pi \tau_\Pi \bigg[(\lambda^\prime - \tau_q) \left(\frac{4}{3}\tilde{\chi}_2 \tau_\Phi + \tilde{\chi}_1 \tau_{\tau_s}\right)\nonumber\\
&~~~~~~~+ c_s^2 \tau_{\tau_s} \tau_\Phi (3\lambda^\prime +\tau_q)\bigg],\nonumber\\
&{d_3 = 8 (\varepsilon_{(0)} + P_{(0)}) (\lambda^\prime - \tau_q) \tau_{\tau_s} \tau_\pi \tau_\Pi \tau_\Phi.}\nonumber
\end{align}

In a similar way, the dispersion relations for the determinant of matrix $\mathbf{B}$ are: 
\begin{align}\label{eq:BC-HM}
    &{\omega_{1} = -\frac{i}{\tau_q} 
    + \mathcal{O}\left(\frac{1}{k_z^2}\right)
    + \mathcal{O}\left(\frac{1}{k_z^4}\right),}\\
    &{\omega_{2} = -\frac{i (\gamma' + \eta)}{\gamma' \tau_\pi + \eta \tau_\phi}
    + \mathcal{O}\left(\frac{1}{k_z^2}\right)+ \mathcal{O}\left(\frac{1}{k_z^4}\right),}\\
    &{\omega_3 = - \frac{i (\tilde{\chi}_2 + \tilde{\chi}_3)}{\tau_{\tau_a}  \tilde{\chi}_2 + \tau_{\tau_s}\tilde{\chi}_3} 
    +\mathcal{O}\left(\frac{1}{k_z^2}\right)
    + \mathcal{O}\left(\frac{1}{k_z^4}\right),}\\
    &\omega_{4,5} = -\frac{i}{2} \left(\frac{1}{\tau_\pi} + \frac{1}{\tau_\phi} + \frac{\lambda'}{\tau_q(\tau_q -\lambda')} - \frac{\gamma' + \eta}{\gamma' \tau_\pi + \eta \tau_\phi}\right)\nonumber\\
    &~~~~~~~~~\pm \sqrt{\frac{\tau_q \left(\gamma' \tau_\pi + \eta \tau_\phi\right)}{(\varepsilon_0 + P_0) (\tau_q - \lambda')\tau_\pi \tau_\phi}}\, k_z + \mathcal{O}\left(\frac{1}{k_z}\right),\\
    &{\omega_{6,7} = -\frac{i (\tau_{\tau_a}^2  \tilde{\chi}_2 + \tau_{\tau_s}^2\tilde{\chi}_3)}{2 \tau_{\tau_a} \tau_{\tau_s} (\tau_{\tau_a}  \tilde{\chi}_2 + \tau_{\tau_s}\tilde{\chi}_3)} \pm \sqrt{- \frac{\tau_{\tau_a}  \tilde{\chi}_2 + \tau_{\tau_s}\tilde{\chi}_3}{\tau_{\tau_a}  \tau_{\tau_s}}}\, k_z + \mathcal{O}\left(\frac{1}{k_z}\right),}\\
    &{\omega_{8,9} = - \frac{i}{2\tau_\Theta} \pm \, \sqrt{\frac{\tilde{\chi}_4}{\tau_\Theta}}\, k_z + \mathcal{O}\left(\frac{1}{k_z}\right).}
\end{align}
We recall that the above dispersion relations are obtained from the condition $\textrm{det}(\mathbf{B})^{2} = 0$, where the square implies that each mode is doubly degenerate.
\medskip 

In the low-wavenumber limit, we found that the solutions for the matrix $\mathbf{D}$ are independent of the wavenumber. Therefore, we do not need to analyze that channel in the high-wavenumber limit. Consequently, we proceed to study the determinant of the matrix $\mathbf{C}$, from which we obtain:
\begin{align}
&{\omega_1 = - \frac{i}{\tau_\phi} 
+ \mathcal{O}\left(\frac{1}{k_z^2}\right)+ \mathcal{O}\left(\frac{1}{k_z^4}\right),}\\
&{\omega_{2,3} = - \frac{i}{2\tau_\Theta} \pm \sqrt{\frac{\tilde{\chi}_4}{\tau_\Theta}} k_z
+ \mathcal{O}\left(\frac{1}{k_z}\right) + \mathcal{O}\left(\frac{1}{k_z^2}\right).}\label{eq:CC-HM}
\end{align}
%
\subsection{Stability check}
To ensure that perturbations of the dynamic variables do not exhibit exponential growth over time, it is essential to ensure the stability of all dispersion relations. Mathematically, this condition is expressed as
\begin{align}
\label{stabilityconditionIS}
\mbox{Im}[\omega(k_{z})]<0.
\end{align}
As discussed in Sec.~\ref{stability}, this requirement ensures that all modes decay, preserving the system's stability.
\subsubsection*{Low-wavenumber limit}
In the sound channel $\mathbf{A}$, the constraint Eq.~\eqref{stabilityconditionIS} leads to the following set of conditions:
\begin{align}
&\Sigma_{\Phi}<0~~\text{and}~~\tau_{\Phi}>\tau_{q}-\lambda^{\prime},\label{con1}\\
&\Lambda_{\pi}<0,\label{con2}\\
&\Lambda_{\Pi}<0,\label{con3}\\
&\Sigma_{\tau_{s}}<0~~\text{and}~~\tau_{\tau_{s}}>\tau_{q}-\lambda^\prime,\label{con4}\\
&\tau_{q}>\lambda^{'},\label{con5}\\
&-1<\sqrt{8D_{b}(\tau_{q}-\lambda^{\prime})+1}<1 \qquad (|(\tau_{q}-\lambda^{\prime})8D_{b}| < 1).\label{con6}
\end{align}
For the shear channel $\mathbf{B}$ we find:
\begin{align}
&\Upsilon_{\Theta}<0,~~\text{and}~~\tau_{\phi}<\tau_{\Theta},\label{con7}\\
&\Sigma_{\tau_{a}}<0~~\text{and}~~\tau_{\tau_{a}}>\tau_{q}-\lambda^{\prime},\label{con8}\\
&-1<\sqrt{1-4D_{s}\tau_{\phi}\beta_{(0)}}<1~~~~(4D_{s}\tau_{\phi}\beta_{(0)} < 1)\label{con9},
\end{align}
and, for the channel $\mathbf{C}$, one find that the stability conditions are
\begin{align}
    -(1-4D_{s}\beta_{(0)}\tau_{\phi})\frac{\tau_{\Theta}}{\tau_{\Theta}-2\tau_{\phi}}<\sqrt{1-4D_{s}\tau_{\phi}\beta_{(0)}}<+(1-4D_{s}\beta_{(0)}\tau_{\phi})\frac{\tau_{\Theta}}{\tau_{\Theta}-2\tau_{\phi}}.\label{con10}
\end{align}
If the conditions in Eqs.~\eqref{con1}-\eqref{con10} are satisfied, the stability of perturbations and, consequently, the stability of the truncated perturbed evolution equations~\eqref{energyperturbation}-\eqref{Thetaperturbation} is ensured in the low-wavenumber limit. Indeed, given the conditions in~\eqref{definition(s)s}, all the above conditions are satisfied. However, the used spin equation of state~\eqref{SOeFHERE} in this analysis plays a crucial role in determining the signs of some of the conditions in~\eqref{definition(s)s} as discussed earlier. If, for example, another spin equation of state, for example~\eqref{XuEOS}, is employed, we can find that some of the above stability conditions are not satisfied. Hence, we conclude that the stability depends on the form of the spin equation of state. 
\medskip 

Within our truncation scheme, we also find the following interesting relations:
\begin{enumerate}
\item Using expressions for the relaxation times $\tau_{\phi}$ and $\tau_{\Theta}$ derived in Eqs.~\eqref{phieq} and \eqref{Thetaeq}, respectively, the condition~\eqref{con7} can be further rewritten as
\begin{align}
\tau_{\phi}<\tau_{\Theta}~\Leftrightarrow~-2a_{5}\gamma<2\tilde{a}_{4}\chi_{4}.
\end{align}
This implies that the transport coefficient $\chi_{4}$ of the spin dissipative current $\Theta^{\lambda\mu\nu}$ in Eq.~\eqref{Thetaeq} is directly related to the rotational viscosity transport coefficient $\gamma$ of the current $\phi^{\mu\nu}$ in Eq.~\eqref{phieq}.\\
\item Similarly, the conditions in Eqs.~\eqref{con1}, \eqref{con4}, and \eqref{con8} imply that the transport coefficients $\chi_{1},\chi_{2}$, and $\chi_{3}$ of the spin dissipative currents $\Phi,\tau_{(s)}^{\mu\nu}$ and $\tau_{(a)}^{\mu\nu}$ (see Eqs.~\eqref{Phieq}, \eqref{tauseq}, and \eqref{tauaeq}, respectively) are related to the transport coefficient $\lambda$ of the dissipative current $q^{\mu}$ in Eq.~\eqref{qeq}.  
\end{enumerate}
%
\subsubsection*{High-wavenumber limit}
As we already discussed in Sec.~\ref{stability}, to conclude about the stability, it is necessary to check the behavior of the perturbations also in the $k_{z}\rightarrow\infty$ limit. Applying the condition~\eqref{stabilityconditionIS} to the dispersion relations in Eqs.~\eqref{eq:AC-HM}-\eqref{eq:CC-HM} results in the following constraints:
\begin{align}
    &\tau_\Phi > 0, \quad \tau_q > 0, \quad \tau_{\tau_a} > 0, \quad \tau_{\tau_s} > 0, \quad \tau_{\pi} > 0, \quad \tau_\Pi > 0, \quad \tau_{\phi}>0, \quad \gamma > 0,\nonumber\\
    &\tau_{\Theta}>0,\quad \tau_{q}>\lambda^{\prime},\label{con1high}\\
    &{\tilde{\chi}_1 < 0, \quad \tilde{\chi}_2 < 0, \quad \tilde{\chi}_3 < 0,}\\
    &{c_{0}>0,\quad d_{0}>0,}\\
    &{\frac{1}{\tau_\pi} + \frac{1}{\tau_\phi} + \frac{\lambda^\prime}{\tau_q(\tau_q -\lambda^\prime)} - \frac{\gamma^\prime + \eta}{\gamma^\prime \tau_\pi + \eta \tau_\phi} >0}\label{con4high}.
\end{align}
Given the conditions in~\eqref{definition(s)s}, we verified that all the above conditions are satisfied. Hence, the stability of perturbations of the dynamical variables, and consequently, the stability of the truncated perturbed evolution equations~\eqref{energyperturbation}-\eqref{Thetaperturbation}, is ensured in the high-wavenumber limit.
\medskip 

While our present investigation has prioritized the stability analysis in the low- and high-wavenumber regimes, there remains a question of whether the stability in these two regimes guarantees that it is also maintained at finite wavenumber values. A deeper examination of this issue can be performed by employing the so-called Routh-Hurwitz criterion~\cite{Bemfica:2019knx,Hoult:2020eho,Bemfica:2020zjp}. We leave this subject for future study.
%
%
%
\subsection{Causality check}

The causality in relativistic hydrodynamics generally refers to the perturbations of dynamical variables propagating within the fluid no faster than the speed of light in a vacuum. Hence, verifying causality is crucial for the evolution equations to admit physically meaningful solutions. As discussed in Sec.~\ref{stability}, to maintain causality, the dispersion relations should satisfy the following constraints:
 \begin{align}\label{CausalitydefIS}
\lim_{k_{z} \to \infty} \bigg|{\rm Re}\frac{w}{k_{z}}\bigg|\leq 1~~\text{and}~~\lim_{k_{z} \to \infty} \bigg|\frac{w}{k_{z}}\bigg|~\text{is bounded.}
\end{align}
Therefore, within our analysis, the truncated evolution equations are causal provided the following conditions hold:
\begin{align}
    &{0 \leq -\frac{\frac{4}{3}\tilde{\chi}_2\tau_{\Phi}+\tilde{\chi}_1\tau_{\tau_{s}}}{\tau_{\tau_{s}}\tau_\Phi}\leq 1,}\label{causality1}\\
    & {0 \leq \frac{\mathcal{A} \pm \mathcal{B}}{2 (\varepsilon_{(0)} + P_{(0)}) (\tau_q - \lambda') \tau_\pi \tau_\Pi} \leq 1,}\label{causality2}\\
    & {0 \leq \frac{\tau_q \left(\gamma' \tau_\pi + \eta \tau_\phi\right)}{(\varepsilon_{(0)} + P_{(0)}) (\tau_q - \lambda')\tau_\pi \tau_\phi} \leq 1,}\label{causality3}\\
    &{0 \leq  -\frac{\tau_{\tau_{a}}\tilde{\chi}_{2}+\tau_{\tau_{s}}\tilde{\chi}_{3}}{\tau_{\tau_{a}}\tau_{\tau_{s}}}\leq 1,}\label{causality4}\\
    &{0 \leq \frac{\tilde{\chi}_4}{\tau_{\Theta}} \leq 1.}\label{causality5}
\end{align}
Recall that, according to the notation in Eq.~\eqref{definition(s)s}, the coefficients $\tilde{\chi}_{1},\, \tilde{\chi}_{2},\,\tilde{\chi}_{3}$ present in the above conditions are all negative, while $\tilde{\chi}_{4}$ is positive. Additionally, $\tau_{q} > \lambda^{\prime}$, as mandated by the stability conditions in Eqs.~\eqref{con5} and \eqref{con1high}. Hence, the causality conditions are satisfied.
\medskip

Since the conditions in Eqs.~\eqref{causality1}-\eqref{causality5} all include relaxation times, their values are critical for ensuring causality. For example, let us focus on Eq.~\eqref{causality1}. We recall that the coefficients $\tilde{\chi}_{1}$ and $\,\tilde{\chi}_{2}$ are defined as,
\begin{align*}
{ \tilde{\chi}_{1}=\frac{2\beta_{(0)}}{\chi_{b}}\chi_{1},~~{\tilde{\chi}_{2}=\frac{\beta_{(0)}}{\chi_{b}}\chi_{2}}},
\end{align*}
and the expression for $\chi_{b}$ was derived in Eq.~\eqref{S11}. Therefore, condition~\eqref{causality1} takes the following form
\begin{align}
{0\leq\,\frac{\pi^{2}}{T^{4} \left(4K_{2}(x)+xK_{1}(x) \right )}\left[\frac{4\chi_{2}}{3\,\tau_{\tau_{s}}}+\frac{2\chi_{1}}{\tau_{\Phi}}\right]\leq\,1.\label{explicitycausality1}}
\end{align}
%
Clearly, the above condition is satisfied if the relaxation times are sufficiently large. The fulfillment of the remaining conditions in Eqs.~\eqref{causality1}-\eqref{causality5}, along with sufficiently large relaxation times, ensures the causality of perturbations in the dynamical variables and, consequently, the causality of the truncated perturbed evolution equations~\eqref{energyperturbation}-\eqref{Thetaperturbation}. 
%
\section{Summary and outlook}
\label{Summary and outlookIS}
In this chapter, we formulated second-order relativistic dissipative spin hydrodynamics hydrodynamics using the Müller-Israel-Stewart approach by deriving 40 coupled evolution equations for 40 dynamical variables. For clarity and tractability, we worked at vanishing chemical potential; its inclusion is left for future study. By ``formulation,'' we mean: \textbf{(i)} deriving the dissipative currents as dynamical variables through relaxation-type dynamical equations, along with their associated transport coefficients, obtained via an entropy-current analysis; and  
\textbf{(ii)} examining the linear stability and causality of a truncated set of the evolution equations.

As in conventional relativistic hydrodynamics, we find that the truncated M\"uller-Israel-Stewart equations for spin hydrodynamics are stable in the low- and high-wavenumber limits in the Landau frame choice $h^{\mu}=0$, however, their stability depends on the chosen spin equation of state. The equations are also found to be causal. A natural next step would be to reexamine stability and causality in a boosted frame and in the generalized Landau frame defined in Eq.~\eqref{generalizedLandueframe}. In addition, a deeper understanding of the emerging spin transport coefficients would improve the physical interpretation of the theory. Finally, a numerical study of the resulting equations, in a way done in the NS limit in Chapter. \ref{Navier-Stokes limit}, would be among interesting future tasks.

\chapter{Relativistic spin hydrodynamics: quantum-statistical formulation}
\label{Quantum-statistical formulation}
In this chapter we develop a first-principles formulation of relativistic dissipative spin hydrodynamics, starting from relativistic quantum-statistical mechanics and thermal quantum-field theory. 
 
We begin in Sec.~\ref{non-relativisticQSM} by revisiting the key concepts of nonrelativistic quantum-statistical mechanics; this review paves the way for Sec.~\ref{Relativistic quantum-statistical mechanicsqm}, where we introduce the \textit{Zubarev} density-operator formalism for conventional relativistic hydrodynamics and extend it to include the spin tensor. In Secs.~\ref{entropycurrentQM}-\ref{Entropy production rateqmmsec} we derive the entropy current and the associated entropy-production rate directly from the underlying statistical ensemble. We then turn to the dissipative currents at the first-order in hydrodynamic gradient expansion: Appendix~\ref{AppendixC} presents a novel method for decomposing an arbitrary rank-$n$ tensor into irreducible components under rotations, allowing us to express the dissipative currents in a compact, irreducible form. Section~\ref{transport} gathers these currents and lists the corresponding transport coefficients, thereby completing the first-order spin-hydrodynamic formulation developed in this chapter.

Our discussion of nonrelativistic quantum-statistical mechanics loosely follows Ref.~\cite{balian2006microphysics}. The \textit{Zubarev} approach to relativistic hydrodynamics builds on the seminal works of Refs.~\cite{Zubarev,ChGvanWeert} and subsequent developments~\cite{Becattini:2014yxa,Hayata:2015lga,Becattini:2019dxo,Becattini:2019poj}; spin degrees of freedom are incorporated following Ref.~\cite{Becattini:2018duy}. The expressions for the entropy current, entropy-production rate, dissipative currents, and transport coefficients in the presence of spin are based on Refs.~\multimyref{AD5, AD8}.

%
\section{Quantum statistical mechanics}
\label{non-relativisticQSM}
In this section, we review the basic concepts of the quantum statistical mechanics of systems of particles in the nonrelativistic limit. This section serves as a foundation for the sections that follow. The discussion is loosely based on Ref.~\cite{balian2006microphysics}.
\medskip

Unlike quantum mechanics, which describes the behavior of a single state $\ket{\psi}$, quantum statistical mechanics focuses on statistical ensembles -- large collections of possible quantum states or \textit{microstates} $\ket{\psi_{i}}$, each associated with a probability $p_{i}$ that the system occupies that state. Such a system can be represented by the set $\{\ket{\psi_{i}}, p_{i}\}$, where each $\ket{\psi_{i}}$ has an associated probability $p_{i} \in [0,1]$, with $\sum p_{i} = 1$.

Recall that a quantum state can be \textit{pure} or \textit{mixed}. A pure state is fully described by a single state $\ket{\psi}$, representing a system with no uncertainty beyond that intrinsic to quantum mechanics. It can be expressed as either a state vector $\ket{\psi}$ or a pure state density operator $\wrho = \ket{\psi}\bra{\psi}$. In contrast, a mixed state arises when our knowledge of the system is incomplete, implying that the system could occupy several possible microstates with specific probabilities. It is described by a \emph{mixed-state density operator}, representing a statistical mixture of pure states:
\begin{align}
\wrho=\sum_{i}p_{i}\ket{\psi_{i}}\bra{\psi_{i}}.
\end{align}
If the system is known to be exactly in one of the \( |\psi_i\rangle \) with \( p_i = 1 \), the state is pure.

The goal of quantum statistical mechanics is to describe the macroscopic physical properties of a system composed of a large number of degrees of freedom. These macroscopic quantities are the expectation values of the corresponding operators. The expectation value of an operator $\mathcal{\widehat{A}}$ is given by: 
\begin{align}
\langle \widehat{\mathcal{A}} \rangle=\sum_{i}p_{i}\bra{\psi_{i}}\widehat{\mathcal{A}}\ket{\psi_{i}}=\Tr(\wrho\widehat{\mathcal{A}}).
\end{align}

The \textit{Von Neumann entropy} quantifies the number of possible microstates a system can have -- or equivalently, how much information we possess about the system's state. It reaches a maximum for a maximally mixed state and a minimum (zero) for a pure state. It is defined as: 
\begin{align}
S = -\Tr(\,\wrho \log \wrho\,).
\end{align}
As the Von Neumann entropy approaches zero, our knowledge of the system increases; conversely, higher entropy indicates greater uncertainty. 

To make predictions -- i.e., to calculate the expectation values of operators -- we must determine the system's state $\wrho$. If we know nothing about a system, the most unbiased assumption is that all microstates are equally probable. This corresponds to a state of maximal entropy or a \emph{maximally mixed state}. In this case, $\wrho$ is the density operator for a system with equiprobable states. We can find its form by maximizing the entropy:
\begin{align}
F[\,\wrho\,]=-\Tr(\,\wrho \log \wrho\,)\qquad \frac{\delta F[\,\wrho\,]}{\delta\wrho}=0.
\end{align}
If we have additional information about the system, such as the expectation value of energy or other observables, we consider all microstates consistent with that data and assume they are equally likely. The best estimate for $\wrho$ is then the one that maximizes entropy while reproducing the known data. This is solved using the method of Lagrange multipliers: 
\begin{align}\label{nonrelconst}
F[\,\wrho\,] = -\Tr(\,\wrho \log \wrho\,) - \sum_i \lambda_i \left[ \Tr(\wrho \widehat{\mathcal{A}}_i) - \mathcal{A}_i \right],
\end{align}
where $\lambda_{i}$ is a Lagrange multiplier for each constraint. Here, the $\mathcal{A}_{i}$ denote the known values of the corresponding observables $\widehat{\mathcal{A}}_i$ (throughout the rest of this Chapter we refer to them as \textit{actual values}). This approach is known as the \emph{maximum entropy principle}, and states determined this way are said to be in thermodynamic equilibrium. Maximizing the functional with respect to $\wrho$ yields the density operator:
\begin{align}
\wrho=\frac{1}{Z}\exp\left(-\sum_{i}\lambda_{i}\mathcal{\widehat{A}}_{i}\right),
\end{align}
where the \textit{partition function} $Z$ is defined as: 
\begin{align}
Z=\Tr\left[\exp\left(-\sum_{i}\lambda_{i}\mathcal{\widehat{A}}_{i}\right)\right].
\end{align}
Different choices of observables and constraints lead to different statistical ensembles. For example, if:
\begin{align}\label{FE0}
F[\,\wrho\,] = -\Tr(\,\wrho \ln \wrho\,) - \beta \left[ \Tr(\wrho \widehat{H}) - E\right] + \zeta \left[\Tr(\wrho \widehat{J}) - J \right],
\end{align}
where \( \widehat{H} \) is the Hamiltonian operator and \( \widehat{J} \) is the particle number operator~\footnote{The operator $\widehat{J}$ can be, in general, any conserved charge in the system. However, throughout this chapter, we will refer to it as a conserved particle number operator.}, and $E$, $J$ are the actual values, then the density operator becomes:
\begin{align}
\wrho = \frac{1}{Z} \exp\left(-\beta \widehat{H} + \zeta \widehat{J}\right),
\end{align}
where \(Z = \Tr \left[\exp\left(-\beta \widehat{H} + \zeta \widehat{J}\right)\right]\). The Lagrange multipliers can be interpreted as $\beta=1/T$ and $\zeta=\mu/T$, where $T$ is the temperature and $\mu$ the chemical potential.  
%

%
%
%
%
%
\section{Relativistic quantum statistical mechanics}
\label{Relativistic quantum-statistical mechanicsqm}
The objective of this section is to extend the quantum statistical mechanics to a \emph{relativistic} framework by deriving the density operator for a relativistic system of particles with spin degrees of freedom. This allows us to characterize the macroscopic properties of the system as those of a relativistic fluid, leading to a formulation of relativistic spin hydrodynamics. The approach without spin degrees of freedom is known as the \textit{Zubarev formulation} of conventional relativistic hydrodynamics. This section is based on Refs.~\cite{Zubarev,ChGvanWeert,Becattini:2014yxa,Hayata:2015lga,Becattini:2019dxo,Becattini:2019poj,Becattini:2018duy}
\medskip

Similar to the nonrelativistic case, the expectation values of operators are defined with respect to a density operator. However, in the relativistic regime, the operators are constructed within the framework of quantum field theory. This distinction arises because particles are now described, loosely speaking, in terms of creation and annihilation operators acting on the vacuum state of the theory. Accordingly, the expectation value of an operator $\mathcal{\widehat{A}}$ is defined as:  
\begin{align}
\langle \widehat{\mathcal{A}}\rangle=\Tr(\wrho\,\widehat{\mathcal{A}})_{\rm ren}=\Tr(\wrho\,\widehat{\mathcal{A}})-\bra{0}\widehat{\mathcal{A}}\ket{0},
\end{align}
where the subscript "ren" denotes \emph{renormalization} by subtraction of the vacuum expectation value. The non-renormalized expectation values are generally divergent. The specific renormalization procedure depends on the details of the underlying QFT and is beyond the scope of this thesis ( see Ref.~\cite{Becattini:2017ljh} and references therein). Moreover, determining the form of $\wrho$ explicitly in terms of the underlying microstates $\ket{\psi_{i}}$ is often intractable or even impossible in QFT. Therefore, as in the nonrelativistic case, see Sec.~\ref{non-relativisticQSM}, we employ a variational approach based on entropy maximization, adapted to a relativistic system with spin. 
%
%
\subsection{Global equilibrium density operator}
As discussed in Chapter~\ref{From fields to fluids}, a relativistic system of particles with spin is invariant under global continuous spacetime symmetries. As a result, the system obeys conservation laws for the energy-momentum, particle number, and total angular momentum tensor operators,  
\begin{align}\label{conservatoperatr}
&\partial_{\mu}\wT^{\mu\nu}(t,\x)=0~,~\partial_{\mu}\widehat{j}^{\mu}(t,\x)=0~,~\partial_{\lambda}\wJ^{\,\lambda\mu\nu}(t,\x)=0.
\end{align}
The total angular momentum tensor can be decomposed into orbital and spin parts,
\begin{align}\label{sandl}
&\wJ^{\,\lambda\mu\nu}=\widehat{L}^{\lambda\mu\nu}+\wspt^{\lambda\mu\nu}, \quad \text{with} \quad \widehat{L}^{\lambda\mu\nu}=2\wT^{\lambda[\nu}x^{\mu]},
\end{align}
and the spin tensor obeys:
\begin{align}\label{continuityopqerrr}
\partial_{\lambda}\wspt^{\lambda\mu\nu}=\wT^{\nu\mu}-\wT^{\mu\nu}.
\end{align}
The corresponding  globally conserved total quantities are defined as integrals of the above tensor densities over a spacelike hypersurface $\Sigma$: 
\begin{align}\label{globalrel}
&\widehat{P}^{\nu}(t)=\int \di \Sigma_{\mu}\wT^{\mu\nu}~,~\widehat{J}(t)=\int \di \Sigma_{\mu}\widehat{j}^{\mu}~,~\widehat{J}^{\mu\nu}(t)=\int \di\Sigma_{\lambda}\left(x^{\mu}\wT^{\lambda\nu}-x^{\nu}\wT^{\lambda\mu}\right)+\wspt^{\lambda\mu\nu},
\end{align}
where $\di\Sigma_{\mu} = \di\Sigma \, n_{\mu}$, with $\Sigma$ being a spacelike three-dimensional hypersurface, $\di\Sigma$ its surface element, and $n^{\mu}$ the future-directed timelike normal vector to the surface. Here, $\widehat{P}^{\nu}$ represents total four-momentum, $\widehat{J}$ is total particle number, and $\widehat{J}^{\mu\nu}$ total angular momentum. More details about the geometry employed will be provided in the next section.

As in the nonrelativistic case, the expectation values of these conserved quantities are constrained with their actual (macroscopic) values, which are here treated as the conserved total hydrodynamic currents. For example, for the four-momentum operator, the constraint reads:
\begin{align}\label{globaleqcost}
\langle \widehat{P}^{\mu}\rangle\equiv\Tr(\wrho\widehat{P}^{\mu})=P^{\mu},
\end{align}
with similar expressions holding for $\langle\widehat{J}\rangle$ and $\langle\wJ^{\mu\nu}\rangle$.

The global equilibrium density operator $\wrho_{\rm GE}$ can be obtained by maximizing the von Neumann entropy subject to these constraints. Analogous to Eq.~\eqref{nonrelconst}, we define:    
\begin{align}
F[\,\wrho_{\rm GE}\,] =& -\Tr(\,\wrho_{\rm GE} \log \wrho_{\rm GE}\,) - b_{\mu} \left[ \Tr(\wrho_{\rm GE} \widehat{P}^{\mu}) - P^{\mu} \right] + \zeta \left[\Tr(\wrho_{\rm GE} \widehat{J}) - J \right]\nonumber\\
&+\varpi_{\mu\nu}\left[ \Tr(\wrho_{\rm GE} \widehat{J}^{\mu\nu}) - J^{\mu\nu} \right], 
\end{align}
where $b_{\mu}, \zeta,$ and $\varpi_{\mu\nu}=-\varpi_{\nu\mu}$ are Lagrange multipliers. Maximizing this functional with respect to $\wrho_{\rm GE}$ yields the \emph{global equilibrium density operator}:
\begin{align}\label{Globalequilibriumdenityoperatorqm}
\wrho_{\rm GE}=\frac{1}{Z_{\rm GE}}\exp\left[-b_{\mu}\widehat{P}^{\mu}+\zeta \widehat{J}+\frac{1}{2}\varpi_{\mu\nu}\wJ^{\mu\nu}\right],
\end{align}
where $Z_{\rm GE}$ is the corresponding partition function. Using Eq.~\eqref{globalrel}, this can be rewritten in terms of the spin tensor operator:
\begin{align}
\wrho_{\rm GE}=\frac{1}{Z_{\rm GE}}\exp\left[-\int \di \Sigma_{\mu}(b_{\nu}+\varpi_{\nu\lambda}x^{\lambda})\wT^{\mu\nu}+\zeta\int\di\Sigma_{\mu}\, \wj^{\mu}+\frac{\varpi_{\mu\nu}}{2}\int \di \Sigma_{\lambda}\wspt^{\lambda\mu\nu}\right].
\end{align}
It is important to emphasize that in global equilibrium, the Lagrange multipliers are constants -- they do not depend on spacetime position. A physical interpretation of these parameters will be provided later in this section. 
%
%
\subsection{Local equilibrium density operator}
\label{Local equilibrium density operator with spin current}
In local equilibrium, the relevant physical input consists of densities, such as energy, momentum, particle number, and spin densities, rather than integrated conserved quantities. Therefore, using only the global expectation values of operators as constraints -- like en Eq.~\eqref{globaleqcost} -- is insufficient to construct the local equilibrium density operator.  
\medskip

To impose density-level constraints, we first define the geometric setting. Let spacetime be foliated into a family of disjoint, three-dimensional spacelike hypersurfaces $\Sigma_{\tau}$, each labeled by a constant value of a scalar function \( \tau(x) \). This foliation of spacetime is commonly known as the \emph{ADM (Arnowitt-Deser-Misner) foliation}. 

At each point $x$, let \( n^\mu(x) \) be the future-directed timelike unit vector normal to the hypersurface $\Sigma_{\tau}$, and define the tangent projector \( \Delta^{\mu\nu} = g^{\mu\nu} - n^\mu n^\nu \). For this foliation to be well-defined, $n^{\mu}$ must be vorticity-free, i.e.,  
\begin{align}
\epsilon^{\mu\nu\rho\sigma} n_\nu \partial_\rho n_\sigma = 0\;,
\end{align}
which is also known as the \emph{Frobenius integrability condition}. The choice of $n^{\mu}$ reflects the freedom to select a reference frame in relativistic hydrodynamics. A common and physically motivated choice is to identify \( n^{\mu} \) with the fluid four-velocity \( u^{\mu} \), as this leads to matching conditions similar to those in Landau and Eckart frames.

In this setup, the renormalized local densities of the  energy-momentum tensor, particle number current, and spin tensor on the hypersurface $\Sigma_{\tau}$, are given by:
\begin{align}
n_{\mu}\Tr(\wrho\widehat{T}^{\mu\nu})_{\rm ren}~,~
n_{\mu}\Tr(\wrho\widehat{j}^{\mu})_{\rm ren}~,~n_{\mu}\Tr(\wrho\wspt^{\lambda\mu\nu})_{\rm ren}.
\end{align}
Accordingly, the constraints that ensure the density operator reproduces the correct local values are:
\begin{align}
&n_{\mu}T^{\mu\nu}_{\rm LE}\equiv n_{\mu}\Tr(\wrho_{\rm LE} \widehat{T}^{\mu\nu})_{\rm ren}=n_{\mu}T^{\mu\nu},\label{const1qm}\\
&n_{\mu}j^{\mu}_{\rm LE}\equiv n_{\mu}\Tr(\wrho_{\rm LE}\widehat{j}^{\mu})_{\rm ren}=n_{\mu} j^{\mu},\label{const2qm}\\
&n_{\mu}\spt^{\mu\lambda\nu}_{\rm LE}\equiv n_{\mu}\Tr(\wrho_{\rm LE}\wspt^{\mu\lambda\nu})_{\rm ren}=n_{\mu}\spt^{\mu\lambda\nu}.\label{const3qm}
\end{align}
Here, the quantities \(T^{\mu\nu}_{\rm LE}, j^{\mu}_{\rm LE}, \spt^{\mu\lambda\nu}_{\rm LE}\) denote the renormalized expectation values in local equilibrium, while \(T^{\mu\nu}, j^{\mu}, \spt^{\mu\lambda\nu}\) are the corresponding physical currents used as constraints. 

To construct the local equilibrium density operator, we maximize the von Neumann entropy over the initial spacelike hypersurface $\Sigma_{\tau_{0}}$, requiring that the local expectation values match the actual hydrodynamic densities. The variational functional $F[\wrho_{\rm LE}]$ becomes: 
\begin{align}
F[\wrho_{\rm LE}]=-\Tr(\wrho_{\rm LE}\log\wrho_{\rm LE})+\int_{\Sigma_{\tau_{0}}}\di \Sigma\,&\bigg[n_{\mu}\left(\Tr(\wrho_{\rm LE}\wT^{\mu\nu})_{\rm ren}-T^{\mu\nu}\right)\beta_{\nu}(x)\nonumber\\
&~~-n_{\mu}\left(\Tr(\wrho_{\rm LE}\wj^{\mu})_{\rm ren}-j^{\mu}\right)\zeta(x)\nonumber\\
&~~-n_{\mu}\left(\Tr(\wrho_{\rm LE}\wspt^{\mu\lambda\nu})_{\rm ren}-\spt^{\mu\lambda\nu}\right)\Omega_{\lambda\nu}(x)\bigg],
\end{align}
where $\di \Sigma_\mu \equiv \di \Sigma \, n_\mu$, and the Lagrange multipliers $\beta_{\nu}(x)$, $\zeta(x)$, $\Omega_{\lambda\nu}(x)$ are now functions of spacetime. Maximizing this functional yields the \emph{local equilibrium density operator}: 
\begin{align}\label{densityoperator}
\wrho_{\rm LE}=\frac{1}{Z_{\rm LE}}\exp\left[-\int_{\Sigma_{\tau_{0}}} \di\Sigma_{\mu}\,\left(\wT^{\mu\nu}\beta_{\nu}
-\zeta \wj^\mu - \frac{1}{2}\Omega_{\lambda\nu}\wspt^{\mu\lambda\nu}\right)\right]\,,
\end{align}
with the corresponding partition function defined as:
\begin{align}\label{partitionfunctionoriginal}
Z_{\rm LE}=\Tr\left\{\exp\left[-\int_{\Sigma_{\tau_{0}}} \di\Sigma_{\mu}\,\left(\wT^{\mu\nu}\beta_{\nu}
-\zeta \wj^\mu - \frac{1}{2}\Omega_{\lambda\nu}\wspt^{\mu\lambda\nu}\right)\right]\right\}.
\end{align}
In Eq.~\eqref{densityoperator}, the fields $\beta_{\nu}(x)$, $\zeta(x)$ 
and $\Omega_{\lambda\nu}(x)$ are the Lagrange multipliers associated with this variational problem. They are interpreted physically as:
\begin{align}\label{lagrangemultipliersRQM}
\beta_{\mu} = \frac{u_{\mu}}{T}\,,  \qquad  \zeta = \frac{\mu}{T}\,, \qquad \Omega_{\mu\nu} = \frac{\omega_{\mu\nu}}{T}\,,
\end{align}
where $T$ is the temperature, $\mu$ the chemical potential, and $\omega_{\mu\nu}=-\omega_{\nu\mu}$ the spin potential (or spin polarization tensor). It is worth emphasizing that these fields are not arbitrary; they must satisfy the constraint equations \eqref{const1qm}-\eqref{const3qm}, and can, in principle, be determined by solving them. For further details, see Refs.~\cite{Becattini:2014yxa,Becattini:2018duy}. 

It is important to point out that the spin tensor density constraint~\eqref{const3qm} originates from the total angular momentum density constraint. In general, one can begin with the density constraint on the total angular momentum; however, since the orbital part is already accounted for in the energy-momentum density constraints~\eqref{const1qm}, the resulting density operator~\eqref{densityoperator} remains unchanged~\cite{Becattini:2018duy}.
%
%
\subsection{Out-of-equilibrium density operator}
We have so far defined a foliation of spacetime into three-dimensional spacelike hypersurfaces and introduced the local equilibrium density operator~\eqref{densityoperator} on the initial-time hypersurface $\Sigma_{\tau_{0}}$. The key observation here is that the local thermodynamic equilibrium density operator~\eqref{densityoperator} does not represent a stationary state. This is because the operators appearing in it -- constructed from quantum fields -- are $\tau$-dependent, and therefore $\wrho_{\rm LE}$ inherits this dependence. 

\begin{figure}[H]
\begin{center}
\includegraphics[width=9cm]{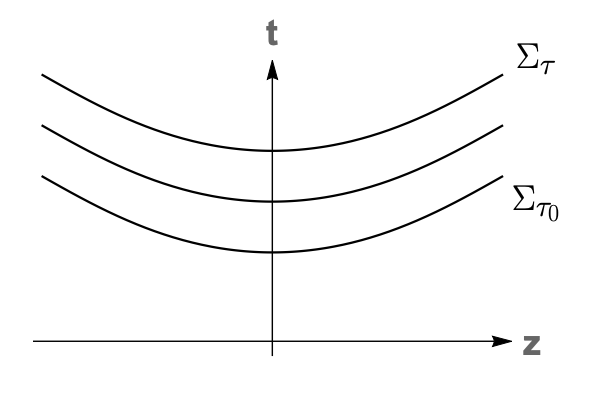}
\caption{A schematic illustration of a family of three-dimensional spacelike hypersurfaces defining a foliation of spacetime, parametrized by the real variable $\tau$, with initial hypersurface $\Sigma_{\tau_{0}}$ and present-time hypersurface $\Sigma_{\tau}$.}
\label{fol}
\end{center}
\end{figure}

To analyze this time evolution, consider a spacetime volume $\Gamma$ bounded by two spacelike hypersurfaces of the foliation: the initial-time hypersurface $\Sigma_{\tau_{0}}$ and the present time hypersurface $\Sigma_{\tau}$, as well as a timelike boundary at spatial infinity connecting them; see also Fig.~\ref{fol}. Assuming that fluxes through the timeline boundary vanish (i.e., the boundary terms at infinity drop out), we can apply the \emph{Gauss (divergence) theorem} to obtain:
\begin{align}
&\int_{\Sigma_{\tau}}\di \Sigma_\mu \;  \left(\wT^{\mu\nu}\beta_{\nu}-\zeta\wj^{\mu}-\frac{1}{2}\Omega_{\lambda\nu}\wspt^{\mu\lambda\nu}\right)- \int_{\Sigma_{\tau_{0}}}\di \Sigma_\mu \;  \left(\wT^{\mu\nu}\beta_{\nu}-\zeta\wj^{\mu}-\frac{1}{2}\Omega_{\lambda\nu}\wspt^{\mu\lambda\nu}\right)\nonumber\\
&=\int_{\Gamma} \di\Gamma\,\partial_{\mu}\left(\wT^{\mu\nu}\beta_{\nu}-\zeta\wj^{\mu}-\frac{1}{2}\Omega_{\lambda\nu}\wspt^{\mu\lambda\nu}\right).
\end{align}

As a result, using Eq.~\eqref{densityoperator}, the generalized density operator evaluated on the present-time hypersurface $\Sigma_{\tau}$ reads: 
\begin{align}\label{outofequilibriumdensityop}
\wrho&=\frac{1}{Z_{\rm LE}}\exp\bigg[-\int_{\Sigma_{\tau}} \di\,\Sigma_{\mu}\,\left(\wT^{\mu\nu}\beta_{\nu}-\zeta \wj^\mu - \frac{1}{2}\Omega_{\lambda\nu}\wspt^{\mu\lambda\nu}\right)\nonumber\\
&~~~~~~~~~~~~~~~~~+\int_{\Gamma} \di\Gamma\,\partial_{\mu}\left(\wT^{\mu\nu}\beta_{\nu}-\zeta\wj^{\mu}-\frac{1}{2}\Omega_{\lambda\nu}\wspt^{\mu\lambda\nu}\right)\bigg].
\end{align}
Note that if the spacetime integral over the volume $\Gamma$ vanishes -- e.g., if the fields $\beta_{\nu}$, $\zeta$ 
and $\Omega_{\lambda\nu}$ are constant in spacetime -- we recover the local equilibrium density operator, now evaluated on a present-time hypersurface $\Sigma_{\tau}$. The second term in Eq.~\eqref{outofequilibriumdensityop}, involving spacetime gradients, represents the out-of-equilibrium correction to the local equilibrium density operator. Importantly, as we will show in Sec.~\ref{Entropy production rateqmmsec}, the volume integral term is also responsible for entropy production in the system.
%
%
\subsection{Global equilibrium conditions}
In a state of \emph{global equilibrium}, the density operator must be independent of $\tau$ -- that is, independent of the choice of spacelike hypersurface. Formally, this means that, starting from the local equilibrium density operator in Eq.~\eqref{densityoperator}, we require the derivative of the integrand to vanish. Imposing this requirement leads to the conditions of global equilibrium. Specifically, the following conditions must hold:
\begin{align}
&\partial_{\mu}\beta_{\nu}+\partial_{\nu}\beta_{\mu}=0,\label{condition1}\\
&\partial_{\mu}\zeta=0,\label{condition2}\\
&\partial_{\mu}\Omega_{\lambda\nu}=0.\label{condition3}
%
\end{align}
The first condition, Eq.~\eqref{condition1}, is the \emph{Killing equation}, which expresses that the thermal vector field $\beta$ must generate a spacetime symmetry. The second condition, Eq.~\eqref{condition2}, requires the ratio of chemical potential to temperature, $\zeta$, to be constant throughout spacetime. Similarly, the third condition, Eq.~\eqref{condition3}, demands that the spin potential divided by temperature, $\Omega$, is also spacetime-independent. 

The general solution to the Killing equation~\eqref{condition1} is:
\begin{align}
\beta_{\nu} (x)=b_{\nu}+\varpi_{\nu\lambda} x^{\lambda},\label{condition4}
\end{align}
where $b_{\nu}$ is a constant four-vector and $\varpi_{\lambda\nu}=-\varpi_{\nu\lambda}$ is a constant antisymmetric tensor known as the \emph{thermal vorticity}. It is related to the derivatives of $\beta_{\nu}$ via:
\begin{align}\label{thermalvoticityqm}
\varpi_{\lambda\nu}=-\frac{1}{2}(\partial_{\lambda}\beta_{\nu}-\partial_{\nu}\beta_{\lambda}),
\end{align}
which is constant at global equilibrium. An additional condition must be imposed to ensure the integrand in the exponent of Eq.~\eqref{densityoperator} remains constant throughout spacetime. This leads to the final global equilibrium condition:
\begin{align}
\varpi^{\lambda\nu}=\Omega^{\lambda\nu}.
\end{align}

Substituting all the above conditions into the local equilibrium density operator~\eqref{densityoperator}, we recover the global equilibrium density operator introduced earlier in Eq.~\eqref{Globalequilibriumdenityoperatorqm}.
%
%
%
%
%
\section{Entropy current}
\label{entropycurrentQM}
The main objective of this section is to derive the local equilibrium entropy current for a relativistic fluid with spin based on the relativistic quantum-statistical formulation developed earlier. We also extend the discussion to the out-of-equilibrium entropy current. The conceptual foundation of the presented material was inspired by Ref.~\cite{Becattini:2019poj}, while the results are based on Ref.~\multimyref{AD5}.
\subsection{Local equilibrium entropy current}
To derive the local equilibrium entropy current, we begin by substituting the local equilibrium density operator from Eq.~\eqref{densityoperator} into the von Neumann entropy:
\begin{align}\label{prelagZ}
S&= -\Tr (\wrho_{\rm LE} \log \wrho_{\rm LE}) \nonumber\\
&= \log Z_{\rm LE} +  \int_\Sigma \di \Sigma_\mu \; 
\left[ \Tr(\wrhol \wT^{\mu\nu}) \beta_{\nu} -\zeta \Tr( \wrhol \wj^\mu) -
\frac{1}{2}\Omega_{\lambda\nu}\Tr( \wrhol \wspt^{\mu\lambda\nu}) \right].
\end{align}
Here, the operator expectation values are defined at spacetime points \( x \in \Sigma \), where \( \Sigma \) denotes a present time hypersurface at which local equilibrium is established—that is, where the constraints given by Eqs.~\eqref{const1qm}–\eqref{const3qm} are satisfied. The key question is whether the partition function, \( \log Z_{\rm LE} \), is \emph{extensive}, i.e., whether it can be written in terms of another thermodynamic field so that the total entropy $S$ can be expressed as an integral over the hypersurface \( \Sigma \)
of a local current $S^\mu$:
\begin{align}
S=\int_\Sigma \di \Sigma_\mu \; S^\mu,
\end{align}
with \( S^\mu \) interpreted as the entropy current. 

This extensivity was demonstrated in the spinless case in Ref.~\cite{Becattini:2019poj}. In Appendix~\ref{Appendix extensivity}, we extend the proof to the presence of spin tensor and show that \( \log Z_{\rm LE} \) can be expressed in terms of the so-called \textit{thermodynamic potential current} $\phi^{\mu}$. Here, we briefly discuss the proof and mention the results.

The initial step consists of modifying the local equilibrium density operator~\eqref{densityoperator} by incorporating a dimensionless parameter \( \lambda \), such that
\begin{align}\label{rholambda}
\wrho_{\rm LE}(\lambda)=\frac{1}{Z_{\rm LE}(\lambda)}\exp\left[-\lambda\int_{\Sigma} \di\Sigma_{\mu}\,\left(\wT^{\mu\nu}\beta_{\nu}
-\zeta \wj^\mu - \frac{1}{2}\Omega_{\lambda\nu}\wspt^{\mu\lambda\nu}\right)\right]\,,
\end{align}
where the partition function $Z_{\rm LE}(\lambda)$ is given by
\begin{align}
Z_{\rm LE}(\lambda)=\Tr\left\{\exp\left[-\lambda\int_{\Sigma} \di\Sigma_{\mu}\,\left(\wT^{\mu\nu}\beta_{\nu}
-\zeta \wj^\mu - \frac{1}{2}\Omega_{\lambda\nu}\wspt^{\mu\lambda\nu}\right)\right]\right\}.
\end{align}
This formulation is constructed such that, when \( \lambda = 1 \), it recovers the original density operator~\eqref{densityoperator} and the corresponding partition function. Based on this, we define the operator $\widehat \Upsilon$ as
\begin{align}
\widehat \Upsilon \equiv \int_{\Sigma} \di\,\Sigma_{\mu}\,\left(\wT^{\mu\nu}\beta_{\nu}
-\zeta \wj^\mu - \frac{1}{2}\Omega_{\lambda\nu}\wspt^{\mu\lambda\nu} \right)~\implies~Z_{\rm LE}(\lambda)=\Tr(e^{-\lambda\widehat \Upsilon }).
\end{align}
It can be shown that if \( \widehat{\Upsilon} \) satisfies the following conditions: \textbf{(i)} it has discrete eigenvalues \( \Upsilon_n \), \textbf{(ii)} these eigenvalues are non-degenerate, and \textbf{(iii)} it is bounded from below, meaning that a lowest non-degenerate eigenvalue \( \Upsilon_0 \) exists with corresponding eigenvector \( \ket{0} \), then the logarithm of \( Z_{\rm LE} \) is \emph{extensive}. In other words, it can be written as an integral over \( \Sigma \) as
\begin{align}\label{extensive}
\log Z_{\rm LE} &= \int_\Sigma \di\Sigma_\mu \; \phi^\mu 
- \bra{0} \widehat \Upsilon \ket{0}\nonumber\\
&= \int_\Sigma \di\Sigma_\mu \; \left[ \phi^\mu 
- \bra{0}(\wT^{\mu\nu}\beta_{\nu} -\zeta \wj^\mu - \frac{1}{2}\Omega_{\lambda\nu}\wspt^{\mu\lambda\nu} 
)\ket{0} \right].
\end{align}
Here, $\phi^{\mu}$ is defined as the \emph{thermodynamic potential current},
\begin{align}\label{phi}
\phi^{\mu}=\int_{1}^{\infty} \di\lambda \; \left( T^{\mu\nu}_{\rm LE}(\lambda) \beta_{\nu}-\zeta j^\mu_{\rm LE}(\lambda) - \frac{1}{2}\Omega_{\lambda\nu} \spt^{\mu\lambda\nu}_{\rm LE}(\lambda)\right),
\end{align}
with its physical interpretation to be discussed shortly. The expectation values $T^{\mu\nu}_{\rm LE}(\lambda)$, $j^\mu_{\rm LE}(\lambda)$, and $\spt^{\mu\lambda\nu}_{\rm LE}(\lambda)$ 
are computed using Eqs.~\eqref{const1qm}-\eqref{const3qm}, but with the modified density operator in Eq.~\eqref{rholambda}. For a complete derivation of Eqs.~\eqref{extensive}, \eqref{phi}, refer to Appendix~\ref{Appendix extensivity}. 

By substituting the expressions for the Lagrange multipliers from Eq.~\eqref{lagrangemultipliersRQM} into Eq.~\eqref{phi}, it becomes clear that $\lambda$ plays the role of a \emph{rescaled inverse 
temperature}. Therefore, one can change the integration variable from $\lambda$ to 
$T'(x) = T(x)/\lambda$, leading to a more appealing form of the thermodynamic potential current: 
\begin{align}\label{phit}
\phi^{\mu}(x) =\int_{0}^{T(x)} \frac{\di T'}{T^{\prime 2}} \; \bigg( &T^{\mu\nu}_{\rm LE}(x)[T',\mu,\omega] 
u_{\nu}(x) -\mu(x) j^\mu_{\rm LE}(x)[T',\mu,\omega]\nonumber\\
&- \frac{1}{2} \omega_{\lambda\nu}(x) 
\spt^{\mu\lambda\nu}_{\rm LE}(x)[T',\mu,\omega] \bigg) \,.
\end{align}
The square brackets indicate functional dependence on the thermodynamic fields. In fact, the local equilibrium values of the currents at a point $x$ depend not only on the local values
of $T',\mu$, and $\omega$ at that point $x$, but on their entire profiles $T'(y),\mu(y),\omega(y)$ over spacetime. Equation \eqref{phit} shows that the thermodynamic 
potential current $\phi^{\mu}$ can be obtained by integrating the local equilibrium values of the relevant currents over temperature. In the special case of homogeneous global equilibrium, which is defined by the condition $\beta$ = const, i.e., vanishing thermal vorticity in Eq.~\eqref{condition4}, and assuming no charges in the system, Eq.~\eqref{phit} reduces to the familiar expression:
\begin{align}
\phi^{\mu}=P\beta^{\mu},
\end{align}
where $P$ is the isotropic pressure and $\beta^{\mu}=u^{\mu}/T$. For full details, see Ref.~\multimyref{AD5}.

Finally, once the thermodynamic potential current $\phi^\mu$ is determined, the \emph{entropy current}~\eqref{prelagZ} can be defined as:
\begin{align}\label{totalent}
S =& -\Tr (\wrho_{\rm LE} \log \wrho_{\rm LE})\nonumber\\
&= \log Z_{\rm LE} +  \int_\Sigma \di \Sigma_\mu \; 
\left[ \Tr(\wrhol \wT^{\mu\nu}) \beta_{\nu} -\zeta \Tr( \wrhol \wj^\mu)-\frac{1}{2}\Omega_{\lambda\nu}\Tr( \wrhol \wspt^{\mu\lambda\nu}) \right]\nonumber\\
&= \int_\Sigma \di \Sigma_\mu \;\left[\phi^\mu + T^{\mu\nu}_{\rm LE}\beta_{\nu}-\zeta j^\mu_{\rm LE}-\frac{1}{2}\Omega_{\lambda\nu}\spt^{\mu\lambda\nu}_{\rm LE} \right].
\end{align}
This expression implies that the local equilibrium entropy current can be defined as:
\begin{align}\label{def1}
S^\mu = \phi^\mu + T^{\mu\nu}_{\rm LE}\beta_{\nu} -\zeta j^\mu_{\rm LE}-\frac{1}{2}\Omega_{\lambda\nu}\spt^{\mu\lambda\nu}_{\rm LE}.
\end{align}

\begin{figure}[H]
\begin{center}
\includegraphics[width=9cm]{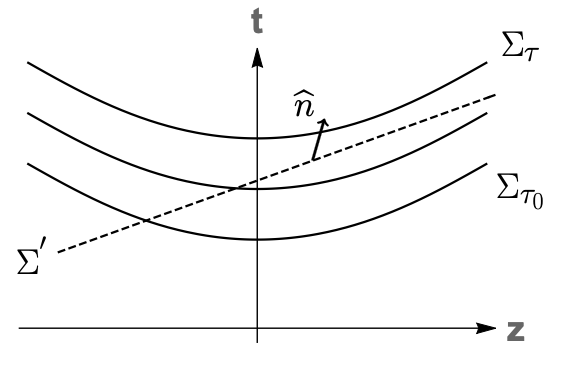}
\caption{An example of a family of three-dimensional spacelike hypersurfaces (solid lines) forming a foliation, which is essential for defining local thermodynamic equilibrium. The hypersurface $\Sigma^{'}$ (shown as a dashed line) does not belong to the foliation.}
\label{sigmaprime}
\end{center}
\end{figure}

Both the thermodynamic potential current $\phi^{\mu}$ Eq.~\eqref{phit} and the entropy current $S^{\mu}$ Eq.~\eqref{def1} explicitly depend on the renormalized expectation values of the quantum operators in local equilibrium. This implies that $\phi^{\mu}$ and $S^{\mu}$ depend not only on the spacetime point $x \in \Sigma$, but also on the choice of foliation used to define the local equilibrium. This foliation dependence leads to a conceptual issue: if the total entropy is computed by integrating the entropy current \eqref{def1} over some hypersurface $\Sigma^{'}$ that does not belong to the foliation (see figure~\ref{sigmaprime}), the result may not correspond to the entropy defined by the von Neumann expression
\begin{align*}\label{sproblem}
\int_{\Sigma^{\prime}} \di\Sigma_\mu \; S^\mu \ne - \Tr(\wrhol(\Sigma^{\prime}) \log \wrhol(\Sigma^{\prime}))\,,
\end{align*}
with equality holding only when $\Sigma^{\prime}$ belongs to the foliation that defines the local equilibrium expectation values.
%
%
\subsection{Out-of-equilibrium entropy current}

The out-of-equilibrium entropy current is defined more generally by expressing it in terms of the actual values of the currents, rather than their local equilibrium counterparts. This is given by,
\begin{equation}
S^{\mu} = \phi^{\mu} + T^{\mu\nu} \beta_{\nu} - \zeta j^{\mu} - \frac{1}{2} \Omega_{\lambda\nu} \spt^{\mu\lambda\nu}, 
\label{eq:entropy_current}
\end{equation}
where the thermodynamic potential current is expressed as:
\begin{align}\label{phi2}
\phi^{\mu}=\int_{0}^{T} \frac{\di T'}{T^{\prime 2}} \; \left( T^{\mu\nu}[T'] u_{\nu}
-\mu j^\mu[T'] - \frac{1}{2} \omega_{\lambda\nu} \spt^{\mu\lambda\nu}[T']
\right),
\end{align}
if and only if the total entropy, integrated over the entire hypersurface, remains invariant. This condition can be written as
\begin{align}
\int_\Sigma \di \Sigma_\mu S^\mu & = \int_\Sigma \di \Sigma \; n_\mu \left[ \phi^\mu + T^{\mu\nu} \beta_{\nu} 
-\zeta j^\mu - \frac{1}{2}\Omega_{\lambda\nu} \spt^{\mu\lambda\nu} \right] \nonumber\\
& = \int_\Sigma \di \Sigma \; n_\mu \bigg[ \int_{0}^{T} \frac{\di T'}{T^{\prime 2}} \; \left( T^{\mu\nu}[T'] 
u_{\nu} -\mu j^\mu[T'] - \frac{1}{2} \omega_{\lambda\nu} \spt^{\mu\lambda\nu}[T'] \right)\nonumber\\
&~~~~~~~~~~~~~~~~~~~~~+ 
T^{\mu\nu} \beta_{\nu} - \zeta j^\mu - \frac{1}{2}\Omega_{\lambda\nu} \spt^{\mu\lambda\nu} \bigg]\nonumber\\
& =\int_\Sigma \di \Sigma \; n_\mu \bigg[ \int_{0}^{T} \frac{\di T'}{T^{\prime 2}} \; 
\left( T^{\mu\nu}_{\rm LE}[T'] u_{\nu} -\mu j^\mu_{\rm LE}[T'] - \frac{1}{2} \omega_{\lambda\nu} 
\spt^{\mu\lambda\nu}_{\rm LE}[T']\right)\nonumber\\
&~~~~~~~~~~~~~~~~~~~~~+ T^{\mu\nu}_{\rm LE} \beta_{\nu} 
- \zeta j^\mu_{\rm LE} - \frac{1}{2}\Omega_{\lambda\nu} \spt^{\mu\lambda\nu}_{\rm LE} \bigg] \nonumber\\
& = - \Tr(\wrhol(\Sigma) \log \wrhol(\Sigma))\,.
\end{align}
In other words, the definition of the out-of-equilibrium currents assumes that the constraints~\eqref{const1qm}-\eqref{const3qm} are satisfied on the hypersurface under consideration. Finally, it is important to point out that Eq.~\eqref{extensive} now holds
\begin{align}\label{extext}
\int \di \Sigma_\mu \phi^\mu =\log Z_{\rm LE}(\Sigma) + \bra{0} \widehat \Upsilon \ket{0},
\end{align}
where the currents represent the actual (not necessarily equilibrium) values. This expression plays an important role in the derivation of the entropy production rate in Sec.~\ref{Entropy production rateqmmsec}.
%
\subsection{Entropy-gauge and pseudo-gauge transformations}
\label{Entropygaugeandpseudogaugetransformations}
The thermodynamic potential current $\phi^{\mu}$ and the entropy current $S^{\mu}$ are not unique. It is well understood that a transformation of $\phi^{\mu}$ by the divergence of an antisymmetric rank-2 tensor $A^{\lambda\mu}$ leads to a transformed entropy current while leaving the total entropy $S$ invariant. Specifically, consider the transformation:
\begin{align}\label{entgauge}
\phi^{\mu}\longrightarrow \phi^{\mu\,\prime} = \phi^\mu + \partial_\lambda A^{\lambda\mu}\,,
\end{align}
where $A^{\lambda\mu}=-A^{\mu\lambda}$ is an arbitrary anti-symmetric tensor. This implies a corresponding transformation of the entropy current
\begin{align}\label{entgauge2}
S^{\mu\,\prime} = S^\mu + \partial_\lambda A^{\lambda\mu}.
\end{align}
Under such a transformation, the total entropy remains invariant by virtue of the \emph{relativistic Stokes theorem}, provided that $A^{\lambda\nu}$ satisfies appropriate boundary conditions: 
\begin{align}
S = \int_\Sigma \di \Sigma_\mu \; \left(S^\mu + \partial_\lambda A^{\lambda\mu}\right)=\int_\Sigma \di \Sigma_\mu \; S^\mu.
\end{align}
Thus, the entropy current \( S^\mu \) is not uniquely defined and may be altered by transformations of this form, which we henceforth refer to as \textit{entropy-gauge transformations}. A specific example of $A^{\lambda \nu}$ and the corresponding boundary conditions will be discussed in future work. 

However, the divergence of the entropy current is invariant under entropy-gauge transformations:
\begin{align}
\partial_\mu S^{\mu\,\prime} = \partial_\mu S^\mu + \partial_\mu \partial_\lambda A^{\lambda\mu}
= \partial_\mu S^\mu\,.
\end{align}

A natural question arises at this point: Is the change in the entropy current induced by a pseudo-gauge transformation of the energy-momentum and spin tensors (discussed in Chapter~\ref{From fields to fluids}) equivalent to an entropy-gauge transformation as in Eq.~\eqref{entgauge2}?

Recall that the pseudo-gauge transformation is given by: 
\begin{align}
&\wT^{\prime \mu\nu}=\wT^{\mu\nu}+\frac{1}{2}\partial_{\lambda}\left(\wPhi^{\lambda\mu\nu}-\wPhi^{\mu\lambda\nu}-\wPhi^{\nu\lambda\mu}\right),\label{pseudogaugeTqm}\\
&\wspt^{\prime \mu\lambda\nu}=\wspt^{\mu\lambda\nu}-\wPhi^{\mu\lambda\nu},
\label{pseudogaugeSqm}
\end{align}
where $\wPhi^{\mu\lambda\nu}$ is an arbitrary rank-3 tensor operator that is anti-symmetric in the last two indices. By inserting Eqs.~\eqref{pseudogaugeTqm}-\eqref{pseudogaugeSqm} into expression for $\phi^\mu$ Eq.~\eqref{phi2}, and assuming the simplifying condition $\Omega=\varpi$, we obtain:
\begin{align}
\phi^{\mu\,\prime }&=\phi^{\mu} +\int_{0}^{T}\frac{\di T^\prime}{T^\prime} \; 
\left[\partial_{\lambda}A^{\lambda\mu}-\Phi^{\lambda\mu\nu} \xi_{\lambda\nu}\right],\label{phipg}\\
S^{\mu\,\prime }&= S^{\mu} + \int_{0}^{T}\frac{\di T^\prime}{T^\prime} \; 
\left[\partial_{\lambda}A^{\lambda\mu}-\Phi^{\lambda\mu\nu} \xi_{\lambda\nu}\right]
+\partial_{\lambda}A^{\lambda\mu}-\Phi^{\lambda\mu\nu}\xi_{\lambda\nu}\label{spg}.
\end{align}
where $A^{\lambda\mu}=(1/2) \beta_{\nu}\left(\Phi^{\lambda\mu\nu}-\Phi^{\nu\lambda\mu}+\Phi^{\mu\nu\lambda}\right)$ 
is an anti-symmetric rank-2 tensor and $\xi_{\lambda\nu}$ is the \emph{thermal shear tensor}, defined by
\begin{align}\label{thermalshearqm}
\xi_{\lambda\nu} = \frac{1}{2} \left( \partial_\lambda \beta_\nu + \partial_\nu \beta_\lambda \right).
\end{align}
Detailed proof of Eqs.~\eqref{phipg}-\eqref{spg} will be presented in the following section. 

It is important to observe that the last term on the right-hand side of equations~\eqref{phipg}-\eqref{spg} cannot, in general, be written 
as the divergence of an anti-symmetric tensor. In essence, this shows that a general pseudo-gauge transformation of the energy-momentum and spin tensors 
\eqref{pseudogaugeTqm}-\eqref{pseudogaugeSqm} does not correspond to an entropy-gauge transformation of the form in Eq.~\eqref{entgauge2}.
%
%
\section{Entropy production rate}
In this section, we derive the entropy production rate within the framework of relativistic spin hydrodynamics. The resulting expression generalizes the formula originally derived by \textit{Zubarev} and \textit{van Weert} in Refs.~\cite{Zubarev, ChGvanWeert} for conventional relativistic hydrodynamics. Furthermore, we examine the behavior of entropy production under pseudo-gauge transformations. The results presented here are based on Ref.~\multimyref{AD5}.
\label{Entropy production rateqmmsec}
\subsection{Derivation of the entropy production rate}
The entropy production rate is obtained by computing the divergence of the out-of-equilibrium entropy current given in Eq.~\eqref{eq:entropy_current}:
\begin{align}\label{diventcur}
\partial_{\mu}S^{\mu}=\partial_{\mu}\phi^{\mu}+\partial_{\mu}T^{\mu\nu}\beta_{\nu}+T^{\mu\nu}\partial_{\mu}\beta_{\nu}-j^{\mu}\partial_{\mu}\zeta-\zeta\partial_{\mu}j^{\mu}-\frac{1}{2}\spt^{\mu\lambda\nu}\partial_{\mu}\Omega_{\lambda\nu}-\frac{1}{2}\Omega_{\lambda\nu}\partial_{\mu}\spt^{\mu\lambda\nu}.
\end{align}
Using the conservation laws for the energy-momentum tensor and particle number current (which hold for their actual values)
\begin{align}\label{conTandjqm}
\partial_{\mu}T^{\mu\nu}=0~~,~~\partial_{\mu}j^{\mu}=0,
\end{align}
equation \eqref{diventcur} simplifies to:
\begin{align}\label{diventcur2}
\partial_{\mu}S^{\mu} & =\partial_{\mu}\phi^{\mu}+T^{\mu\nu}\partial_{\mu}\beta_{\nu}-j^\mu \partial_\mu \zeta 
-\frac{1}{2} \spt^{\mu\lambda\nu}\partial_{\mu} \Omega_{\lambda\nu}- \frac{1}{2}\Omega_{\lambda\nu} 
\partial_{\mu}\spt^{\mu\lambda\nu}.
\end{align}
Furthermore, using the continuity equation for the spin tensor, valid for the actual values:
\begin{align}
\partial_{\lambda}S^{\lambda\mu\nu}=-2T^{\mu\nu}_{(A)},
\end{align}\label{spincontinuityeqfinalqmforp}
we can rewrite Eq.~\eqref{diventcur2} as
\begin{align}\label{entropyproductionQMnotfinal}
\partial_{\mu}S^{\mu}= \partial_{\mu}\phi^{\mu}+T_{(S)}^{\mu\nu}\xi_{\mu\nu} -j^\mu \partial_\mu \zeta + 
 T_{(A)}^{\mu\nu} \left(\Omega_{\mu\nu} - \varpi_{\mu\nu}\right)
-\frac{1}{2} \spt^{\mu\lambda\nu}\partial_{\mu} \Omega_{\lambda\nu},
\end{align}
where the energy-momentum tensor is decomposed into symmetric and anti-symmetric parts, $T^{\mu\nu}=T^{\mu\nu}_{(S)}+T^{\mu\nu}_{(A)}$. Moreover, here $\xi_{\mu\nu}$ is the thermal shear~\eqref{thermalshearqm} and $\varpi_{\mu\nu}$~\eqref{thermalvoticityqm} is the thermal vorticity.

\begin{figure}[H]
\begin{center}
 \includegraphics[width=9cm]{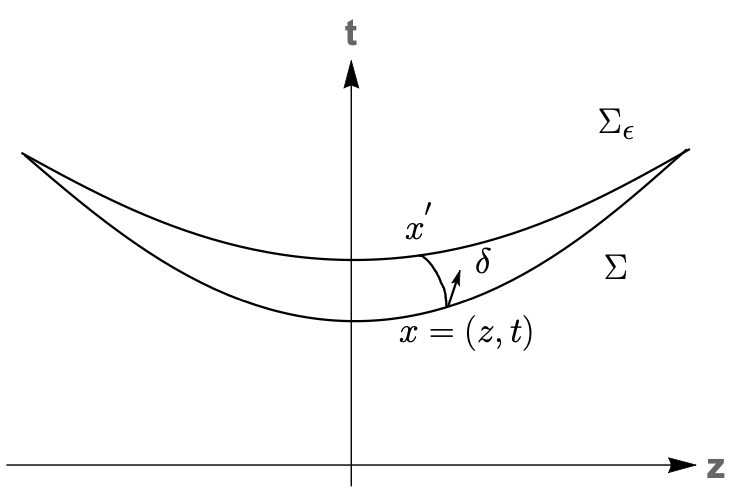}
 \caption{The hypersurface $\Sigma$ is mapped to $\Sigma_{\epsilon}$
 via an infinitesimal diffeomorphism. The deformation vector field is 
 $\di x'(x,\epsilon)/\di \epsilon|_{\epsilon=0}$.}
 \label{Diffeomorphic hypersurfaces}
\end{center}
\end{figure}

The next step is to compute the divergence $\partial_{\mu}\phi^{\mu}$ appearing in Eq~\eqref{entropyproductionQMnotfinal}. From Eq.~\eqref{extext}, $\phi^{\mu}$ is related to $\log Z_{\rm LE}$. To derive $\partial_{\mu}\phi^{\mu}$, we examine how $\log Z_{\rm LE}$ changes under an infinitesimal change of the hypersurface. An infinitesimal change of the hypersurface $\Sigma_{\epsilon}$ can be seen as the result of a local infinitesimal deformation such that each point $x \in \Sigma$ is mapped to a nearby point $x^{\prime}(x, \epsilon) \in \Sigma_{\epsilon}$ by an amount $\epsilon \delta^{\mu}$ (see Figure~\ref{Diffeomorphic hypersurfaces}). Formally, this defines a one-parameter family of diffeomorphisms:
\begin{align}
x^{\mu}\rightarrow x^{\mu\,\prime}(x,\epsilon)&=x^{\mu\,\prime}(x,0)+\epsilon\frac{\di x^{\mu\,\prime}(x,\epsilon)}{\di\epsilon}\bigg|_{\epsilon=0}\nonumber\\
&=x^{\mu}+\epsilon\delta^{\mu}(x),
\end{align}
where 
\begin{align}
\frac{\di x^{\mu\,\prime}(x,\epsilon)}{\di \epsilon}\bigg|_{\epsilon=0}&=\lim_{\epsilon \to 0}\frac{x^{\mu\,\prime}(x,\epsilon)-x^{\mu\,\prime}(x,0)}{\epsilon}\nonumber\\
&=\lim_{\epsilon \to 0}\frac{x^{\mu\,\prime}(x,\epsilon)-x^{\mu}}{\epsilon}\nonumber\\
&\equiv \delta^{\mu}(x).
\end{align}
The vector field $\delta^{\mu}(x)$ represents the direction in which the hypersurface is deformed, and $\epsilon$ is a finite (real) parameter, which measures how far along that direction we move.
\noindent
To compute how integrals over hypersurfaces change, we use the relation~\cite{Becattini:2023ouz}
\begin{align}\label{dsigma}
\lim_{\epsilon\to\ 0} \frac{1}{\epsilon} \left( \int_{\Sigma_\epsilon} \di \Sigma_{\mu}V^{\mu}-
\int_{\Sigma} \di \Sigma_{\mu} \, V^{\mu} \right) = \int_{\partial \Sigma} \di \tilde S_{\mu\nu}\, 
\delta^{\mu} V^{\nu} +\int_{\Sigma} \di \Sigma \cdot \delta \, \partial_{\mu} V^{\mu}\,,
\end{align}
where $\partial \Sigma$ is the two-dimensional boundary surface and $V^\mu$ is a vector field. 

Assuming that the boundary term vanishes, we apply the above formula to Eq.~\eqref{extext}, getting:

\resizebox{\textwidth}{!}{
\begin{minipage}{\textwidth}
\begin{align}\label{liederiv1}
& \lim_{\epsilon\to\ 0} \frac{1}{\epsilon} \left[ \log Z_{\rm LE}(\Sigma_\epsilon) - \log Z_{\rm LE}(\Sigma) \right] \nonumber\\
&= \int_{\Sigma} \di \Sigma \cdot \delta \, \partial_{\mu} \left( \phi^{\mu} - \bra{0} \wT^{\mu\nu} \beta_{\nu} - 
 \zeta \wj^\mu - \frac{1}{2}\Omega_{\lambda\nu} \wspt^{\mu\lambda\nu} \ket{0} \right) \nonumber \\
 & =  \int_{\Sigma} \di \Sigma \cdot \delta \left(\partial_{\mu} \phi^{\mu} - 
\bra{0} \wT_{(S)}^{\mu\nu} \xi_{\mu\nu} - 
   \wj^\mu \partial_\mu \zeta + \wT_{(A)}^{\mu\nu} (\Omega_{\mu\nu} - \varpi_{\mu\nu}) - 
  \frac{1}{2} \wspt^{\mu\lambda\nu} \partial_\mu \Omega_{\lambda\nu} \ket{0}\right).
\end{align}
\end{minipage}
}
\smallskip

\noindent In the last step, we have used Eqs.~\eqref{conservatoperatr} and~\eqref{continuityopqerrr}.  

On the other hand, the logarithm of the partition function can be calculated via its definition as a trace:

\resizebox{\textwidth}{!}{
\begin{minipage}{\textwidth}
\begin{align}
& Z_{\rm LE}(\Sigma_\epsilon) = \Tr \left( \exp\left[ - \int_{\Sigma_\epsilon}\di \Sigma_{\mu}
\left(\wT^{\mu\nu}\beta_{\nu}- \zeta \wj^\mu - \frac{1}{2}\Omega_{\lambda\nu}\wspt^{\mu\lambda\nu}\right)\right]\right)
\nonumber\\
&~~~~~~~~~~~\simeq \Tr\bigg( \exp\bigg[ - \int_{\Sigma} \di \Sigma_{\mu}\left(\wT^{\mu\nu}\beta_{\nu} 
- \zeta \wj^\mu - \frac{1}{2}\Omega_{\lambda\nu}\wspt^{\mu\lambda\nu}\right)\nonumber\\
&~~~~~~~~~~~~~~-\epsilon \int_{\Sigma} \di \Sigma \cdot \delta
\nabla_{\mu}\left(\wT^{\mu\nu}\beta_{\nu} -\zeta \wj^\mu - \frac{1}{2}\Omega_{\lambda\nu}\wspt^{\mu\lambda\nu}
\right)\bigg] \bigg)\nonumber \\
&~~~~~~~~~~~ = \Tr\left( \exp\left[ - \int_{\Sigma} \di \Sigma_{\mu} \left( \wT^{\mu\nu}\beta_{\nu} 
- \zeta \wj^\mu - \frac{1}{2}\Omega_{\lambda\nu}\wspt^{\mu\lambda\nu} \right) \right. \right.\nonumber \\
&~~~~~~~~~~~~~~ - \left. \left. \epsilon \int_{\Sigma} \di \Sigma \cdot \delta \; \left( \wT_S^{\mu\nu}\xi_{\mu\nu} -\wj^\mu \partial_\mu \zeta + 
 \wT_A^{\mu\nu} (\Omega_{\mu\nu} - \varpi_{\mu\nu}) - \frac{1}{2} \wspt^{\mu\lambda\nu}\partial_{\mu} \Omega_{\lambda\nu} 
 \right) \right] \right)\,.
\end{align}
\end{minipage}
}
\smallskip

\noindent Here we used Eq.~\eqref{dsigma} assuming that the boundary term vanishes and, again,~Eqs.~\eqref{conservatoperatr} and~\eqref{continuityopqerrr}. 

Expanding the trace to first order in $\epsilon$, and applying Eq.~\eqref{densityoperator}, we obtain:
\begin{align}
 Z_{\rm LE}(\Sigma_\epsilon) & \simeq  Z_{\rm LE}(\Sigma) - \epsilon Z_{\rm LE}(\Sigma)\times \int_{\Sigma} \di \Sigma \cdot \delta \; \bigg[ \Tr(\wrhol \wT_S^{\mu\nu}) \xi_{\mu\nu}\nonumber \\
 &~~~ -\Tr(\wrhol \wj^\mu) \partial_\mu \zeta + \Tr(\wrhol \wT_A^{\mu\nu}) (\Omega_{\mu\nu} - \varpi_{\mu\nu})\nonumber\\
&~~~-\frac{1}{2} \Tr(\wrhol \wspt^{\mu\lambda\nu}) \partial_\mu \Omega_{\lambda\nu} \bigg]\,.
\end{align}
Therefore:
\begin{align}\label{liederiv2}
& \lim_{\epsilon\to\ 0} \frac{1}{\epsilon} \left[ \log Z_{\rm LE}(\Sigma_\epsilon) - \log Z_{\rm LE}(\Sigma) \right]\nonumber \\
\nonumber
& = - \int_{\Sigma} \di \Sigma \cdot \delta \; \bigg[ \Tr(\wrhol \wT_S^{\mu\nu}) \xi_{\mu\nu} -
   \Tr(\wrhol \wj^\mu) \partial_\mu \zeta + \Tr(\wrhol \wT_A^{\mu\nu}) (\Omega_{\mu\nu} - \varpi_{\mu\nu})\nonumber\\
&~~~~~~~~~~~~~~~~~~~~~-\frac{1}{2} \Tr(\wrhol \wspt^{\mu\lambda\nu}) \partial_\mu \Omega_{\lambda\nu} \bigg]\,. 
\end{align}
Comparing Eqs.~\eqref{liederiv1} and~\eqref{liederiv2}, and noting that $\Sigma$ and $\delta$ are arbitrary, we obtain:

\resizebox{\textwidth}{!}{
\begin{minipage}{\textwidth}
\begin{align}\label{phidiverg}
\partial_\mu \phi^{\mu} &= - \bigg[ \left( 
  \Tr(\wrhol \wT_S^{\mu\nu}) -\bra{0} \wT_S^{\mu\nu} \ket{0} \right) \xi_{\mu\nu} - 
 \left( \Tr(\wrhol \wj^\mu) -  \bra{0} \wj^{\mu} \ket{0} \right) \partial_\mu \zeta \nonumber \\
& ~~~+\left( \Tr(\wrhol \wT_A^{\mu\nu}) - \bra{0} \wT_A^{\mu\nu} \ket{0} \right) (\Omega_{\mu\nu} - \varpi_{\mu\nu}) 
 - \frac{1}{2} \bigg( \Tr(\wrhol \wspt^{\mu\lambda\nu})\nonumber\\
 &~~~- \bra{0} \wspt^{\mu,\lambda\nu} \ket{0} \bigg)
 \partial_\mu \Omega_{\lambda\nu} \bigg] \nonumber \\
 &=  - \left( T^{\mu\nu}_{S\rm (LE)} \xi_{\mu\nu} - 
 j^{\mu}_{\rm LE} \partial_\mu \zeta + T^{\mu\nu}_{A \rm (LE)} (\Omega_{\mu\nu} - \varpi_{\mu\nu}) 
 - \frac{1}{2} \spt^{\mu\lambda\nu}_{\rm LE} \partial_\mu \Omega_{\lambda\nu} \right)\,,
\end{align}
\end{minipage}
}
\smallskip 

\noindent where, in the last step, we have used the definition of local equilibrium values.
\medskip

Substituting Eq.~\eqref{phidiverg} into Eq.~\eqref{entropyproductionQMnotfinal}, we obtain the final expression for the \emph{entropy production rate}:
\begin{align}\label{entropyproductionrate}
\partial_{\mu}S^{\mu}=&\left( T_{(S)}^{\mu\nu}-T^{\mu\nu}_{(S)\,\rm LE}\right) \xi_{\mu\nu}
- \left( j^\mu-j^\mu_{\rm LE}\right) \partial_\mu \zeta  + 
\left( T^{\mu\nu}_{(A)}-T^{\mu\nu}_{(A)\,\rm LE}\right) (\Omega_{\mu\nu}-\varpi_{\mu\nu})\nonumber\\
&- \frac{1}{2}\left( \spt^{\mu\lambda\nu}-\spt^{\mu\lambda\nu}_{\rm LE} \right) \partial_{\mu}\Omega_{\lambda\nu} 
\geq 0.
\end{align}
The above expression is a generalization of the entropy production rate found by \textit{Zubarev} and \textit{van Weert}~\cite{Zubarev,ChGvanWeert} in the quantum-statistical formulation of a relativistic fluid without spin tensor. It expresses entropy production in terms of all possible gradients of the system’s fields, which we will refer to as the \textit{thermo-hydrodynamic fields}. The quantities in parentheses denote the dissipative currents. In particular, the symmetric and antisymmetric dissipative components of the energy-momentum tensor are given by
\begin{align}\label{symmiantid}
&\delta T^{\mu\nu}_{(S)}=\left( T_{(S)}^{\mu\nu}-T^{\mu\nu}_{(S)\,\rm LE}\right),\\
&\delta T^{\mu\nu}_{(A)}=\left( T^{\mu\nu}_{(A)}-T^{\mu\nu}_{(A)\,\rm LE}\right),
\end{align}  
whereas the dissipative current associated with the particle number is  
\begin{align}\label{pd}
\delta j^{\mu}=\left( j^\mu-j^\mu_{\rm LE}\right).
\end{align}  
Finally, the dissipative current for the spin tensor takes the form  
\begin{align}\label{sd}
\delta S^{\mu\lambda\nu}=\left( \spt^{\mu\lambda\nu}-\spt^{\mu\lambda\nu}_{\rm LE} \right).  
\end{align}  

The expression for entropy production rate in Eq.~\eqref{entropyproductionrate} allows us to establish the hierarchy of hydrodynamic gradient ordering for the relevant terms:
\begin{align}
\xi_{\mu\nu}\sim\mathcal{O}(\partial)~,~~\partial_{\mu}\zeta\sim\mathcal{O}(\partial)~,~~(\Omega_{\mu\nu}-\varpi_{\mu\nu})\sim\mathcal{O}(\partial)~,~~\partial_{\mu}\Omega_{\lambda\nu}\sim\mathcal{O}(\partial).
\end{align}
This implies that the spin potential itself is a zeroth-order hydrodynamic field
\begin{align}
\Omega_{\mu\nu}\sim\mathcal{O}(1).
\end{align}
The treatment of the hydrodynamic gradient counting for the spin potential distinguishes the approach in this chapter from the formalism in Chapters~\ref{Navier-Stokes limit} and \ref{MISMIS}, where the spin potential was considered a first-order quantity, \(\mathcal{O}(\partial)\).
%
%
\subsection{Pseudo-gauge transformation}
\label{Entropy production rate and pseudo-gauge transformation}
In Sec.~\ref {Entropygaugeandpseudogaugetransformations}, we demonstrated that the entropy production rate is invariant under entropy-gauge transformations. We also clarified that entropy-gauge transformations are conceptually distinct from pseudo-gauge transformations. In this section, we examine the effect of pseudo-gauge transformations on the entropy production rate. 

To explore this, we start with the entropy current given in Eq.~\eqref{eq:entropy_current}, and apply the pseudo-gauge transformation to the following combination of terms,  
\begin{align}\label{partofentropycurrent}
T^{\mu\nu} \beta_\nu - \frac{1}{2} \varpi_{\lambda\nu} \spt^{\mu\lambda\nu},
\end{align}
where we restrict to the special case $\Omega_{\lambda\nu}=\varpi_{\lambda\nu}$. This simplification is motivated by calculational ease. If the entropy production rate were invariant under pseudo-gauge in this special case, one would then proceed to investigate the general expression~\eqref{eq:entropy_current}. However, if invariance already fails in this simplified setting, the result is sufficient to draw meaningful conclusions.

By substituting Eqs.~\eqref{pseudogaugeTqm}-\eqref{pseudogaugeSqm} into Eq.~\eqref{partofentropycurrent}, we obtain 
\begin{align}\label{D1}
T^{\mu\nu\,\prime} \beta_\nu - \frac{1}{2} \varpi_{\lambda\nu} \spt^{\mu\lambda\nu\,\prime}
= &T^{\mu\nu} \beta_\nu - \frac{1}{2} \varpi_{\lambda\nu} \spt^{\mu\lambda\nu} 
+ \frac{1}{2}\partial_{\lambda}(\Phi^{\lambda\mu\nu}-\Phi^{\mu\lambda\nu}-\Phi^{\nu\lambda\mu})
\beta_{\nu}\nonumber\\
&+\frac{1}{2}\varpi_{\lambda\nu}\Phi^{\mu\lambda\nu}.
\end{align}
The last two terms can be transformed as follows,
\begin{align}\label{D2}
& \frac{1}{2}\partial_{\lambda}(\Phi^{\lambda\mu\nu}-\Phi^{\mu\lambda\nu}-\Phi^{\nu\lambda\mu})\beta_{\nu}
 +\frac{1}{2}\varpi_{\lambda\nu}\Phi^{\mu\lambda\nu}\nonumber \\
 = & \frac{1}{2}\partial_{\lambda}(\Phi^{\lambda\mu\nu}-\Phi^{\mu\lambda\nu}-\Phi^{\nu\lambda\mu})\beta_{\nu}
 +\frac{1}{2}\partial_{\nu}\beta_{\lambda}\Phi^{\mu\lambda\nu},\nonumber\\
= &\frac{1}{2}\partial_{\lambda}(\Phi^{\lambda\mu\nu}-\Phi^{\mu\lambda\nu}-\Phi^{\nu\lambda\mu})\beta_{\nu}
+\frac{1}{2}\partial_{\nu}(\beta_{\lambda}\Phi^{\mu\lambda\nu})-\frac{1}{2}\beta_{\lambda}\partial_{\nu}\Phi^{\mu\lambda\nu},
\nonumber \\
= & \frac{1}{2}(\partial_{\lambda}\Phi^{\lambda\mu\nu}-\partial_{\lambda}\Phi^{\nu\lambda\mu})\beta_{\nu}
+\frac{1}{2}\partial_{\lambda}(\beta_{\nu}\Phi^{\mu\nu\lambda}).
\end{align}
In the above, we have used the definition of thermal vorticity, the antisymmetry of
$\Phi^{\mu\nu\lambda}$ in the last two indices, and in the last step, swapped the saturated indices 
$\lambda$ and $\nu$ appropriately. Using the same methods,  
Eq.~\eqref{D2} can be further transformed into,
\begin{align}\label{D4}
& \frac{1}{2}(\partial_{\lambda}\Phi^{\lambda\mu\nu}-\partial_{\lambda}\Phi^{\nu\lambda\mu})\beta_{\nu}
+\frac{1}{2}\partial_{\lambda}(\beta_{\nu}\Phi^{\mu\nu\lambda}) \nonumber \\
= & \frac{1}{2}\partial_{\lambda}\left(\beta_{\nu}\Phi^{\lambda\mu\nu}-\Phi^{\nu\lambda\mu}\beta_{\nu}+\beta_{\nu}\Phi^{\mu\nu\lambda}\right)-\frac{1}{2}\Phi^{\lambda\mu\nu}\partial_{\lambda}\beta_{\nu}+\frac{1}{2}\Phi^{\nu\lambda\mu}
\partial_{\lambda}\beta_{\nu}\nonumber\\
= & \frac{1}{2}\partial_{\lambda}\left(\beta_{\nu}\Phi^{\lambda\mu\nu}-\Phi^{\nu\lambda\mu}\beta_{\nu}+\beta_{\nu}\Phi^{\mu\nu\lambda}\right)-\frac{1}{2}\Phi^{\lambda\mu\nu}
\partial_{\lambda}\beta_{\nu}-\frac{1}{2}\Phi^{\lambda\mu\nu}\partial_{\nu}\beta_{\lambda} \nonumber \\
= & \frac{1}{2}\partial_{\lambda}\left(\beta_{\nu}\Phi^{\lambda\mu\nu}-\Phi^{\nu\lambda\mu}\beta_{\nu}+\beta_{\nu}\Phi^{\mu\nu\lambda}\right)-\Phi^{\lambda\mu\nu}\xi_{\lambda\nu}.
\end{align}
Let us now define the anti-symmetric tensor $A^{\lambda\nu}$ as,
\begin{align}
A^{\lambda\mu}=\frac{1}{2}\left(\beta_{\nu}\Phi^{\lambda\mu\nu}-\Phi^{\nu\lambda\mu}\beta_{\nu}+\beta_{\nu}\Phi^{\mu\nu\lambda}\right).
\end{align}
Then, combining Eqs.~\eqref{D1}-\eqref{D4}, we arrive at the result
\begin{align}\label{D5}
T^{\mu\nu\,\prime} \beta_\nu - \frac{1}{2} \varpi_{\lambda\nu} \spt^{\mu\lambda\nu\,\prime}= T^{\mu\nu} \beta_\nu - \frac{1}{2} \varpi_{\lambda\nu} \spt^{\mu\lambda\nu}  + \nabla_\lambda A^{\lambda\mu}
 -\Phi^{\lambda\mu\nu}\xi_{\lambda\nu}.
\end{align}
Substituting this into the expressions for the thermodynamic potential~\eqref{phi2} and the entropy current~\eqref{eq:entropy_current}, we find the transformed versions:
\begin{align}
\phi^{\mu\,\prime}&=\phi^{\mu} +\int_{0}^{T}\frac{\di T^\prime}{T^\prime} \; 
\left[\partial_{\lambda}A^{\lambda\mu}-\Phi^{\lambda\mu\nu} \xi_{\lambda\nu}\right],\label{phipgg}\\
S^{\mu\,\prime}&= S^{\mu} + \int_{0}^{T}\frac{\di T^\prime}{T^\prime} \; 
\left[\partial_{\lambda}A^{\lambda\mu}-\Phi^{\lambda\mu\nu} \xi_{\lambda\nu}\right]
+\partial_{\lambda}A^{\lambda\mu}-\Phi^{\lambda\mu\nu}\xi_{\lambda\nu}\label{spgg}.
\end{align}
Therefore, it becomes evident that the divergence 
of the entropy current is not invariant under a pseudo-gauge transformation. Moreover, a simultaneous
entropy-gauge transformation cannot generally restore this invariance. This conclusion differs from that reported in Ref.~\cite{Li:2020eon}.
\medskip

It is important to note that this result is not unexpected. A pointed out in Ref.~\cite{Becattini:2018duy}, the density operator describing initial local equilibrium is not invariant under pseudo-gauge transformations for a system that is not in global equilibrium -- such as the quark-gluon plasma (QGP) throughout its evolution. Consequently, physical measurements may, in principle, depend on the pseudo-gauge choice if the initial quantum state lacks pseudo-gauge invariance. 
%
\section{First-order dissipative currents and transport coefficients}
\label{transport}
The objective of this section is to derive the first-order dissipative currents of the system — namely, \(\delta T^{\mu\nu}_{(S)}\), \(\delta T^{\mu\nu}_{(A)}\), \(\delta j^{\mu}\), and \(\delta \spt^{\mu\lambda\nu}\) — in a form that is irreducible under rotations. 
These irreducible decompositions allow for a systematic identification of the corresponding transport coefficients. The method used to obtain these results is using the irreducible representations of the rotation group $\mathrm{SO}(3,\mathbb{R})$ and is detailed in Appendix~\ref{AppendixC}. The results of this section are based exclusively on Ref.~\multimyref{AD8}.
%
%
\subsection{Dissipative currents and their irreducible decompositions under rotation}

In general, without imposing physical constraints, the dissipative currents $\delta T^{\mu\nu}_{(S)}$, $\delta T^{\mu\nu}_{(A)}$, $\delta j^{\mu}$, and $\delta \spt^{\mu\lambda\nu}$ can be expressed as linear combinations of all possible thermo-hydrodynamic gradients, contracted with \textit{tensor coefficients}. Explicitly, we can write:
\begin{align}
&\delta T^{\mu\nu}_{(S)} = H^{\mu\nu\rho\sigma} \; \xi_{\rho\sigma} + \frac{1}{T} \; K^{\mu\nu\rho} \; \partial_{\rho} \zeta + L^{\mu\nu\rho\sigma} \; \left(\Omega_{\rho\sigma} - \varpi_{\rho\sigma}\right) + \frac{1}{T} \; M^{\mu\nu\rho\sigma\tau} \; \partial_{\rho} \Omega_{\sigma\tau}\label{EMTsymmetricgradientdecomposition}\,, \\[1.5em]
&\delta T^{\mu\nu}_{(A)} = N^{\mu\nu\rho\sigma} \; \xi_{\rho\sigma} + \frac{1}{T} \; P^{\mu\nu\rho} \; \partial_{\rho} \zeta + Q^{\mu\nu\rho\sigma} \; \left(\Omega_{\rho\sigma} - \varpi_{\rho\sigma}\right) + \frac{1}{T} \; R^{\mu\nu\rho\sigma\tau} \; \partial_{\rho} \Omega_{\sigma\tau}\label{EMTantisymmetricgradientdecomposition}\,, \\[1.5em]
&T \, \delta j^{\mu} = G^{\mu\rho\sigma} \; \xi_{\rho\sigma} + \frac{1}{T} \; I^{\mu\rho} \; \partial_{\rho} \zeta + O^{\mu\rho\sigma} \; (\Omega_{\rho\sigma} - \varpi_{\rho\sigma}) + \frac{1}{T} \; F^{\mu\rho\sigma\tau} \; \partial_{\rho} \Omega_{\sigma\tau}\label{Particlenumbergradientdecomposition}\,, \\[1.5em]
&T \, \delta \spt^{\mu\lambda\nu} = T^{\mu\lambda\nu\rho\sigma} \; \xi_{\rho\sigma} + \frac{1}{T} \; U^{\mu\lambda\nu\rho} \; \partial_{\rho} \zeta + V^{\mu\lambda\nu\rho\sigma} \; \left(\Omega_{\rho\sigma} - \varpi_{\rho\sigma}\right) + \frac{1}{T} \; W^{\mu\lambda\nu\rho\sigma\tau} \; \partial_{\rho} \Omega_{\sigma\tau}\label{Spingradientdecomposition}\,.
\end{align}
Each of the tensors $H, K, L, M, \dots, W$, carries a specific rank and index symmetries appropriate to the contraction. Their structures are detailed in Tables~\ref{SummaryofIndexSymmetryProperties underExchangeTS}-\ref{SummaryofIndexSymmetryProperties underExchangeS} in Appendix~\ref{AppendixC}. The extra factors of $T$ or $1/T$ in Eqs.~\eqref{Particlenumbergradientdecomposition} and~\eqref{Spingradientdecomposition} are introduced to ensure that all tensor coefficients have the same energy dimension in natural units, i.e., $[H]=[N]=\dots=[W]=[E]^4$.

The task is now reduced to determining the explicit forms of these tensor coefficients, which in turn gives the explicit forms of the dissipative currents. To achieve this, we proceed as follows:
\begin{enumerate}
\item We assume isotropic in the local rest frame of the fluid element. We also consider equilibrium with a constant four-temperature field $\beta = u/T$, so that the thermal vorticity vanishes, $\varpi = 0$. This avoids any anisotropies that could otherwise be introduced by $\varpi$. As a result, the basis for decomposition of any tensor are:
\begin{align}\label{basisrep}
u^{\mu}~~,~~\Delta^{\mu\nu}=g^{\mu\nu}-u^{\mu}u^{\nu}~~,~~\epsilon^{\mu\nu\lambda\gamma}u_\gamma.
\end{align}
These are rotationally invariant in the rest frame associated to $u_{\mu}$.\\
\item We calculate all tensor coefficients in this equilibrium setup. The minimal physical requirement is that they be invariant under rotation in the fluid's local comoving frame.\\
\item To satisfy rotational invariance, we irreducibly decompose each tensor coefficient under the rotation group $\mathrm{SO}(3,\mathbb{R})$ and retain only the scalar (invariant) parts. The decomposition is constructed using the basis from step 1. The method used to irreducibly decompose the tensors is explained in full detail in Appendix~\ref{AppendixC}~\cite{Daher:2025pfq}.   
\end{enumerate}

Below, we list the irreducible rotation-invariant components for each tensor coefficient, as summarized in Tables \ref{Table-G-I-O-F}-\ref{Table-W}. These components are constructed such that they are invariant under rotation in the comoving frame. Each term is labeled with lowercase Latin letters with numbers as subscripts (e.g., $h_1$,  $k_2$,  $l_3$).
\begin{table}[H]
\centering
\resizebox{\textwidth}{!}{
\renewcommand{\arraystretch}{1.7}
\begin{tabular}{@{\extracolsep{0.5em}}|c|c||c|c||c|c||c|c||c|c|}
\hline
\multicolumn{2}{|c||}{$H^{\mu\nu\rho\sigma}$} & \multicolumn{2}{c||}{$K^{\mu\nu\rho}$} & \multicolumn{2}{c||}{$L^{\mu\nu\rho\sigma}$} & \multicolumn{4}{c|}{$M^{\mu\nu\rho\sigma\tau}$}\tabularnewline
\hline
$h_{1}$ & $u^{(\mu}\Delta^{\nu)(\rho}u^{\sigma)}$ & $k_{1}$ & $u^{(\mu}\Delta^{\nu)\rho}$ & $l_{1}$ & $u^{(\mu}\Delta^{\nu)[\rho}u^{\sigma]}$ & $m_{1}$ & $u^{(\mu}\Delta^{\nu)[\sigma}u^{\tau]}u^{\rho}$ & $m_{7}$ & $u^{\mu}u^{\nu}u^{[\sigma}\Delta^{\tau]\rho}$\tabularnewline
\hline
$h_{2}$ & $\Delta^{\mu\nu}\Delta^{\rho\sigma}$ & $k_{2}$ & $\Delta^{\mu\nu}u^{\rho}$ & $l_{2}$ & $u^{(\mu}\epsilon^{\nu)\rho\sigma\gamma}u_{\gamma}$ & $m_{2}$ & $u^{(\mu}\epsilon^{\nu)\sigma\tau\gamma}u_{\gamma}u^{\rho}$ & $m_{8}$ & $u^{\mu}u^{\nu}\epsilon^{\sigma\tau\gamma\rho}u_{\gamma}$\tabularnewline
\hline
$h_{3}$ & $\Delta^{\mu\nu,\rho\sigma}$ & $k_{3}$ & $u^{\mu}u^{\nu}u^{\rho}$ &  &  & $m_{3}$ & $\Delta^{\mu\nu}u^{[\sigma}\Delta^{\tau]\rho}$ & $m_{9}$ & $(\epsilon^{\sigma\tau\gamma(\mu}\Delta^{\nu)\rho}+\frac{1}{3}\Delta^{\mu\nu}\epsilon^{\rho\sigma\tau\gamma})u_{\gamma}$\tabularnewline
\hline
$h_{4}$ & $\Delta^{\mu\nu}u^{\rho}u^{\sigma}$ &  &  &  &  & $m_{4}$ & $\Delta^{\mu\nu}\epsilon^{\sigma\tau\gamma\rho}u_{\gamma}$ & $m_{10}$ & $\Delta^{\mu\nu,\rho[\tau}u^{\sigma]}$\tabularnewline
\hline
$h_{5}$ & $u^{\mu}u^{\nu}\Delta^{\rho\sigma}$ &  &  &  &  & $m_{5}$ & $2u^{(\mu}\Delta^{\nu)[\sigma}\Delta^{\tau]\rho}$ &  & \tabularnewline
\hline
$h_{6}$ & $u^{\mu}u^{\nu}u^{\rho}u^{\sigma}$ &  &  &  &  & $m_{6}$ & $u^{(\mu}\epsilon^{\nu)\rho\gamma[\sigma}u^{\tau]}u_{\gamma}$ &  & \tabularnewline
\hline
\end{tabular}
}
\caption{Irreducible rotation-invariant decompositions of $H^{\mu\nu\rho\sigma}$, $K^{\mu\nu\rho}$, $L^{\mu\nu\rho\sigma}$, and $M^{\mu\nu\rho\sigma\tau}$.}
\label{Table-G-I-O-F}
\end{table}
\begin{table}[H]
\centering
\resizebox{\textwidth}{!}{
\renewcommand{\arraystretch}{1.7}
\begin{tabular}{@{\extracolsep{0.5em}}|c|c||c|c||c|c||c|c||c|c|}
\hline
\multicolumn{2}{|c||}{$N^{\mu\nu\rho\sigma}$} & \multicolumn{2}{c||}{$P^{\mu\nu\rho}$} & \multicolumn{2}{c||}{$Q^{\mu\nu\rho\sigma}$} & \multicolumn{4}{c|}{$R^{\mu\nu\rho\sigma\tau}$}\tabularnewline
\hline
$n_{1}$ & $u^{[\mu}\Delta^{\nu](\rho}u^{\sigma)}$ & $p_{1}$ & $\Delta^{\rho[\mu}u^{\nu]}$ & $q_{1}$ & $u^{[\mu}\Delta^{\nu][\rho}u^{\sigma]}$ & $r_{1}$ & $u^{[\mu}\Delta^{\nu][\sigma}u^{\tau]}u^{\rho}$ & $r_{5}$ & $\epsilon^{\mu\nu\gamma[\sigma}u^{\tau]}u_{\gamma}u^{\rho}$\tabularnewline
\hline
$n_{2}$ & $\epsilon^{\mu\nu\gamma(\rho}u^{\sigma)}u_{\gamma}$ & $p_{2}$ & $\epsilon^{\mu\nu\rho\gamma}u_{\gamma}$ & $q_{2}$ & $2\Delta^{\mu[\rho}\Delta^{\sigma]\nu}$ & $r_{2}$ & $u^{[\mu}\epsilon^{\nu]\sigma\tau\gamma}u_{\gamma}u^{\rho}$ & $r_{6}$ & 2$\Delta^{\mu[\sigma}\Delta^{\tau]\nu}u^{\rho}$\tabularnewline
\hline
 &  &  &  & $q_{3}$ & $u^{[\rho}\epsilon^{\sigma]\mu\nu\gamma}u_{\gamma}$ & $r_{3}$ & 2$u^{[\mu}\Delta^{\nu][\sigma}\Delta^{\tau]\rho}$ & $r_{7}$ & $2\Delta^{\rho[\nu}\epsilon^{\mu]\sigma\tau\gamma}u_{\gamma}$\tabularnewline
\hline
 &  &  &  & $q_{4}$ & $u^{[\mu}\epsilon^{\nu]\rho\sigma\gamma}u_{\gamma}$ & $r_{4}$ & $u^{[\mu}\epsilon^{\nu]\gamma\rho[\sigma}u^{\tau]}u_{\gamma}$ & $r_{8}$ & $2\Delta^{\rho[\mu}\Delta^{\nu][\sigma}u^{\tau]}$\tabularnewline
\hline
\end{tabular}
}
\caption{Irreducible rotation-invariant decompositions of $N^{\mu\nu\rho\sigma}$, $P^{\mu\nu\rho}$, $Q^{\mu\nu\rho\sigma}$, and $R^{\mu\nu\rho\sigma\tau}$.}\label{Table-N-P-Q-R}
\end{table}
\begin{table}[H]
\centering
\resizebox{\textwidth}{!}{
\renewcommand{\arraystretch}{1.7}
\begin{tabular}{@{\extracolsep{0.5em}}|c|c||c|c||c|c||c|c||c|c|}
\hline
\multicolumn{2}{|c||}{$G^{\mu\rho\sigma}$} & \multicolumn{2}{c||}{$I^{\mu\rho}$} & \multicolumn{2}{c||}{$O^{\mu\rho\sigma}$} & \multicolumn{4}{c|}{$F^{\mu\rho\sigma\tau}$}\tabularnewline
\hline
$g_{1}$ & $\Delta^{\mu(\rho}u^{\sigma)}$ & $i_{1}$ & $\Delta^{\mu\rho}$ & $o_{1}$ & $\Delta^{\mu[\rho}u^{\sigma]}$ & $f_{1}$ & $\Delta^{\mu[\sigma}u^{\tau]}u^{\rho}$ & $f_{4}$ & $\epsilon^{\mu\rho\gamma[\sigma}u^{\tau]}u_{\gamma}$\tabularnewline
\hline
$g_{2}$ & $u^{\mu}\Delta^{\rho\sigma}$ & $i_{2}$ & $u^{\mu}u^{\rho}$ & $o_{2}$ & $\epsilon^{\mu\rho\sigma\gamma}u_{\gamma}$ & $f_{2}$ & $\epsilon^{\mu\sigma\tau\gamma}u^{\rho}u_{\gamma}$ & $f_{5}$ & $u^{\mu}\Delta^{\rho[\sigma}u^{\tau]}$\tabularnewline
\hline
$g_{3}$ & $u^{\mu}u^{\rho}u^{\sigma}$ &  &  &  &  & $f_{3}$ & $2\Delta^{\mu[\sigma}\Delta^{\tau]\rho}$ & $f_{6}$ & $u^{\mu}\epsilon^{\rho\sigma\tau\gamma}u_{\gamma}$\tabularnewline
\hline
\end{tabular}
}
\caption{Irreducible rotation-invariant  decompositions of $G^{\mu\rho\sigma}$, $I^{\mu\rho}$, $O^{\mu\rho\sigma}$, and $F^{\mu\rho\sigma\tau}$.}
\end{table}
\begin{table}[H]
\centering
\resizebox{\textwidth}{!}{
\renewcommand{\arraystretch}{1.7}
\begin{tabular}{@{\extracolsep{0.5em}}|c|c||c|c||c|c||c|c|}
\hline
\multicolumn{4}{|c||}{$T^{\mu\lambda\nu\rho\sigma}$} & \multicolumn{2}{c||}{$U^{\mu\lambda\nu\rho}$} & \multicolumn{2}{c|}{$V^{\mu\lambda\nu\rho\sigma}$}\tabularnewline
\hline
$t_{1}$ & $\Delta^{\mu[\nu}u^{\lambda]}u^{\rho}u^{\sigma}$ & $t_{6}$ & $u^{[\lambda}\epsilon^{\nu]\mu\gamma(\rho}u^{\sigma)}u_{\gamma}$ & $u_{1}$ & $\Delta^{\mu[\nu}u^{\lambda]}u^{\rho}$ & $v_{1}$ & $2\Delta^{\mu[\rho}\Delta^{\sigma][\nu}u^{\lambda]}$\tabularnewline
\hline
$t_{2}$ & $\Delta^{\mu[\nu}u^{\lambda]}\Delta^{\rho\sigma}$ & $t_{7}$ & $u^{[\lambda}\Delta^{\nu]\mu,\rho\sigma}$ & $u_{2}$ & $2\Delta^{\mu[\nu}\Delta^{\lambda]\rho}$ & $v_{2}$ & $2\Delta^{\mu[\rho}\epsilon^{\sigma]\nu\lambda\gamma}u_{\gamma}$\tabularnewline
\hline
$t_{3}$ & $2\Delta^{\mu[\nu}\Delta^{\lambda](\rho}u^{\sigma)}$ & $t_{8}$ & $(\Delta^{\mu(\rho}\epsilon^{\sigma)\nu\lambda\gamma}-\Delta^{\rho\sigma}\epsilon^{\mu\nu\lambda\gamma}/3)u_{\gamma}$ & $u_{3}$ & $\epsilon^{\mu\nu\lambda\gamma}u_{\gamma}u^{\rho}$ & $v_{3}$ & $2\Delta^{\mu[\lambda}\Delta^{\nu][\rho}u^{\sigma]}$\tabularnewline
\hline
$t_{4}$ & $\epsilon^{\mu\nu\lambda\gamma}u_{\gamma}u^{\rho}u^{\sigma}$ & $t_{9}$ & $u^{\mu}u^{[\lambda}\Delta^{\nu](\rho}u^{\sigma)}$ & $u_{4}$ & $\epsilon^{\mu\rho\gamma[\nu}u^{\lambda]}u_{\gamma}$ & $v_{4}$ & $u^{[\rho}\epsilon^{\sigma]\mu\gamma[\lambda}u^{\nu]}u_{\gamma}$\tabularnewline
\hline
$t_{5}$ & $\epsilon^{\mu\nu\lambda\gamma}u_{\gamma}\Delta^{\rho\sigma}$ & $t_{10}$ & $u^{\mu}\epsilon^{\lambda\nu\gamma(\rho}u^{\sigma)}u_{\gamma}$ & $u_{5}$ & $u^{\mu}\Delta^{\rho[\nu}u^{\lambda]}$ & $v_{5}$ & $u^{\mu}u^{[\lambda}\Delta^{\nu][\rho}u^{\sigma]}$\tabularnewline
\hline
 &  &  &  & $u_{6}$ & $u^{\mu}\epsilon^{\nu\lambda\rho\gamma}u_{\gamma}$ & $v_{6}$ & $u^{\mu}u^{[\nu}\epsilon^{\lambda]\rho\sigma\gamma}u_{\gamma}$\tabularnewline
\hline
 &  &  &  &  &  & $v_{7}$ & $u^{\mu}\epsilon^{\nu\lambda\gamma[\rho}u^{\sigma]}u_{\gamma}$\tabularnewline
\hline
 &  &  &  &  &  & $v_{8}$ & $u^{\mu}\Delta^{\lambda[\rho}\Delta^{\sigma]\nu}$\tabularnewline
\hline
\end{tabular}
}
\caption{Irreducible rotation-invariant decompositions of $T^{\mu\lambda\nu\rho\sigma}$, $U^{\mu\lambda\nu\rho}$, and $V^{\mu\lambda\nu\rho\sigma}$.}\label{Table-T-U-V}
\end{table}
\begin{table}[H]
\centering
\resizebox{\textwidth}{!}{
\renewcommand{\arraystretch}{1.7}
\begin{tabular}{@{\extracolsep{0.5em}}|c|c||c|c||c|c|}
\hline
\multicolumn{6}{|c|}{$W^{\mu\lambda\nu\rho\sigma\tau}$}\tabularnewline
\hline
$w_{1}$ & $\Delta^{\mu[\nu}u^{\lambda]}u^{[\sigma}\Delta^{\tau]\rho}$ & $w_{7}$ & $2\Delta^{\mu[\nu}\Delta^{\lambda][\tau}u^{\sigma]}u^{\rho}$ & $w_{13}$ & $u^{\mu}u^{\rho}\Delta^{\lambda[\sigma}\Delta^{\tau]\nu}$\tabularnewline
\hline
$w_{2}$ & $\epsilon^{\mu\lambda\nu\gamma}u_{\gamma}\epsilon^{\rho\sigma\tau\kappa}u_{\kappa}$ & $w_{8}$ & $2\Delta^{\mu[\tau}\epsilon^{\sigma]\lambda\nu\gamma}u_{\gamma}u^{\rho}$ & $w_{14}$ & $u^{\mu}\epsilon^{\lambda\nu\gamma[\sigma}u^{\tau]}u^{\rho}u_{\gamma}$\tabularnewline
\hline
$w_{3}$ & $\Delta^{\mu[\nu}u^{\lambda]}\epsilon^{\sigma\tau\gamma\rho}u_{\gamma}+\epsilon^{\mu\lambda\nu\gamma}u_{\gamma}\Delta^{\rho[\sigma}u^{\tau]}$ & $w_{9}$ & $4\Delta^{\mu[\lambda}\Delta^{\nu][\sigma}\Delta^{\tau]\rho}$ & $w_{15}$ & $u^{\mu}u^{[\nu}\epsilon^{\lambda]\rho\gamma[\sigma}u^{\tau]}u_{\gamma}$\tabularnewline
\hline
$w_{4}$ & $\Delta^{\mu[\nu}u^{\lambda]}\epsilon^{\sigma\tau\gamma\rho}u_{\gamma}-\epsilon^{\mu\lambda\nu\gamma}u_{\gamma}\Delta^{\rho[\sigma}u^{\tau]}$ & $w_{10}$ & $u^{[\lambda}\Delta^{\nu]\mu,\rho\tau}u^{\sigma]}$ & $w_{16}$ & $u^{\mu}\Delta^{\rho[\nu}\Delta^{\lambda][\sigma}u^{\tau]}$\tabularnewline
\hline
$w_{5}$ & $u^{[\nu}\epsilon^{\lambda]\mu\gamma[\sigma}u^{\tau]}u_{\gamma}u^{\rho}$ & $w_{11}$ & $u^{\mu}u^{[\lambda}\Delta^{\nu][\sigma}u^{\tau]}u^{\rho}$ & $w_{17}$ & $u^{\mu}u^{[\nu}\Delta^{\lambda][\tau}\Delta^{\sigma]\rho}$\tabularnewline
\hline
$w_{6}$ & $2\Delta^{\mu[\sigma}\Delta^{\tau][\nu}u^{\lambda]}u^{\rho}$ & $w_{12}$ & $u^{\mu}u^{[\lambda}\epsilon^{\nu]\sigma\tau\gamma}u_{\gamma}u^{\rho}$ & $w_{18}$ & $u^{\mu}\Delta^{\rho[\sigma}\epsilon^{\tau]\lambda\nu\gamma}u_{\gamma}$\tabularnewline
\hline
$w_{19}$ & \multicolumn{5}{c|}{$\Delta^{\mu\rho}u^{[\tau}\Delta^{\sigma][\lambda}u^{\nu]}-\Delta^{\mu[\tau}u^{\sigma]}u^{[\lambda}\Delta^{\nu]\rho}$}\tabularnewline
\hline
$w_{20}$ & \multicolumn{5}{c|}{$\Delta^{\mu[\nu}\epsilon^{\lambda]\rho\gamma[\tau}u^{\sigma]}u_{\gamma}+(\Delta^{\rho[\lambda}u^{\nu]}\epsilon^{\mu\sigma\tau\gamma}+\Delta^{\mu\rho}u^{[\lambda}\epsilon^{\nu]\sigma\tau\gamma})u_{\gamma}/2$}\tabularnewline
\hline
$w_{21}$ & \multicolumn{5}{c|}{$\Delta^{\mu[\nu}\epsilon^{\lambda]\rho\gamma[\tau}u^{\sigma]}u_{\gamma}-(\Delta^{\rho[\lambda}u^{\nu]}\epsilon^{\mu\sigma\tau\gamma}+\Delta^{\mu\rho}u^{[\lambda}\epsilon^{\nu]\sigma\tau\gamma})u_{\gamma}/2$}\tabularnewline
\hline
$w_{22}$ & \multicolumn{5}{c|}{$\Delta^{\mu\rho}\epsilon^{\nu\lambda\gamma[\sigma}u^{\tau]}u_{\gamma}-\Delta^{\mu[\nu}\epsilon^{\lambda]\rho\gamma[\tau}u^{\sigma]}u_{\gamma}-\epsilon^{\mu\nu\lambda\gamma}u_{\gamma}u^{[\sigma}\Delta^{\tau]\rho}/3$}\tabularnewline
\hline
$w_{23}$ & \multicolumn{5}{c|}{$(\Delta^{\rho[\lambda}u^{\nu]}\epsilon^{\mu\sigma\tau\gamma}-\Delta^{\mu\rho}u^{[\lambda}\epsilon^{\nu]\sigma\tau\gamma})u_{\gamma}/2-\Delta^{\mu[\nu}u^{\lambda]}\epsilon^{\sigma\tau\gamma\rho}u_{\gamma}/3$}\tabularnewline
\hline
$w_{24}$ & \multicolumn{5}{c|}{$\Delta^{\mu\rho}\Delta^{\lambda[\sigma}\Delta^{\tau]\nu}+\epsilon^{\mu\lambda\nu\gamma}u_{\gamma}
\epsilon^{\rho\sigma\tau\kappa}u_{\kappa}/6-\Delta^{\mu[\nu}\Delta^{\lambda][\sigma}\Delta^{\tau]\rho}$}
\tabularnewline
\hline
\end{tabular}
}
\caption{Irreducible rotation-invariant decompositions of $W^{\mu\lambda\nu\rho\sigma\tau}$.}\label{Table-W}
\end{table}
\noindent
For example, from Table \ref{Table-G-I-O-F}, the irreducible decomposition of $H$, $K$, and $L$ reads:
\begin{align}
&H^{\mu\nu\rho\sigma}=h_1 u^{(\mu}\Delta^{\nu)(\rho}u^{\sigma)}+h_2\Delta^{\mu\nu}\Delta^{\rho\sigma}+h_3 \Delta^{\mu\nu,\rho\sigma}
+h_4\Delta^{\mu\nu}u^{\rho}u^{\sigma}+h_5u^{\mu}u^{\nu}\Delta^{\rho\sigma}\nonumber\\
&~~~~~~~~~~~+ h_6u^{\mu}u^{\nu}u^{\rho}u^{\sigma}\;,\label{H123}\\
&K^{\mu\nu\rho}=k_{1}u^{(\mu}\Delta^{\nu)\rho}+k_{2}\Delta^{\mu\nu}u^{\rho}+k_{3}u^{\mu}u^{\nu}u^{\rho}\;,\label{K123}\\
&L^{\mu\nu\rho\sigma}=l_{1}u^{(\mu}\Delta^{\nu)[\rho}u^{\sigma]}+l_{2}u^{(\mu}\epsilon^{\nu)\rho\sigma\gamma}u_{\gamma}.\label{L123}
\end{align}
The remaining coefficients $M, N, P, \dots, W$ can be constructed likewise. Above we introduced the traceless symmetric projector $\Delta_{\mu\nu,\alpha\beta}\equiv(\Delta_{\mu\alpha}\Delta_{\nu\beta}+\Delta_{\nu\alpha}\Delta_{\mu\beta})/2-\Delta_{\mu\nu}\Delta_{\alpha\beta}/3$.  

By substituting the decomposed tensors from Tables~\ref{Table-G-I-O-F}-\ref{Table-W} into the general forms of the dissipative currents given in Eqs.~\eqref{EMTsymmetricgradientdecomposition}-\eqref{Spingradientdecomposition}, we obtain their complete irreducible expressions under rotation. This leads to a total of 98 undetermined transport coefficients, resulting in highly complex expressions. This naturally leads us to the next section, where we discuss the physical constraints that can be applied to systematically reduce the number of independent coefficients by eliminating unphysical terms.
%
%
\subsection{Matching conditions, parity-evenness, and semi-positivity constraints}
Tables~\ref{Table-G-I-O-F}--\ref{Table-W} present 98 initially undetermined coefficients, resulting in highly complex expressions for the dissipative currents. However, many of these coefficients can be systematically eliminated by applying the following three types of physical constraints: \textbf{(i)} matching conditions (frame choice), \textbf{(ii)} parity-even structure of the entropy production rate, and \textbf{(iii)} the second law of thermodynamics -- i.e., the requirement that the entropy production rate satisfies the semi-positivity condition, $\partial_\mu S^\mu \geq 0$. These constraints reduce the number of free coefficients from 98 to 23.
\subsubsection*{Matching conditions}
As discussed in Sec.~\ref{Local equilibrium density operator with spin current}, the local equilibrium expectation values, $T^{\mu\nu}_{\rm LE}$, $j^{\mu}_{\rm LE}$, and $\spt^{\lambda\mu\nu}_{\rm LE}$, are constrained by Eqs.~\eqref{const1qm}-\eqref{const3qm} to reproduce the actual hydrodynamic currents $T^{\mu\nu}$, $j^{\mu}$, and $\spt^{\lambda\mu\nu}$ projected along the timelike unit vector $n^{\mu}$ normal to the hypersurface. Choosing $n^{\mu} = u^{\mu}$, the fluid four-velocity, recovers matching conditions consistent with Landau or Eckart frames in conventional formulations of dissipative hydrodynamics.
Let $X_{\rm LE}$ denote the local equilibrium expectation values of the operators $T^{\mu\nu}_{\rm LE}$, $j^{\mu}_{\rm LE}$, and $\spt^{\lambda\mu\nu}_{\rm LE}$, and $X$ the corresponding actual values. Then, for $n^{\mu} = u^{\mu}$, the matching conditions~\eqref{const1qm}-\eqref{const3qm} read:
\begin{align}
u_\mu X^{\mu}_{\rm LE} \equiv u_{\mu}\Tr(\wrho_{\rm LE}\widehat{X})=u_\mu X^{\mu}~~\implies~~ u_{\mu}\delta X^{\mu}=0,
\end{align}
where the actual values are decomposed into equilibrium and dissipative parts $X^{\mu}=X_{0}^{\mu}+\delta X^{\mu}$. This implies that Eqs.~\eqref{const1qm}-\eqref{const3qm} read: 
\begin{align}
&u_{\mu}\left(\delta T^{\mu\nu}_{(S)}+\delta T^{\mu\nu}_{(A)}\right)=0,\label{TStAmat}\\
&u_{\mu}\delta j^{\mu}=0,\label{matparti}\\
&u_{\mu}\delta \spt^{\mu\lambda\nu}=0.\label{spinmatt}
\end{align}
Note that the first matching condition corresponds to the generalized Landau frame, or the Landau frame in spin hydrodynamics, as also discussed in Chapter~\ref{Navier-Stokes limit}.

Employing the decompositions of the tensor coefficients from Tables~\ref{Table-G-I-O-F}-\ref{Table-W} and substituting Eqs.~\eqref{EMTsymmetricgradientdecomposition}-\eqref{Spingradientdecomposition} into the matching conditions~\eqref{TStAmat}-\eqref{spinmatt} yields a system of constraints on the transport coefficients. These are:
\begin{align}
u_{\mu}\left(\delta T^{\mu\nu}_{(S)}+\delta T^{\mu\nu}_{(A)}\right)=0&\quad\implies\quad
\begin{cases}
     &h_{1}+n_{1}=0,\quad h_{5}=h_{6}=0,\quad k_{1}-p_1=0,\\
     &k_{3}=0,\quad l_{1}+q_{1}=0,\quad l_{2}+{q_{4}}=0,\\
     &m_{1}+r_{1}=0,\quad m_{2}+r_2=0,\\
     &m_5+r_3=0,\quad m_6-r_4=0,\\
     &m_7=m_8=0.
\end{cases}\label{matchingconditions-1}
\end{align}
\begin{align}
u_{\mu}\delta j^{\mu}=0&\quad \implies\quad 
\hspace{0.65cm} g_2=g_3=0,\quad i_2=0,\quad f_5=f_6=0.\label{matchingconditions-2}
\end{align}
\begin{align}
u_{\mu}\delta \spt^{\mu\lambda\nu}=0&\quad\implies\quad
\begin{cases}
  &t_9=t_{10}=0,\quad u_{5}=u_{6}=0,\quad v_{5}=v_{6}=0,\\
  &v_{7}=v_{8}=0,\quad w_{11}=w_{12}=0,\\
  &w_{13}=w_{14}=0,\quad w_{15}=w_{16}=0,\\ &w_{17}=w_{18}=0.
\end{cases}\label{matchingconditions-3}
\end{align}
These impose 34 constraints, reducing the number of coefficients from $98$ to $64$.
%
%
\subsubsection{Parity-even entropy production rate}
Another set of physical constraints on the dissipative currents originates from the entropy production rate given in Eq.~\eqref{entropyproductionrate}. The first of these constraints is that the entropy production rate, $\partial_{\mu}S^{\mu}$, is a Lorentz scalar which is invariant under spatial reflections as well as spatial rotations in the rest frame defined by the fluid four-velocity $u^\mu$. In other words, $\partial_{\mu}S^{\mu}$ must be parity-even. Consequently, we must ensure that all terms contributing to the entropy production rate are parity-even or scalars, which provides a systematic means of eliminating certain undetermined coefficients. 

To implement this condition, we decompose the thermo-hydrodynamic gradients appearing in the entropy production rate~\eqref{entropyproductionrate} into irreducible components under spatial rotations. This decomposition follows the same procedure described in the previous section, which led to the classifications in Tables~\ref{Table-G-I-O-F}--\ref{Table-W}. However, unlike before, we do not limit ourselves to the decomposition of scalar components only (for the full derivation, see Appendix~\ref{AppendixC}). After this, we contract these gradient terms with the tensor coefficients in the dissipative currents to obtain the complete irreducible decomposition of the entropy production rate. This approach enables us to systematically discard terms that violate the parity-even condition.
\medskip 

We begin with the symmetric shear tensor: 
\begin{align}\label{thermalsheardecomposition}
\xi^{\mu\nu}=(D\beta)u^{\mu}u^{\nu}+\frac{1}{3T}\theta\Delta^{\mu\nu}+\frac{1}{T}\mathcal{J}_h^{\alpha}\left(\Delta^\mu_\alpha u^{\nu}+\Delta^\nu_\alpha u^\mu \right)+\frac{1}{T}\sigma^{\alpha\beta} \Delta^{\mu\nu}_{\alpha\beta},
\end{align}
where we use the comoving and spatial derivatives $D\equiv u_\mu\partial^\mu$ and $\nabla^\mu\equiv\Delta^{\mu\nu}\partial_\nu$, the expansion scalar $\theta\equiv \partial_\mu u^\mu$, the heat flow $\mathcal{J}_h^\mu\equiv D u^\mu-(1/T)\nabla^\mu T$, and the shear tensor $\sigma^{\mu\nu}\equiv \Delta^{\mu\nu}_{\alpha\beta}\partial^\alpha u^\beta$.

Next, we consider the difference between the spin potential and thermal vorticity, which is antisymmetric and can be decomposed as,
\begin{align}\label{Omega-omegadecomposition}
\Omega^{\mu\nu}-\varpi^{\mu\nu}=\frac{1}{T}\mathcal{E}^{\alpha}\left(\Delta^\nu_\alpha u^{\mu}-\Delta^\mu_\alpha u^\nu \right)-\frac{1}{2T}\epsilon^{\mu\nu\alpha\beta}u_{\alpha}\mathcal{B}_{\beta},
\end{align}
where the electric-like part is $\mathcal{E}^{\mu}\equiv T u_{\nu}(\Omega^{\nu\mu}-\varpi^{\nu\mu})$ and the magnetic-like part reads $\mathcal{B}^{\mu}\equiv T\epsilon^{\alpha\beta\gamma\mu}u_{\gamma}(\Omega_{\alpha\beta}-\varpi_{\alpha\beta})$. 

The gradient of the chemical potential-to-temperature ratio is decomposed as: 
\begin{align}\label{dzetadecomposition}
\partial_{\mu}\zeta=\left(D\zeta\right)u_{\mu}+\left(\nabla_{\alpha}\zeta\right) \Delta^\alpha_\mu \;.
\end{align}
Finally, the derivative of the spin potential is expressed as:
\begin{align}\label{spinpotentialgradientdecompositionnn}
 \partial_{\mu}\Omega_{\lambda\nu}=&\ 2\mathcal{X}^{\gamma}u_{\mu}u_{[\lambda}\Delta_{\nu]\gamma}-
\frac12\mathcal{Y}^{\gamma}u_{\mu}\epsilon_{\lambda\nu\sigma\gamma}u^{\sigma}
+\frac12\mathcal{Z}u_{[\lambda}\Delta_{\nu]\mu}+\mathcal{T}^{\gamma}u_{[\lambda}\epsilon_{\nu]\mu\alpha\gamma}u^{\alpha}+2\mathcal{F}^{\rho\sigma}u_{[\lambda}\Delta_{\nu]\sigma}\Delta_{\mu \rho}\nonumber\\
 &\ -\frac18\mathcal{H}\epsilon_{\lambda\nu\alpha\mu}u^{\alpha}-\frac12\mathcal{G}_{[\lambda}\Delta_{\nu]\mu}-\frac12\mathcal{I}^{\rho\sigma}\Delta_{\mu \rho}\epsilon_{\lambda\nu\tau \sigma}u^{\tau}\;,
\end{align}
where the scalars, vectors, and tensors appearing in the expression above are defined as follows,
\begin{align}
&\mathcal{X}^{\gamma}=u^{[\rho}\Delta^{\sigma]\gamma}D\Omega_{\rho\sigma},\quad \mathcal{Y}^{\gamma}=\epsilon^{\rho\sigma\tau\gamma}u_{\tau}D\Omega_{\rho\sigma},\quad
\mathcal{Z}=u^{[\rho}\nabla^{\sigma]}\Omega_{\rho\sigma},\quad
\mathcal{T}^{\gamma}=u^{[\rho}\epsilon^{\sigma]\lambda\theta\gamma}u_{\lambda}\partial_{\theta}\Omega_{\rho\sigma},     \nonumber\\
&\mathcal{F}^{\rho\sigma}=u^{[\gamma}\Delta^{\theta](\rho}\nabla^{\sigma)}\Omega_{\gamma\theta}-\Delta^{\rho\sigma}\mathcal{Z}/4
,\quad
\mathcal{H}=\epsilon^{\rho\sigma\lambda\gamma}u_{\lambda}\partial_{\gamma}\Omega_{\rho\sigma},\quad
\mathcal{G}^{\gamma}=2\Delta^{\gamma\tau}\nabla^{\rho}\Omega_{\rho\tau},    \nonumber\\
&\mathcal{I}^{\rho\sigma}=u_{\gamma}\epsilon^{\theta\lambda\gamma(\rho}\nabla^{\sigma)}\Omega_{\theta\lambda}-\Delta^{\rho\sigma}\mathcal{H}/4.
\end{align}
Note that the tensors $\mathcal{F}^{\mu\nu}$ and $\mathcal{I}^{\mu\nu}$ are symmetric, but not traceless, with their traces being related to $\mathcal{Z}$ and $\mathcal{H}$, respectively. They can be further written in terms of traceless symmetric tensors $\mathcal{F}^{\mu\nu}_{S}$ and $\mathcal{I}^{\mu\nu}_{S}$ as
\begin{align}
    \mathcal{F}^{\mu\nu}=\mathcal{F}^{\mu\nu}_{S}+\frac{1}{12}\Delta^{\mu\nu}\mathcal{Z},\quad
    \mathcal{I}^{\mu\nu}=\mathcal{I}^{\mu\nu}_{S}+\frac{1}{12}\Delta^{\mu\nu}\mathcal{H}.
\end{align}

By contracting the decomposed gradients~\eqref{thermalsheardecomposition}-\eqref{spinpotentialgradientdecompositionnn} with the corresponding tensor coefficients in the dissipative currents~\eqref{EMTsymmetricgradientdecomposition}-\eqref{Spingradientdecomposition}, and applying the matching conditions \eqref{matchingconditions-1}-\eqref{matchingconditions-3}, we obtain the full irreducible form of the entropy production rate~\eqref{entropyproductionrate},
\begin{align}\label{contraction2}
&T^2\xi_{\mu\nu}\delta T^{\mu\nu}_{S}=\,h_1 \mathcal{J}_h^\mu \mathcal{J}_{h,\mu}+h_2\theta^2+h_3\sigma^{\mu\nu}\sigma_{\mu\nu}+\left(h_4 T D\beta+k_2D\zeta+m_3\mathcal{Z}+m_4\mathcal{H}\right)\theta
\nonumber\\
&~~\hspace{2.1cm}-\mathcal{J}^h_\mu\left(- k_1\nabla^\mu \zeta+l_1 \mathcal{E}^\mu  +l_2 \mathcal{B}^\mu +m_1\mathcal{X}^\mu  +m_2\mathcal{Y}^\mu +m_5\mathcal{G}^\mu +m_6\mathcal{T}^\mu \right)
\nonumber\\
&~~\hspace{2.1cm}+\sigma_{\mu\nu}\left(m_9\mathcal{I}_S^{\mu\nu}+m_{10}\mathcal{F}_S^{\mu\nu}\right) \,,\\ 
&T^2\partial_{\mu}\zeta\delta j^{\mu}=\,i_1(\nabla^\mu\zeta)(\nabla_\mu\zeta)(-g_1\mathcal{J}_h^{\mu}+o_1\mathcal{E}^{\mu}+o_2\mathcal{B}^{\mu}+f_1\mathcal{X}^{\mu}+f_2\mathcal{Y}^{\mu}+f_3\mathcal{G}^{\mu}\nonumber\\
&~~~~~~~~~~~~~~~\,+f_4\mathcal{T}^{\mu})\nabla_{\mu}\zeta\,,\\
&T^2(\Omega_{\mu\nu}-\varpi_{\mu\nu})\delta T^{\mu\nu}_{A}=\,-q_{1}\mathcal{E}^\mu \mathcal{E}_\mu-q_{2}\mathcal{B}^\mu \mathcal{B}_\mu-(q_3+q_4)\mathcal{E}^{\mu}\mathcal{B}_{\mu}\nonumber\\
&~~~\,\hspace{3.6cm}-\mathcal{E}_{\mu}\left(-n_1\mathcal{J}_h^{\mu}+p_1\nabla^{\mu}\zeta+r_1\mathcal{X}^{\mu}+r_2\mathcal{Y}^{\mu}+r_3\mathcal{G}^{\mu}-r_4\mathcal{T}^{\mu}\right)
\nonumber\\
&~~~\,\hspace{3.6cm}-\mathcal{B}_{\mu}\left(-n_{2}\mathcal{J}_h^{\mu}+p_2\nabla^{\mu}\zeta+r_5\mathcal{X}^{\mu}+r_6\mathcal{Y}^{\mu}+r_7\mathcal{G}^{\mu}-r_8\mathcal{T}^{\mu}\right)\,,\\
&T^2\partial_{\mu}\Omega_{\lambda\nu}\delta \spt^{\mu\lambda\nu}=\,(t_1 TD\beta+t_2\theta+u_1 D\zeta)\mathcal{Z}+(t_4 TD\beta+t_5\theta+u_3 D\zeta)\mathcal{H}
+w_1\mathcal{Z}^{2}
\nonumber\\
&~~~\hspace{2.7cm}+w_2\mathcal{H}^{2}+2w_3\mathcal{Z}\mathcal{H}-(t_3\mathcal{J}^h_{\mu}+u_2\nabla_{\mu}\zeta+v_2\mathcal{B}_{\mu}+v_3\mathcal{E}_{\mu}+w_7\mathcal{X_{\mu}}
\nonumber\\
&~~~\hspace{2.7cm}+w_8\mathcal{Y}_{\mu})\mathcal{G}^{\mu}-(t_6\mathcal{J}^h_{\mu}+u_4\nabla_{\mu}\zeta+v_1\mathcal{B}_{\mu}+v_4\mathcal{E}_{\mu}+w_5\mathcal{X_{\mu}}
\nonumber\\
&~~~\hspace{2.7cm}+w_6\mathcal{Y}_{\mu})\mathcal{T}^{\mu}-w_9\mathcal{G}^{\mu}\mathcal{G}_{\mu}-w_{19}\mathcal{T}^{\mu}\mathcal{T}_{\mu}-w_{21}\mathcal{G}^{\mu}\mathcal{T}_{\mu}\nonumber\\
&~~~\hspace{2.7cm}+(t_7\mathcal{F}^{\mu\nu}_{S}+t_8I^{\mu\nu}_{S})\sigma_{\mu\nu}+w_{10}\mathcal{F}^{\mu\nu}_{S}\mathcal{F}_{S,\mu\nu}\nonumber\\
&~~~\hspace{2.7cm}+(w_{22}+w_{23})\mathcal{I}^{\mu\nu}_{S}\mathcal{F}_{S,\mu\nu}-\frac{1}{2}w_{24}\mathcal{I}^{\mu\nu}_{S}\mathcal{I}_{S,\mu\nu}\,.
\label{contraction4}
\end{align}

As discussed earlier, the entropy production rate $\partial_{\mu}S^{\mu}$ is a Lorentz scalar, invariant under reflections as well as spatial rotations in the local rest frame associated with $u^\mu$. To eliminate the terms that violate the latter condition, we may first classify all the terms of the irreducible entropy production rate appearing in Eqs.~\eqref{contraction2}-\eqref{contraction4} according to their transformations properties under such operations, as summarized in Table~\ref{Table-classification}.
\begin{table}[H]
\renewcommand{\arraystretch}{1.5}
\begin{tabularx}{1\textwidth} { 
| >{\raggedright\arraybackslash}X 
| >{\raggedright\arraybackslash}X
| >{\raggedright\arraybackslash}X
| >{\raggedright\arraybackslash}X
| >{\centering\arraybackslash}X 
| >{\raggedleft\arraybackslash}X | }
\hline
\textbf{Scalars} & \textbf{Pseudo-scalars} & \textbf{Vectors} & \textbf{Pseudo-vectors} & \textbf{Tensors} & \textbf{Pseudo-tensors} \\
 
\hline
$\theta$  & $\mathcal{H}$  &  $\mathcal{J}_h^\mu$ & $\mathcal{B}^{\mu}$ & $\sigma^{\mu\nu}$ & $\mathcal{I}^{\mu\nu}_{S}$ \\
\hline
$TD\beta$ & --- & $\nabla^{\mu}\zeta$  & $\mathcal{Y}^{\mu}$ & $\mathcal{F}^{\mu\nu}_{S}$ & --- \\
\hline
$ D\zeta$ & --- &$ \mathcal{X}^{\mu} $& $\mathcal{T}^{\mu}$& --- &---\\
\hline
$\mathcal{Z}$ & --- & $\mathcal{E}^{\mu}$ & --- & ---& ---\\
\hline
--- & --- &  $\mathcal{G}^{\mu}$ & --- & --- & ---\\
\hline
\end{tabularx}
\caption{Classification of the thermo-hydrodynamic gradient terms based on their properties under reflections and rotations.}
\label{Table-classification}
\end{table}
\noindent
To ensure that the entropy production rate is parity-even, terms with odd parity in Eqs. \eqref{contraction2}-\eqref{contraction4} must be accompanied by parity-odd coefficients to preserve overall evenness. For simplicity, we set all parity-odd coefficients to zero. This eliminates 28 additional coefficients, reducing the total number of undetermined coefficients from $64$ to $36$.

We then use the information from Table~\ref{Table-classification} to reorganize the entropy production rate $\partial_{\mu}S^{\mu}$ as a sum of six bilinear contributions:
\begin{align}\label{partsofentropyproduction}
\partial_{\mu}S^{\mu}= \partial_{\mu}S^{\mu}_{S.S}+ \partial_{\mu}S^{\mu}_{PS.PS}+\partial_{\mu}S^{\mu}_{V.V}+ \partial_{\mu}S^{\mu}_{PV.PV}+\partial_{\mu}S^{\mu}_{T.T}+ \partial_{\mu}S^{\mu}_{PT.PT}\;.
\end{align}
Here, the labels follow the notation: 
$S$: scalar, $PS$: pseudo-scalar, $V$: vector, $PV$: pseudo-vector, $T$: tensor, and $PT$: pseudo-tensor. 

Explicit forms of these contributions in matrix form read:
\begin{align}
&\partial_\mu S_{S.S}^\mu=\frac{1}{T^2}
\resizebox{0.6\textwidth}{!}{$
\begin{pmatrix} 
\theta \\ TD\beta \\ D\zeta \\ \mathcal{Z}
\end{pmatrix}^\text{T}
\left(
\begin{array}{cccc}
 h_2 & h_4/2 & k_2/2 &(2m_3-t_2)/4 \\
 h_4/2 & 0 & 0 & -t_1/4 \\
 k_2/2 & 0 & 0 & -u_1/4\\
 (2m_3-t_2)/4 & -t_1/4& -u_1/4 & -w_1/2 \\
\end{array}
\right)
\begin{pmatrix} 
\theta \\ TD\beta \\ D\zeta \\ \mathcal{Z}
\end{pmatrix}$}\;,
\\
~~~~~~~&\partial_\mu S_{PS.PS}^\mu=-\frac{1}{2T^2}w_2 \mathcal{H}^2\;,\\
&\partial_\mu S_{V.V}^\mu=-\frac{1}{T^2}
\resizebox{0.65\textwidth}{!}{$
\begin{pmatrix} 
\mathcal{J}_h^{\mu}\\ \nabla^{\mu}\zeta\\ \mathcal{E}^{\mu}\\ \mathcal{G}^{\mu}\\ \mathcal{X}^{\mu}
\end{pmatrix}^\text{T}
\left(
\begin{array}{ccccc}
 -h_{1} &(g_1-k_1)/2 & (l_1-n_1)/2& (2m_5-t_3)/4 & m_{1}/2 \\
 (g_1-k_1)/2& i_1 & (p_1-o_1)/2  &-(2f_3+u_2)/4& -f_1/2 \\
 (l_1-n_1)/2 & (p_1-o_1)/2 & q_1& (2r_3-v_3)/4 & r_1/2  \\
 (2m_5-t_3)/4 & -(2f_3+u_2)/4 & (2r_3-v_3)/4 & -w_9/2 & -w_7/4 \\
  m_1/2 & -f_1/2 & r_1/2 & -w_7/4  & 0\\
\end{array}
\right)
\begin{pmatrix} 
\mathcal{J}^h_{\mu}\\ \nabla_{\mu}\zeta\\ \mathcal{E}_{\mu}\\ \mathcal{G}_{\mu}\\ \mathcal{X}_{\mu}
\end{pmatrix}$}\;,\\
&\partial_\mu S_{PV.PV}^\mu=-\frac{1}{T^2}
\resizebox{0.54\textwidth}{!}{$
\begin{pmatrix}
\mathcal{B}^{\mu}\\ \mathcal{T}^{\mu}\\ \mathcal{Y}^{\mu}
\end{pmatrix}^\text{T}
\left(
\begin{array}{ccc}
 q_2  & -(2r_8+v_1)/4& r_6/2 \\
 -(2r_8+v_1)/4& -w_{19}/2 & -w_6/4  \\
  r_6/2 & -w_6/4 & 0 \\
\end{array}
\right)
\begin{pmatrix}
\mathcal{B}^{\mu}\\ \mathcal{T}^{\mu}\\ \mathcal{Y}^{\mu}
\end{pmatrix}$},
\\
&\partial_\mu S_{T.T}^\mu=\frac{1}{T^2}
\resizebox{0.54\textwidth}{!}{$
\begin{pmatrix}
\sigma^{\mu\nu}\\ \mathcal{F}^{\mu\nu}_{S}
\end{pmatrix}^\text{T}
\left(
\begin{array}{ccc}
 h_3 & (2m_{10}-t_7)/4 \\
 (2m_{10}-t_7)/4& -w_{10}/2\\
\end{array}
\right)
\begin{pmatrix}
\sigma^{\mu\nu}\\ \mathcal{F}^{\mu\nu}_{S}
\end{pmatrix}$}\;,
\\
&\partial_\mu S_{PT.PT}^\mu=\frac{1}{4T^2}w_{24}\mathcal{I}^{\mu\nu}_{S}\mathcal{I}_{\mu\nu\, S}\;.
\end{align}

In summary, enforcing the parity-even nature of the entropy production rate allows us to \textbf{(i)} reduce the number of undetermined coefficients from $64$ to $36$ and \textbf{(ii)} organize the entropy production rate into six structurally distinct contributions, simplifying further analysis. However, despite this reduction, the dissipative current expressions remain highly involved and require additional constraints, such as those imposed by the second law of thermodynamics, to further constrain the theory.
%
%
\subsubsection{Semi-positivity of the entropy production rate}
The dissipative currents are constrained by the second law of thermodynamics, which requires the entropy production rate in Eq.~\eqref{entropyproductionrate} to be semi-positive. This means that each individual contribution in Eq.~\eqref{partsofentropyproduction} must be non-negative. There are two equivalent approaches to verify this semi-positivity: (1) diagonalize the corresponding matrices and express the results in terms of sums of perfect squares, or (2) use the fact that all eigenvalues of a matrix are non-negative if and only if all principal minors of the matrix are non-negative. In what follows, we adopt the second approach, which yields the following conditions:
\begin{align}
\label{eq:divS_s}
&\partial_{\mu}S^{\mu}_{S.S}\geq0\quad\implies\quad\begin{cases}
        &h_2\geq0,\quad w_1\leq 0,\quad h_4=k_2=t_1=u_1=0,\\
        &-8h_2w_1\geq (2m_3-t_2)^2,
\end{cases}
\\
&\partial_{\mu}S^{\mu}_{PS.PS}\geq0\quad\implies\quad 
\hspace{0.7cm}w_2\leq 0,
\\
&\partial_{\mu}S^{\mu}_{V.V}\geq0\quad\implies\quad\begin{cases}
    &{m_1=f_1=r_1=w_7=0},\quad h_1\leq 0,~i_1\geq 0,~w_9\leq 0,\\
       &q_1+h_1=0,\quad g_1-o_1=0,\quad t_3+v_3=0,\\
       &-4h_1 i_1\geq(k_1-g_1)^2,\quad 8h_1w_9\geq(2m_5-t_3)^2,\\
       &-8i_1w_9\geq (2f_3+u_2)^2,\\
       & h_1(2f_3+u_2)^2-i_1(2m_5-t_3)^2+2w_9(k_1-g_1)^2\\
       &+(2f_3+u_2)(2m_5-t_3)(k_1-g_1)+8h_1i_1w_9\geq0,
\end{cases}
\\
&\partial_{\mu}S^{\mu}_{PV.PV}\geq0\quad\implies\quad\begin{cases}
&{q_2\geq 0},\quad {w_{19}\leq0},\quad {r_6=w_6=0},\\
&-8q_2w_{19}\geq (2r_8+v_1)^2,
\end{cases}
\\
&\partial_{\mu}S^{\mu}_{T.T}\geq0\quad\implies\quad\begin{cases}
&{h_{3}\geq 0},\quad {w_{10}\leq 0},\\
&-8h_3w_{10}\geq (2m_{10}-t_7)^2,
\end{cases}
\\
\label{eq:divS_PT}
&\partial_{\mu}S^{\mu}_{PT.PT}\geq0\quad\implies\quad\hspace{0.7cm}{w_{24}\geq 0}.
\end{align}

These constraints include 13 equalities, which reduce the number of independent coefficients from $36$ to $23$. The remaining inequalities serve as additional constraints on those 23 coefficients. Specifically, the inequalities in Eqs.~\eqref{eq:divS_s}-\eqref{eq:divS_PT} can be classified as follows: 11 inequalities impose sign conditions on individual coefficients, 6 inequalities set bounds on products of two coefficients, and 1 inequality involves a bound on the product of three coefficients. Taking all of these constraints into account allows us to construct the final form of the dissipative currents, as will be presented in the next section.
%
%
\subsection{Summary of dissipative currents and transport coefficients}
In this section, we present the final expressions for the dissipative currents at first-order in gradients and tabulate the associated transport coefficients, along with their physical interpretations as understood at this stage. In total, we identify 23 \emph{dissipative transport coefficients} that relate the dissipative currents to gradients of the thermo-hydrodynamic fields: 4 for the symmetric part of the energy-momentum tensor, 5 for the antisymmetric part, 3 for the conserved vector current, and 11 for the spin tensor.
\medskip 
\subsubsection{Symmetric part of the energy-momentum tensor}

The symmetric dissipative part of the energy-momentum tensor takes the form,
\begin{align}
\delta T^{\mu\nu}_{S}=&2u^{(\mu}\left[-\kappa_{h}(\mathcal{J}_h^{\nu)}-\mathcal{E}^{\nu)})+\kappa_{\zeta}\,\nabla^{\nu)}\zeta+\kappa_{\mathcal{G}}\,\mathcal{G}^{\nu)}\right]
+\left(\zeta_b\theta+\zeta_{\mathcal{Z}}\,\mathcal{Z}\right)\Delta^{\mu\nu}\nonumber\\
&+2\left(\eta\sigma^{\mu\nu}
+\eta_{\mathcal{F}}\,\mathcal{F}_{S}^{\mu\nu}\right),
\label{EMTsymmetricdissipativecurrent}
\end{align}
where the thermo-hydrodynamic fields, the corresponding transport coefficients, their constraints, and physical interpretations are given in Table~\ref{Tablesymmetric}.
%
%
\begin{table}[H]\label{symmset}
\renewcommand{\arraystretch}{2.3}
\begin{tabularx}{1\textwidth} { 
| >{\raggedright\arraybackslash}X
| >{\raggedright\arraybackslash}X
| >{\raggedright\arraybackslash}X
| >{\centering\arraybackslash}X 
| >{\raggedleft\arraybackslash}X | }
\hline
\textbf{Thermo-hydrodynamic field} &  \textbf{Coefficient} & \textbf{Constraint} & \textbf{Interpretation}\\
\hline
$\theta=\nabla^{\mu}u_{\mu}$&$\zeta_b={h_2}/{T}$&$\zeta_b \geq 0$& Bulk viscosity\\
\hline
$\sigma^{\mu\nu}=\nabla^{(\mu}u^{\nu)}-\frac{1}{3}\theta\Delta^{\mu\nu}$&$\eta={h_{3}}/{(2T)}$&$\eta\geq 0$&Shear viscosity\\
\hline
$\mathcal{Z}=u^{[\mu}\nabla^{\nu]}\Omega_{\mu\nu}$  &  $\zeta_{\mathcal{Z}}=m_3/T$ &$4\zeta\chi_\mathcal{Z}\geq(\zeta_\mathcal{Z}+\chi_{\theta})^{2}$&{Gyro-bulk viscosity}\\
\hline
$\mathcal{F}^{\mu\nu}_{S}={-u_\rho \nabla^{(\mu}\Omega^{\nu)\rho}-\frac{1}{3}\Delta^{\mu\nu}\mathcal{Z}}$ & $\eta_{\mathcal{F}}={m_{10}}/{(2T)}$ &$16\eta\chi_\mathcal{F}\geq(4\eta_\mathcal{F}+\chi_{\sigma})^{2}$&Gyro-shear viscosity\\
\hline
\end{tabularx}
\caption{Summary of contributions to \( \delta T^{\mu\nu}_{S} \) in Eq.~\eqref{EMTsymmetricdissipativecurrent}, along with the corresponding transport coefficients. The 
first three coefficients in Eq.~\eqref{EMTsymmetricdissipativecurrent} are not independent, as they also appear in 
\(\delta T^{\mu\nu}_{A}\) in Eq.~\eqref{EMTantisymmetricdissipativecurrent} due to the matching 
condition~\eqref{TStAmat}; they are listed in Table \ref{Tableantisymmetric}.}
\label{Tablesymmetric}
\end{table}
\noindent
In Table \ref{Tablesymmetric}, the coefficients $\eta$ and $\zeta_b$ represent the shear and bulk viscosities in conventional relativistic hydrodynamics. Two new transport coefficients, $\zeta_{\mathcal{Z}}$ and $\eta_{\mathcal{F}}$, are identified, corresponding to the gradients of the spin potential. These two transport coefficients are of particular interest because their current structures are the same as those of the bulk and shear tensors, apart from the type of the gradient involved. Note that the first three coefficients in Eq. \eqref{EMTsymmetricdissipativecurrent}, which multiply
vector terms, also appear in $\delta T^{\mu\nu}_{A}$, according to
Eq.~\eqref{EMTantisymmetricdissipativecurrent} because they are involved in the matching 
condition $u_\mu (T^{\mu\nu}_S+T^{\mu\nu}_A)=0$. For this reason, they have
been omitted from Table \ref{Tablesymmetric} and are listed in Table \ref{Tableantisymmetric}.
\medskip 
\subsubsection{Antisymmetric part of the energy-momentum tensor}

The anti-symmetric contribution to the energy-momentum tensor reads:
\begin{align}
\delta T^{\mu\nu}_{A}=&-2u^{[\mu}\left[-\kappa_{h}(\mathcal{J}_h^{\nu]}-\mathcal{E}^{\nu]})+\kappa_{\zeta}\,\nabla^{\nu]}\zeta+\kappa_{\mathcal{G}}\,\mathcal{G}^{\nu]}\right]-2\gamma_{\phi}\phi^{\mu\nu}-2\gamma_{\Xi}\Xi^{\mu\nu}\;,
\label{EMTantisymmetricdissipativecurrent}
\end{align}
and the relevant transport coefficients are summarized in Table~\ref{Tableantisymmetric}.  
%
%
\begin{table}[H]
\renewcommand{\arraystretch}{2.1}
\begin{tabularx}{1\textwidth} { 
| >{\raggedright\arraybackslash}X 
| >{\raggedright\arraybackslash}X
| >{\raggedright\arraybackslash}X
| >{\centering\arraybackslash}X 
| >{\raggedleft\arraybackslash}X | }
\hline
\textbf{Thermo-hydrodynamic field} &  \textbf{Coefficient} & \textbf{Constraint} & \textbf{Interpretation}\\
\hline
$\mathcal{J}_h^{\mu}-\mathcal{E}^{\mu}= Du^{\mu}-\frac{1}{T}\nabla^{\mu}T+T(\Omega^{\mu\nu}-\varpi^{\mu\nu})u_{\nu}$&$\kappa_{h}=-h_{1}/(2T)$&$\kappa_{h}\geq 0$& Boost heat conductivity\\
\hline
$\nabla^{\mu}\zeta$ &  $\kappa_{\zeta}={k_1}/{(2T)}$ &$8T\kappa_h\mathcal{K}_\zeta\geq(T\mathcal{K}_h-2\kappa_\zeta)^2$& Chemical-boost heat conductivity\\
\hline
$\mathcal{G}^{\mu}=2\Delta^{\mu\tau}\nabla^{\rho}\Omega_{\rho\tau}$&$\kappa_{\mathcal{G}}=-m_5/(2T)$ &$16\kappa_h\chi_\mathcal{G}\geq(8\kappa_\mathcal{G}-\chi_h)^2$& Gyro-boost heat conductivity\\
\hline
$\phi^{\mu\nu}=T\Delta^{\mu[\sigma}\Delta^{\rho]\nu}(\Omega_{\rho\sigma}-\varpi_{\rho\sigma})$ &  $\gamma_{\phi}=q_2/T$ &$\gamma_{\phi}\geq0$& Rotational viscosity\\
\hline
$\Xi^{\mu\nu}=-u_\rho\nabla^{[\mu}\Omega^{\nu]\rho}$&$\gamma_{\Xi}={r_{8}}/{T}$&
$16\gamma_\phi\chi_\Xi\geq (4\gamma_\Xi+\chi_\phi)^2$&
Gyro-rotation viscosity\\
\hline
\end{tabularx}
\caption{Summary of contributions to $\delta T^{\mu\nu}_{A}$ in Eq.~\eqref{EMTantisymmetricdissipativecurrent}, along with the corresponding transport coefficients.}
\label{Tableantisymmetric}
\end{table}
\noindent
The coefficients $\kappa_{h}$ and $\gamma_{\phi}$ were previously obtained 
in Ref.~\cite{Hattori:2019lfp} and discussed in detail in Chapter~\ref{Navier-Stokes limit} where they were referred to as the boost heat conductivity 
and rotational viscosity, denoted as $\lambda$ and $\gamma$, 
respectively. It should be noted that the current associated with 
$\kappa_{h}$ does not exactly match that of $\lambda$ (see Eq.~\ref{phemenologicalq} in Chapter~\ref{Navier-Stokes limit})
due to the choice of matching condition. Nevertheless, they share the same underlying physical 
meaning. Further studies on the rotational viscosity coefficient have also been conducted in Refs.~\cite{Hongo:2022izs,Hidaka:2023oze}. In addition, we identify three new transport coefficients: $\kappa_{\zeta}$, $\kappa_{\mathcal{G}},$ and $\gamma_{\Xi}$, which arise from the inclusion of the gradient of the spin 
potential and the presence of a finite chemical potential.
\medskip

\subsubsection{Vector current}
For the conserved vector current, we obtain the following expression:
\begin{align}
\delta j^{\mu}=\mathcal{K}_{\zeta}\nabla^{\mu}\zeta+\mathcal{K}_{h}\left(\mathcal{J}_h^{\mu}-\mathcal{E}^{\mu}\right)+\mathcal{K}_{\mathcal{G}}\,\mathcal{G}^{\mu}.
\label{particlenodissipativecurrent}
\end{align}
A summary of the corresponding transport coefficients, along with their respective constraints and physical interpretations, is provided in Table~\ref{taparticlenumber}. 
%
%
\begin{table}[H]
\renewcommand{\arraystretch}{2.9}
\begin{tabularx}{1\textwidth} { 
| >{\raggedright\arraybackslash}X 
| >{\raggedright\arraybackslash}X
| >{\raggedright\arraybackslash}X
| >{\centering\arraybackslash}X 
| >{\raggedleft\arraybackslash}X | }
\hline
\textbf{Thermo-hydrodynamic field} &  \textbf{Coefficient} & \textbf{Constraint} & \textbf{Interpretation}\\
\hline
$\nabla^{\mu}\zeta$ & $\mathcal{K}_{\zeta}={i_1}/{T^{2}}$& $\mathcal{K}_{\zeta}\geq0$& Diffusion coefficient\\
 \hline
$\mathcal{J}_h^{\mu}-\mathcal{E}^{\mu}=Du^{\mu}-\frac{1}{T}\nabla^{\mu}T+T(\Omega^{\mu\nu}-\varpi^{\mu\nu})u_{\nu}$ &$\mathcal{K}_{h}={g_1}/{T^2}$&$8T\kappa_h\mathcal{K}_\zeta\geq(T\mathcal{K}_h-2\kappa_\zeta)^2$ &Diffusion-boost heat conductivity\\
\hline
$\mathcal{G}^{\mu}=2\Delta^{\mu\tau}\nabla^{\rho}\Omega_{\rho\tau}$ & $\mathcal{K}_{\mathcal{G}}=-{f_3}/{T^{2}}$&$8T\chi_G\mathcal{K}_\zeta\geq(2T\mathcal{K}_\mathcal{G}+\chi_\zeta)^2$&Diffusion-gyro-boost heat conductivity\\
\hline
\end{tabularx}
\caption{Summary of various transport coefficients for \( \delta j^{\mu} \) in Eq.~\eqref{particlenodissipativecurrent}.}
\label{taparticlenumber}
\end{table}
\noindent
The coefficient $\mathcal{K}_\zeta$ relates the dissipative current to the gradient of the chemical
potential and can thus be identified as the diffusion coefficient, as in conventional relativistic hydrodynamics. However, in the context of relativistic spin hydrodynamics, we observe the emergence of two additional transport coefficients, $\mathcal{K}_{h}$ and $\mathcal{K}_{\mathcal{G}}$.
\medskip 

\subsubsection{Spin tensor}
Finally, we report the expression for the spin dissipative current at first-order in the hydrodynamic gradient expansion, 
\begin{align}
\delta \spt^{\mu\lambda\nu}=&\frac{1}{T}\bigg[\Delta^{\mu[\lambda}\left(\chi_{h}(\mathcal{J}_h^{\nu]}-\mathcal{E}^{\nu]})+\chi_{\zeta}\nabla^{\nu]}\zeta+\chi_{\mathcal{G}}\mathcal{G}^{\nu]}\right)+\Delta^{\mu[\lambda}u^{\nu]}\left(\chi_{\theta}\theta+\chi_{\mathcal{Z}}\mathcal{Z}\right)\nonumber\\
&+\left(\chi_{\sigma}\sigma^{\mu[\lambda}u^{\nu]}+\chi_{\mathcal{F}}\mathcal{F}_{S}^{\mu[\lambda}u^{\nu]}\right)+\left(\chi_{\phi}\phi^{\mu[\lambda}u^{\nu]}+\chi_{\Xi}\Xi^{\mu[\lambda}u^{\nu]}\right)\nonumber\\
&+\left(\chi_{\mathcal{H}}\epsilon^{\mu \lambda\nu\rho}u_{\rho}\mathcal{H}+\chi_{\mathcal{I}}\mathcal{I}^{\mu}_{\rho,S}\epsilon^{\lambda\nu\rho\gamma}u_{\gamma}\right)\bigg]\;,
\label{Spindissipativecurrent}
\end{align}
with the transport coefficients and their respective constraints listed in Table~\ref{Tablespin}. 
%
\begin{table}[H]
\renewcommand{\arraystretch}{2.3}
\begin{tabularx}{1\textwidth} { 
| >{\raggedright\arraybackslash}X 
| >{\raggedright\arraybackslash}X
| >{\raggedright\arraybackslash}X
| >{\centering\arraybackslash}X 
| >{\raggedleft\arraybackslash}X | }
\hline
\textbf{Thermo-hydrodynamic field}& \textbf{Coefficient} & \textbf{Constraint}& \textbf{Interpretation}\\
 \hline
$(\mathcal{J}_h^{\nu}-\mathcal{E}^{\nu})=D u^{\nu}-\frac{1}{T}\nabla^{\nu}T+T(\Omega^{\nu\lambda}-\varpi^{\nu\lambda})u_{\lambda}$& $\chi_{h}=-t_3/T$& $16\kappa_{h}\chi_\mathcal{G}\geq(8\kappa_\mathcal{G}-\chi_{h})^{2}$& Spin-boost heat conductivity\\
\hline
$\nabla^{\nu}\zeta$&$\chi_{\zeta}=-u_2/T$&$8T\chi_\mathcal{G}\mathcal{K}_{\zeta}\geq(2T\mathcal{K}_\mathcal{G}+\chi_{\zeta})^{2}$& Spin-chemical-boost heat conductivity\\
\hline
$\mathcal{G}^{\nu}=2\Delta^{\nu\tau}\nabla^{\rho}\Omega_{\rho\tau}$& $\chi_{\mathcal{G}}=-w_9/T$&$\chi_{\mathcal{G}}\geq 0$&  Spin-gyro-boost heat conductivity\\
\hline
$\theta=\nabla^{\mu}u_{\mu}$ & $\chi_{\theta}=-t_2/(2T)$&$4\zeta\chi_\mathcal{Z}\geq(\zeta_\mathcal{Z}+\chi_{\theta})^{2}$ &{  Spin-bulk viscosity}\\
\hline
$\mathcal{Z}=u^{[\mu}\nabla^{\nu]}\Omega_{\mu\nu}$ & $\chi_{\mathcal{Z}}=-w_1/(2T)$&$\chi_{\mathcal{Z}}\geq 0$&{ Spin-gyro-bulk viscosity}\\
\hline
$\sigma^{\mu\lambda}=\nabla^{(\mu}u^{\lambda)}-\frac{1}{3}\theta\Delta^{\mu\lambda}$& $\chi_{\sigma}=-t_7/T$&$16\eta\chi_\mathcal{F}\geq(4\eta_\mathcal{F}+\chi_{\sigma})^{2}$&{ Spin-shear viscosity}\\
\hline
$\mathcal{F}^{\mu\lambda}_{S}=-u_{\rho} \nabla^{(\mu}\Omega^{\lambda)\rho}-\frac{1}{3}\Delta^{\mu\lambda}\mathcal{Z}$ & $\chi_{\mathcal{F}}=-w_{10}/T$& $\chi_{\mathcal{F}}\geq 0$&{ Spin-gyro-shear viscosity}\\
\hline
$\phi^{\mu\lambda}=T\Delta^{\mu[\sigma}\Delta^{\rho]\lambda}(\Omega_{\rho\sigma}-\varpi_{\rho\sigma})$& $\chi_{\phi}=2v_1/T$& $16\gamma_{\phi}\chi_{\Xi}\geq(4\gamma_{\Xi}+\chi_{\phi})^{2}$& { Spin-rotation viscosity}\\
\hline
$\Xi^{\mu\lambda}=-u_\rho\nabla^{[\mu}\Omega^{\lambda]\rho}$& $\chi_{\Xi}=-2w_{19}/T$&$\chi_{\Xi}\geq 0$ &{ Spin-gyro-rotation viscosity}\\
\hline
$\mathcal{H}=\epsilon^{\rho\sigma\lambda\gamma}u_{\lambda}\partial_{\gamma}\Omega_{\rho\sigma}$& $\chi_{\mathcal{H}}=-w_2/T$& $\chi_{\mathcal{H}}\geq 0$& { Spin-gyro-pseudo bulk viscosity}\\
\hline
$\mathcal{I}^{\mu\rho}_{ S}=u_{\gamma}\epsilon^{\theta\lambda\gamma(\mu}\nabla^{\rho)}\Omega_{\theta\lambda}-\Delta^{\mu}_{\rho}\mathcal{H}/3$& $\chi_{\mathcal{I}}=w_{24}/(2T)$& $\chi_{\mathcal{I}}\geq 0$&{ Spin-gyro-pseudo shear viscosity}\\
\hline
\end{tabularx}
\caption{Summary of various transport coefficients for \( \delta S^{\mu\lambda\nu} \) in Eq.~\eqref{Spindissipativecurrent}.}
\label{Tablespin}
\end{table}
\noindent
The coefficients \( \chi_{Z}, \chi_{F}\), and \( \chi_{\Xi} \) were identified in Refs.~\cite{She:2021lhe,Biswas:2023qsw,Drogosz:2024gzv}, while most of the coefficients related to derivatives
of the spin potential were obtained in Refs.~\cite{Gallegos:2020otk,Dey:2024cwo}. In addition to the 
constraints given in Tables~\ref{Tablesymmetric}-\ref{Tablespin}, an additional condition must be satisfied:
\begin{align}
&16T\kappa_h \mathcal{K}_\zeta \chi_\mathcal{G}\geq T\mathcal{K}_\zeta(8\kappa_\mathcal{G}-\chi_h)^2+2\kappa_h(2T\mathcal{K}_\mathcal{G}+\chi_\zeta)^2+2\chi_\mathcal{G}(2\chi_\zeta-T\mathcal{K}_h)^2\nonumber\\
&-(8\kappa_\mathcal{G}-\chi_h)(2T\mathcal{K}_\mathcal{G}+\chi_\zeta)(2\chi_\zeta-T\mathcal{K}_h)\;.
\end{align}

It is worth emphasizing that the names assigned to the transport coefficients in Tables~\ref{Tablesymmetric}-\ref{Tablespin} reflect the current physical understanding of their roles. A more detailed exploration of their microscopic origins could lead to a more precise interpretation.
\medskip

With these results, we can formulate the evolution equations for the energy-momentum tensor, the particle number 
current, and the spin tensor at first order in gradients, commonly referred to as the relativistic 
Navier-Stokes limit. However, the presence of 23 transport coefficients - each of which must be either derived from a microscopic theory or fitted to experimental data - may pose considerable challenges for future numerical simulations. 
%
\section{Summary and outlook}
In this chapter, we developed a first-principles formulation of relativistic dissipative spin hydrodynamics based on relativistic quantum-statistical mechanics. 
Our work yielded the entropy current, the entropy production rate, the full set of dissipative currents at the first-order of hydrodynamic gradient expansion, and their associated transport coefficients. 
Moreover, we also introduced, in Appendix~\ref{AppendixC}, 
a general method for the irreducible decomposition of rank-$n$ tensors under rotation -- a tool that should prove useful even beyond the scope of hydrodynamics.

In the future, we aim to analyze each transport coefficient in detail and identify parameter regimes in which the theory is stable. Moreover, motivated by causality considerations, we also plan to develop a second-order extension of this framework. However, the presence of twenty-three independent transport coefficients -- each of which must ultimately be fixed by microscopic theory or empirical input -- poses a significant challenge for numerical simulations. Hence, a detailed study of the model’s numerical performance is therefore deferred to future work.

\chapter{Summary and outlook}
\label{summaryandoutlook}
\section{Summary}
In this thesis, we formulated relativistic dissipative spin hydrodynamics using two approaches. The first, presented in Chapters~\ref{Navier-Stokes limit}-\ref{MISMIS}, is grounded in covariant thermodynamics and extends the Navier-Stokes and Müller-Israel-Stewart frameworks of conventional relativistic hydrodynamics by incorporating a spin tensor. The second approach discussed in Chapter~\ref{Quantum-statistical formulation}, is derived from first principles of relativistic quantum statistical mechanics and thermal quantum field theory, building upon the foundational Zubarev formalism for relativistic fluids.
\medskip

In the Navier–Stokes limit (Chapter~\ref{Navier-Stokes limit}) we obtained ten coupled evolution equations for ten dynamical variables in the zero chemical potential case. These variables are characterized by energy density, momentum density, and spin density. The procedure involved: \textbf{(i)} establishing the thermodynamics of the system including the spin tensor; \textbf{(ii)} identifying the dissipative currents and transport coefficients using conservation laws (derived in Chapter~\ref{From fields to fluids}) and entropy-current analysis; \textbf{(iii)} analyzing the linear stability of the evolution equations; and \textbf{(iv)} solving the resulting equations under Bjorken flow conditions. Apart from their inherent acausal nature due to their parabolic structure, we found that the evolution equations are generally unstable. Interestingly, we found that in the rest frame, at low-wavenumber, and under the Landau frame choice $h^{\mu} = 0$, the stability depends sensitively on the form of the spin equation of state. Similarly, under Bjorken flow, the obtained solutions are strongly influenced by the spin equation of state.
\medskip

In Chapter~\ref{MISMIS}, we extended the formulation to the Müller-Israel-Stewart limit by deriving 40 coupled evolution equations for 40 dynamical variables: 16 corresponding to the independent components of the energy-momentum tensor and 24 to the spin tensor. By ``formulation'', we mean the process of \textbf{(i)} determining the transport coefficients and deriving the relaxation-type dynamical equations of the dissipative currents via entropy-current analysis, and \textbf{(ii)} investigating the linear stability and causality properties of a simplified version of the evolution equations. As in conventional second-order relativistic hydrodynamics, the truncated evolution equations are stable in both the low- and high-wavenumber limits when evaluated in the Landau frame ($h^{\mu} = 0$); however, the stability again depends on the spin equation of state. Importantly, the equations were found to be causal.
\medskip

In Chapter~\ref{Quantum-statistical formulation}, we presented a first-principles derivation of first-order relativistic spin hydrodynamics using quantum-statistical methods. This formulation included: \textbf{(i)} the construction of a density operator incorporating the spin tensor, \textbf{(ii)} the derivation of the entropy current and \textbf{(iii)} the entropy production rate, \textbf{(iv)} the determination of the structure of dissipative currents at the first-order of hydrodynamic gradient expansion, and \textbf{(v)} the identification of the associated transport coefficients. To support this, in Appendix~\ref{AppendixC} we introduced a general method for performing the irreducible decomposition of tensors of arbitrary rank under spatial rotations, providing a powerful and broadly applicable tool for analyzing dissipative structures.
%
%
\section{Future plans}
A potential application of the theory developed in this thesis is the analysis of global and local spin polarization measurements in relativistic high-energy nuclear collisions. The two approaches we have presented offer distinct perspectives on this problem. To make contact with experiment, it is essential to develop a framework that translates the macroscopic outputs of relativistic spin hydrodynamics into particle-level observables at freeze-out. In conventional hydrodynamic simulations, such observables are typically extracted using the Cooper–Frye prescription. Extending this method to include spin requires a generalized version of the Cooper–Frye formula that allows the calculation of the average spin polarization vector of particles emitted at freeze-out -- a directly measurable quantity. 
\medskip

Several theoretical developments remain to be pursued. For the first (thermodynamic) approach developped in Chapters~\eqref{Navier-Stokes limit}–\eqref{MISMIS}, a key next step is to validate the thermodynamic relations using an underlying microscopic theory. Additionally, the hydrodynamic gradient ordering scheme for the spin potential remains unresolved due to the absence of a microscopic foundation for any specific counting method. Further progress also requires a detailed analysis of the stability and causality of the second-order evolution equations, both in a boosted frame and in the generalized Landau frame. A numerical
study of the resulting MIS equations would be among interesting future tasks.
\medskip

For the second (quantum-statistical) approach developed in Chapter~\eqref{Quantum-statistical formulation}, the next major task is to investigate all 23 transport coefficients in detail, ideally deriving them from a microscopic theory. Special attention should be given to those coefficients that govern spin dissipation, as they play a central role in the dynamics and phenomenology of spin hydrodynamics. Perhaps the most demanding theoretical goal is the systematic development of a second-order spin hydrodynamic framework using the second method, extending beyond the current formulation. 
%


\appendix
\chapter{Decomposition of an asymmetric rank-2 tensor with respect to the flow velocity}
%
%
\label{Appendix A}
In this appendix, we present a method to decompose an arbitrary asymmetric rank-2 tensor along and orthogonal to the flow four-velocity. It is important to note that the derivation shown here holds only 
when the entropy production of the theory is only driven by the gradients of flow four-velocity. In other words, it holds in the absence of the gradients of the spin potential. This is justified in the studies shown in Chapter~\ref{Navier-Stokes limit} due to the chosen counting scheme, where the spin potential is at least first-order in hydrodynamic gradient expansion. In such a case, its gradients are considered to be of higher order (see Eq.~\eqref{Xuentropyproduction}). Moreover, the decomposition strictly holds in the case of zero chemical potential, $\mu=0$, i.e., when there are no gradients of the chemical potential in the system. In Chapter~\ref{Quantum-statistical formulation} and related Appendix~\ref{AppendixC}, we go beyond these assumptions and irreducibly decompose the currents, considering all possible gradients in the system.
\medskip

In general, the energy-momentum tensor $T^{\mu\nu}$ is an asymmetric rank-2 tensor. By asymmetric, we mean that it can be decomposed into symmetric, $T^{\mu\nu}_{(S)}=(T^{\mu\nu}+T^{\nu\mu})/2\equiv T^{(\mu\nu)}$, and antisymmetric, $T^{\mu\nu}_{(A)}=(T^{\mu\nu}-T^{\nu\mu})/2\equiv T^{[\mu\nu]}$, part as  
 \begin{align}
T^{\mu\nu}=T^{\mu\nu}_{(S)}+T^{\mu\nu}_{(A)},
\end{align}
with both parts being non-zero. Moreover, in hydrodynamics, the energy-momentum tensor can be decomposed into parts of different orders in gradient expansion. Hence, the above equation  can be written as
\begin{align}
T^{\mu\nu}&=T^{\mu\nu}_{(0)}+T^{\mu\nu}_{(1)}=T^{\mu\nu}_{(0)}+T^{\mu\nu}_{(1S)}+T^{\mu\nu}_{(1A)},
\end{align}
where $T^{\mu\nu}_{(0)}$ denotes the leading order (or equilibrium) part, while $T^{\mu\nu}_{(1)}=T^{\mu\nu}_{(1S)}+T^{\mu\nu}_{(1A)}$ contains the symmetric and antisymmetric parts which are of the first order in gradients. 

Using the projector on the space orthogonal to the four-velocity $\Delta^{\mu\nu}=g^{\mu\nu}-u^{\mu}u^{\nu}$, we perform decomposition of the first and second index of $T^{\mu\nu}$ into the parts orthogonal and parallel to four-velocity,
\begin{align}
T^{\mu\nu}&=T_{\alpha\beta}g^{\alpha\mu}g^{\beta\nu}\nonumber\\
&=T_{\alpha\beta}\left(\Delta^{\alpha\mu}+u^{\alpha}u^{\mu}\right)(\Delta^{\beta\nu}+u^{\beta}u^{\nu})\nonumber\\
&=T_{\alpha\beta}u^{\alpha}u^{\beta}u^{\mu}u^{\nu}+T_{\alpha\beta}u^{\alpha}u^{\mu}\Delta^{\beta\nu}+T_{\alpha\beta}u^{\beta}u^{\nu}\Delta^{\alpha\mu}+T_{\alpha\beta}\Delta^{\alpha\mu}\Delta^{\beta\nu}\nonumber\\
&=\left(T_{(0)\alpha\beta}+T_{(1S)\alpha\beta}+T_{(1A)\alpha\beta}\right)u^{\alpha}u^{\beta}u^{\mu}u^{\nu}+\left(T_{(0)\alpha\beta}+T_{(1S)\alpha\beta}+T_{(1A)\alpha\beta}\right)u^{\alpha}u^{\mu}\Delta^{\beta\nu}\nonumber\\
&~~~+\left(T_{(0)\alpha\beta}+T_{(1S)\alpha\beta}+T_{(1A)\alpha\beta}\right)u^{\beta}u^{\nu}\Delta^{\alpha\mu}+\left(T_{(0)\alpha\beta}+T_{(1S)\alpha\beta}+T_{(1A)\alpha\beta}\right)\Delta^{\alpha\mu}\Delta^{\beta\nu}.
\end{align}
Rearranging terms in the above equation, we can rewrite the last equation as
\begin{align}
T^{\mu\nu}&=T_{\alpha\beta (0)}u^{\alpha}u^{\beta}u^{\mu}u^{\nu}+u^{\mu}\left(T_{\alpha\beta (1S)}+T_{\alpha\beta (1A)}\right)u^{\alpha}\Delta^{\beta\nu}+u^{\nu}\left(T_{\alpha\beta (1S)}+T_{\alpha\beta (1A)}\right)u^{\beta}\Delta^{\alpha\mu}\nonumber\\
&~~~~+\left(T_{\alpha\beta (0)}+T_{\alpha\beta (1S)}+T_{\alpha\beta (1A)}\right)\Delta^{\alpha\mu}\Delta^{\beta\nu}.
\end{align}
Above, we assumed that the first-order terms are orthogonal to the flow four-velocity,
\begin{align}
&T_{\alpha\beta(1S)}u^{\alpha}u^{\beta}=T_{\alpha\beta(1A)}u^{\alpha}u^{\beta}=0,
\end{align}
as well as 
\begin{align}
u^{\mu}T_{\alpha\beta(0)}u^{\alpha}\Delta^{\beta\nu}=u^{\nu}T_{\alpha\beta(0)}u^{\beta}\Delta^{\alpha\mu}=0.
\end{align}
The expressions appearing above define some fundamental quantities, which are listed in Table
\ref{tab:EMTdecompSymbols}.
\begin{table}[H]
    \centering
    \renewcommand{\arraystretch}{1.5} 
    \begin{tabular}{|c|c|}
        \hline
        \textbf{Symbol} & \textbf{Expression} \\ \hline
        $\varepsilon$ & $\left(T_{\alpha\beta(0)} u^{\alpha} u^{\beta}\right) u^{\mu} u^{\nu} = \varepsilon u^{\mu} u^{\nu}$ \\ \hline
        $P$ & $T_{\alpha\beta(0)} \Delta^{\alpha\mu} \Delta^{\beta\nu} = -P \Delta^{\mu\nu}$ \\ \hline
        $h^{\mu}$ & $u^{\nu} T_{\alpha\beta(1S)} u^{\beta} \Delta^{\alpha\mu} = u^{\nu} h^{\mu}$ \\ \hline
        $h^{\nu}$ & $u^{\mu} T_{\alpha\beta(1S)} u^{\alpha} \Delta^{\beta\nu} = u^{\mu} h^{\nu}$ \\ \hline
        $q^{\mu}$ & $u^{\nu} T_{\alpha\beta(1A)} u^{\beta} \Delta^{\alpha\mu} = u^{\nu} T_{\alpha\beta(1A)} u^{\beta} g^{\alpha\mu} = u^{\nu} q^{\mu}$ \\ \hline
        $q^{\nu}$ & $u^{\mu} T_{\alpha\beta(1A)} u^{\alpha} \Delta^{\beta\nu} = -u^{\mu} T_{\alpha\beta(1A)} u^{\beta} g^{\alpha\nu} = -u^{\mu} q^{\nu}$ \\ \hline
        $\phi^{\mu\nu}$ & $\frac{1}{2}T_{\alpha\beta(1A)} \Delta^{[\alpha\mu} \Delta^{\beta]\nu} = \phi^{\mu\nu}$ \\ \hline
        $\pi^{\mu\nu}+\Delta^{\mu\nu}\Pi$ & $T_{\alpha\beta(1S)} \Delta^{\alpha\mu} \Delta^{\beta\nu} = 
        \left[
            \frac{T_{\alpha\beta(1S)}}{2}
            \left( 
                \Delta^{\alpha\mu} \Delta^{\beta\nu} + \Delta^{\alpha\nu} \Delta^{\beta\mu} 
            \right) 
            - \frac{\Delta^{\mu\nu}}{3} \Delta_{\alpha\beta} T^{\alpha\beta}_{(1S)}
        \right] + \frac{\Delta^{\mu\nu}}{3} \Delta_{\alpha\beta} T^{\alpha\beta}_{(1S)}$ \\ \hline
    \end{tabular}
    \caption{Decomposition of the energy-momentum tensor with symbolic labels, in absence of gradients of spin potential and in the vanishing chemical potential limit.}
    \label{tab:EMTdecompSymbols}
\end{table}
\noindent
Consequently, the energy-momentum tensor can be expressed as, 
\begin{align}\label{EMTGENREALDECOM}
T^{\mu\nu}=&T^{\mu\nu}_{(0)}+T^{\mu\nu}_{(1S)}+T^{\mu\nu}_{(1A)}\nonumber\\
=&\varepsilon u^{\mu}u^{\nu}-P\Delta^{\mu\nu}+2h^{(\mu}u^{\nu)}+\pi^{\mu\nu}+\Pi\Delta^{\mu\nu}+2q^{[\mu}u^{\nu]}+\phi^{\mu\nu},
\end{align}
where 
\begin{align} 
& T^{\mu\nu}_{(1S)} = 2h^{(\mu}u^{\nu)}+\pi^{\mu\nu}+\Pi\Delta^{\mu\nu},\label{}\\ 
&T^{\mu\nu}_{(1A)} = 2q^{[\mu}u^{\nu]}+\phi^{\mu\nu}\label{}.
\end{align}
Finally, in the absence of dissipative currents, the energy-momentum reduces to its equilibrium form, 
\begin{align}
T^{\mu\nu}=T^{\mu\nu}_{(0)}=\varepsilon u^{\mu}u^{\nu}-P\Delta^{\mu\nu}.
\end{align}

\chapter{Decomposition of a rank-3 tensor, antisymmetric in the last two indices, with respect to the flow velocity}
\label{Appendix B}
In this appendix, we demonstrate how to decompose a rank-3 tensor that is antisymmetric in its last two indices into parts orthogonal and parallel to the four-velocity. The obtained result is used to express the spin tensor \( S^{\lambda\mu\nu}_{(1)} \) in Eq.~\eqref{eq17} in terms of lower-rank tensors with well-defined properties regarding index symmetries. We note that, although the following decomposition holds and is consistent with our assumptions, it is, as we shall show later, not fully irreducible.
\medskip

Consider an arbitrary rank-3 tensor $\phi^{\lambda\mu\nu}$ antisymmetric in the last two indices. Employing the decomposition of its first index into the parts orthogonal and parallel to four-velocity, we obtain 
\begin{align}
\phi^{\lambda\mu\nu}& =g^{\lambda}_{~\alpha} \phi^{\alpha\mu\nu} 
 = u^{\lambda}\gamma^{\mu\nu}+\Delta^{\lambda}_{~\alpha}\phi^{\alpha\mu\nu}  = u^{\lambda}\gamma^{\mu\nu}+\phi^{\langle\lambda\rangle\mu\nu}.
\label{equ88new1}
\end{align}
Here, we have used the projector $\Delta^{\lambda}_{~\alpha}$ that satisfies $g^{\lambda}_{~\alpha}=\Delta^{\lambda}_{~\alpha}+u^{\mu}u_{\lambda}$, and introduced the following notation: 
\begin{align}
&\gamma^{\mu\nu}\equiv u_{\alpha}\phi^{\alpha\mu\nu},\\
&\phi^{\langle\lambda\rangle\mu\nu}=\Delta^{\lambda}_{~\alpha}\phi^{\alpha\mu\nu},
\end{align}
where $\gamma^{\mu\nu}$ is an antisymmetric tensor. This immediately implies that $F^{\nu}\equiv u_{\mu}\gamma^{\mu\nu}$ satisfies $F\cdot u=0$. In the next step, we proceed with the decomposition of $\gamma^{\mu\nu}$ as follows, 
\begin{align}
\gamma^{\mu\nu}& =g^{\mu}_{~\rho}\gamma^{\rho\nu}\nonumber\\
&=(u^{\mu}u_{\rho}+\Delta^{\mu}_{~\rho})\gamma^{\rho\nu}\nonumber\\
&=u^{\mu}F^{\nu}+\gamma^{\langle\mu\rangle\nu}\nonumber\\
& = u^{\mu}F^{\nu}+g^{\nu}_{~\rho}\gamma^{\langle\mu\rangle\rho}\nonumber\\
&=u^{\mu}F^{\nu}+(u^{\nu}u_{\rho}+\Delta^{\nu}_{~\rho})\gamma^{\langle\mu\rangle\rho}\nonumber\\
& = u^{\mu}F^{\nu}+u^{\nu}u_{\rho}\gamma^{\langle\mu\rangle\rho}+\gamma^{\langle\mu\rangle\langle\nu\rangle}.
\label{equ89new1}
\end{align}
It can be easily shown that $u^{\nu}u_{\rho}\gamma^{\langle\mu\rangle\rho}=-u^{\nu}F^{\mu}$. Therefore, $\gamma^{\mu\nu}$ has the form, 
\begin{align}\label{B5}
\gamma^{\mu\nu}=u^{\mu}F^{\nu}-u^{\nu}F^{\mu}+\gamma^{\langle\mu\rangle\langle\nu\rangle}.
\end{align}
Subsequently, let us consider the last term in Eq.~\eqref{equ88new1} and employ the decomposition of its second index into the parts orthogonal and parallel to four-velocity,
\begin{align}
\phi^{\langle\lambda\rangle\mu\nu}&=g^{\mu}_{~\rho}\phi^{\langle\lambda\rangle\rho\nu}\nonumber\\
&=(u^{\mu}u_{\rho}+\Delta^{\mu}_{~\rho})\phi^{\langle\lambda\rangle\rho\nu}\nonumber\\
&=u^{\mu}u_{\rho}\phi^{\langle\lambda\rangle\rho\nu}+\phi^{\langle\lambda\rangle\langle\mu\rangle\nu}.
\end{align}
Defining $u_{\rho}\phi^{\langle\lambda\rangle\rho\nu}\equiv-\Sigma^{\lambda\nu}$, implies $u_{\lambda}\Sigma^{\lambda\nu}=0$. Therefore, the above equation further reads, 
\begin{align}
\phi^{\langle\lambda\rangle\mu\nu} & = \phi^{\langle\lambda\rangle\langle\mu\rangle\nu}-u^{\mu}\Sigma^{\lambda\nu}.
\end{align}
Finally, we employ the decomposition of the last index of $\phi^{\langle\lambda\rangle\langle\mu\rangle\nu}$ into the parts orthogonal and parallel to four-velocity,
\begin{align}
\phi^{\langle\lambda\rangle\mu\nu} & = g^{\nu}_{~\alpha}\phi^{\langle\lambda\rangle\langle\mu\rangle\alpha}-u^{\mu}\Sigma^{\lambda\nu}\nonumber\\
& = \phi^{\langle\lambda\rangle\langle\mu\rangle\langle\nu\rangle}+u^{\nu}u_{\alpha}\phi^{\langle\lambda\rangle\langle\mu\rangle\alpha}-u^{\mu}\Sigma^{\lambda\nu}\nonumber\\
& = \phi^{\langle\lambda\rangle\langle\mu\rangle\langle\nu\rangle}+u^{\nu}\Sigma^{\lambda\mu}-u^{\mu}\Sigma^{\lambda\nu}.
\label{equ92new1}
\end{align}
Substituting Eqs.~\eqref{B5} and \eqref{equ92new1} into Eq.~\eqref{equ88new1}, the decomposition of the tensor $\phi^{\lambda\mu\nu}$ reads, 
\begin{align}\label{phibeforefinal}
\phi^{\lambda\mu\nu}=u^{\lambda}\left(u^{\mu}F^{\nu}-u^{\nu}F^{\mu}+\gamma^{\langle\mu\rangle\langle\nu\rangle}\right)+u^{\nu}\Sigma^{\lambda\mu}-u^{\mu}\Sigma^{\lambda\nu}+\phi^{\langle\lambda\rangle\langle\mu\rangle\langle\nu\rangle}.
\end{align}
Here, we can introduce the following notation $S^{\mu\nu}\equiv u^{\mu}F^{\nu}-u^{\nu}F^{\mu}+\gamma^{\langle\mu\rangle\langle\nu\rangle}$. 
Since $\Sigma^{\mu\nu}$ is asymmetric and orthogonal to $u^{\mu}$ it can also be decomposed into a symmetric part denoted by $\Sigma_{(s)}^{\mu\nu}$ and an antisymmetric part denoted by $\Sigma_{(a)}^{\mu\nu}$. The symmetric part can be decomposed into a trace $\Sigma$ and a traceless part $\Sigma_{(s)}^{\langle\mu\nu\rangle}$. Therefore, Eq.~\eqref{phibeforefinal} can be further rewritten as follows,
\begin{align}
\phi^{\lambda\mu\nu}=& u^{\lambda}{S}^{\mu\nu}+\left(u^{\nu}\Delta^{\lambda\mu}-u^{\mu}\Delta^{\lambda\nu}\right)\Sigma+\left(u^{\nu}\Sigma_{(s)}^{\langle\lambda\mu\rangle}-u^{\mu}\Sigma_{(s)}^{\langle\lambda\nu\rangle}\right)+\left(u^{\nu}\Sigma_{(a)}^{\lambda\mu}-u^{\mu}\Sigma_{(a)}^{\lambda\nu}\right)\nonumber\\
&+\phi^{\langle\lambda\rangle\langle\mu\rangle\langle\nu\rangle}.
\end{align}
The first term in the expression above is parallel to the four-velocity, whereas the remaining terms are orthogonal. The presence of the last term $\phi^{\langle\lambda\rangle\langle\mu\rangle\langle\nu\rangle}$ in the decomposition suggests that it cannot be regarded as fully irreducible; however, when considered as a method to simplify the analysis of the tensor structure of the equations of motion, the decomposition remains valid.
\medskip

As a consistency check, let us count the number of independent components on both sides of the decomposition. Given that \( \phi^{\lambda\mu\nu} \) is a rank-3 tensor that is antisymmetric in its last two indices, it follows that it has 24 independent components or degrees of freedom. Therefore, for consistency, the total number of independent components of all terms on the right-hand side must sum to 24. The tensor ${S}^{\mu\nu}$ has 6 DOF, and $\Sigma$ is a scalar, hence, it has only one DOF. $\Sigma_{(s)}^{\langle\mu\nu\rangle}$ is symmetric, traceless, and orthogonal to the fluid flow vector, hence, it has 5 DOF. The tensor $\Sigma_{(a)}^{\mu\nu}$ is antisymmetric and transverse to the fluid flow; hence, it has 3 DOF. Finally, $\phi^{\langle\lambda\rangle\langle\mu\rangle\langle\nu\rangle}$ is antisymmetric in the last two indices and orthogonal to flow vector in all indices; hence, it has only 9 DOF. 
%
%
%
%
%
%
%
%
%
%
%
%
%
\chapter{Extensivity of the logarithm of the partition function in the presence of a spin tensor}
%
%
\label{Appendix extensivity}
Here, we extend the proof of extensivity of the logarithm of the partition function $\log Z_{\rm LE}$ and the expression for the thermodynamic potential current $\phi^{\mu}$, originally derived in Ref.~\cite{Becattini:2019poj}, by incorporating the contribution of the spin tensor~\multimyref{AD5}. 
\medskip 

The first step involves modifying the local equilibrium density operator~\eqref{densityoperator} by introducing a dimensionless parameter \( \lambda \), defined as:
\begin{align}\label{rholambdaapp}
&\wrho_{\rm LE}(\lambda)=\frac{1}{Z_{\rm LE}(\lambda)}\exp\left[-\lambda\int_{\Sigma} \di\,\Sigma_{\mu}\,\left(\wT^{\mu\nu}\beta_{\nu}
-\zeta \wj^\mu - \frac{1}{2}\Omega_{\lambda\nu}\wspt^{\mu\lambda\nu}\right)\right]\,,\\
&Z_{\rm LE}(\lambda)=\Tr\left\{\exp\left[-\lambda\int_{\Sigma} \di\,\Sigma_{\mu}\,\left(\wT^{\mu\nu}\beta_{\nu}
-\zeta \wj^\mu - \frac{1}{2}\Omega_{\lambda\nu}\wspt^{\mu\lambda\nu}\right)\right]\right\}.
\end{align}
This is deliberately constructed to recover the original density operator~\eqref{densityoperator} and partition function~\eqref{partitionfunctionoriginal} when \( \lambda = 1 \). Differentiating the logarithm of the \( \lambda \)-dependent partition function with respect to \( \lambda \) yields:
%
%
%
%

\begin{align}
\frac{\partial \log Z_{\rm LE}(\lambda)}{\partial \lambda}
&= \int_{\Sigma} \di\Sigma_{\mu} \Big[ 
    -\beta_{\nu} \Tr(\wrho_{\rm LE}(\lambda) \wT^{\mu\nu}) 
    + \zeta \Tr(\wrho_{\rm LE}(\lambda) \wj^{\mu}) \notag \\
&\qquad\qquad\quad
    + \frac{1}{2} \Omega_{\lambda\nu} \Tr(\wrho_{\rm LE}(\lambda) \wspt^{\mu\lambda\nu}) 
\Big]
\end{align}

\noindent On the other hand, integrating this equation  from a lower bound \( \lambda_0 \) to \( \lambda = 1 \) gives:
\begin{align}\label{intpar}
\int_{\lambda_{0}}^{1} \frac{\partial\log Z_{\rm LE}(\lambda)}{\partial \lambda}\di\lambda=\log Z_{\rm LE}(1)-\log Z_{\rm LE}(\lambda_{0}).
\end{align}
To determine $\lambda_{0}$, we introduce the operator:
\begin{align}
\widehat \Upsilon \equiv \int_{\Sigma} \di\,\Sigma_{\mu}\,\left(\wT^{\mu\nu}\beta_{\nu}
-\zeta \wj^\mu - \frac{1}{2}\Omega_{\lambda\nu}\wspt^{\mu\lambda\nu} \right)~\implies~Z_{\rm LE}(\lambda)=\Tr(e^{-\lambda\widehat \Upsilon }).
\end{align}
Suppose that $\widehat \Upsilon$ has discrete, non-degenerate eigenvalues \( \Upsilon_{n} \), with degeneracy \( g_n=1 \), and is bounded from below -- i.e., there exists a unique lowest eigenvalue $\Upsilon_0$ and corresponding eigenvector $\ket{0}$. Then we can write:
\begin{align}\label{expansionqm}
Z_{\rm LE}(\lambda)&=\Tr(e^{-\lambda \widehat \Upsilon}) = \sum_n e^{-\lambda \Upsilon_n} 
=e^{-\lambda\Upsilon_{0}}\left(1+e^{-\lambda(\Upsilon_{1}-\Upsilon_{0})}+e^{-\lambda(\Upsilon_{2}-\Upsilon_{0})}+....\right).
\end{align}
If we further assume $\Upsilon_{0}=0$ and take the limit $\lambda\rightarrow +\infty$, we find: 
\begin{align}\label{resulttt}
\lim_{\lambda \to \infty}Z_{\rm LE}(\lambda)=1~\implies~\lim_{\lambda \to \infty}\log Z_{\rm LE}(\lambda)=0.
\end{align}
Therefore, choosing $\lambda_{0}=+\infty$, Eq.~\eqref{intpar} becomes
\begin{align}\label{D8}
\int_{+\infty}^{1} \frac{\partial\log Z_{\rm LE}(\lambda)}{\partial \lambda}\di\lambda=\log Z_{\rm LE}.
\end{align}
To proceed, we use the fact that adding non-operator constants to the exponent of the density operator does not affect its physical content, as long as the same constant is subtracted from the partition function. For instance
\begin{align}
\wrho_{\rm LE}^{~\prime}&=\frac{\exp\left[-\int_{\Sigma} \di\,\Sigma_{\mu}\,\left(\wT^{\mu\nu}- \bra{0} \widehat{T}^{\mu\nu} \ket{0}\right)\beta_{\nu}\right]}{\Tr\left(\exp\left[-\int_{\Sigma} \di\,\Sigma_{\mu}\,\left(\wT^{\mu\nu}- \bra{0} \widehat{T}^{\mu\nu} \ket{0}
\right)\beta_{\nu}\right]\right)}\,,\nonumber\\
&=\frac{\exp\left[-\int_{\Sigma} \di\,\Sigma_{\mu}\,\left(\wT^{\mu\nu}\beta_{\nu}\right)\right]}{\Tr\left(\exp\left[-\int_{\Sigma} \di\,\Sigma_{\mu}\,\left(\wT^{\mu\nu}\beta_{\nu}
\right)\right]\right)},\nonumber\\
&=\wrho_{\rm LE}.
\end{align}
For simplicity, in the above, we omitted the parts related to $\wj^{\mu}$ and $\wspt^{\mu\lambda\nu}$. Thus, we define the shifted operator:
\begin{align}
\widehat \Upsilon \rightarrow  \widehat \Upsilon^{\prime} &= \widehat \Upsilon - \Upsilon_{0} \nonumber\\
&= \widehat \Upsilon - \bra{0}\widehat \Upsilon\ket{0} \nonumber\\
&= \int_{\Sigma} \mathrm{d}\Sigma_{\mu} \left[ \left(\widehat{T}^{\mu\nu} - \bra{0} \widehat{T}^{\mu\nu} \ket{0} \right) \beta_{\nu} 
- \zeta \left(\widehat{j}^{\mu} - \bra{0} \widehat{j}^{\mu} \ket{0}\right) \right. \nonumber\\
&\quad~~~~~~~~~~~~\left. - \frac{1}{2} \Omega_{\lambda\nu} \left(\widehat{S}^{\mu\lambda\nu} - \bra{0} \widehat{S}^{\mu\lambda\nu} \ket{0} \right) \right].
\end{align}
Accordingly, the $\lambda$-dependent partition function becomes:
\begin{align}
Z_{\rm LE}(\lambda)&=\Tr[e^{-\lambda \widehat \Upsilon}]\nonumber\\
&=\Tr\bigg(\exp\bigg[
-\lambda\int_{\Sigma} \mathrm{d}\Sigma_{\mu}\left(\widehat{T}^{\mu\nu} - \bra{0} \widehat{T}^{\mu\nu} \ket{0} \right) \beta_{\nu} 
- \zeta \left(\widehat{j}^{\mu} - \bra{0} \widehat{j}^{\mu} \ket{0}\right)\nonumber\\
&~~~~~~~~~~~~~~~~~- \frac{1}{2} \Omega_{\lambda\nu} \left(\widehat{S}^{\mu\lambda\nu} - \bra{0} \widehat{S}^{\mu\lambda\nu} \ket{0} \right) \bigg]\bigg).
\end{align}
Repeating the steps from Eqs.~\eqref{expansionqm}-\eqref{D8}, we arrive at:
\begin{align}
\log Z_{\rm LE}&=-\int^{+\infty}_{1}\frac{\partial \log Z_{\rm LE}(\lambda)}{\partial \lambda}\di\lambda\nonumber\\
&= \int_\Sigma \di \Sigma_\mu \; \phi^\mu 
- \bra{0} \widehat \Upsilon \ket{0}\nonumber\\
&= \int_\Sigma \di \Sigma_\mu \; \left[ \phi^\mu 
- \bra{0}(\wT^{\mu\nu}\beta_{\nu} -\zeta \wj^\mu - \frac{1}{2}\Omega_{\lambda\nu}\wspt^{\mu\lambda\nu})\ket{0} \right].
\end{align}
Here, the thermodynamic potential current is given by:
\begin{align}
\phi^{\mu}=\int_{1}^{\infty} \di\lambda \; \left( T^{\mu\nu}_{\rm LE}(\lambda) \beta_{\nu}
 -\zeta j^\mu_{\rm LE}(\lambda) - \frac{1}{2}\Omega_{\lambda\nu} \spt^{\mu\lambda\nu}_{\rm LE}(\lambda)\right),
\end{align} 
which completes the proof that $\log Z_{\rm LE}$ is an extensive quantity. 
Further discussion on the thermodynamic potential current is provided in Chapter~\ref{Quantum-statistical formulation}.

%
\chapter{Decomposition of rank-$n$ tensors into irreducible components under $\mathrm{SO}(3,\mathbb{R})$}
\vspace{0.5em}

%
\label{AppendixC}
Experts in representation theory may proceed directly to Sec.~\ref{Application in relativistic hydrodynamics}. The primary goal of this appendix is to introduce a method for decomposing arbitrary rank-n tensors into:
\begin{enumerate}
\item Irreducible components under rotations, as in Eqs.~\eqref{thermalsheardecomposition}–\eqref{spinpotentialgradientdecompositionnn}. This is the focus of Sec.~\ref{ApplicationI}.
\item Irreducible components under rotations, restricted to rotation-invariant terms, as exemplified by the tensor coefficients in Tables~\ref{Table-G-I-O-F}–\ref{Table-W}. This is the subject of Sec.~\ref{Application II}.
\end{enumerate}
Moreover, beyond its direct application in Chapter~\ref{Quantum-statistical formulation}, we apply the method in Sec.~\ref{applicationnn} to conventional relativistic hydrodynamics, where it is used to reproduce the well-known dissipative currents in the Landau frame.

The foundational concepts in the prerequisites section are loosely based on Refs.~\cite{gilmore2006lie,tung1985group} and the references therein. The method presented in Sec.~\ref{Application in relativistic hydrodynamics} is based exclusively on Ref.~\multimyref{AD8}.
%
%
\section{Representation theory prerequisites}
\label{Non-relativistic representations of SO(3) group} 
In this section, we introduce fundamental concepts from the representation theory of the group $\text{SO}(3, \mathbb{R})$, providing the necessary mathematical tools for the following section. The group $\text{SO}(3, \mathbb{R})$ consists of all real \(3 \times 3\) orthogonal matrices with determinant one; it describes the set of all proper rotations in three-dimensional Euclidean space.
\medskip 

A group \(\mathrm{G}\) is said to have a (linear) \emph{representation} \(\mathbf{T}\) on an \(n\)-dimensional vector space \(\mathcal{V}\) if \(\mathbf{T}\) is a homomorphism that assigns to each element \(\mathbf{g} \in \mathrm{G}\) an \(n \times n\) invertible linear transformation in the general linear group \(\mathrm{GL}(\mathcal{V})\), which acts on \(\mathcal{V}\). Formally, this can be written as:
\begin{align*}
\mathbf{T}: & \quad \mathrm{G} \rightarrow \mathrm{GL}(\mathcal{V}) \\
& \quad \mathbf{g} \mapsto \mathbf{T}(\mathbf{g})
\end{align*}
such that \(\mathbf{T}(\mathbf{g}_1 \mathbf{g}_2) = \mathbf{T}(\mathbf{g}_1) \mathbf{T}(\mathbf{g}_2)\) for all \(\mathbf{g}_1, \mathbf{g}_2 \in \mathrm{G}\), and \(\mathbf{T}(\mathbf{e}) = \mathbf{I}\), where \(\mathbf{e}\) is the identity elements of the group and \(\mathbf{I}\in\mathrm{GL}(\mathcal{V})\) is the identity transformation on \(\mathcal{V}\). The dimension of the vector space \(\mathcal{V}\) is referred to as the \emph{dimension} or \emph{degree} of the representation.

A representation is said to be \emph{reducible} if there exists a proper, nontrivial subspace of \(\mathcal{V}\) that is invariant under the action of the representation. By a \emph{proper, nontrivial} subspace, we mean one that is neither the \emph{zero subspace} \(\{0\}\) nor the entire space \(\mathcal{V}\). 


When we say that a subspace is invariant under a representation, we do not mean that the individual elements of the subspace remain unchanged under the group action. Rather, we mean that the subspace is mapped into itself — that is, the action of the group sends any element of the subspace to another element within the same subspace.
For example, if \(\mathcal{V}\) is a space of tensors and $\mathbf{A}\in\mathcal{V}$ is an antisymmetric tensor prior to the transformation, then $\mathbf{A}^{\prime}$ will also be antisymmetric afterward. 

A representation is said to be \emph{irreducible} if the only invariant subspaces of \(\mathcal{V}\) are the zero subspace and \(\mathcal{V}\) itself. 
%
%
\subsection{Three-dimensional (or vector) representation}
\label{Vector(3)-representation}
The most commonly used representation of $\mathrm{SO}(3, \mathbb{R})$ is the three-dimensional representation, also known as the \emph{vector representation}. In this case, an element \({\theta} \in\)  $\mathrm{SO}(3, \mathbb{R})$ is mapped to a $3 \times 3$ rotation matrix that acts on vectors in three-dimensional real vector space $\mathbb{R}^{3}$, such that:
\begin{align}
\textbf{R}:~ &\mathrm{SO}(3, \mathbb{R})\longrightarrow \mathrm{GL}(3, \mathbb{R}) \nonumber\\
&~~~~~~~~~~\theta \longrightarrow  \textbf{R}(\theta)\quad \text{such that}~~\textbf{R}(\theta) \mathbf{V} = \mathbf{V}^{'}
\end{align}
Here, \(\textbf{R}(\theta)\) is an orthogonal matrix with determinant one, and $\mathbf{V}\in\mathbb{R}^{3}$. 

What if the underlying vector space is not three-dimensional? In that case, the representation space is no longer composed of \(3 \times 3\) matrices. This highlights the fact that a given group may admit multiple, inequivalent representations -- a topic explored in the following subsections.
%
%
\subsection{One-dimensional (or scalar) representation}
\label{Scalar(1)-representation}
The simplest representation of \(\mathrm{SO}(3, \mathbb{R})\) is the one-dimensional, or \emph{scalar representation} (also known as the \emph{trivial representation}). It acts on the one-dimensional real vector space \( \mathbb{R} \) by mapping every element \( \theta \in \mathrm{SO}(3, \mathbb{R}) \) to the identity transformation. Formally, it is defined as:
\begin{align}\label{scalaringeneral_unique}
\textbf{R}:~ &\mathrm{SO}(3, \mathbb{R}) \longrightarrow \mathrm{GL}(1, \mathbb{R}) \nonumber \\
&~~~~~~~~~~\theta \longrightarrow  \mathbf{R}(\theta) = 1,
\end{align}
such that for all \( \mathbf{V} \in \mathbb{R} \),
\begin{equation}
\textbf{R}(\theta) \mathbf{V} = \mathbf{V}.
\end{equation}
That is, the scalar representation leaves all elements invariant.
\subsection{Nine-dimensional (or rank-2 tensor) representation}
\label{Matrix (9)-representation}
The nine-dimensional representation of $\mathrm{SO}(3, \mathbb{R})$, often referred to as the \emph{rank-2 tensor representation},  acts on the space of rank-2 tensors $\mathbf{X}_{ij}$ (or equivalently, on a nine-dimensional vector space \( \mathcal{V} \)). Each element element \( \theta \in\) $\mathrm{SO}(3, \mathbb{R})$ is mapped to a \( 9 \times 9\) matrix whose components can be expressed in terms of products \( \textbf{R}_{ai} \textbf{R}_{bj} \), corresponding to the transformation of rank-2 tensors\footnote{This nine-dimensional space corresponds to the tensor product $\mathbb{R}^3\otimes \mathbb{R}^3$, on which $\mathrm{SO}(3)$ acts via \( \textbf{R}_{ai} \textbf{R}_{bj} \).}. This representation acts on a rank-2 tensor \( \textbf{X}_{ij} \) as follows
\begin{align}
\textbf{X}_{ab}^\prime &
= \sum_{i,j=1}^{3} \textbf{R}_{ai} \textbf{R}_{bj} \textbf{X}_{ij}.
\end{align}
This approach can be extended to define higher-dimensional \emph{tensor representations} of \(\mathrm{SO}(3, \mathbb{R})\), which act on higher-rank tensors or higher-dimensional vector spaces. However, in what follows, we restrict our attention to rank-2 tensors, as this level of generality is sufficient for the purposes of the next sections. 
%
%
\subsection{Irreducible nine-dimensional (or rank-2 tensor) representation}
%
A pertinent question arises: Is the nine-dimensional representation reducible? According to the discussion above, we should check whether the space of rank-2 tensors has any proper, nontrivial subspaces that are invariant under the action of the representation. To do this, recall that any rank-2 tensor (or \( 3 \times 3\) matrix) can be decomposed into its symmetric and antisymmetric parts:
\begin{align}\label{decompositioneuclidean}
\textbf{X}_{ij}=\frac{1}{2}(\textbf{X}_{ij}+\textbf{X}_{ji})+\frac{1}{2}(\textbf{X}_{ij}-\textbf{X}_{ji})=\textbf{S}_{ij}+\textbf{A}_{ij},
\end{align}
where $\textbf{S}_{ij}$ and $\textbf{A}_{ij}$ are symmetric and antisymmetric parts, respectively. This decomposition defines two subspaces. To determine whether the representation is reducible, we must check whether these subspaces are invariant under the group action. Consider the antisymmetric part:
\begin{align}\label{Aabinvariant}
\textbf{A}_{ab}^{\prime}&=\frac{1}{2}(\textbf{X}_{ab}^{\prime}-\textbf{X}_{ba}^{\prime})\nonumber\\
&=\frac{1}{2}\sum_{i,j=1}^{3}(\textbf{R}_{ai}\textbf{R}_{bj}\textbf{X}_{ij}-\textbf{R}_{bi}\textbf{R}_{aj}\textbf{X}_{ij})\nonumber\\
&=\frac{1}{2}\sum_{i,j=1}^{3}\textbf{R}_{ai}\textbf{R}_{bj}(\textbf{X}_{ij}-\textbf{X}_{ji})=\sum_{i,j=1}^{3}\textbf{R}_{ai}\textbf{R}_{bj}\textbf{A}_{ij}.
\end{align}
Thus, the antisymmetric part $\textbf{A}_{ij}$ transforms into another antisymmetric tensor, showing that the subspace of antisymmetric tensors is invariant. A similar argument applies to the symmetric part \( \textbf{S}_{ij} \), confirming that it also forms an invariant subspace. Therefore, the nine-dimensional representation acting on the space of rank-2 tensors is reducible since it preserves the decomposition into symmetric and antisymmetric subspaces. 

To reduce the nine-dimensional representation into less dimensional representations, we examine the dimensions of the invariant subspaces. The space of symmetric $3\times 3$ matrices \( \textbf{S}_{ij} \) has 6 independent components (degrees of freedom), while the antisymmetric part \( \textbf{A}_{ij} \) has 3. Thus, the nine-dimensional representation decomposes as a direct sum:
\begin{align}
    \textbf{9}=\textbf{6} \oplus \textbf{3} 
\end{align}
Here, the notation \(\textbf{9}=\textbf{6} \oplus \textbf{3} \) indicates that the original nine-dimensional representation decomposes into six-dimensional subrepresentation acting on \( \textbf{S}_{ij} \) and a three-dimensional subrepresentation acting on \( \textbf{A}_{ij} \).

Are these subrepresentations themselves reducible? To answer this, consider the trace of the symmetric tensor and check if it forms a subspace that is invariant under the corresponding representation: 
\begin{align}\label{traceeuclidean}
\textbf{S}^{\prime}&=\sum_{a=1}^{3}\textbf{S}_{aa}^{\prime}\nonumber\\
&=\sum_{a=1}^{3}\sum_{i,j=1}^{3}\textbf{R}_{ai}\textbf{R}_{aj}\textbf{S}_{ij}\nonumber\\
&=\sum_{i,j=1}^{3}\left(\sum_{a=1}^{3}\textbf{R}^{\mathrm{T}}_{ia}\textbf{R}_{aj}\right)\textbf{S}_{ij}\nonumber\\
&=\sum_{i,j=1}^{3}\delta_{ij}\textbf{S}_{ij}=\textbf{S}.
\end{align}
Hence, the trace is invariant under the action of the group. This scalar forms a one-dimensional invariant subspace. The remaining part of \( \textbf{S}_{ij} \) -- its traceless symmetric part -- then spans a five-dimensional invariant subspace. As a result, the symmetric representation can be further decomposed:
\begin{align}
\textbf{6} = \textbf{5} \oplus \textbf{1}.  
\end{align}
Thus, the full decomposition of the original nine-dimensional representation is
\begin{align}\label{9=5x3x1}
\textbf{9} = (\textbf{1} \oplus \textbf{3} \oplus \textbf{5}).
\end{align}

The subspaces correspond respectively to:
\begin{itemize}
\item the scalar (trace part of \( \textbf{S}_{ij} \)),
\item the antisymmetric tensors \( \textbf{A}_{ij} \), and
\item the symmetric traceless tensors.
\end{itemize}

An alternative notation is sometimes used to label the degree of the irreducible representations by the angular momentum quantum number \(\mathbf{j}\)  (being a non-negative integer), where the dimension of each irreducible representation is given by \(\mathbf{2j+1}\). Using this, the decomposition~\eqref{9=5x3x1} becomes:
\begin{align}\label{secondnotation}
\textbf{9} = (\textbf{1} \oplus \textbf{3} \oplus \textbf{5})\equiv (\mathbf{0} \oplus \mathbf{1} \oplus \mathbf{2}).
\end{align}
where:
\begin{itemize}
\item $\textbf{j}=\textbf{0}$ corresponds to dimension  $\mathbf{2(0)+1 = 1}$, 
\item $\textbf{j}=\textbf{1}$  corresponds to dimension $\mathbf{2(1)+1 = 3}$,
\item $\textbf{j}=\textbf{2}$ corresponds to dimension $\mathbf{2(2)+1 = 5}$.
\end{itemize} 

%
%
For now, we simply treat this as a change in notation, and we leave aside the underlying representation theory behind the \textbf{2j+1} rule. 
%
\subsection{Irreducible representations: Minkowski spacetime}
\label{Irriduciblerepresentationandhydrodynamics}
%
%
%

In relativistic theories formulated in Minkowski spacetime, the three-dimensional and nine-dimensional representations of $\mathrm{SO}(3,\mathrm{R})$ discussed previously are limited in scope, as they act solely on spatial components. While we do not consider the full Lorentz group here, it is important to understand how representations of $\mathrm{SO}(3,\mathbb{R})$ operate within the broader structure of Minkowski space. 
\medskip

Consider a four-dimensional vector space equipped with the Minkowski metric. We can define a four-dimensional representation in which \({\theta} \in \mathrm{SO}(3, \mathbb{R})\) is mapped to a \(4 \times 4\) matrix \(\mathbf{R}(\theta)\), acting on four-vectors in Minkowski spacetime as:
\begin{align}\label{abstractdefofff}
\textbf{R}^{\mu}_{~\nu}\textbf{V}^{\nu}=\textbf{V}^{\mu\,\prime}.
\end{align}
We now ask: Is this representation reducible? That is, does the space of four-vectors $\textbf{V}^{\mu}$ contain a proper, nontrivial subspace that remains invariant under the group action? 

From the transformation law
\begin{align}\label{abstractdefof}
\textbf{V}^{0\,\prime}=\textbf{V}^{0}~~,~~\textbf{V}^{i\prime}=\textbf{R}^{i}_{~j}\textbf{V}^{j},
\end{align}
we see that the time component $\mathbf{V}^{0}$ remains unchanged, and the spatial components $\mathbf{V}^{i}$ transform under the three-dimensional rotation representation. Thus, the subspaces spanned by  $\mathbf{V}^{0}$ and by the spatial components $\mathbf{V}^{i}$ are invariant under the group action. Therefore, the four-dimensional representation is reducible.

Based on the number of degrees of freedom in each invariant subspace, the four-dimensional representation can be decomposed into a one-dimensional (or scalar) representation and a three-dimensional (or vector) representation:
\begin{align}
\textbf{4}=\textbf{1}\oplus\textbf{3},
\end{align}
or, using the $\mathbf{j}$-notation from Eq.~\eqref{secondnotation},
\begin{align}\label{vectorused}
\textbf{0} \oplus \textbf{1}.
\end{align}

Next, consider the space of \(4\times4\) matrices (or 2-tensors), forming a sixteen-dimensional vector space. We define a representation in which \({\theta} \in\) $\mathrm{SO}(3, \mathbb{R})$ is mapped to a $16 \times 16$ matrix with components {\(\textbf{R}^{\mu}_{~\lambda}\textbf{R}^{\nu}_{~\delta}\)}, acting on rank-2 tensors (e.g., the energy-momentum tensor $\textbf{T}^{\lambda\delta}$) as 
\begin{align}
\textbf{T}^{\mu\nu\,\prime}= \mathbf{R}^{\mu}_{\ \lambda} \mathbf{R}^{\nu}_{\delta} \textbf{T}^{\lambda\delta}.
\end{align}
 
To determine whether this representation is reducible, we as whether the space of \(4\times4\) tensors admits proper invariant subspaces. As with the previous cases, any rank-2 tensor can be decomposed into symmetric and antisymmetric parts,
\begin{align}
\textbf{T}^{\mu\nu} = \frac{1}{2}(\textbf{T}^{\mu\nu} + \textbf{T}^{\nu\mu}) + \frac{1}{2}(\textbf{T}^{\mu\nu} - \textbf{T}^{\nu\mu}) = \textbf{S}^{\mu\nu} + \textbf{A}^{\mu\nu},
\end{align}
Using arguments analogous to Eq.~\eqref{Aabinvariant}, one can show that both $\textbf{S}^{\mu\nu}$ and $\textbf{A}^{\mu\nu}$ form invariant subspaces.

Based on the number of degrees of freedom, the dimension of the invariant subspaces provided by $\textbf{S}^{\mu\nu}$ and $\textbf{A}^{\mu\nu}$ are ten and six, respectively, hence 
\begin{align}
\textbf{16}=\textbf{10}\oplus\textbf{6}.
\end{align}

The symmetric tensor $\textbf{S}^{\mu\nu}$ has ten independent components. This space can be further decomposed. Although the trace $\textbf{S}=g_{\mu\nu}\textbf{S}^{\mu\nu}$ is a Lorentz scalar, in the context of $\mathrm{SO}(3, \mathbb{R})$, it can be expressed as a sum of contributions from temporal and spatial components:
\begin{align}
\textbf{S}&=g_{\mu\nu}\textbf{S}^{\mu\nu} =g_{00}\textbf{S}^{00}+g_{ii}\textbf{S}^{ii} =\textbf{S}_{(1)}+\textbf{S}_{(2)}.
\end{align}
Both $\textbf{S}_{(1)}$ and $\textbf{S}_{(2)}$ are invariant under the action of the group, as shown using a procedure analogous to Eq.~\eqref{traceeuclidean}. Therefore, we can decompose the ten-dimensional space as:
\begin{align}
\textbf{10}=\textbf{1}\oplus\textbf{1}\oplus\textbf{8}.  
\end{align}

The remaining traceless components of $\textbf{S}^{\mu\nu}$ include the traceless symmetric spatial part $\textbf{S}^{\mu\nu}_{\text{traceless}}=\textbf{S}^{ij}_{\text{traceless}}$, and the vector-like components $\textbf{S}^{\mu0}$. Using the same procedure as in Eq.~\eqref{Aabinvariant}, one can show that both $\textbf{S}^{ij}_{\text{traceless}}$ and $\textbf{S}^{\mu0}$ form subspaces that are invariant under the action of $\mathrm{SO}(3, \mathbb{R})$. The number of degrees of freedom of $\textbf{S}^{ij}_{\text{traceless}}$ is five while $\textbf{S}^{\mu0}$ has three degrees of freedom. Therefore, the full decomposition of the ten-dimensional symmetric representation is:
\begin{align}
\textbf{10}=\textbf{1}\oplus\textbf{1}\oplus\textbf{3}\oplus\textbf{5},
\end{align}
or in $\textbf{j}$ notation,
\begin{align}\label{symmetricused}
\textbf{0}\oplus\textbf{0}\oplus\textbf{1}\oplus\textbf{2}.
\end{align}

Finally, we consider the antisymmetric part $\mathbf{A}^{\mu\nu}$, which spans a six-dimensional space.  The components of $\mathbf{A}^{\mu\nu}$ split into the spatial antisymmetric components $\mathbf{A}^{ij}$ and the time-space components $\mathbf{A}^{0i}$. Using the procedure from in Eq.~\eqref{Aabinvariant} as before, one may show that both $\mathbf{A}^{ij}$ and $\mathbf{A}^{0i}$ form proper nontrivial subspaces that are invariant under the action of the group. Based on the number of degrees of freedom (both $\mathbf{A}^{ij}$ and $\mathbf{A}^{0i}$ have three), the decomposition of the six-dimensional representation into irreducible parts reads
\begin{align}
\textbf{6}=\textbf{3}\oplus\textbf{3},
\end{align}
or in $\mathbf{j}$ notation,
\begin{align}\label{antisymmused}
\textbf{1} \oplus \textbf{1}.
\end{align}
To summarize, the sixteen-dimensional representation decomposes into irreducible 
representations as $\textbf{16}=(\textbf{1}\oplus\textbf{1}\oplus\textbf{3}\oplus\textbf{5})\oplus(\textbf{3}\oplus\textbf{3})$, or equivalently as $(\textbf{0}\oplus\textbf{0}\oplus\textbf{1}\oplus\textbf{2})\oplus(\textbf{1}\oplus\textbf{1})$.
\section{The method}
\label{Application in relativistic hydrodynamics}
%
%
%
%
%
%


%
%
\subsection{Basis and building blocks}
To this end, we consider a fluid element that is isotropic in the local rest frame, while assuming that the thermal velocity $\beta^\mu = u^\mu / T$ is constant and that the thermal vorticity $\varpi = 0$, to avoid any potential anisotropy that could be induced by thermal vorticity. In this scenario, the basis for decomposing tensors consists of the fluid four-velocity $u^\mu$, the spatial projector $\Delta^{\mu\nu}$, and the rank-3 pseudotensor constructed from the Levi-Civita tensor and the four-velocity, $\epsilon^{\mu\nu\lambda\gamma}u_\gamma$. Namely, our basis is
\begin{align}\label{basisrepresentation}
u^{\mu}~~,~~\Delta^{\mu\nu}=g^{\mu\nu}-u^{\mu}u^{\nu}~~,~~\epsilon^{\mu\nu\lambda\gamma}u_\gamma.
\end{align}
These quantities are rotationally invariant in the local rest frame associated with $u^{\mu}$, where $u^{\mu}=(1,0,0,0)$ and $\Delta^{\mu\nu}=g^{\mu\nu}-u^\mu u^\nu=\text{diag}(0,-1,-1,-1)$.

Given this physical setup, and to better understand the derivation of irreducible decompositions of arbitrary $n$-rank tensors under spatial rotations (as detailed in the next subsections), we proceed as follows:
\begin{enumerate}
\item Express the irreducible representations acting on four-vectors $\textbf{V}^{\mu}$~\eqref{vectorused}, symmetric tensors $\textbf{S}^{\mu\nu}$~\eqref{symmetricused}, and antisymmetric tensors $\textbf{A}^{\mu\nu}$~\eqref{antisymmused}, in terms of the hydrodynamic basis. The resulting components are referred to as \emph{building blocks}. 
\item Demonstrate how to obtain the irreducible decomposition under spatial rotations for  $\textbf{V}^{\mu}$,  $\textbf{S}^{\mu\nu}$, and  $\textbf{A}^{\mu\nu}$, using the  building blocks. 
\item Discuss the equilibrium (rotationally invariant) limit of the various irreducible decompositions of $\textbf{V}^{\mu}$, $\textbf{S}^{\mu\nu}$, and $\textbf{A}^{\mu\nu}$.
\end{enumerate}

The irreducible representation acting on $\textbf{V}^{\mu}$, as in Eq.~\eqref{vectorused}, reads
\begin{align}\label{vecdeco}
\textbf{V}^{\mu}: (\textbf{0}\oplus \textbf{1}) \to (u^{\mu}\oplus \Delta^{\mu}_{\alpha}).
\end{align}
Here, we associate the four-vector $u^\mu$ with the scalar representation $\textbf{j}=\textbf{0}$, and the projector 
$\Delta^\mu_\alpha$ with the vector representation $\textbf{j}=\textbf{1}$. In practical terms, this means:
\begin{align}
&u_{\mu}\textbf{V}^{\mu}=u_{0}\mathbf{V}^{0}=a_{1},\\
&\Delta^{\alpha}_{\mu}\textbf{V}^{\mu}=a_{2}X^{\beta}\Delta_{\beta}^{\alpha},
\end{align}
where $a_{1}$, $a_{2}$, and $X^{\alpha}$ are to be determined for each considered quantity represented by four-vector $\textbf{V}^{\mu}$. This leads to the following irreducible decomposition under spatial rotations:
\begin{align}\label{lost0}
\textbf{V}^{\mu}=a_{1}u^{\mu}+a_{2}X^{\alpha}\Delta_{\alpha}^{\mu}.
\end{align}
In equilibrium — that is, in the rotationally invariant local rest frame — the vector $\textbf{V}^{\mu}$ must remain invariant under spatial rotations. To ensure this condition, only the scalar component of the decomposition is allowed. Consequently, the rotationally invariant irreducible decomposition of $\textbf{V}^{\mu}$ reduces to:
\begin{align}
\textbf{V}^{\mu}=a_{1}u^{\mu},
\end{align}

The irreducible representation acting on symmetric tensors $\mathbf{S}^{\mu\nu}$, as given in Eq.~\eqref{symmetricused}, can be expressed 
as follows,
\begin{align}
&\mathbf{S}^{\mu\nu}: (\textbf{0}\oplus\textbf{0}\oplus\textbf{1}\oplus\textbf{2}) \to \left(u^{\mu}u^{\nu}\oplus \Delta^{\mu\nu}\oplus u^{(\mu}\Delta^{\nu)}_{\alpha}\oplus \Delta^{(\mu}_{\alpha}\Delta^{\nu)}_{\beta}-\frac{1}{3}\Delta_{\alpha\beta}\Delta^{\mu\nu}\right).\label{Stensdeco}
\end{align}
Here, the additional indices $\alpha$ and $\beta$ indicate that the terms possessing them act as projectors (or are proportional to projectors) when applied to tensors.  In practice, the above means, 
\begin{align}
    &u_{\mu}u_{\nu}\textbf{S}^{\mu\nu}=b_{1},\\
    &\Delta_{\mu\nu}\textbf{S}^{\mu\nu}=3b_{2},\\
    &u_{(\mu}\Delta_{\nu)\alpha}\textbf{S}^{\mu\nu}=\frac{1}{2}b_{3}\Delta_{\alpha\beta}Y^{\beta},\\
    &{\left(\Delta^{\alpha}_{(\mu}\Delta^{\beta}_{\nu)}-\frac{1}{3}\Delta^{\alpha\beta}\Delta_{\mu\nu}\right)\mathbf{S}^{\mu\nu}=\frac
    {b_{4}}{2}\left(\Delta^{\alpha}_{\gamma}\Delta^{\beta}_{\delta}+\Delta^{\beta}_{\gamma}\Delta^{\alpha}_{\delta}\right)Z^{\gamma\delta}-\frac{2b_{4}}{3}\Delta_{\gamma\delta}\Delta^{\alpha\beta}Z^{\gamma\delta}}\nonumber\\
    &~~~~~~~~~~~~~~~~~~~~~~~~~~~~~~~~~~~~~~~+\frac{b_{4}}{3}\Delta_{\gamma\delta}\Delta^{\alpha\beta}Z^{\gamma\delta}.
\end{align}
Here, the coefficients $b_{i}\, (\text{for} \,\,i=1,2,3,4)$, along with $Y^{\beta}$ and $Z^{\alpha\beta}$, depend on the specific tensor $\textbf{S}^{\mu\nu}$. This leads to the irreducible decomposition of a symmetric tensor $\textbf{S}^{\mu\nu}$ as, 
\begin{align}\label{lost-1}
{\textbf{S}^{\mu\nu}=b_{1}u^{\mu}u^{\nu}+b_{2}\Delta^{\mu\nu}+b_{3}u^{(\mu}\Delta^{\nu)}_{\alpha}Y^{\alpha}+b_{4}\left(\Delta^{(\mu}_{\alpha}\Delta^{\nu)}_{\beta}-\frac{1}{3}\Delta_{\alpha\beta}\Delta^{\mu\nu}\right)Z^{\alpha\beta}.}
\end{align}

In equilibrium, the tensor $\textbf{S}^{\mu\nu}$ must remain invariant under spatial rotations in the local co-moving frame of the fluid. To satisfy this condition, only the scalar components are permitted. Therefore, the rotationally invariant irreducible decomposition of $\textbf{S}^{\mu\nu}$ is
\begin{align}
\textbf{S}^{\mu\nu}=b_{1}u^{\mu}u^{\nu}+b_{2}\Delta^{\mu\nu}.
\end{align}

The irreducible representation acting on antisymmetric tensors $\mathbf{A}^{\mu\nu}$, as in Eq.~\eqref{Atensdeco}, reads
\begin{align}
&\mathbf{A}^{\mu\nu}: (\textbf{1}\oplus\textbf{1}) \to \left(u^{[\mu}\Delta^{\nu]\alpha}\oplus \epsilon^{\mu\nu\tau\alpha}u_{\tau}\right).\label{Atensdeco}
\end{align}
Here again, the extra indices $\alpha$ and $\beta$ signify that these expressions act as projectors (or are proportional to projectors) when applied to tensors. Note that the Levi-Civita tensor is required to generate an independent vector projector for antisymmetric tensors. In practice, this gives
\begin{align}
    &{u_{[\mu}\Delta_{\nu]\alpha}\textbf{A}^{\mu\nu}=\frac{c_{1}}{2}\Delta^{\beta}_{\alpha}B_{\beta}}\\
    &{\epsilon_{\mu\nu\tau\alpha}u^{\tau}\textbf{A}^{\mu\nu}=-c_{2}\Delta^{\beta}_{\alpha}D_{\beta},}
\end{align}
where $c_{1}$, $c_{2}$, $B_{\beta}$, and $D_{\beta}$ depend on the particular tensor $\textbf{A}^{\mu\nu}$ considered. This leads to the irreducible decomposition of antisymmetic tensors $\textbf{A}^{\mu\nu}$ as
\begin{align}\label{lost}
\mathbf{A}^{\mu\nu}=c_{1}u^{[\mu}\Delta^{\nu]\alpha}B_{\alpha}+c_{2}\epsilon^{\mu\nu\tau\alpha}u_{\tau}D_{\alpha}.
\end{align}
%
%
%
\subsection{Irreducible decomposition of a rank-$n$ tensor under rotations}
\label{ApplicationI}

By using the above irreducible representations~\eqref{vecdeco}\eqref{Stensdeco}\eqref{Atensdeco} as building blocks, we can determine irreducible representation acting on n-rank tensors even if n$>$2, leading to the determination of their irreducible decomposition under rotation. 

For example, let us consider the gradient of the spin potential, \( \partial_{\rho}\Omega_{\lambda\nu} \)~\eqref{spinpotentialgradientdecompositionnn}, which is a rank-3 tensor antisymmetric in the last two indices. Using Eqs.~\eqref{vecdeco} and \eqref{Atensdeco}, we can express the irreducible representation under rotation acting on \( \partial_{\rho}\Omega_{\lambda\nu} \)  as follows,
\begin{align}\label{spinpotentialdecom}
\partial_{\rho}\Omega_{\sigma\tau}: (\textbf{0}\oplus\textbf{1})\otimes(\textbf{1}\oplus\textbf{1})
\to (u_{\rho}\oplus\Delta_{\rho\alpha})\otimes\left(\frac{1}{2}(u_{\sigma}\Delta_{\tau\beta}-u_{\tau}\Delta_{\sigma\beta})\oplus\epsilon_{\sigma\tau\gamma\beta}u^{\gamma}\right).
\end{align}
Here, the first factor $(\textbf{0} \oplus \textbf{1})$ refers to the gradient vector $\partial_\rho$ and the 
second, i.e. $(\textbf{1}\oplus\textbf{1})$, to the spin potential tensor $\Omega_{\sigma\tau}$. Before expanding the above tensor product we recall the rules of tensor multiplication of representations,
\begin{align}\label{rulesoftensormultii}
&\textbf{0}\,\otimes\,\textbf{0} = \textbf{0} \qquad \textbf{0}\, \otimes\, \textbf{1}= \textbf{1} \qquad \textbf{1}\, \otimes\, \textbf{1} = (\textbf{0}, \textbf{1}, \textbf{2}),  \qquad \textbf{1}\, \otimes\, \textbf{2} = (\textbf{1}, \textbf{2}, \textbf{3}),\nonumber\\
&\textbf{2}\, \otimes\, \textbf{2} = (\textbf{0}, \textbf{1}, \textbf{2},\textbf{3},\textbf{4}).
\end{align}
Therefore, by taking into account the above rules and expanding Eq.~\eqref{spinpotentialdecom} we get,
\begin{align}\label{listrep}
    &(\textbf{0}\otimes\textbf{1})= \textbf{1} \to u_{\rho}u_{[\sigma}\Delta_{\tau]\beta},\nonumber\\
    &(\textbf{0}\otimes\textbf{1})= \textbf{1} \to  u_{\rho}\epsilon_{\sigma\tau\gamma\beta}u^{\gamma},\nonumber\\
    &(\textbf{1}\otimes\textbf{1}) \to \Delta_{\rho}^{\alpha}u_{[\sigma}\Delta_{\tau]\beta},\nonumber\\
    &(\textbf{1}\otimes\textbf{1})  \to \Delta^{\alpha}_{\rho}\epsilon_{\sigma\tau\gamma\beta}u^{\gamma}.
\end{align}
Now we can reduce the $(\mathbf{1} \otimes \mathbf{1})$ representation into $\mathbf{j} = \mathbf{0}, \mathbf{1}, \mathbf{2}$, and associate to each one a corresponding projector.

To obtain the correct form of the scalar representation $\mathbf{j} = \mathbf{0}$, we have to contract the indices $\alpha, \beta$ in~\eqref{listrep}. For the vector representation $\mathbf{j} = \mathbf{1}$, we antisymmetrize them using a Levi-Civita projector $\epsilon^{\alpha\beta\lambda\,\delta} u_{\lambda}$. Finally, for the rank-2 tensor representation $\mathbf{j} = \mathbf{2}$, we symmetrize the indices $\alpha, \beta$ and remove the trace. Therefore:
\begin{align}
& (\textbf{1}\otimes\textbf{1}) \to \Delta_{\rho}^{\alpha}u_{[\sigma}\Delta_{\tau]\beta}=\begin{cases}
        \textbf{0} \to u_{[\sigma}\Delta_{\tau]\rho}\nonumber\\
        \textbf{1} \to \epsilon^{\alpha\beta\lambda\,\delta}u_{\lambda}\Delta_{\rho\alpha}u_{[\sigma}\Delta_{\tau]\beta}\\
        \textbf{2} \to \Delta_{\rho(\alpha}u_{[\sigma}\Delta_{\tau]\beta)}-\frac{1}{3}
        \Delta_{\alpha\beta}u_{[\sigma}\Delta_{\tau]\rho}
    \end{cases}\nonumber\\
    &(\textbf{1}\otimes\textbf{1}) \to \Delta^{\alpha}_{\rho}\epsilon_{\sigma\tau\gamma\beta}u^{\gamma}=\begin{cases}
        \textbf{0} \to \epsilon_{\sigma\tau\gamma\rho}u^{\gamma}\nonumber\\
        \textbf{1} \to \epsilon^{\alpha\beta\lambda\,\delta}u_{\lambda}\Delta_{\rho\alpha}
        \epsilon_{\sigma\tau\gamma\beta}u^{\gamma}\nonumber\\
        \textbf{2} \to \Delta_{\rho(\alpha}\epsilon_{\sigma\tau\gamma\beta)}u^{\gamma}-\frac{1}{3}\Delta_{\alpha\beta}\epsilon_{\sigma\tau\gamma\rho}u^{\rho}
    \end{cases}\nonumber
\end{align}
Altogether, and following the same terminology used in Eqs.~\eqref{lost0}\eqref{lost-1}\eqref{lost}, the tensor \( \partial_{\rho}\Omega_{\lambda\nu} \) can be decomposed into eight irreducible components under rotations, and expressed as
\begin{align}  
\partial_{\rho}\Omega_{\sigma\tau}=&c_1\mathcal{X}^{\gamma}u_{\rho}u_{[\sigma}\Delta_{\tau]\gamma}
+c_2\mathcal{Y}^{\gamma}u_{\rho}\epsilon_{\sigma\tau\lambda\gamma}u^{\lambda}
+c_3\mathcal{Z}u_{[\sigma}\Delta_{\tau]\rho}
+c_4\mathcal{T}^{\gamma}u_{[\sigma}\epsilon_{\rho\tau]\lambda\gamma}u^{\lambda}
+c_5\mathcal{F}^{\alpha\beta}\Delta_{\rho(\alpha}u_{[\sigma}\Delta_{\tau]\beta)}\nonumber\\
&+c_6\mathcal{H}\epsilon_{\sigma\tau\gamma\rho}u^{\gamma}
+c_7\mathcal{G}_{\gamma}\epsilon_{\rho}^{~~\beta\lambda\gamma}u_{\lambda}\epsilon_{\sigma\tau\alpha\beta}u^{\alpha}
+c_8\mathcal{I}^{\alpha\beta}\Delta_{\rho(\alpha}\epsilon_{\sigma\tau\gamma\beta)}u^{\gamma}.
\end{align}
Here, $c_i,\ i=1,2,\dots,8$ are arbitrary factors whose setting amounts to a rescaling of
the scalar, vector and tensor coefficients. In the expression above we have merged the trace term of the $\textbf{j}=\textbf{2}$ representations into the $\textbf{j}=\textbf{0}$ representations, without 
loss of generality. To determine the s $c_i,\ i=1,2,\dots,8$ as well as the coefficients $\mathcal{X}^{\gamma},\mathcal{Y}^{\gamma},\mathcal{Z}....,\mathcal{I}^{\alpha\beta}$ associated with each term, we contract on both sides its corresponding term. For example, 
\begin{align}\label{spinpotentialgradientdecomposition}
 \partial_{\mu}\Omega_{\lambda\nu}=&\ 2\mathcal{X}^{\gamma}u_{\mu}u_{[\lambda}\Delta_{\nu]\gamma}-
\frac12\mathcal{Y}^{\gamma}u_{\mu}\epsilon_{\lambda\nu\sigma\gamma}u^{\sigma}
+\frac12\mathcal{Z}u_{[\lambda}\Delta_{\nu]\mu}+\mathcal{T}^{\gamma}u_{[\lambda}\epsilon_{\nu]\mu\alpha\gamma}u^{\alpha}+2\mathcal{F}^{\rho\sigma}u_{[\lambda}\Delta_{\nu]\sigma}\Delta_{\mu \rho}\nonumber\\
 &\ -\frac18\mathcal{H}\epsilon_{\lambda\nu\alpha\mu}u^{\alpha}-\frac12\mathcal{G}_{[\lambda}\Delta_{\nu]\mu}-\frac12\mathcal{I}^{\rho\sigma}\Delta_{\mu \rho}\epsilon_{\lambda\nu\tau \sigma}u^{\tau}\;.
\end{align}
where,
\begin{align}
&\mathcal{X}^{\gamma}=u^{[\rho}\Delta^{\sigma]\gamma}D\Omega_{\rho\sigma},\quad \mathcal{Y}^{\gamma}=\epsilon^{\rho\sigma\tau\gamma}u_{\tau}D\Omega_{\rho\sigma},\quad
\mathcal{Z}=u^{[\rho}\nabla^{\sigma]}\Omega_{\rho\sigma},\quad
\mathcal{T}^{\gamma}=u^{[\rho}\epsilon^{\sigma]\lambda\theta\gamma}u_{\lambda}\partial_{\theta}\Omega_{\rho\sigma},     \nonumber\\
&\mathcal{F}^{\rho\sigma}=u^{[\gamma}\Delta^{\theta](\rho}\nabla^{\sigma)}\Omega_{\gamma\theta}-\Delta^{\rho\sigma}\mathcal{Z}/4
,\quad
\mathcal{H}=\epsilon^{\rho\sigma\lambda\gamma}u_{\lambda}\partial_{\gamma}\Omega_{\rho\sigma},\quad
\mathcal{G}^{\gamma}=2\Delta^{\gamma\tau}\nabla^{\rho}\Omega_{\rho\tau},    \nonumber\\
&\mathcal{I}^{\rho\sigma}=u_{\gamma}\epsilon^{\theta\lambda\gamma(\rho}\nabla^{\sigma)}\Omega_{\theta\lambda}-\Delta^{\rho\sigma}\mathcal{H}/4.
\end{align}
The thermal shear $\xi^{\mu\nu}$~\eqref{thermalsheardecomposition}, the difference between spin potential and thermal vorticity $\Omega^{\mu\nu}-\varpi^{\mu\nu}$~\eqref{Omega-omegadecomposition}, and the derivative of the chemical potential by temperature $\partial_{\mu}\zeta$~\eqref{dzetadecomposition} follows the same procedure as $\partial_{\lambda}\Omega_{\mu\nu}$.
%
%
\subsection{Irreducible decomposition of an rank-$n$ tensor under rotations with constraint to retain only rotation-invariant terms}  
\label{Application II}
At this point, as discussed in the previous section, we know how to determine the irreducible representations acting on rank-n tensors, even for $n >$ 2, which allows us to identify the irreducible decomposition of such tensors under rotation. In this section, however, we go a step further and demonstrate how to isolate those irreducible components that are invariant under rotations.

As a direct application, we consider the tensor coefficients \( H^{\mu\nu\rho\sigma}, K^{\mu\nu\rho}, \dots, W^{\mu\lambda\nu\rho\sigma\tau} \) listed in Tables~\ref{Table-G-I-O-F}--\ref{Table-W}. These tensors are expanded in the isotropic local rest frame of the fluid element and are therefore required to be invariant under spatial rotations in the local comoving frame. To ensure this invariance, we first perform a general irreducible decomposition of the tensors (as described in the previous subsection) and then retain only the scalar components.

Let us focus on the tensor coefficient \( H^{\mu\nu\rho\sigma}\) which is symmetric in the first two indices and the last two indices. The irreducible representation of \( H^{\mu\nu\rho\sigma}\) can be directly constructed from the irreducible representations building blocks mainly in Eq.\eqref{Stensdeco} such that 
\begin{align}\label{HdecompositionApp}
H^{\mu\nu\rho\sigma}=\textbf{S}^{\mu\nu}\otimes \textbf{S}^{\rho\sigma}: (\textbf{0} \oplus \textbf{0} \oplus \textbf{1} \oplus \textbf{2}) \otimes (\textbf{0} \oplus \textbf{0} \oplus \textbf{1} \oplus \textbf{2}).
\end{align}
According to the rules in~\eqref{rulesoftensormultii}, the irreducible representation acting on the tensor $H^{\mu\nu\rho\sigma}$ reads,
$$
H^{\mu\nu\rho\sigma}: 2( \textbf{0} \oplus \textbf{0} \oplus \textbf{1} \oplus \textbf{2}) \oplus \textbf{1} \oplus \textbf{1} \oplus \textbf{0} \oplus \textbf{1} \oplus \textbf{2}
\oplus \textbf{1} \oplus \textbf{2} \oplus \textbf{3} \oplus \textbf{2} \oplus \textbf{2} \oplus \textbf{1} \oplus \textbf{2} \oplus \textbf{3} \oplus \textbf{0} \oplus \textbf{1} \oplus \textbf{2} \oplus \textbf{3} \oplus \textbf{4}
$$
As it can be inferred from the above expression, we have six scalar representations and so the tensor \(H^{\mu\nu\rho\sigma}\) has six independent scalar components. Hence, following the same terminology in Eqs.~\eqref{lost0}\eqref{lost-1}\eqref{lost}, we infer that 
\begin{equation}
H^{\mu\nu\rho\sigma}=h_1 u^{(\mu}\Delta^{\nu)(\rho}u^{\sigma)}+h_2\Delta^{\mu\nu}\Delta^{\rho\sigma}+h_3 \Delta^{\mu\nu,\rho\sigma}
+h_4\Delta^{\mu\nu}u^{\rho}u^{\sigma}+h_5 u^{\mu}u^{\nu}\Delta^{\rho\sigma}+ h_6u^{\mu}u^{\nu}u^{\rho}u^{\sigma}\;.
\end{equation}
Here, the traceless symmetric projector $\Delta_{\mu\nu,\alpha\beta}\equiv(\Delta_{\mu\alpha}\Delta_{\nu\beta}+\Delta_{\nu\alpha}\Delta_{\mu\beta})/2-\Delta_{\mu\nu}\Delta_{\alpha\beta}/3$ was used, and $h_i,\ i=1,2,\dots,6$ are arbitrary coefficients to be determined according to the underlying physical system. The remaining tensor coefficients are treated using the same strategy.

The tables below summarize the symmetry properties under index exchange of all tensor coefficients appearing in Eqs.~\eqref{EMTsymmetricgradientdecomposition}-\eqref{Spingradientdecomposition},
\begin{table}[H]
\renewcommand{\arraystretch}{2}
\begin{tabularx}{1\textwidth} { 
| >{\raggedright\arraybackslash}X 
| >{\raggedright\arraybackslash}X
| >{\raggedright\arraybackslash}X
| >{\raggedright\arraybackslash}X
| >{\centering\arraybackslash}X 
| >{\raggedleft\arraybackslash}X | }
\hline
\textbf{Tensor coefficient} & \textbf{Symmetric exchange}& \textbf{Antisymmetric exchange} & \textbf{No symmetry}\\
\hline
$H^{\mu\nu\rho\sigma}$& $\mu$-$\nu$ \& $\rho$-$\sigma$ & - & -\\
\hline
$K^{\mu\nu\rho}$& $\mu$-$\nu$ & - & $\rho$\\
\hline 
$L^{\mu\nu\rho\sigma}$& $\mu$-$\nu$ & $\rho$-$\sigma$ & -\\
\hline
$M^{\mu\nu\rho\sigma\tau}$& $\mu$-$\nu$ & $\sigma$-$\tau$ & $\rho$\\
\hline
\end{tabularx}
\caption{Summary of symmetry properties under index exchange for the tensor coefficients in $\delta T^{\mu\nu}_{(S)}$.}
\label{SummaryofIndexSymmetryProperties underExchangeTS}
\end{table}
\begin{table}[H]
\renewcommand{\arraystretch}{2}
\begin{tabularx}{1\textwidth} { 
| >{\raggedright\arraybackslash}X 
| >{\raggedright\arraybackslash}X
| >{\raggedright\arraybackslash}X
| >{\raggedright\arraybackslash}X
| >{\centering\arraybackslash}X 
| >{\raggedleft\arraybackslash}X | }
\hline
\textbf{Tensor coefficient} & \textbf{Symmetric exchange}& \textbf{Antisymmetric exchange} & \textbf{No symmetry}\\
\hline
$N^{\mu\nu\rho\sigma}$ & $\rho$-$\sigma$ & $\mu$-$\nu$ & -\\
\hline
$P^{\mu\nu\rho}$ & - & $\mu$-$\nu$ & $\rho$\\
\hline
$Q^{\mu\nu\rho\sigma}$ & - &$\mu$-$\nu$ \& $\rho$-$\sigma$ & -\\
\hline
$R^{\mu\nu\rho\sigma\tau}$& - & $\mu$-$\nu$ \& $\sigma$-$\tau$ & $\rho$\\
\hline
\end{tabularx}
\caption{Summary of symmetry properties under index exchange for the tensor coefficients in $\delta T^{\mu\nu}_{(A)}$.}
\label{SummaryofIndexSymmetryProperties underExchangeTA}
\end{table}
\begin{table}[H]
\renewcommand{\arraystretch}{2}
\begin{tabularx}{1\textwidth} { 
| >{\raggedright\arraybackslash}X 
| >{\raggedright\arraybackslash}X
| >{\raggedright\arraybackslash}X
| >{\raggedright\arraybackslash}X
| >{\centering\arraybackslash}X 
| >{\raggedleft\arraybackslash}X | }
\hline
\textbf{Tensor coefficient} & \textbf{Symmetric exchange}& \textbf{Antisymmetric exchange} & \textbf{No symmetry}\\
\hline
$G^{\mu\rho\sigma}$& $\rho$-$\sigma$ & - & $\mu$\\
\hline
$I^{\mu\rho}$& - & - & $\mu$ \& $\rho$\\
\hline
$O^{\mu\rho\sigma}$& - & $\rho$-$\sigma$ & $\mu$\\
\hline
$F^{\mu\rho\sigma\tau}$& - & $\sigma$-$\tau$ & $\mu$ \& $\rho$\\
\hline
\end{tabularx}
\caption{Summary of symmetry properties under index exchange for the tensor coefficients in $\delta j^{\mu}$.}
\label{SummaryofIndexSymmetryProperties underExchangeJ}
\end{table}
\begin{table}[H]
\renewcommand{\arraystretch}{2}
\begin{tabularx}{1\textwidth} { 
| >{\raggedright\arraybackslash}X 
| >{\raggedright\arraybackslash}X
| >{\raggedright\arraybackslash}X
| >{\raggedright\arraybackslash}X
| >{\centering\arraybackslash}X 
| >{\raggedleft\arraybackslash}X | }
\hline
\textbf{Tensor coefficient} & \textbf{Symmetric exchange}& \textbf{Antisymmetric exchange} & \textbf{No symmetry}\\
\hline
$T^{\mu\lambda\nu\rho\sigma}$ & $\rho$-$\sigma$ & $\lambda$-$\nu$ & $\mu$\\
\hline
$U^{\mu\lambda\nu\rho}$& - &  $\lambda$-$\nu$ & $\mu$ \& $\rho$\\
\hline
$V^{\mu\lambda\nu\rho\sigma}$& - &$\lambda$-$\nu$ \& $\rho$-$\sigma$ & $\mu$\\
\hline
$W^{\mu\lambda\nu\rho\sigma\tau}$& - & $\lambda$-$\nu$ \& $\sigma$-$\tau$ & $\mu$ \& $\rho$\\
\hline
\end{tabularx}
\caption{Summary of symmetry properties under index exchange for the tensor coefficients in $\delta \spt^{\mu\lambda\nu}$.}
\label{SummaryofIndexSymmetryProperties underExchangeS}
\end{table}
%
%
\subsection{Application}
\label{applicationnn}
In this subsection, we apply the above method to reproduce the well-known dissipative currents and transport coefficients in conventional relativistic hydrodynamics. The motivation is to confirm that the method applied in Chapter~\ref{Quantum-statistical formulation} holds true in absence of spin tensor.

%
\medskip 

For relativistic hydrodynamic without spin, the entropy production rate can be obtained as a limiting case of equation~\eqref{entropyproductionrate} by neglecting terms proportional to the spin potential and its gradients. This results in the following expression
\begin{align}\label{entropyratewithoutspin}
\partial_{\mu}S^{\mu}=\left( T_{(S)}^{\mu\nu}-T^{\mu\nu}_{(S)\,\rm LE}\right) \xi_{\mu\nu}
- \left( j^\mu-j^\mu_{\rm LE}\right) \partial_\mu \zeta. 
\end{align}
Here, $ T_{(S)}^{\mu\nu}-T^{\mu\nu}_{(S)\,\rm LE}\equiv\delta T^{\mu\nu}_{(S)}$ is the dissipative part of the energy-momentum tensor which is symmetric and $ j^\mu-j^\mu_{\rm LE}\equiv\delta j^{\mu} $ is the dissipative part of the particle number current. The term $\xi_{\mu\nu}=\partial_{(\mu}\beta_{\nu)}$ is the thermal shear tensor and $\zeta$ is the chemical potential by temperature. 

As discussed in Chapter~\ref{Quantum-statistical formulation}, and without imposing any physical constraints, the dissipative currents can be decomposed in terms of all the thermo-hydrodynamic gradients in the system such that,
\begin{align}
&\delta T^{\mu\nu}_{(S)}=H^{\mu\nu\rho\sigma}\xi_{\rho\sigma}+\frac{1}{T}K^{\mu\nu\rho}\partial_{\rho}\zeta,\label{TSSS}\\
&T\delta j^{\mu}=G^{\mu\rho\sigma}\xi_{\rho\sigma}+\frac{1}{T}I^{\mu\rho}\partial_{\rho}\zeta.\label{JSS}
\end{align}
Here, we introduce tensor coefficients $H, K, G, I$ each of which has a specific rank and index symmetry based on the respective contraction. The additional factors $T$ and $1/T$ in $\delta j^{\mu}$ are introduced to ensure that all tensor coefficients have the same energy dimension in natural units, i.e., $[H]=[K]=[G]=[I]=[E]^4$.

Using the method described in Sec.~\eqref{Application II}, the rotation invariant irreducible components of the tensor coefficients reads,
\begin{align}
&H^{\mu\nu\rho\sigma}=h_1 u^{(\mu}\Delta^{\nu)(\rho}u^{\sigma)}+h_2\Delta^{\mu\nu}\Delta^{\rho\sigma}+h_3 \Delta^{\mu\nu,\rho\sigma}
+h_4\Delta^{\mu\nu}u^{\rho}u^{\sigma}\nonumber\\
&~~~~~~~~~~+h_5 u^{\mu}u^{\nu}\Delta^{\rho\sigma}+ h_6u^{\mu}u^{\nu}u^{\rho}u^{\sigma},\\
&K^{\mu\nu\rho}=k_{1}u^{(\mu}\Delta^{\nu)\rho}+k_{2}\Delta^{\mu\nu}u^{\rho}+k_{3}u^{\mu}u^{\nu}u^{\rho}\;,\\
&G^{\mu\rho\sigma}=g_{1}\Delta^{\mu(\rho}u^{\sigma)}+g_{2}u^{\mu}\Delta^{\rho\sigma}+g_{3}u^{\mu}u^{\rho}u^{\sigma},\\
&I^{\mu\rho}=i_{1}\Delta^{\mu\rho}+i_{2}u^{\mu}u^{\rho}.
\end{align}
Here, the traceless symmetric projector $\Delta_{\mu\nu,\alpha\beta}\equiv(\Delta_{\mu\alpha}\Delta_{\nu\beta}+\Delta_{\nu\alpha}\Delta_{\mu\beta})/2-\Delta_{\mu\nu}\Delta_{\alpha\beta}/3$ was used.

We obtain the necessary matching conditions by neglecting those related to the dissipative part of the antisymmetric energy-momentum tensor and the spin dissipative tensors in Eqs.~\eqref{TStAmat} and \eqref{spinmatt}. This leads to:
\begin{align}
&u_{\mu}\delta T^{\mu\nu}_{(S)}=0,\label{TSmatt}\\
&u_{\mu}\delta j^{\mu}=0.\label{matparti}
\end{align}
The first of these is equivalent to the Landau frame choice. Imposing these relations on Eqs.~\eqref{TSSS}--\eqref{JSS} yields the following constraints:
\begin{align}
&h_{1}=h_{5}=h_{6}=0,\\
&k_{1}=k_{3}=0,\\
&g_{2}=g_{3}=0,\\
&i_{2}=0.
\end{align}
The above conditions reduces the forms of the dissipative currents to, 
\begin{align}
&\delta T^{\mu\nu}_{(S)}=\left(h_{2}\Delta^{\mu\nu}\Delta^{\rho\sigma}+h_{3}\Delta^{\mu\nu,\rho\sigma}+h_{4}\Delta^{\mu\nu}u^{\rho}u^{\sigma}\right)\xi_{\rho\sigma}+\frac{k_{2}}{T}\Delta^{\mu\nu}u^{\rho}\partial_{\rho}\zeta,\\
&T\delta j^{\mu}=g_{1}\Delta^{\mu(\rho}u^{\sigma)}\xi_{\rho\sigma}+\frac{i_{1}}{T}\Delta^{\mu\rho}\partial_{\rho}\zeta.
\end{align}

The dissipative currents are also subject to the second law of thermodynamics as appearing in the entropy production rate $\partial_{\mu}S^{\mu}\geq0$. To study this condition, we decompose the gradient terms appearing in the entropy production rate using the irreducible components under rotation. However, unlike the tensor coefficients, we do not restrict the decomposition to scalar terms only (see Sec~\ref{ApplicationI}). Subsequently, we contract the gradients with the tensor coefficients to obtain the complete irreducible decomposition of the entropy production rate. This allows us to eliminate terms that do not respect the second law of thermodynamics. 
\medskip 

We start by the symmetric shear tensor,  
\begin{align}\label{thermalsheardecompositionapp}
\xi^{\mu\nu}=(D\beta)u^{\mu}u^{\nu}+\frac{1}{3T}\theta\Delta^{\mu\nu}+\frac{1}{T}\mathcal{J}_h^{\alpha}\left(\Delta^\mu_\alpha u^{\nu}+\Delta^\nu_\alpha u^\mu \right)+\frac{1}{T}\sigma^{\alpha\beta} \Delta^{\mu\nu}_{\alpha\beta},
\end{align}
where we have used the comoving derivative $D\equiv u_\mu\partial^\mu$, spatial gradient $\nabla^\mu\equiv\Delta^{\mu\nu}\partial_\nu$, the expansion scalar $\theta\equiv \partial_\mu u^\mu$, the heat flow $\mathcal{J}_h^\mu\equiv D u^\mu-(1/T)\nabla^\mu T$. The next gradient term, the derivative of the of the chemical potential over temperature, is given by 
\begin{align}\label{dzetadecompositionapp}
\partial_{\mu}\zeta=\left(D\zeta\right)u_{\mu}+\left(\nabla_{\alpha}\zeta\right) \Delta^\alpha_\mu \;.
\end{align}

By contracting the above gradient terms~\eqref{thermalsheardecompositionapp}-\eqref{dzetadecompositionapp} with the tensor coefficients in the dissipative currents~\eqref{TSSS}-\eqref{JSS} we arrive at the final irreducible form of the entropy production rate, 
\begin{align}\label{hopefullyfinee}
\partial_{\mu}S^{\mu}=\partial_{\mu}S^{\mu}_{S.S}+\partial_{\mu}S^{\mu}_{V.V}+\partial_{\mu}S^{\mu}_{T.T},
\end{align}
where, 
\begin{align}
\partial_\mu S_{S.S}^{\mu}&\sim\begin{pmatrix}
\theta\\ D\beta\\ D\zeta
\end{pmatrix}^\text{T}
\left(
\begin{array}{ccc}
 h_2  & h_{4}/2& k_{2}/2 \\
 h_{4}/2& 0 & 0 \\
  k_{2}/2 & 0 & 0 \\
\end{array}
\right)
\begin{pmatrix}
\theta\\ D\beta\\ D\zeta
\end{pmatrix}\;,
\\
\partial_\mu S_{V.V}^{\mu}&\sim\begin{pmatrix}
\mathcal{J}^{\alpha}\\ \nabla^{\alpha}\zeta
\end{pmatrix}^\text{T}
\left(
\begin{array}{ccc}
 0 & -g_{1}/2 \\
-g_{1}/2& -i_{1}\\
\end{array}
\right)
\begin{pmatrix}
\mathcal{J}^{\alpha}\\ \nabla^{\alpha}\zeta
\end{pmatrix}\;,
\\
\partial_\mu S_{T.T}^{\mu}&\sim h_{3}\sigma_{\alpha\beta}\sigma^{\alpha\beta}.
\end{align}
We note that, for simplicity, we used $\sim$ rather than strict equality, as we eliminate some positive constants and temperature scaling factor. However, they will be included appropriately in the final forms. The semi-positivity of $\partial_\mu S^\mu$ requires each of the parts in Eq.~\eqref{hopefullyfinee} to be semi-positive. One possible way to show that a matrix is semi-positive is by showing that each of its principal minors are semi-positive. Hence the semi-positivity constraints leads to, 
\begin{align}
&h_{2}\geq0~,~h_{4}=0~,~k_{2}=0,\\
&g_{1}=0~,~i_{1}\leq0,\\
&h_{3}\geq0.
\end{align}
Therefore the final forms of the dissipative currents reads, 
\begin{align}
    &\delta T^{\mu\nu}_{(S)}=\frac{h_{2}}{T}\theta\Delta^{\mu\nu}+\frac{h_{3}}{T}\sigma^{\mu\nu},\\
    &T\delta j^{\mu}=\frac{i_{1}}{T}\nabla^{\mu}\zeta.
\end{align}
The above formulas represent the dissipative currents in relativistic hydrodynamics in the Landau frame if and only if \( h_2 \) is identified as the bulk viscosity, \( h_3 \) as the shear viscosity, and \( i_1 \) as the diffusion coefficient.


\defbibheading{bibintoc}{\chapter*{#1}\addcontentsline{toc}{backmatter}{\refname}}

\printbibliography[title={\refname},heading=bibintoc]

\end{document}